\documentclass{article}
\usepackage{arxiv}

\RequirePackage{amsthm,amsmath,amsfonts,amssymb}
\usepackage[numbers]{natbib}
\RequirePackage[colorlinks,citecolor=blue,urlcolor=blue,backref=page,backref=page]{hyperref}
\RequirePackage{graphicx}

\usepackage{amsfonts, amsmath, amssymb}
\usepackage{bm}
\usepackage{subfigure}
\usepackage{subcaption}
\usepackage{float}
\usepackage{bigints}
\usepackage{xcolor}
\usepackage{soul}
\usepackage{relsize}

\usepackage{mathtools}
\newcommand{\defeq}{\vcentcolon=}

\DeclareMathOperator*{\argmin}{arg\,min}
\newcommand\norm[1]{\left\lVert#1\right\rVert}
\newcommand{\inv}{^{\raisebox{.2ex}{$\scriptscriptstyle-1$}}}

\usepackage{enumitem}
\usepackage{algorithm}
\usepackage{algorithmic}
\newlist{steps}{enumerate}{1}
\setlist[steps, 1]{label = Step \arabic*:}

\usepackage{mathtools,tikz,caption}
\DeclareRobustCommand\sampleline[1]{%
  \tikz\draw[#1] (0,0) (0,\the\dimexpr\fontdimen22\textfont2\relax)
  -- (2em,\the\dimexpr\fontdimen22\textfont2\relax);%
}

\theoremstyle{plain}

\newtheorem{theorem}{Theorem}[section]

\title{Validating uncertainty propagation approaches for two-stage Bayesian spatial models using simulation-based calibration}

\author{
 Stephen Jun Villejo \\
  School of Mathematics and Statistics/School of Statistics\\
  University of Glasgow/University of the Philippines\\
  \texttt{stephen.villejo@glasgow.ac.uk} \\
  orcid.org/0000-0002-0510-3143\\
   \And
 Sara Martino \\
  Department of Mathematical Sciences\\
  Norwegian University of Science and Technology\\
\texttt{sara.martino@ntnu.no} \\
orcid.org/0000-0003-4326-9029\\
  \And
 Janine Illian \\
  School of Mathematics and Statistics\\
  University of Glasgow\\
  \texttt{janine.illian@glasgow.ac.uk} \\
  orcid.org/0000-0002-6130-2796\\
  \And
  William Ryan \\
  School of Mathematics and Statistics\\
  University of Glasgow\\ \texttt{william.ryan@glasgow.ac.uk} \\
  orcid.org/0000-0002-6379-2186
  \And
  Finn Lindgren \\
  School of Mathematics \\
  University of Edinburgh \\  \texttt{finn.lindgren@ed.ac.uk} \\
  orcid.org/0000-0002-5833-2011\\
}

\begin{document}

\maketitle

\begin{abstract}
This work tackles the problem of uncertainty propagation in two-stage Bayesian models, with a focus on spatial applications. A two-stage modeling framework has the advantage of being more computationally efficient than a fully Bayesian approach when the first-stage model is already complex in itself, and avoids the potential problem of unwanted feedback effects. Two ways of doing two-stage modeling are the crude plug-in method and the posterior sampling method. The former ignores the uncertainty in the first-stage model, while the latter can be computationally expensive. This paper validates the two aforementioned approaches and  proposes a new approach to do uncertainty propagation, which we call the $\mathbf{Q}$ uncertainty method, implemented using the Integrated Nested Laplace Approximation (INLA). We validate the different approaches using the simulation-based calibration method, which tests the self-consistency property of Bayesian models. Results show that the crude plug-in method underestimates the true posterior uncertainty in the second-stage model parameters, while the resampling approach and the proposed method are correct. We illustrate the approaches in a real life data application which aims to link relative humidity and Dengue cases in the Philippines for August 2018. 
\end{abstract}


\section{Introduction}\label{sec:intro}

A two-stage modeling framework is commonly used in different areas of statistics such as longitudinal data analysis, survival analysis, and spatial statistics. As an example, in survival analysis, it is common practice to first fit a model for longitudinal markers and then use the estimated trends in the biomarker as input to a survival model (\citet{rustand2024fast, ye2008semiparametric}). In the area of spatial statistics, a two-stage framework is often used to address spatial misalignment problems, e.g., when the response variable and covariates have different spatial supports (\citet{szpiro2011efficient, gryparis2009measurement}). 

\begin{figure}[t]
    \centering    
    \includegraphics[trim={15mm 4mm 10mm 20mm},clip,scale=.39]{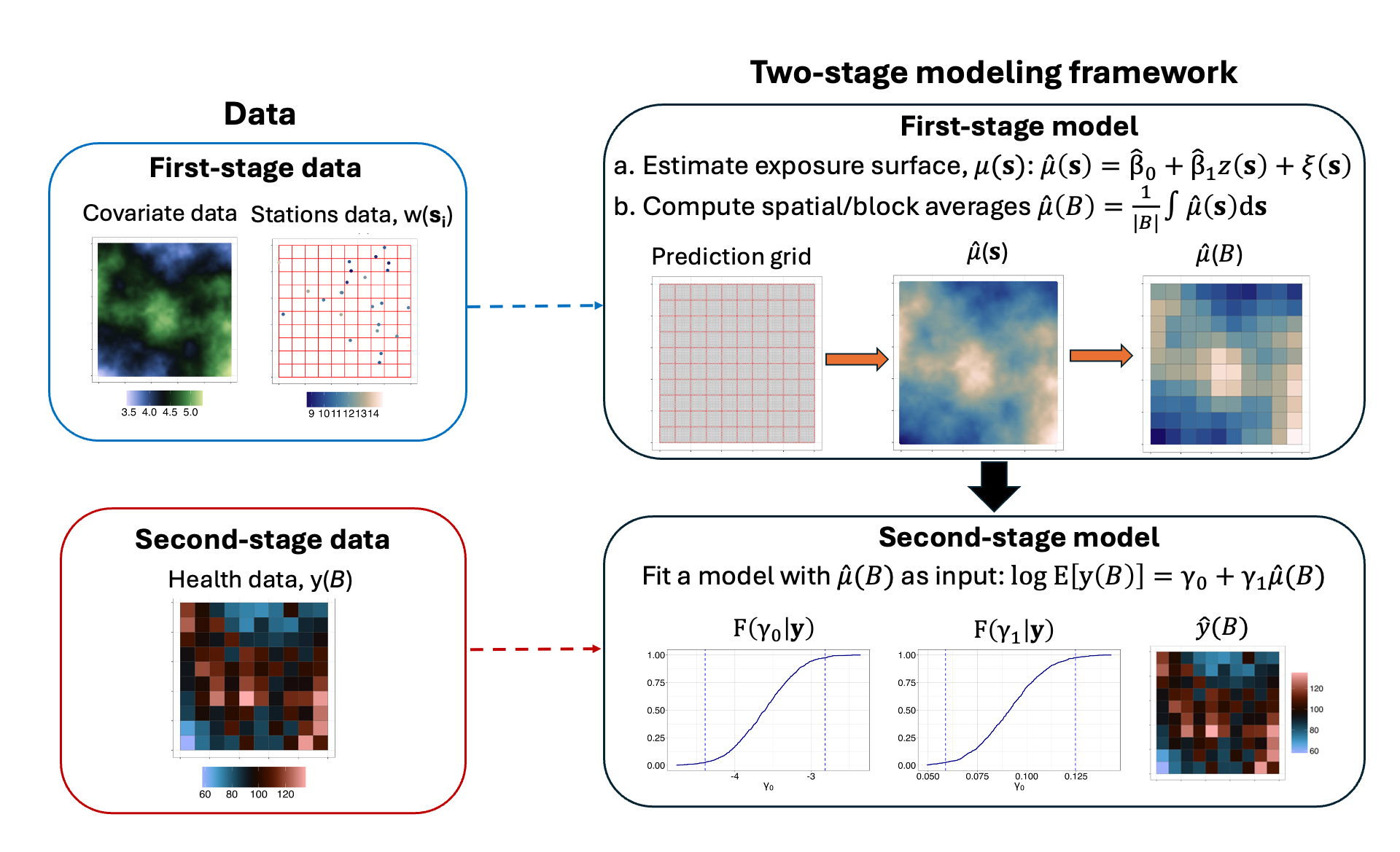}
    \caption{A two-stage modelling framework in spatial epidemiology: linking health outcomes, such as areal data on case counts of a disease, and pollution and/or meteorological variables observed at finite number of spatial locations or stations. \textit{First stage}: fitting a spatial model for the true exposure surface. \textit{Second stage}: fitting the health model.}
    \label{fig:motivation}
\end{figure}

An example in spatial epidemiology is shown in Figure \ref{fig:motivation} (\citet{blangiardo2016two, lee2017rigorous,cameletti2019bayesian, liu2017incorporating}). Here, the aim is to understand the link between case counts of a disease and some exposure variable, for example pollution level or meteorological variables. Data on case counts are areal, i.e., aggregated quantities over regions or blocks; while the exposure variable is spatially continuous and  observed at a finite number of spatial points (e.g.  weather stations). The first step consists of fitting a spatial model (\textit{first-stage model}), which is then used to predict the exposure surface on a fine grid. Spatial averages of the predicted surface for each area/block are then computed. The next step is to fit a model (\textit{second-stage model}) to link the case counts and the block estimates of exposures. A simple formulation of this two-stage model is as follows:
\begin{align}
    &\textcolor{blue}{\text{\textbf{First stage}}:}\;\;\; \mu(\mathbf{s})=\beta_0 + \beta_1z(\mathbf{s})+\xi(\mathbf{s})\label{eq:motivatingeq1}\\
   &\;\;\;\;\;\;\;\;\;\;\;\;\;\;\;\;\;\;\;\;\;\;\;\;\;\;\;\;\;
   \text{w}(\mathbf{s}_i) = \mu(\mathbf{s}_i) + \epsilon(\mathbf{s}_i), \;\;\; \epsilon(\mathbf{s}_i) \overset{\text{iid}}{\sim}\mathcal{N}(0,\sigma^2)\;\;\; i=1,\ldots,n \label{eq:motivatingeq2} \\
    &\textcolor{blue}{\text{\textbf{Second stage}}:} \;\;\; \log\Big(\mathbb{E}\big[\text{y}(B)\big]\Big) = \gamma_0+\gamma_1\mu(B) \label{eq:motivatingeq3}\\
    &\;\;\;\;\;\;\;\;\;\;\;\;\;\;\;\;\;\;\;\;\;\;\;\;\;\;\;\;\; \mu(B) = \dfrac{1}{|B|}\int_B\mu(\mathbf{s})d\mathbf{s}, \label{eq:motivatingeq4} 
\end{align}
where $\mu(\mathbf{s})$ is the true exposure surface, $\xi(\mathbf{s})$ is a spatially correlated random field, $z(\mathbf{s})$ is a known covariate, $\text{w}(\mathbf{s}_i)$ is the observed value of $\mu(\cdot)$ at a spatial location $\mathbf{s}_i$, $\epsilon(\mathbf{s}_i)$ is a measurement error term. $\text{y}(B)$ denotes the observed count of the disease in block $B$ and is assumed to follow the  Poisson distribution. The above model assumes that the counts $\text{y}(B)$ are linked to $\mu(\mathbf{s})$ via the spatial averages $\mu(B)=\dfrac{1}{|B|}\bigintsss_B\mu(\mathbf{s})d\mathbf{s},$ where $|B|$ is the size of block $B$ (Equation \eqref{eq:motivatingeq4}). The model unknowns are the fixed effects $\{\gamma_0,\gamma_1,\beta_0,\beta_1\}$, and the hyperparameters which include the error variance $\sigma^2$, and the parameters of the random field $\xi(\mathbf{s})$.

Simultaneous fitting of Equations \eqref{eq:motivatingeq1}-\eqref{eq:motivatingeq4}, also referred to as a \textit{joint modeling approach}, can be computationally challenging and expensive (\citet{gryparis2009measurement,liu2017incorporating}). Moreover, the first-stage model is typically the most computationally demanding; thus, joint modeling can be inconvenient if there are several epidemiological models of interest (\citet{liu2017incorporating, blangiardo2016two}). Equations \eqref{eq:motivatingeq1} -- \eqref{eq:motivatingeq4} can be viewed as a model with covariate measurement error (\citet{berry2002bayesian}). \citet{berry2002bayesian} proposed to fit the equations simultaneously, which they refer to as a \textit{fully Bayesian approach}. However, a potential problem with this approach is that it could cause potential `feedback' effects, wherein the data $\text{y}(B)$ influence and distort the model for $\mu(\mathbf{s})$, which consequently may compromise the $\big(\mu(B),\text{y}(B)\big)$ relationship (\citet{wakefield2006health,shaddick2002modelling,gryparis2009measurement}). This could happen when  data to inform about $\mu(\mathbf{s})$ are sparse (\citet{gryparis2009measurement}) or due to model misspecification (\citet{yucel2005imputation}). One way to `cut' the feedback between the two stages is by introducing a cut function in an MCMC algorithm (\citet{plummer2015cuts}). The cut function essentially simplifies the full conditional distribution of a graphical model into smaller modules that interact more weakly than in a full Bayesian analysis (\citet{bayarri2009modularization, spiegelhalter2003winbugs}). However, this approach may not converge to a well-defined limiting distribution unless tempered transitions are introduced (\citet{plummer2015cuts}).  In addition, it is well-known to be difficult to implement and computationally expensive (\citet{chakraborty2023modularized}). Thus, a fully Bayesian approach may not be practical, especially with the increase in the volume of available and accessible data nowadays, which implies that fitting the first-stage model can be complex in itself. This problem has been approached using frequentist estimation methods as well (\citet{lopiano2013estimated,szpiro2013measurement}).

Hence, a two-stage modeling framework, as illustrated in Figure \ref{fig:motivation}, has often been used. Here, the first stage would fit Equations \eqref{eq:motivatingeq1} and \eqref{eq:motivatingeq2}, providing estimates for $\mu(\mathbf{s})$, denoted as $\hat{\mu}(\mathbf{s})$. Such estimates can be used to compute  the spatial averages $\hat{\mu}(B) = \dfrac{1}{|B|}\bigintssss_B\hat{\mu}(\mathbf{s})$. In the second stage, $\hat{\mu}(B)$ serves as input in Equation \eqref{eq:motivatingeq3}, yielding posterior estimates of $\gamma_0$ and $\gamma_1$. However, the predictions $\hat{\mu}(\mathbf{s})$ are subject to uncertainty from estimation error or model misspecification, which must be appropriately propagated to the second stage.

We argue that the two-stage modeling framework is often more practical or appropriate in several scenarios due to three main reasons. First, it offers an intuitive physical interpretation, as there is a unidirectional relationship between $\mu(\cdot)$ and $\text{y}(\cdot)$ (e.g., climate and pollution levels affect disease risks but not vice versa). Second, it is computationally efficient, particularly when the first-stage model is already complex. Third, it avoids potential feedback issues that can arise in fully Bayesian approaches.

While uncertainty propagation is intrinsic to a fully Bayesian approach, it must be explicitly accounted for in a two-stage modeling framework. The novelty of this work lies in validating different two-stage modeling approaches using the self-consistency property of Bayesian algorithms. This property holds if the data-averaged posterior is equal to the prior distribution (\citet{geweke2004getting}). To test this, we use the simulation-based calibration (SBC) method (\citet{talts2018validating}), discussed in Section \ref{sec:SBC}. The SBC method allows us to assess the correctness of a Bayesian algorithm and has a frequentist interpretation, since the results are viewed as the expected behavior averaged over all potential data outcomes.
 We also introduce a variation of SBC tailored for situations where some first-stage model parameters violate self-consistency, but the second-stage parameters are of primary interest. This is detailed in Section \ref{subsec:sbcvariation}. Both the original and modified SBC methods are implemented in our experiments. As noted by \citet{talts2018validating}, SBC is a crucial part of a robust Bayesian workflow that involves model building, inference, and model checking/improvement (\citet{gelman2020bayesian}).

Another significant contribution of this work is a new approach for uncertainty propagation in two-stage Bayesian models, referred to as the $\mathbf{Q}$ uncertainty method. This method introduces an error component in the second-stage model, which is assigned a Gaussian prior with zero mean and covariance matrix $\mathbf{Q}^{-1}$. It encodes the full uncertainty from the first-stage model. While this method has similarities to the prior exposure method (\citet{cameletti2019bayesian}), it incorporates the full covariance structure of the first-stage latent parameters. Unlike previous approaches (\citet{chang2011estimating, peng2010spatial}), which use first-stage posterior results as priors in the second stage, our method explicitly adds a new model component to capture first-stage uncertainty. Additionally, we explore a low-rank approximation of the error component to handle high-dimensional spatial models, such as large spatio-temporal datasets. This results in two variants: the full $\mathbf{Q}$ uncertainty approach and the low-rank $\mathbf{Q}$ uncertainty approach. Both methods are particularly convenient in the framework of latent Gaussian models and when inference is performed using INLA.


Section \ref{sec:framework} formally discusses the two-stage modeling framework and the uncertainty propagation problem. Section \ref{subsec:currentapproaches} reviews two current approaches: the crude plug-in approach and the resampling approach. Section \ref{subsec:proposedmethods} presents the proposed methods within the INLA framework, applied to spatial modeling contexts. Section \ref{sec:SBC} elaborates on the self-consistency property, the SBC method, and our proposed SBC variant. In Section \ref{sec:simulationstudy}, we validate four uncertainty propagation approaches: the crude plug-in approach, the resampling approach, the full $\mathbf{Q}$ approach, and the low rank $\mathbf{Q}$ approach through simulation experiments. These include a two-stage spatial model with Gaussian observations and point referenced data in the second stage (Section \ref{subsec:twostage_spatial_Gaussian}), and one with Poisson observations and areal data in the second stage (Section \ref{subsec:spatial_twostage_poisson}). We focus on the SBC results for the second-stage parameters, $\gamma_0$ and $\gamma_1$, which are most affected by potential underestimation of posterior uncertainty in a two-stage framework. Finally, we demonstrate the proposed methods in a real-world application, which aims to link relative humidity (a climate variable) and Dengue case counts in the Philippines in August 2018 (Section \ref{sec:data_application}). The paper concludes with insights and future directions in Section \ref{sec:conclusions}.

\section{Uncertainty propagation problem}\label{sec:framework}

\begin{figure}[H]
\centering
\includegraphics[scale=.48]{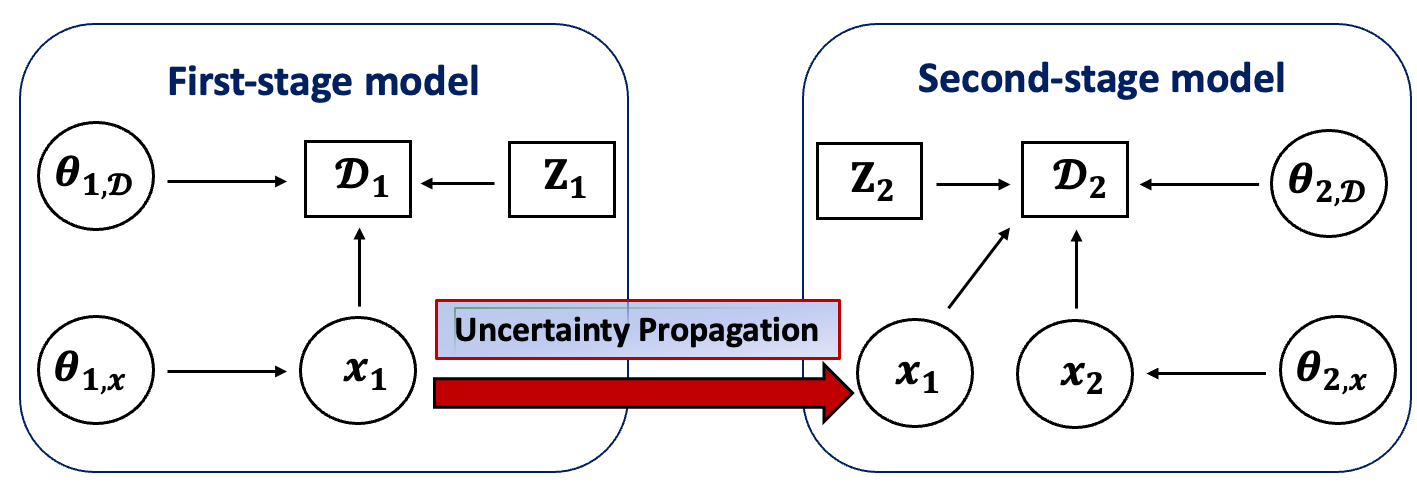}
\caption{Two-stage modelling framework for uncertainty propagation}
\label{fig:framework}
\end{figure}

\vspace{-5mm}

We assume that model inference for a physical process of interest is performed in two stages, as shown in Figure \ref{fig:framework}. The observed data is $\bm{\mathcal{D}} = \{\bm{\mathcal{D}}_1,\bm{\mathcal{D}}_2\}$, partitioned into the first-stage data and second-stage data, respectively. First-stage inference is performed using $\bm{\mathcal{D}}_1$ only and without looking at $\bm{\mathcal{D}}_2$, since for instance the health counts are not intended to inform the estimation of exposures or the health biomarkers causes the probability of survival and not the other way around. We model such process using a Bayesian hierarchical model. Let $\bm{x}_1$ and $\bm{\theta}_1$ be the latent parameters and hyperparameters linked to $\bm{\mathcal{D}}_1$, respectively. We partition $\bm{\theta}_1$ into $\{\bm{\theta}_{1,\bm{x}}, \bm{\theta}_{1,\bm{\mathcal{D}}}\}$, for which $\bm{\theta}_{1,\bm{x}}$ are the hyperparameters linked to $\bm{x}_1$, while $ \bm{\theta}_{1,\bm{\mathcal{D}}}$ are the hyperparameters linked to $\bm{\mathcal{D}}_1$. Also, let $\mathbf{Z}_1$ be a set of fixed inputs/covariates. We assume that $\bm{\mathcal{D}}_1 \sim \mathcal{F}_1(\bm{\mathcal{D}}_1|\bm{x}_1,\bm{\theta}_1,\mathbf{Z}_1).$ Similarly, let $\bm{x}_2$ and $\bm{\theta}_2 = \{\bm{\theta}_{2,\bm{x}}, \bm{\theta}_{2,\bm{\mathcal{D}}}\}$ be the latent parameters and hyperparameters linked to $\bm{\mathcal{D}}_2$, respectively, and $\mathbf{Z}_2$ be a set of fixed inputs. The modeling framework in Figure \ref{fig:framework} assumes that $\bm{x}_1$ from the first-stage model is also linked to $\bm{\mathcal{D}}_2$, so that we have $\bm{\mathcal{D}}_2 \sim \mathcal{F}_2(\bm{\mathcal{D}}_2|\bm{x}_1,\bm{x}_2,\bm{\theta}_2,\mathbf{Z}_2).$ This gives us the full data model: $\{\bm{\mathcal{D}}_1,\bm{\mathcal{D}}_2\} \sim \mathcal{F}_2(\bm{\mathcal{D}}_2|\bm{x}_1,\bm{x}_2,\bm{\theta}_2,\mathbf{Z}_2)\mathcal{F}_1(\bm{\mathcal{D}}_1|\bm{x}_1,\bm{\theta}_1,\mathbf{Z}_1).$

Figure \ref{fig:framework} considers $\bm{x}_1$, or some function of it, as an input when fitting the second-stage model. However, in practice, $\bm{x}_1$ is unknown and needs to be estimated in the first stage. Hence, its uncertainty due to estimation error or model misspecification error needs to be correctly propagated into the second stage; otherwise, the standard errors of the second-stage model parameters may be underestimated. The end goal, therefore, is to correctly estimate the following posterior distributions:\begin{itemize}
    \item The posterior distribution of the first-stage parameters, given by $\pi(\bm{x}_1,\bm{\theta}_1|\bm{\mathcal{D}}_1)$.
    \item The posterior distribution of the second-stage parameters. For the plug-in method, this is given by $\pi(\bm{x}_2,\bm{\theta}_2|\bm{\mathcal{D}}_2,\bm{x}_1^*)$, were  $\bm{x}_1^*$ denotes the posterior mean of $\bm{x}_1$ from the first-stage model results. For the resampling method, this is given by  $\int\pi(\bm{x}_2,\bm{\theta}_2|\bm{\mathcal{D}}_2,\bm{x}_1)\pi(\bm{x}_1|\bm{\mathcal{D}}_1)d\bm{x}_1$.
\end{itemize}
When estimating the posterior distribution of the second-stage model, the uncertainty in $\bm{x}_1^*$ needs to be accounted for, which is fundamentally the \textit{uncertainty propagation problem}. 

\subsection{Current approaches}\label{subsec:currentapproaches}
This section discusses two existing approaches to fit the two-stage model in Figure \ref{fig:framework}. Although not exhaustive, these approaches serve as benchmarks for comparison with our proposed approaches. Related methods, which use first-stage posteriors as priors for the second stage, are discussed in Section \ref{subsec:proposedmethods}.


\begin{enumerate}
    \item \textbf{Plug-in Method} -- Let $\hat{\bm{\mu}}_{\bm{x}_1} = \mathbb{E}[\bm{x}_1|\bm{\mathcal{D}}_1]$ be the posterior mean of $\bm{x}_1$ estimated from the first-stage model. The crude plug-in method simply uses this as an input to the second stage. The linear predictor of the second-stage model is then: \begin{align}\label{eq:plugin}
    g\Big(\mathbb{E}[\bm{\mathcal{D}}_2|\cdots]\Big) = \gamma_0\bm{1} + \gamma_1\mathbf{h}(\hat{\bm{\mu}}_{\bm{x}_1},\cdot) + \mathbf{Z}_2\gamma_2,
\end{align} where $g(\cdot)$ is the link function, $\{\gamma_0, \gamma_1, \gamma_2\}$ are model fixed effects, and $\mathbf{h}(\cdot)$ is a vector-valued linear function  $\mathbf{h}\text{:} \; \bm{x}_1\text{-space} \rightarrow \mathbb{R}^{\text{dim}(\bm{\mathcal{D}}_2)}$. Note that Equation \eqref{eq:plugin} can also include random effects. The estimated uncertainty in the second-stage posterior distribution, $\pi(\gamma_0, \gamma_1, \bm{\gamma}_2,\bm{\theta}_2|\bm{\mathcal{D}}_2,\hat{\bm{\mu}}_{\bm{x}_1})$, is possibly underestimated since it fails to account for the uncertainty in $\hat{\bm{\mu}}_{\bm{x}_1}$.


\item \textbf{Resampling method} -- The resampling method, described in  Algorithm \ref{alg:resampling}, accounts for the uncertainty in the first-stage model in a natural way, but can be computationally expensive since it requires fitting the second-stage model several times. 
\begin{algorithm}[H]
    \caption{Implementation of the resampling method}\label{alg:resampling}
    \flushleft
    \text{Repeat steps 1--2 for $j=1,2,\ldots,J:$}
    \begin{steps}
    \item Sample $\tilde{\bm{\mu}}^{(j)}_{\bm{x}_1} \sim \hat{\pi}(\bm{x}_1|\bm{\mathcal{D}}_1)$.
    \item Plug-in the sampled values in the second stage model, i.e., plug-in $\tilde{\bm{\mu}}^{(j)}_{\bm{x}_1}$ instead of $\hat{\bm{\mu}}_{\bm{x}_1}$ in Equation \eqref{eq:plugin}. Store all posterior marginals, such as $\pi\big(\bm{\gamma}_1^{(j)}|\bm{\mathcal{D}}_2,\tilde{\bm{\mu}}^{(j)}_{\bm{x}_1}\big)$.
    \item All $J$ results are then combined using model averaging, e.g., $\hat{\pi}(\gamma_1|\bm{\mathcal{D}}_2) = \dfrac{1}{J}\sum_{j=1}^J \pi\big(\bm{\gamma}_1^{(j)}|\bm{\mathcal{D}}_2,\tilde{\bm{\mu}}^{(j)}_{\bm{x}_1}\big).$
\end{steps}
\end{algorithm}

\end{enumerate}

This approach was adopted in \citet{blangiardo2016two, liu2017incorporating}, and \citet{zhu2003hierarchical}. A related approach was also implemented in \citet{lee2017rigorous}, where a new value $\tilde{\bm{\mu}}^{(j)}_{\bm{x}_1}$ is sampled at each iteration of the MCMC algorithm and then the second-stage model is fitted for each sample. 
\subsection{Proposed method -- $\mathbf{Q}$ uncertainty}\label{subsec:proposedmethods}

In this section, we present the proposed $\mathbf{Q}$ uncertainty method for uncertainty propagation. This approach avoids multiple runs of the Bayesian algorithm for the second-stage model, offering potential computational efficiency over the resampling method. The method shares similarities with MCMC algorithms used by \citet{chang2011estimating}, \citet{peng2010spatial}, and \citet{gryparis2009measurement}, where the second-stage model is fitted using first-stage results as an informative prior. Two implementation approaches are noted in the literature: one that allows feedback by updating the prior distribution with second-stage data (\citet{gryparis2009measurement,chang2011estimating}), and another that cuts feedback by fixing the prior at each iteration (\citet{peng2010spatial}). Our method aligns with the latter, as we also cut feedback. It is also related to \citet{cameletti2019bayesian}, but uniquely accounts for the full uncertainty in first-stage latent parameters via the $\mathbf{Q}$ matrix.
We propose two versions: the \textit{full $\mathbf{Q}$ uncertainty method} and the \textit{low rank $\mathbf{Q}$ uncertainty method}, the latter being an approximation useful for large $\mathbf{Q}$  matrices, such as in spatio-temporal applications. Both methods are implemented within the INLA framework (\citet{rue2009approximate, van2023new}) and demonstrated in spatial applications.

\subsubsection{Full Q uncertainty method}

In order to formally describe the proposed method, we first briefly describe the INLA methodology, with particular focus on how the $\mathbf{Q}$ matrix is computed. This is important since $\mathbf{Q}^{-1}$ encodes the uncertainty in the latent parameters of the first-stage model, i.e., the uncertainty in $\bm{x}_1$ (see Figure \ref{fig:framework}), which is then propagated to the second-stage model. The first-stage Bayesian hierarchical model is as follows:
\begin{align*}
\bm{\mathcal{D}}_1|\bm{x}_1,\bm{\theta}_{1,\bm{\mathcal{D}}} &\sim \prod_{i=1}^{n_1} \pi(\bm{\mathcal{D}}_{1i}|\bm{x}_1,\bm{\theta}_{1,\bm{\mathcal{D}}})\\   \bm{x}_1|\bm{\theta}_{1,\bm{x}} &\sim \mathcal{N}\Big(\bm{0},\mathbf{Q}_{\text{prior}}^{-1}(\bm{\theta}_{1,\bm{x}})\Big)\\
    \bm{\theta}_1=\{\bm{\theta}_{1,\bm{\mathcal{D}}},\bm{\theta}_{1,\bm{x}}\} &\sim \pi(\bm{\theta}_1)
\end{align*}
The above is a latent Gaussian model since a Gaussian prior is assumed for the latent parameters $\bm{x}_1$. For inference, the needed quantities are the posteriors $\pi(\bm{\theta}_1|\bm{\mathcal{D}}_1)$ and $\pi(\bm{x}_1|\bm{\mathcal{D}}_1)$. The estimated posterior marginal of $\bm{\theta}$, $\hat{\pi}(\bm{\theta}_1|\bm{\mathcal{D}}_1)$, is
\begin{align*}
   \hat{\pi}(\bm{\theta}_1|\bm{\mathcal{D}}_1) = \dfrac{\pi(\bm{x}_1,\bm{\theta}_1|\bm{\mathcal{D}}_1)}{\pi_G(\bm{x}_1|\bm{\theta}_1,\bm{\mathcal{D}}_1)}\Bigg|_{\bm{x}_1=\hat{\bm{\mu}}_{\bm{x}_1}(\bm{\theta}_1)}, 
\end{align*}
where $\pi_G(\bm{x}_1|\bm{\theta}_1,\bm{\mathcal{D}}_1)$ is the Gaussian approximation of $\pi(\bm{x}_1|\bm{\theta}_1,\bm{\mathcal{D}}_1)$,  computed from a second-order expansion of the log-posterior density around its mode. In particular, $\pi_G(\bm{x}_1|\bm{\theta}_1,\bm{\mathcal{D}}_1)$ is given by
\begin{equation}    \label{eq:latentapprox}\bm{x}_1|\bm{\theta}_1,\bm{\mathcal{D}}_1 \approx \mathcal{N}\Big(\hat{\bm{\mu}}_{\bm{x}_1}(\bm{\theta}_1),\mathbf{Q}_{\bm{x}_1}^{-1}(\bm{\theta}_1)  \Big),
\end{equation}
where $\hat{\bm{\mu}}_{\bm{x}_1}(\bm{\theta}_1)$ is the mean of the Gaussian approximation for a given $\bm{\theta}_1$, and $\mathbf{Q}_{\bm{x}_1}(\bm{\theta}_1)$ is a sparse precision matrix which primarily depends on two components: the graph obtained from the prior of $\bm{x}_1$ and the graph based on the mapping from $\bm{x}_1$ to the linear predictors (\citet{van2023new}), and is also computed given $\bm{\theta}_1$. The marginal posterior of each element of $\bm{x}_1$ are then calculated by integrating out $\bm{\theta}_1$ using numerical integration. 


The variance-covariance matrix $\mathbf{Q}_{\bm{x}_1}^{-1}(\bm{\theta}_1)$ in Equation \eqref{eq:latentapprox} encodes the uncertainty in the latent parameters $\bm{x}_1$. Its inverse is what we refer to as the $\mathbf{Q}$ matrix, i.e., $\mathbf{Q}\equiv \mathbf{Q}_{\bm{x}_1}(\bm{\theta}_1)$. We then use this information when fitting the second-stage model. In particular, in the second-stage hierarchical model, we introduce a new model component $\bm{\epsilon}$, which we call an \textit{error component}. Its prior model is derived from Equation \eqref{eq:latentapprox}, i.e., $\bm{\epsilon} \sim \mathcal{N}\Big(\mathbf{0},\mathbf{Q}_{\bm{x}_1}^{-1}(\bm{\theta}_1)\Big)$. In practice, we propose to evaluate $\mathbf{Q}_{\bm{x}_1}(\bm{\theta}_1)$ at the mode of $\hat{\pi}(\bm{\theta}_1|\mathcal{D}_1)$. The  predictor in the second stage is then given by
\begin{align}\label{eq:fullQ}
    g\Big(\mathbb{E}[\bm{\mathcal{D}}_2|\cdots]\Big) = \gamma_0\bm{1} + \gamma_1\mathbf{h}\Big(\hat{\bm{\mu}}_{\bm{x}_1}(\bm{\theta}_1)+\bm{\epsilon},\cdot\Big) + \mathbf{Z}_2\bm{\gamma}_2,
\end{align}
We call this approach the \textit{full $\mathbf{Q}$ uncertainty method}.

 In fitting Equation \eqref{eq:fullQ}, the quantity $\hat{\bm{\mu}}_{\bm{x}_1}(\bm{\theta}_1)$ is assumed to be fixed and known. Since $\mathbf{h}(\cdot)$ is a linear function of $\hat{\bm{\mu}}_{\bm{x}_1}(\bm{\theta}_1)+\bm{\epsilon}$, it follows that $\mathbf{h}\big(\hat{\bm{\mu}}_{\bm{x}_1}(\bm{\theta}_1)+\bm{\epsilon},\cdot\big)$ is also Gaussian. However, $\gamma_1$ is also unknown and assigned a Gaussian prior; thus, Equation \eqref{eq:fullQ} is not a latent Gaussian model since the  predictor involves a product of two components, each with a Gaussian prior. One way to fit the model is to linearize the predictor in Equation \eqref{eq:fullQ} using a first-order Taylor approximation as implemented in the \texttt{R} library \texttt{inlabru} (\citet{lindgren2024inlabru}).  Another approach is to specify a grid of values for $\gamma_1$, and then fit Equation \eqref{eq:fullQ} conditional on each $\gamma_1$. All estimates are then combined using model averaging (\citet{gomez2020bayesian}). A third approach is to fit the model using a hybrid INLA with MCMC or importance sampling approach. Here $\gamma_1$ is estimated using sampling, while the rest of the parameters are estimated using INLA (\citet{gomez2018markov, berild2022importance}). 
 
 In the INLA framework, the model component $\bm{\epsilon}$ in Equation \eqref{eq:fullQ} involves a scaling or precision parameter, say $\tau_{\bm{\epsilon}}$, for $\mathbf{Q}_{\bm{x}_1}(\bm{\theta}_1)$. We propose fixing the value of this scaling parameter at $\tau_{\bm{\epsilon}}=1$. Note that fixing this value to higher (lower) values implies a reduction (increase) in the uncertainty carried over from the first-stage model to the second-stage model. 
 An approach for determining the optimal value of $\tau_{\bm{\epsilon}}$ could be pursued in  future work. 

\textit{An application to spatial models}

In spatial applications, the linear predictor in the first-stage model is a combination of fixed effects and a random field. In these scenarios, a common specification for the function $\mathbf{h}(\cdot)$ in Equation \eqref{eq:fullQ} is as follows:
\begin{equation}
    \label{eq:stageonelinearpred}
    \mathbf{h}\big(\bm{x}_1\big) = \mathbf{Z}_1\bm{\beta}+\bm{\xi},
\end{equation}
where $\bm{\beta}$ are fixed effects and $\bm{\xi}$ is a random field. An efficient method for estimating the random field is the stochastic partial differential equations (SPDE) approach (\citet{lindgren2011explicit}). This approach provides a finite-dimensional but continuously-indexed approximation of Gaussian fields with Mat\'ern covariance function. The discretization is defined on a mesh and expresses the approximation as 
\begin{equation}\label{eq:spde}
    \xi(\mathbf{s}) \approx \sum_{k=1}^K\psi_k \omega_k,
\end{equation}
where $K$ is the number of mesh nodes or vertices, $\{\psi_k\}$ are basis functions chosen to be piecewise linear in each triangle, i.e., $\psi_k=1$ at vertex $k$ and $\psi_k=0$ otherwise, and $\{\omega_k\}$ are Gaussian-distributed weights. The approximation in Equation \eqref{eq:spde} is fully specified by the probability distribution of the weights $\bm{\omega}\sim \mathcal{N}\big(\bm{0}, \mathbf{Q}^{-1}_{\bm{\omega}}\big)$, where $\mathbf{Q}_{\bm{\omega}}$ is a sparse precision matrix. The speed of the computation depends on the resolution of the mesh, i.e., the number of mesh nodes $K$. Using the SPDE approach, we can write Equation \eqref{eq:stageonelinearpred} as 
\begin{equation*}
    \mathbf{h}\big(\bm{x}_1\big) = \mathbf{Z}_1\bm{\beta}+\mathbf{A}\bm{\omega},
\end{equation*}
where the latent parameter vector is $\bm{x}_1=\begin{pmatrix}
    \bm{\beta} & \omega_1 & \ldots & \omega_K
\end{pmatrix}^\intercal$, and $\mathbf{A}$ is the mapping matrix from the mesh nodes to the observed data points. With the full $\mathbf{Q}$ uncertainty method, we use the posterior mean $\hat{\bm{\mu}}_{\bm{x}_1} = \begin{pmatrix}
    \hat{\bm{\beta}} & \hat{\bm{\omega}}
\end{pmatrix}^\intercal$ and the precision matrix $\mathbf{Q}_{\bm{x}_1}^{-1}(\bm{\theta}_1)$ from the Gaussian approximation in Equation \eqref{eq:latentapprox}. The linear predictor in the second-stage model is then specified as:
\begin{align}
    \label{eq:fullQuncertainty}
    g\Big(\mathbb{E}[\bm{\mathcal{D}}_2|\cdots]\Big) &= \gamma_0\bm{1} + \gamma_1\Bigg\{ \begin{bmatrix}
        \mathbf{Z}_1 & \mathbf{A}
    \end{bmatrix}\bigg(\begin{bmatrix}
        \hat{\bm{\beta}} \\ \hat{\bm{\omega}}
    \end{bmatrix} + \bm{\epsilon} \bigg) \Bigg\} + \mathbf{Z}_2\bm{\gamma}_2,
\end{align}
where $\gamma_0,\gamma_1$, and $\bm{\gamma}_2$ are the model parameters; $\mathbf{Z}_1$,$\mathbf{Z}_2$ and $\mathbf{A}$ are known matrices; and $\bm{\epsilon}$ is the error component with prior given by  $\bm{\epsilon}\sim\mathcal{N}\Big(\mathbf{0},\mathbf{Q}_{\bm{x}_1}^{-1}(\bm{\theta}_1)\Big)$.

\subsubsection{Low rank Q uncertainty method}

A potential problem with the specification in Equation \eqref{eq:fullQuncertainty} is that when extending this to large spatio-temporal data, the dimension of $\hat{\bm{\omega}}$ scales linearly as the number of time points, which in effect also applies to the dimension of the error component $\bm{\epsilon}$. Thus, we propose a low rank approximation of $\mathbf{Q}$, which then expresses Equation \eqref{eq:fullQuncertainty} as follows:
\begin{align}
    \label{eq:lowrankQuncertainty}
      g\Big(\mathbb{E}[\bm{\mathcal{D}}_2|\cdots]\Big) &= \gamma_0\bm{1} + \gamma_1\Bigg\{ \begin{bmatrix}
        \mathbf{Z}_1 & \mathbf{A}
    \end{bmatrix}\begin{bmatrix}
        \hat{\bm{\beta}} + \bm{\epsilon}_{\beta} \\ \hat{\bm{\omega}} + \mathbf{B}\bm{\epsilon}_{\omega}^*
    \end{bmatrix} \Bigg\} + \mathbf{Z}_2\bm{\gamma}_2 \\ \label{eq:lowrankQuncertainty2}
    &=\gamma_0\bm{1} + \gamma_1\Bigg\{ \begin{bmatrix}
        \mathbf{Z}_1 & \mathbf{A}
    \end{bmatrix}\bigg(\begin{bmatrix}
        \hat{\bm{\beta}} \\ \hat{\bm{\omega}}
    \end{bmatrix} + \begin{bmatrix}
        \mathbb{I} & \mathbf{0} \\ \mathbf{0} & \mathbf{B}
    \end{bmatrix} \begin{bmatrix}
        \bm{\epsilon}_{\beta} \\ \bm{\epsilon}_{\omega}^*
    \end{bmatrix} \bigg) \Bigg\} + \mathbf{Z}_2\bm{\gamma}_2,
\end{align}
where we have implicitly partitioned $\bm{\epsilon}$ into $\begin{pmatrix}
    \bm{\epsilon}_{\beta} & \bm{\epsilon}_{\omega}
\end{pmatrix}^\intercal$, with $\bm{\epsilon}_{\beta}$ being the fixed effects error component which is time-invariant, and  $\bm{\epsilon}_{\omega}$ is the spatial error component which varies in time. In Equation \eqref{eq:lowrankQuncertainty}, $\mathbf{B}\bm{\epsilon}_{\omega}^*$ is used to approximate $\bm{\epsilon}_{\omega}$, where $\bm{\epsilon}_{\omega}^*$ is defined on a coarser mesh compared to the mesh used for $\hat{\bm{\omega}}$, and $\mathbf{B}$ is the appropriate projection matrix from the coarse mesh to the fine mesh. 

The probability model for $\bm{\epsilon}_{\omega}^*$ depends on the distribution of the weights at the coarser mesh, say $\bm{\phi}\in \mathbb{R}^M$, $M \ll K$, $K$ being the dimension of $\hat{\bm{\omega}}$. The probability model for $\bm{\epsilon}_{\omega}^*$ is given by $\bm{\epsilon}_{\omega}^* \sim \mathcal{N}\Big(\mathbf{0},\big(\mathbf{B}^\intercal \mathbf{Q}_{\bm{\omega}}\mathbf{B}\big)^{-1}\Big)$ as stated in Theorem \ref{thm:lowrankQ}.  
\begin{theorem}\label{thm:lowrankQ}
Suppose $\bm{\omega}$ is defined on a discretization of dimension $\mathbb{R}^K$, i.e. $\bm{\omega}\in \mathbb{R}^K$, with probability model $\bm{\omega}\sim \mathcal{N}\big(\mathbf{0},\mathbf{Q}_{\bm{\omega}}^{-1}\big)$. Suppose we define a coarser discretization specified via the weights $\bm{\phi}\in \mathbb{R}^M, M \ll K$, such that $\bm{\omega}=\mathbf{B}\bm{\phi}$. Then, the precision matrix of $\bm{\phi}$ is given by $\mathbf{Q}_{\bm{\phi}}=\mathbf{B}^\intercal \mathbf{Q}_{\bm{\omega}}\mathbf{B}$ and the probability model of $\bm{\phi}$ is given by
\begin{align*}
    \bm{\phi}\sim \mathcal{N}\Big((\mathbf{B}^\intercal\mathbf{Q}_{\bm{\omega}}\mathbf{B})^{\inv}\mathbf{B}^\intercal\mathbf{Q}_{\bm{\omega}}\bm{\omega}, \mathbf{Q}_{\bm{\phi}}^{-1}\Big).
\end{align*}
\end{theorem}
\begin{proof}
This is simply a generalized least squares problem, i.e., $\hat{\bm{\phi}} \defeq \argmin_{\bm{\phi}} \norm{\bm{\omega} - \mathbf{B}\bm{\phi}}^2_{\mathbf{Q}_{\bm{\omega}}^{\inv}}$. This yields $\mathbb{E}[\hat{\bm{\phi}}]=(\mathbf{B}^\intercal\mathbf{Q}_{\bm{\omega}}\mathbf{B})^{\inv}\mathbf{B}^\intercal\mathbf{Q}_{\bm{\omega}}\bm{\omega}$ and $\mathbb{V}[\hat{\bm{\phi}}]=\mathbf{Q}_{\bm{\phi}}=\big(\mathbf{B}^\intercal \mathbf{Q}_{\bm{\omega}}\mathbf{B}\big)^{-1}.$
\end{proof}

\section{Simulation-based calibration for model validation}\label{sec:SBC}

We validate the uncertainty propagation methods discussed in Sections \ref{subsec:currentapproaches} and \ref{subsec:proposedmethods} using the simulation-based calibration (SBC) approach, originally proposed by \citet{talts2018validating} and based on  ideas from \citet{cook2006validation}. The SBC method tests for the self-consistency property of Bayesian models, which states that the posterior distribution, averaged over all possible outcomes from the full generative model, is equal to the prior distribution. Formally, suppose $\pi(\bm{\theta})$ is the prior model, $\pi(\mathbf{y}|\bm{\theta})$ is the observation density or probability mass function, and $\pi(\bm{\theta}|\mathbf{y})$ is the posterior distribution. The self-consistency property is stated as:
\begin{equation*}
    \pi(\bm{\theta}') =  \int \pi(\bm{\theta}'|\mathbf{y})\pi(\mathbf{y}|\bm{\theta})\pi(\bm{\theta}) d\mathbf{y}d\bm{\theta}.
\end{equation*}
Any discrepancy between the prior model and the data-averaged posterior indicates an error in the Bayesian algorithm. The SBC method tests this property using rank statistics. It involves sampling from the data's generative model and applying the Bayesian algorithm to each data replicate. Specifically, consider the following sequence of samples drawn from the Bayesian model:

\begin{align*}
    \tilde{\bm{\theta}}&\sim \pi(\bm{\theta})\\
    \tilde{\mathbf{y}}&\sim\pi(\mathbf{y}|\tilde{\bm{\theta}}) \\
    \{\bm{\theta}_1,\ldots,\bm{\theta}_L\} &\overset{iid}{\sim}\pi(\bm{\theta}|\tilde{\mathbf{y}}). 
\end{align*}
If the Bayesian algorithm is correct, then for any one-dimensional function of the parameters, $f:\Theta \rightarrow \mathbb{R}$, where $\Theta$ is the $\bm{\theta}$-space, the \textit{rank statistic} of the prior sample relative to the posterior sample, given by
\textsmaller[.5]{\begin{equation}\label{eq:rankstatistic}
    r\Big( \big\{f(\bm{\theta}_1),\ldots,f(\bm{\theta}_L)\big\}, f(\tilde{\bm{\theta}}) \Big) = \sum_{\ell=1}^L \mathbb{I}\big[f(\bm{\theta}_{\ell})<f(\tilde{\bm{\theta}})\big], \;\;\; \mathbb{I}\big[f(\bm{\theta}_{\ell})<f(\tilde{\bm{\theta}})\big]=\begin{cases}
        1 & \text{if } f(\bm{\theta}_{\ell})< f(\tilde{\bm{\theta}})\\
        0 & \text{if } f(\bm{\theta}_{\ell})\geq f(\tilde{\bm{\theta}})
    \end{cases}
\end{equation}} 
should be uniformly distributed across the integers $\{0,1,\ldots,L\}$. 

Deviations from uniformity in the rank distribution provide insights into errors in the posteriors. A $\cap$-shaped rank distribution suggests that the data-averaged posterior is overdispersed compared to the prior, meaning uncertainty is overestimated, violating Bayesian self-consistency. Conversely, a $\cup$-shaped rank distribution indicates underdispersion, where the estimated posterior underestimates the true uncertainty. Asymmetry in the rank distribution reveals bias in the data-averaged posterior, deviating in the opposite direction relative to the prior distribution.


\subsection{Implementation of SBC for the two-stage model}

This section discusses how to implement the SBC in a two-stage modelling framework following Figure \ref{fig:framework}. We assume that $\bm{\theta}_1\in \bm{\Theta}_1, \bm{\theta}_2\in \bm{\Theta}_2, \bm{x}_1\in \bm{\chi}_1$, and $\bm{x}_2\in \bm{\chi}_2$, and that $\bm{\Theta}_1, \bm{\Theta}_2, \bm{\chi}_1,$ and $\bm{\chi}_2$ are continuous spaces. The assumption of continuous spaces for the model parameters are crucial for the SBC method in \citet{talts2018validating} to work.  For cases when some of the spaces are discrete, an SBC variant is proposed in \citet{modrak2023simulation}. 

Following Equations \eqref{eq:plugin}, \eqref{eq:fullQuncertainty}, and \eqref{eq:lowrankQuncertainty}, we are interested in the posterior marginals of $\gamma_0$ and $\gamma_1$, since these are the parameters which are potentially underestimated when the uncertainty in the first stage model is not properly propagated to the second stage. We use the individual parameters as test quantities to check that their uncertainty is correctly calibrated. Testing individual parameters allows the diagnosis of a large number of problems with posterior approximation (\citet{modrak2023simulation}). However, \citet{modrak2023simulation} also recommended the use of test functions which are data-dependent, such as the joint likelihood of the data, since there are large classes of problems which cannot be detected when the test quantities are functions only of the parameters. We have not considered such test functions in this work, but  plan to do so in  future work.

Algorithm \ref{alg:sbc} shows the steps to implement the SBC for a two-stage Bayesian model in Figure \ref{fig:framework}, 
where the test quantities are $\gamma_0$ and $\gamma_1$. To test for the uniformity of the rank statistic in Equation \eqref{eq:rankstatistic}, we primarily use the graphical approach in \citet{sailynoja2022graphical}, which generates simultaneous confidence bands for the difference between the empirical cumulative distribution function (ECDF) and the uniform CDF. The method is not sensitive to binning, does not require smoothing, and provides intuitive visual interpretation.

\begin{algorithm}[hbt!]
    \caption{Implementing SBC for Figure \ref{fig:framework} with $\gamma_0$ and $\gamma_1$ as test quantities}\label{alg:sbc}
    \flushleft
    \text{Do for $k=1,2,\ldots,K:$}
    \begin{steps}
    
    \item Sample hyperparameter values:  $\tilde{\bm{\theta}}_{1}^{(k)} \sim \pi(\bm{\theta}_{1}), \tilde{\bm{\theta}}_{2}^{(k)} \sim \pi(\bm{\theta}_{2})$.
    \item Sample latent parameter values: $\tilde{\bm{x}}_{1}^{(k)} \sim \pi(\bm{x}_{1}|\tilde{\bm{\theta}}_{1}^{(k)}), \tilde{\bm{x}}_{2}^{(k)} \sim \pi(\bm{x}_{2}|\tilde{\bm{\theta}}_{2}^{(k)})$.
    \item Sample observed data values: 
    \begin{equation*}
        \tilde{\bm{\mathcal{D}}}_1^{(k)} \sim \pi_1(\bm{\mathcal{D}}_1|\tilde{\bm{x}}_1^{(k)},\tilde{\bm{\theta}}_1^{(k)}), \tilde{\bm{\mathcal{D}}}_2^{(k)} \sim \pi_2(\bm{\mathcal{D}}_2|\tilde{\bm{x}}_1^{(k)}, \tilde{\bm{x}}_2^{(k)},\tilde{\bm{\theta}}_2^{(k)}).
    \end{equation*}
    \item Perform inference in order to obtain estimated posteriors: $\hat{\pi}^{(k)}(\bm{\theta}_1,\bm{x}_1|\bm{\mathcal{D}}_1)$ and $\hat{\pi}^{(k)}(\bm{\theta}_2,\bm{x}_2|\bm{\mathcal{D}}_2)$. 
    \item Generate $L$ samples from the estimated posterior distributions of $\gamma_0$ and $\gamma_1$:
    \begin{align*}
        \gamma_{0,1}^{(k)}, \gamma_{0,2}^{(k)},\ldots, \gamma_{0,L}^{(k)} \sim \hat{\pi}(\gamma_0|\bm{\mathcal{D}}_2) \\
        \gamma_{1,1}^{(k)}, \gamma_{1,2}^{(k)},\ldots, \gamma_{1,L}^{(k)} \sim \hat{\pi}(\gamma_1|\bm{\mathcal{D}}_2)
    \end{align*}
    \item Compute the ranks: 
    \begin{align*}
        &r\Big(\big\{\gamma_{0,1}^{(k)}, \gamma_{0,2}^{(k)},\ldots, \gamma_{0,L}^{(k)}\big\},\tilde{\gamma}_0^{(k)}\Big) = \sum_{\ell=1}^L\mathbb{I}\big[\gamma_{0,\ell}^{(k)}<\tilde{\gamma}_0^{(k)}\big], \;\;\;\; \mathbb{I}\big[\gamma_{0,\ell}^{(k)}<\tilde{\gamma}_0^{(k)}\big] = \begin{cases} 
      1 & \gamma_{0,\ell}^{(k)}< \tilde{\gamma}_0^{(k)} \\
      0 & \gamma_{0,\ell}^{(k)}\geq \tilde{\gamma}_0^{(k)} 
   \end{cases}\\
   &r\Big(\big\{\gamma_{1,1}^{(k)}, \gamma_{1,2}^{(k)},\ldots, \gamma_{1,L}^{(k)}\big\},\tilde{\gamma}_1^{(k)}\Big) = \sum_{\ell=1}^L\mathbb{I}\big[\gamma_{1,\ell}^{(k)}<\tilde{\gamma}_1^{(k)}\big], \;\;\;\; \mathbb{I}\big[\gamma_{1,\ell}^{(k)}<\tilde{\gamma}_1^{(k)}\big] = \begin{cases} 
      1 & \gamma_{1,\ell}^{(k)}< \tilde{\gamma}_1^{(k)} \\
      0 & \gamma_{1,\ell}^{(k)}\geq \tilde{\gamma}_1^{(k)} 
   \end{cases},
    \end{align*}
    where $\tilde{\gamma}_0^{(k)}$ and $\tilde{\gamma}_1^{(k)}$ are prior samples. The ranks are normalized by computing
    \begin{equation}\label{eq:pkformula}
        p_k=\dfrac{1}{L}\sum_{\ell=1}^L\mathbb{I}\big[\gamma_{0,\ell}^{(k)}<\tilde{\gamma}_0^{(k)}\big] \;\;\;\; \text{and} \;\;\;\; p_k=\dfrac{1}{L}\sum_{\ell=1}^L\mathbb{I}\big[\gamma_{1,\ell}^{(k)}<\tilde{\gamma}_1^{(k)}\big] 
    \end{equation}
\end{steps}
\end{algorithm}

\subsection{Variation in the SBC}\label{subsec:sbcvariation}

In this section, we propose a variation of the SBC approach in a two-stage modeling framework. The motivation here is that the parameters of primary interest are the fixed effects of the second-stage model, namely $\gamma_0$ and $\gamma_1$, and we want to avoid the influence of certain parameters of the first-stage model which may violate the self-consistency property for specific priors or model specification. As an example, some parameters in $\bm{\theta}_{1,\bm{x}}$ (Figure \ref{fig:framework}) may violate the self-consistency property. Hence, we propose Theorem \ref{thm:SBChierarhicalmodels1}, which derives the distribution of the rank statistic for an arbitrary unidimensional test function conditional on $\bm{\theta}_{1,\bm{x}}$. This is then used as the theoretical basis for doing the SBC conditional on $\bm{\theta}_{1,\bm{x}}$. 


\begin{theorem}\label{thm:SBChierarhicalmodels1}
    Let $\pi(\bm{\theta}_{1})=\pi(\bm{\theta}_{1,\bm{x}})\pi(\bm{\theta}_{1,\bm{\mathcal{D}}})$ be the prior model, $\pi(\bm{x}_1|\bm{\theta}_1)$ be the latent model, and $\pi(\bm{\mathcal{D}}_1|\bm{x}_1,\bm{\theta}_1)$ be the observation density or probability mass function. Let $\bm{\chi}_1$ be the latent space, $\bm{\Theta}_{1,\bm{x}}$ be the $\bm{\theta}_{1,\bm{x}}$-space, and $\bm{\Theta}_{1,\bm{\mathcal{D}}}$ be the $\bm{\theta}_{1,\bm{\mathcal{D}}}$-space, all continuous. Suppose $\bm{\theta}_{1,\bm{x}}\in \bm{\Theta}_{1,\bm{x}}$ is fixed. Let $\tilde{\bm{\theta}}_{1,\bm{\mathcal{D}}}$ be a sample from the prior, i.e., $\tilde{\bm{\theta}}_{1,\bm{\mathcal{D}}}\sim \pi(\bm{\theta}_{1,\bm{\mathcal{D}}})$, $\tilde{\bm{x}}_1$ be a sample from the latent model, i.e., $\tilde{\bm{x}}_1\sim \pi(\bm{x}_1|\bm{\theta}_{1,\bm{x}},\tilde{\bm{\theta}}_{1,\bm{\mathcal{D}}})$, and $\tilde{\bm{\mathcal{D}}}_1$ a sample from the observation model, i.e., $\tilde{\bm{\mathcal{D}}}_1 \sim \pi(\bm{\mathcal{D}}_1|\tilde{\bm{x}}_1,\bm{\theta}_{1,\bm{x}},\tilde{\bm{\theta}}_{1,\bm{\mathcal{D}}})$. Suppose that the approximate posteriors from applying the Bayesian algorithm are $\hat{\pi}(\bm{x}_1|\tilde{\bm{\mathcal{D}}}_1)$ and $\hat{\pi}(\bm{\theta}_{1}|\tilde{\bm{\mathcal{D}}}_1)$. Let $\{\bm{x}_{1,\ell}\}$ and  $\{\bm{\theta}_{1,\bm{\mathcal{D}},\ell}\}, \ell=1,\ldots,L$ be independent samples from the posteriors, i.e., $\begin{pmatrix}
    \mathbf{x}_{1,1} & \bm{x}_{1,2} & \ldots \bm{x}_{1,L}
\end{pmatrix} \overset{\text{iid}}{\sim}\hat{\pi}(\mathbf{x}_1|\tilde{\bm{\mathcal{D}}}_1)$, and $\begin{pmatrix}
    \bm{\theta}_{1,\bm{\mathcal{D}},1} & \bm{\theta}_{1,\bm{\mathcal{D}},2} & \ldots \bm{\theta}_{1,\bm{\mathcal{D}},L}
\end{pmatrix} \overset{\text{iid}}{\sim}\hat{\pi}(\bm{\theta}_{1,\bm{\mathcal{D}}}|\tilde{\bm{\mathcal{D}}}_1)$. Then, we have the following results:
 \begin{enumerate}[label=(\arabic*)]
        \item For any uni-dimensional function $f: \bm{\chi}_1\rightarrow \mathbb{R}$, the distribution of the rank statistic 
\begin{equation*}
    r = \sum_{\ell=1}^L \mathbb{I}\big[f(\bm{x}_{1,\ell}) < f(\tilde{\bm{x}}_1)\big], \;\;\; \mathbb{I}\big[f(\bm{x}_{1,\ell}) < f(\tilde{\bm{x}}_1)\big]=
    \begin{cases}
        1 & \text{if } f(\mathbf{x}_{1,\ell})< f(\tilde{\bm{x}}_1)\\
        0 & \text{if } f(\bm{x}_{1,\ell})\geq f(\tilde{\bm{x}}_1)
    \end{cases}
\end{equation*}
is $\mathcal{U}(0,1,\ldots,L)$.
        \item  For any uni-dimensional function $f: \bm{\Theta}_{1,\bm{\mathcal{D}}}\rightarrow \mathbb{R}$, the distribution of the rank statistic
\begin{equation*}
    r = \sum_{\ell=1}^L \mathbb{I}\big[f(\bm{\theta}_{1,\bm{\mathcal{D}},\ell}) < f(\tilde{\bm{\theta}}_{1,\bm{\mathcal{D}}})\big], \;\;\; \mathbb{I}\big[f(\bm{\theta}_{1,\bm{\mathcal{D}},\ell}) < f(\tilde{\bm{\theta}}_{1,\bm{\mathcal{D}}})\big]=
    \begin{cases}
        1 & \text{if } f(\bm{\theta}_{1,\bm{\mathcal{D}},\ell})< f(\tilde{\bm{\theta}}_{1,\bm{\mathcal{D}}})\\
        0 & \text{if } f(\bm{\theta}_{1,\bm{\mathcal{D}},\ell})\geq f(\tilde{\bm{\theta}}_{1,\bm{\mathcal{D}}})
    \end{cases}
\end{equation*}
is $\mathcal{U}(0,1,\ldots,L)$.
    \end{enumerate}

\end{theorem}

The proof of Theorem \ref{thm:SBChierarhicalmodels1} is in Section 1.1 of the Supplementary Material. Theorem \ref{thm:SBChierarhicalmodels1} implies a variation in the implementation of the original SBC. Instead of sampling from the full data generative model, we fix the value of $\bm{\theta}_{1,\bm{x}}$. In particular, the changes in Algorithm \ref{alg:sbc} only apply to Steps 1 -- 3, which are the steps for generating data from the model. The changes are formalized in Algorithm \ref{alg:sbcconditional}. The remaining steps for doing the SBC conditional on $\bm{\theta}_{1,\bm{x}}$ are the same as the original SBC, i.e., the model inference is done without knowledge of $\bm{\theta}_{1,\bm{x}}$ and the test quantities for the SBC are $\gamma_0$ and $\gamma_1$. Moreover, we extend Theorem \ref{thm:SBChierarhicalmodels1} to the case where we condition on the entire hyperaparameter vector $\bm{\theta}_1$. This is formalized in Section 1.2 of the Supplementary Material, and which we refer to as Theorem 3.2.

\begin{algorithm}[hbt!]
\algsetup{linenosize=\tiny}
    \caption{Data generation mechanism for the SBC conditional on $\bm{\theta}_{1,\bm{x}}$}\label{alg:sbcconditional}
    \flushleft
    Fix $\bm{\theta}_{1,\bm{x}}\in\Theta_{1,\bm{x}}$. \text{Do for $k=1,2,\ldots,K:$}
    \begin{steps}
    
    \item Sample hyperparameter values:  $\tilde{\bm{\theta}}_{1,\bm{\mathcal{D}}}^{(k)} \sim \pi(\bm{\theta}_{1,\bm{\mathcal{D}}}), \tilde{\bm{\theta}}_{2}^{(k)} \sim \pi(\bm{\theta}_{2})$.
    \item Sample latent parameter values: $\tilde{\bm{x}}_{1}^{(k)} \sim \pi\big(\bm{x}_{1}|\bm{\theta}_{1,\bm{x}}\big), \tilde{\bm{x}}_{2}^{(k)} \sim \pi\big(\bm{x}_{2}|\tilde{\bm{\theta}}_{2}^{(k)}\big)$.
    \item Sample observed data values: 
    \begin{equation*}
        \tilde{\bm{\mathcal{D}}}_1^{(k)} \sim \pi_1(\bm{\mathcal{D}}_1|\tilde{\bm{x}}_1^{(k)},\tilde{\bm{\theta}}_{1,\bm{\mathcal{D}}}^{(k)},\bm{\theta}_{1,\bm{x}}), \tilde{\bm{\mathcal{D}}}_2^{(k)} \sim \pi_2(\bm{\mathcal{D}}_2|\tilde{\bm{x}}_1^{(k)}, \tilde{\bm{x}}_2^{(k)},\tilde{\bm{\theta}}_2^{(k)}).
    \end{equation*}
\end{steps}
\end{algorithm}

\section{Simulation experiments}\label{sec:simulationstudy}
We perform simulation experiments where we compare the 
different uncertainty propagation approaches on two two-stage models: one with Gaussian observations (Section \ref{subsec:twostage_spatial_Gaussian}) and  one with Poisson observations (Section \ref{subsec:spatial_twostage_poisson}). For each model, we highlight the SBC results for $\gamma_0$ and $\gamma_1$ using both Algorithms \ref{alg:sbc} and \ref{alg:sbcconditional}.

\subsection{A two-stage spatial model with Gaussian observations}\label{subsec:twostage_spatial_Gaussian}

\begin{figure}[t]
    \centering
   \subfigure[Spatial locations of data]{\includegraphics[width=.27\linewidth]{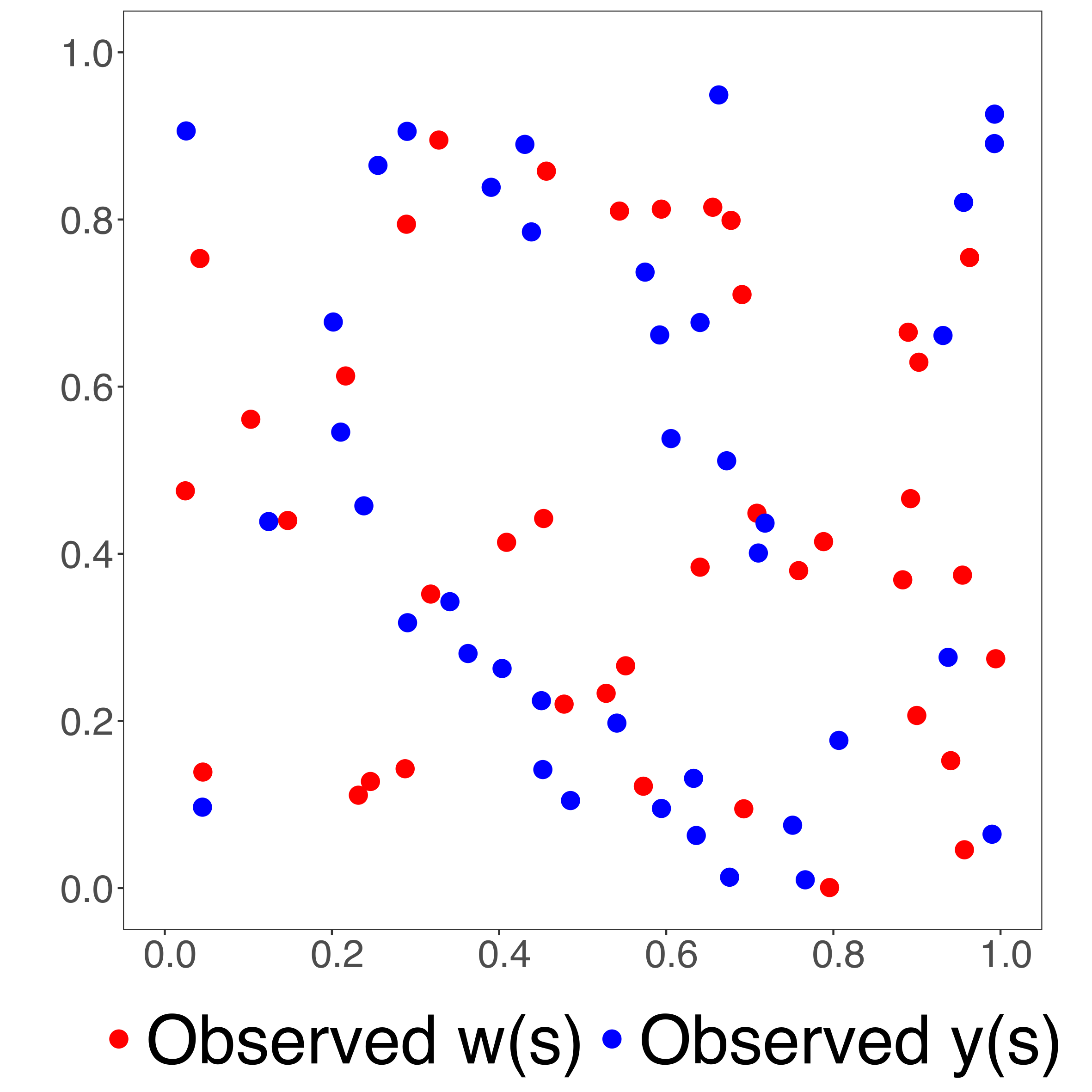}}\hspace{5mm}
   \subfigure[$\mu(\mathbf{s})$]{\includegraphics[width=.27\linewidth]{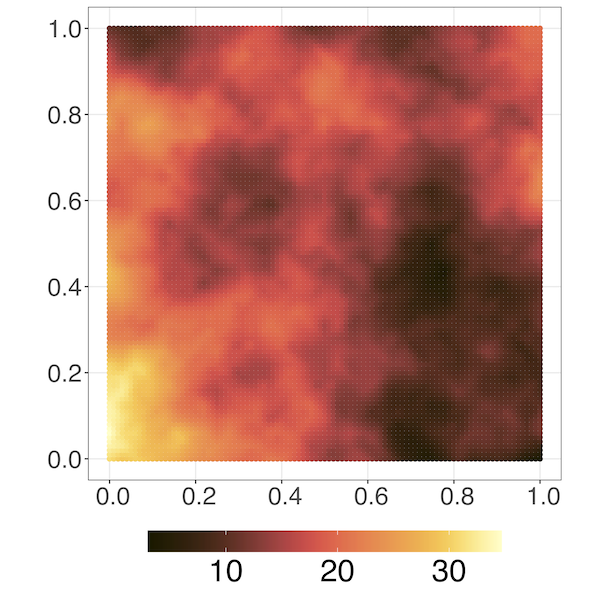}}\hspace{5mm} \subfigure[$\mathbb{E}\big(\text{y}(\mathbf{s})|\mu(\mathbf{s})\big)$]{\includegraphics[width=.27\linewidth]{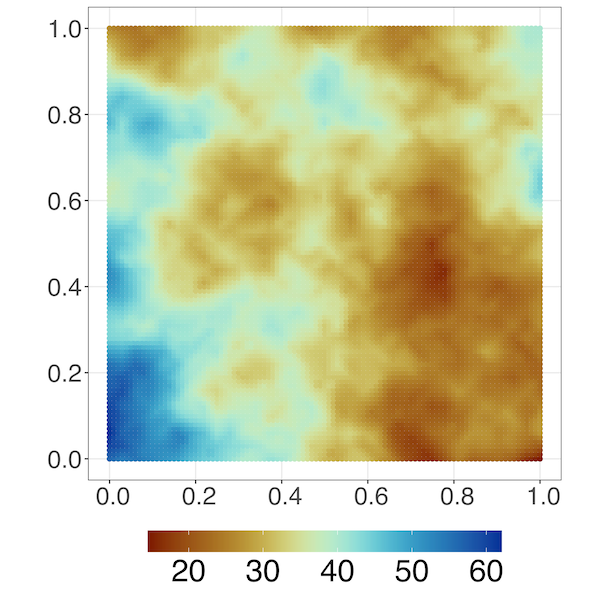}} 
    \caption{Simulated data for the two-stage model in Section \ref{subsec:twostage_spatial_Gaussian}: (a) spatial locations of the data (b) simulated $\mu(\mathbf{s})$ (c) simulated second-stage field $\mathbb{E}\big[\text{y}(\mathbf{s})|\mu(\mathbf{s})\big]$}
\label{fig:res_twostage_gaussian_simdata}
\end{figure}

In the first experiment, the first-stage latent process $\mu(\mathbf{s})$ is given by $\mu(\mathbf{s}) = \beta_0 + \beta_1 z(\mathbf{s}) + \xi(\mathbf{s})$,
where $\beta_0$ and $\beta_1$ are fixed effects, $z(\mathbf{s})$ is a known covariate, and $\xi(\mathbf{s})$ is a Gaussian field with Mat\'ern covariance function. The error-prone observed outcomes are $\bm{\mathcal{D}}_1\equiv \big\{\text{w}(\mathbf{s}_i), i=1,\ldots,n_{{\text{w}}}\big\}$, which follows the classical error model, i.e.,
\begin{align}\label{eq:sec4.1firststage}
    \text{w}(\mathbf{s}_i) = \mu(\mathbf{s}_i) + e_1(\mathbf{s}_i), \;\;\; e_1(\mathbf{s}_i) \overset{\text{iid}}{\sim} \mathcal{N}(0,\sigma^2_{e_1}), \;\;\; i=1,\ldots,n_{\text{w}}
\end{align}
 The latent process $\mu(\mathbf{s})$ is an input in the second-stage model, i.e.,
\begin{align}\label{eq:sec4.1secondstage}
    \text{y}(\mathbf{s}_j) = \gamma_0 + \gamma_1\mu(\mathbf{s}_j) + e_2(\mathbf{s}_j), \;\;\; e_2(\mathbf{s}_j) \overset{\text{iid}}{\sim} \mathcal{N}(0,\sigma^2_{e_2}),\;\;\; j=1,\ldots,n_{\text{y}},
\end{align}
Here, $\bm{\mathcal{D}}_2\equiv \big\{\text{y}(\mathbf{s}_j),j=1,\ldots,n_{{\text{y}}}\big\}$  represents observations at locations different from where $\text{w}(\mathbf{s}_i)$ is measured.
This model belongs to  the class of measurement error models because  
$\mu(\mathbf{s}_j)$ in Equation \eqref{eq:sec4.1secondstage} is unobserved and  needs to be estimated using Equation \eqref{eq:sec4.1firststage} (\citet{berry2002bayesian, banerjee2002prediction, madsen2008regression}). 
Figure \ref{fig:res_twostage_gaussian_simdata} shows a simulated observed locations for $\text{w}(\mathbf{s}_i)$ and $\text{y}(\mathbf{s}_j)$ where $n_{\text{w}}=n_{\text{y}}=80$. Note that $\mathbb{E}\big[\text{y}(\mathbf{s})|\mu(\mathbf{s})\big] = \gamma_0+\gamma_1\mu(\mathbf{s})$ is also another latent field of interest.

We use INLA with SPDE representations of the spatial fields to fit this model. The first- and second-stage latent parameters are $\bm{x}_1=\big\{\beta_0,\beta_1,\omega_1,\omega_2,\ldots,\omega_K\big\}$ and $\bm{x}_2=\big\{\gamma_0,\gamma_1\big\}$, respectively. The $\big\{\omega_1,\omega_2,\ldots,\omega_K\big\}$ are the Gaussian weights of the SPDE approximation (\citet{lindgren2011explicit}), i.e., $\xi(\mathbf{s}) \approx \sum_{i=1}^K\psi_i\omega_i$ where $\psi_i, i=1,\ldots,K$, are basis functions as discussed in Section \ref{subsec:proposedmethods}. Moreover, the first-stage hyperparameters are $\bm{\theta}_1 = \big\{\sigma_{e_1},\rho_{\xi},\sigma_{\xi}\big\}$, where $\sigma^2_{\xi}$ and $\rho_{\xi}$ are the marginal variance and range parameter of the random field $\xi(\mathbf{s})$, respectively. The second-stage hyperparameter is $\bm{\theta}_2 = \big\{\sigma_{e_2}\big\}$. 

We used Gaussian priors for the fixed effects: $\beta_0 \sim \mathcal{N}\big(0,10^2\big), \beta_1 \sim \mathcal{N}\big(0,5^2\big), \gamma_0 \sim \mathcal{N}\big(0,10^2\big), \gamma_1 \sim \mathcal{N}\big(0,3^2\big)$; and penalized-complexity (PC) prior for $\sigma_{e_1}$ and $\sigma_{e_2}$ (\citet{fuglstad2019constructing, simpson2017penalising}). A PC prior for $\sigma_{e_1}$ and $\sigma_{e_2}$ penalizes deviation from the base model of zero variance. The formulation requires the user to specify a constant $\sigma_{\text{0}}$ and a probability value $\alpha$ such that $\mathbb{P}(\sigma_{e_i}>\sigma_{\text{0}})=\alpha, i=1,2$. This is equivalent to the prior $\sigma_{e_i}\sim\text{Exp}\Big(\lambda=-(\ln\alpha)/\sigma_{\text{0}}\Big)$, where the rate parameter $\lambda$ determines the magnitude of the penalty, with higher values corresponding to higher penalty. In particular, the PC priors for $\sigma_{e_1}$ and $\sigma_{e_2}$ are specified as $\mathbb{P}(\sigma_{e_1} > 1)=0.5$ and $ 
    \mathbb{P}(\sigma_{e_2} > 1)=0.5$. For the Mat\'ern parameters, we used a joint normal prior for $\log(\tau)$ and $\log(\kappa)$ (\citet{lindgren2015bayesian}), where 
\begin{align} \label{eq:jointnormalpriorA}
    \log(\tau) &= \dfrac{1}{2}\log\Big( \dfrac{1}{4\pi}\Big)-\log(\sigma_{\xi}) - \log(\kappa) \\
     \label{eq:jointnormalpriorB}
   \log(\kappa) &= \dfrac{\log(8)}{2} - \log(\rho_{\xi}).
\end{align}
In particular, we have $\begin{bmatrix}
    \log(\tau) \\ \log(\kappa) 
\end{bmatrix} \sim \mathcal{N}\Bigg(\begin{bmatrix}
    -0.7547-\log(\kappa) \\ 1.0397
\end{bmatrix}, \begin{bmatrix}
    20.67 & 0 \\ 0 & 8.67
\end{bmatrix}\Bigg)$. This prior specification implies that the plausible values for the Mat\'ern parameters are $0.3 \leq \sigma_{\xi} \leq 0.8$ and $0.6\leq\rho_{\xi}\leq1.5$. 

 \begin{figure}[t]
    \centering
    \subfigure[Full $\mathbf{Q}$ mesh]{\includegraphics[trim={3cm 2cm 3 3},clip, width=.28\linewidth]{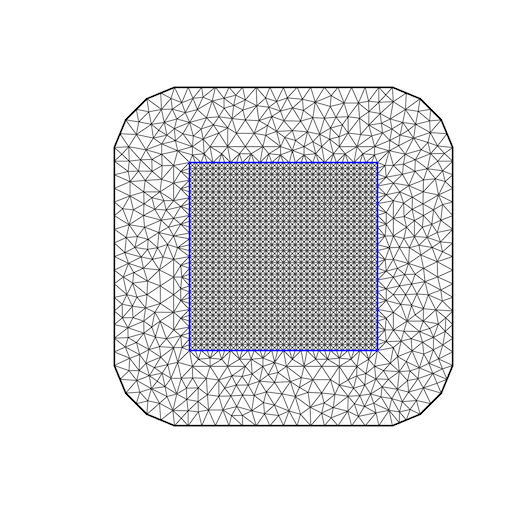}} \hspace{5mm}
   \subfigure[Low rank  $\mathbf{Q}$ mesh A]{\includegraphics[trim={3cm 2cm 3 3},clip, width=.28\linewidth]{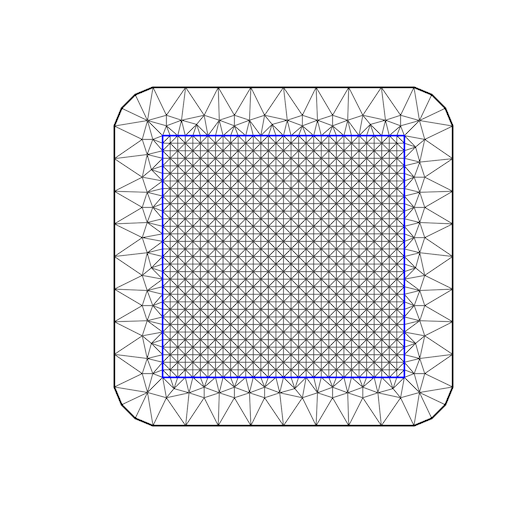}}\hspace{5mm}
   \subfigure[Low rank  $\mathbf{Q}$ mesh B]{\includegraphics[trim={3cm 2cm 3 3},clip, width=.28\linewidth]{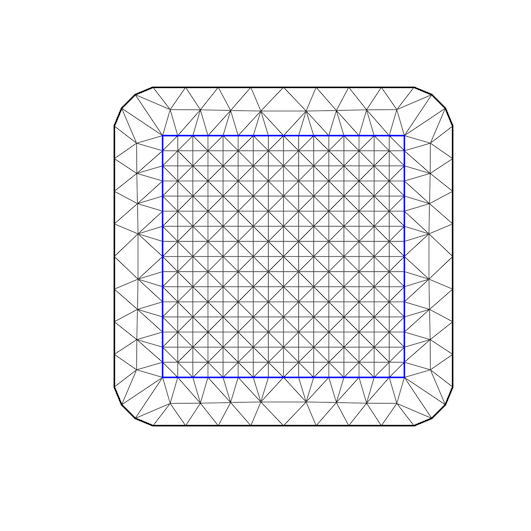}}
    \caption{Meshes used for the simulation experiments: (a) mesh for the full $\mathbf{Q}$ (b) slightly coarser mesh for the low rank $\mathbf{Q}$ method (c) very coarse mesh for the low rank $\mathbf{Q}$ method}
\label{fig:res_twostage_spatial_Gaussian_meshes}
\end{figure}
    
In doing the SBC, we fixed the spatial locations of $\text{w}(\mathbf{s}_i)$ and $\text{y}(\mathbf{s}_j)$ for all the data replicates which is shown in Figure \ref{fig:res_twostage_gaussian_simdata}a. The covariate field $z(\mathbf{s})$ was simulated from a Mat\'ern process with range of 0.6, standard deviation of 2, and mean-squared differentiability parameter of 1. This is fixed for all data replicates since $z(\mathbf{s})$ is a known quantity in the model. The random field $\xi(\mathbf{s})$ was also simulated from a Mat\'ern process, with range $\rho_{\xi}=4$ and marginal standard deviation $\sigma_{\xi}=0.6$. It varies for the different data replicates since this is an unknown quantity which needs to be estimated. 

\begin{figure}[t]
    \centering
    \subfigure[Plug-in method]{\includegraphics[width=.32\linewidth]{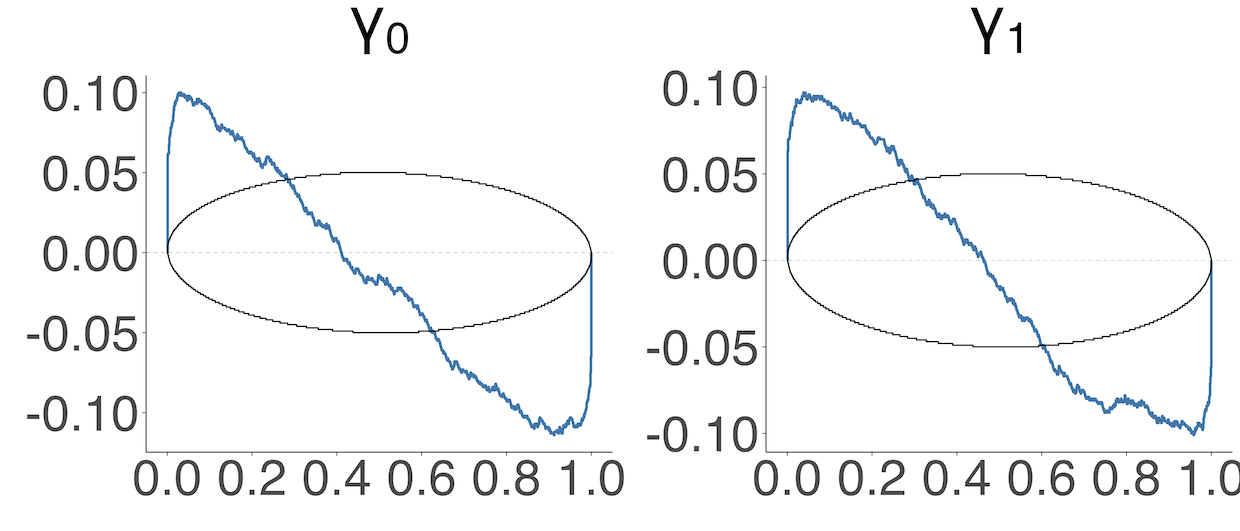}}     
    \subfigure[Resampling method]{\includegraphics[width=.32\linewidth]{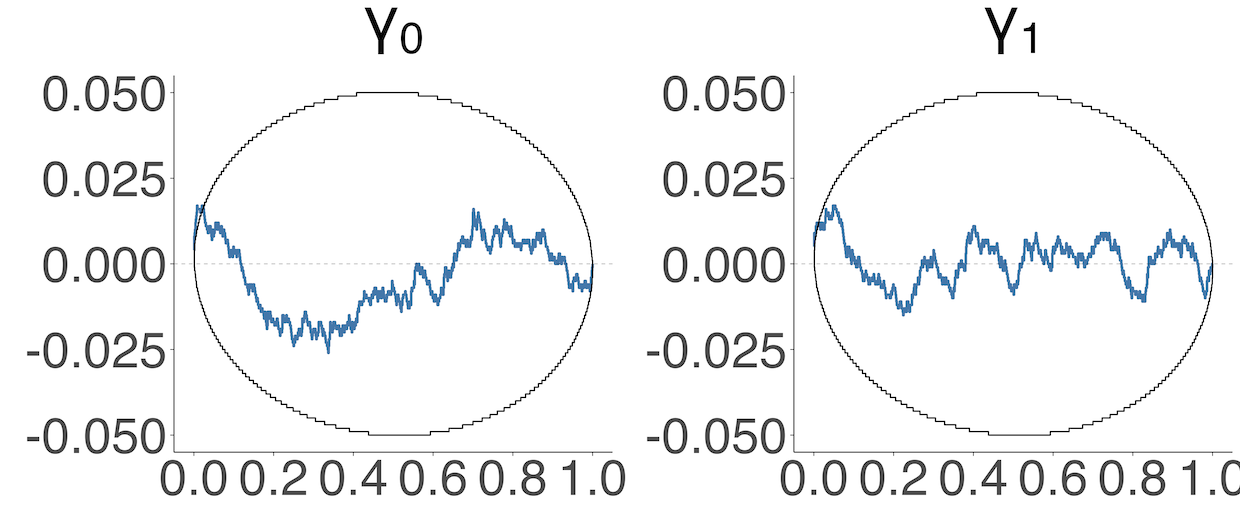}}  
    
    \subfigure[Full $\mathbf{Q}$ uncertainty method]{\includegraphics[width=.32\linewidth]{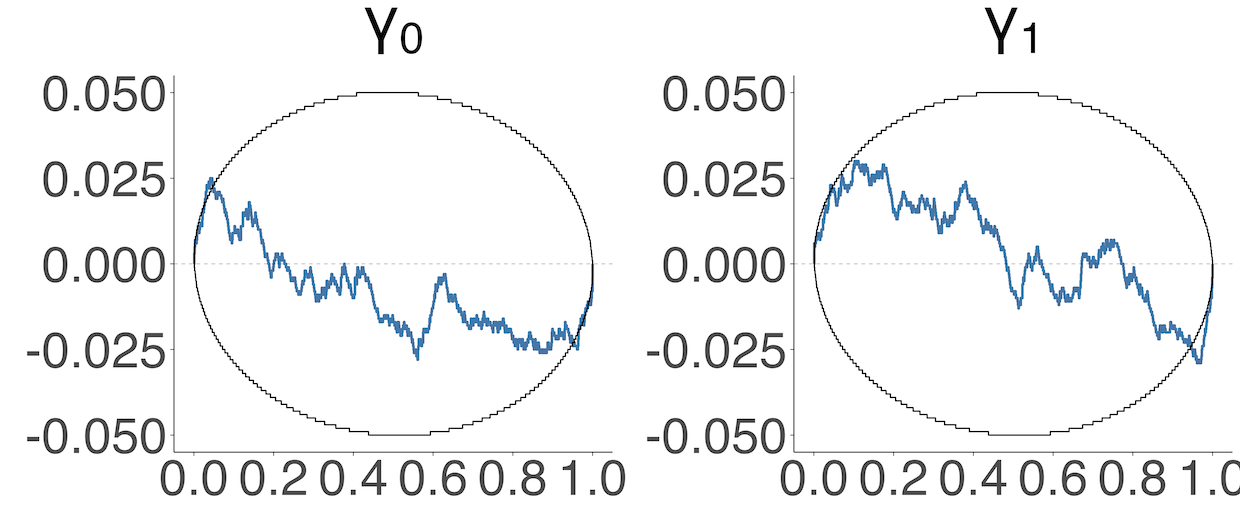}}
    \subfigure[Low rank $\mathbf{Q}$ (mesh A)]{\includegraphics[width=.32\linewidth]{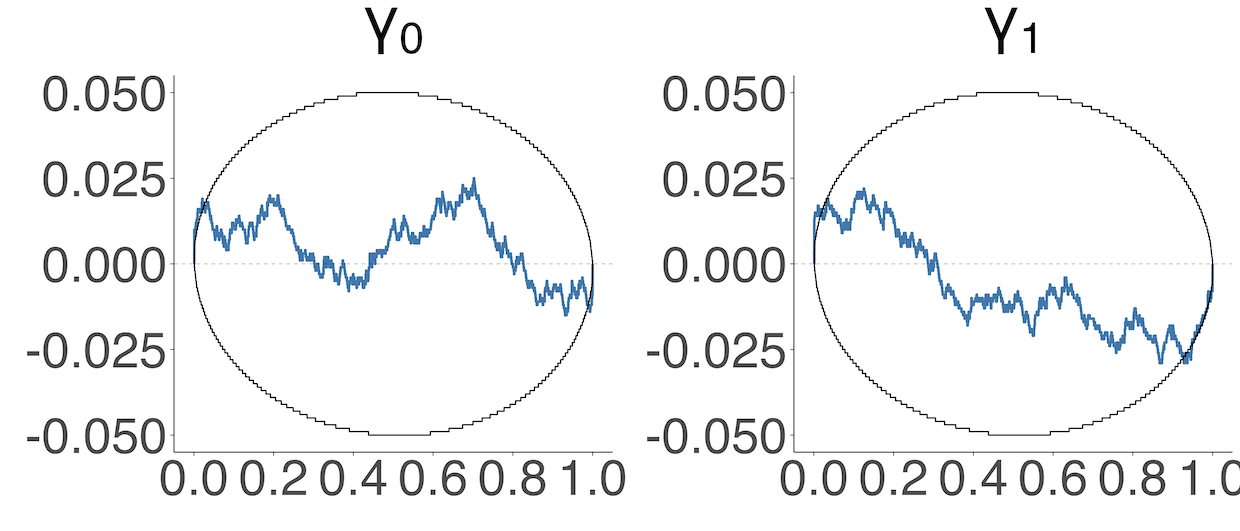}}
    \subfigure[Low rank $\mathbf{Q}$ (mesh B)]{\includegraphics[width=.32\linewidth]{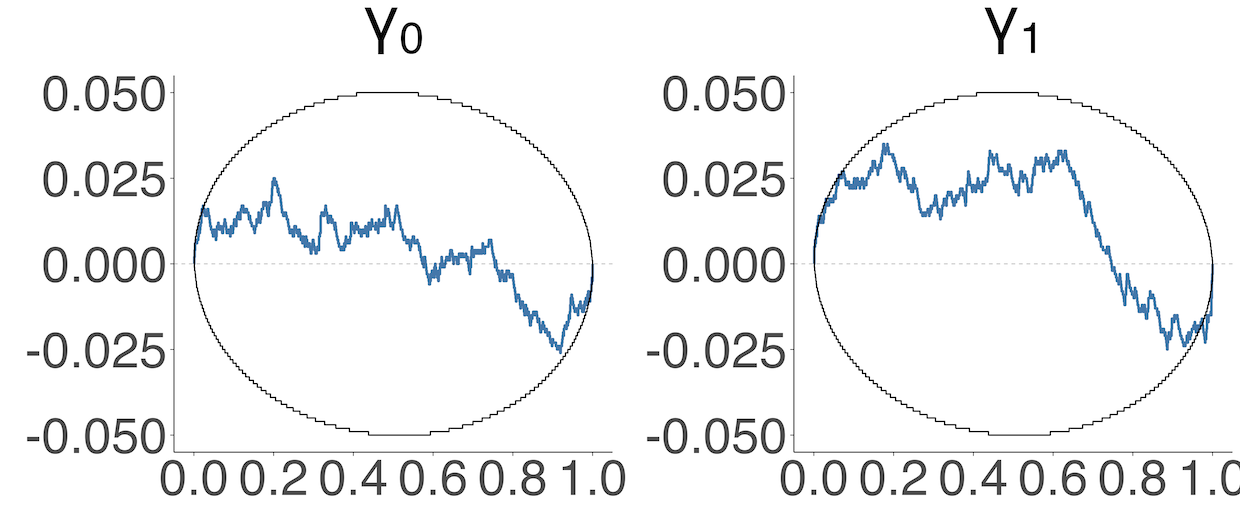}}
    \caption{ECDF difference plot of $p_k$ for $\gamma_0$ and $\gamma_1$ using Algorithm \ref{alg:sbc} out of 1000 data replicates for the two-stage Gaussian spatial model (Section \ref{subsec:twostage_spatial_Gaussian}) using different approaches: (a) plug-in method (b) resampling method (c) full $\mathbf{Q}$ method (d) low rank $\mathbf{Q}$ (mesh A) (e) low rank $\mathbf{Q}$ (mesh B)}
\label{fig:res_twostage_spatial_Gaussian_p_k}
\end{figure}

We compare four uncertainty propagation approaches: plug-in, resampling (with $J=30$ resamples), full $\mathbf{Q}$, and low rank $\mathbf{Q}$ uncertainty. The mesh used to fit the full $\mathbf{Q}$ uncertainty method is shown in Figure \ref{fig:res_twostage_spatial_Gaussian_meshes}a. Here, the maximum triangle edge lengths for the inner domain and the outer extension are 0.04 and 0.1 units, respectively. We use the same mesh to simulate and estimate $\xi(\mathbf{s})$. For the low rank $\mathbf{Q}$ approach, we explored two meshes (A and B) with different level of coarseness: for mesh A (Figure \ref{fig:res_twostage_spatial_Gaussian_meshes}b), the maximum edge lengths are 0.05 and 0.2 units, while for the coarser mesh B (Figure \ref{fig:res_twostage_spatial_Gaussian_meshes}c) (low rank $\mathbf{Q}$ mesh B), they are
0.1 and 0.25, respectively. The number of nodes is $2672$ for the full $\mathbf{Q}$ mesh,  $1228$ for mesh A, and $365$ for mesh B.

 Figure \ref{fig:res_twostage_spatial_Gaussian_p_k} shows the ECDF difference plot of the normalized ranks $p_k$ for $\gamma_0$ and $\gamma_1$ from 1000 data replicates using Algorithm \ref{alg:sbc} (corresponding histograms are  in Figure 7 of the Supplementary Material). Results show that for the plug-in method, the hypotheses of uniform distribution of the ranks $p_k$ is rejected for both $\gamma_0$ and $\gamma_1$ (Figure \ref{fig:res_twostage_spatial_Gaussian_p_k}a). In addition, the $\cup$-shaped histogram (Supplementary Material) reveals an underestimation of the posterior variance. The resampling method and the two proposed methods do not show deviations from uniformity, not even with the low rank $\mathbf{Q}$ approach using the  coarser mesh B. This suggests that both  the resampling approach and the proposed methods correctly capture the posterior uncertainty of $\gamma_0$ and $\gamma_1$. 
 
 Although the primary focus is on these parameters,  we also examined the histogram and ECDF difference plot of the normalized ranks $p_k$ for all first-stage model parameters. Section 2.1 in the Supplementary Material presents the SBC results for the first-stage model using  Algorithm \ref{alg:sbc}. The results show a uniform distribution of $p_k$ for all first-stage model hyperparameters, except for $\sigma_{\xi}$ which is slightly $\cap$-shaped. This motivates the use of Algorithm \ref{alg:sbcconditional},  based on Theorem \ref{thm:SBChierarhicalmodels1}, where SBC is applied conditional on $\bm{\theta}_{1,\bm{x}}=\{\sigma_{\xi},\rho_{\xi}\}$. Moreover, three mesh nodes fail the Kolmogorov-Smirnov test for uniformity at 10\% significance level (see Figure 1 in Section 2.1 of the Supplementary Material).

The results from  Algorithm \ref{alg:sbcconditional} are shown in Sections 2.2 and 2.4 in the Supplementary Material. The conclusions are consistent with those from Algorithm \ref{alg:sbc}, i.e., the plug-in method 
underestimates the posterior uncertainty of $\gamma_0$ and $\gamma_1$, while the resampling method and the proposed methods are also correct, althought there may be slight underestimation with the low rank $\mathbf{Q}$ approach. Finally, we also attempted to perform the SBC on a non-spatial two-stage model, i.e., similar to Equations \eqref{eq:sec4.1firststage} and \eqref{eq:sec4.1secondstage} but without the spatial field $\xi(\mathbf{s})$. We used both the INLA and no U-turn sampler (NUTS) (\citet{hoffman2014no}). The results, which are given in Section 5.1 of the Supplementary Material, also show that the plug-in method underestimates the posterior uncertainty of both $\gamma_0$ and $\gamma_1$, while the resampling and the $\mathbf{Q}$ methods are correct. Note that the $\mathbf{Q}$ method is implemented only using the INLA method.

\subsubsection{Illustration with simulated data}

\begin{figure}[t]
    \centering
    \subfigure[Posterior mean of $\mathbb{E}(\text{y}(\mathbf{s})|\mu(\mathbf{s}))$]{\includegraphics[width=.46\linewidth]{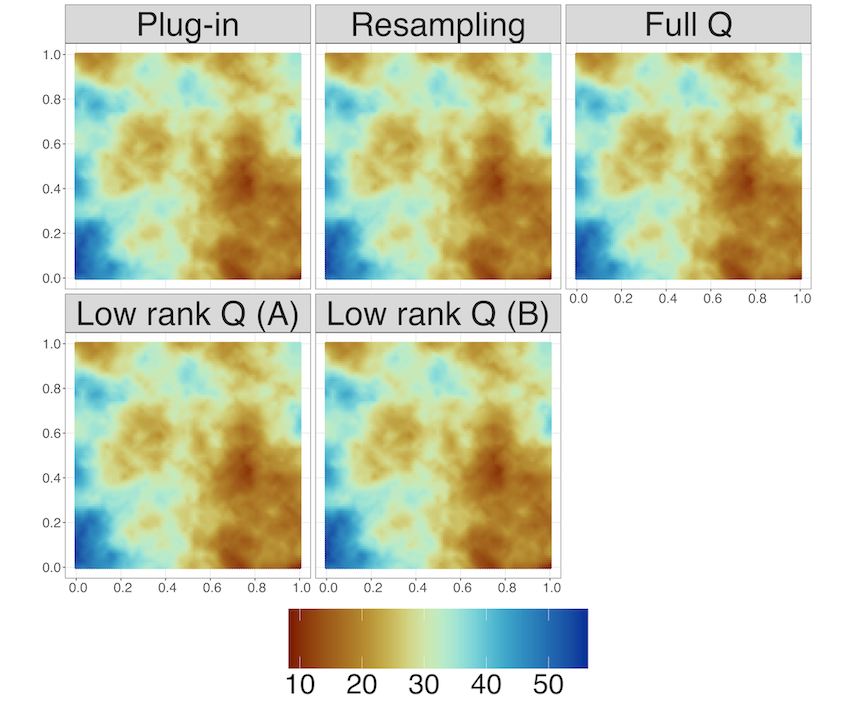}} 
   \subfigure[Posterior SD of $\mathbb{E}(\text{y}(\mathbf{s})|\mu(\mathbf{s}))$]{\includegraphics[width=.46\linewidth]{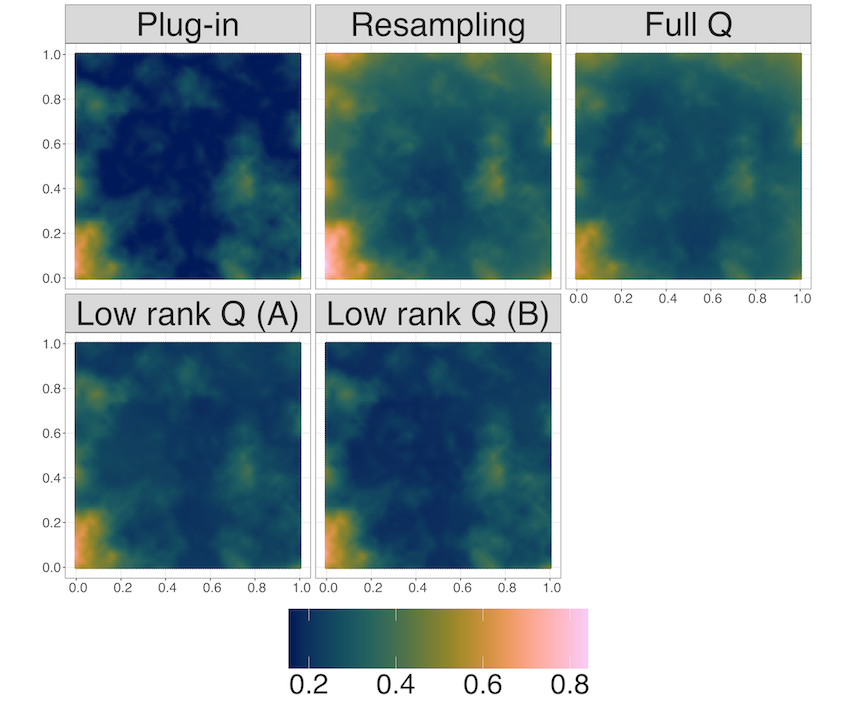}}
    \caption{Comparison of the posterior mean and posterior SD of $\mathbb{E}\big[\text{y}(\mathbf{s})|\mu(\mathbf{s})\big]=\gamma_0+\gamma_1\mu(\mathbf{s})$ for the two-stage Gaussian model in Section \ref{subsec:twostage_spatial_Gaussian} from different approaches: plug-in method, resampling method, full $\mathbf{Q}$ method, low rank $\mathbf{Q}$ (mesh A), low rank $\mathbf{Q}$ (mesh B)}
\label{fig:res_spatial_twostage_gaussian_predfield_s2}
\end{figure}
To gain additional insights, we analyze in detail one simulated data from the previous section with true values of the parameters as follows: $\beta_0=10, \beta_1=3, \gamma_0=10, \gamma_1=1.5,\sigma_{e_1}^2=1,\sigma_{e_2}^2=1, \sigma_{\xi}=4$, and  $\rho_{\xi}=0.6$. 
Figure \ref{fig:res_twostage_gaussian_simdata} shows: (a) the spatial locations for the data $\text{w}(\mathbf{s}_i)$ and $\text{y}(\mathbf{s}_j)$, (b) the simulated field $\mu(\mathbf{s})$, and (c) the simulated field $\mathbb{E}\big[\text{y}(\mathbf{s})|\mu(\mathbf{s})\big]$ which we estimate using the different uncertainty propagation approaches. 
The posterior means of $\mathbb{E}\big[\text{y}(\mathbf{s})|\mu(\mathbf{s})\big]$ (Figure \ref{fig:res_spatial_twostage_gaussian_predfield_s2}a) are all very similar and close to the truth (Figure \ref{fig:res_twostage_gaussian_simdata}c). Posterior standard deviations (Figures \ref{fig:res_spatial_twostage_gaussian_predfield_s2}b) are smallest, as expected, for the plug-in method. The resampling method resulted in the highest overall uncertainty, while the full $\mathbf{Q}$ uncertainty method produced posterior uncertainties nearly identical to those of the resampling method. The posterior uncertainty from the low-rank $\mathbf{Q}$ method is lower than that of the full $\mathbf{Q}$ method, but higher than the uncertainty from the plug-in method.

\begin{figure}[t]
    \centering    \includegraphics[scale=.2]{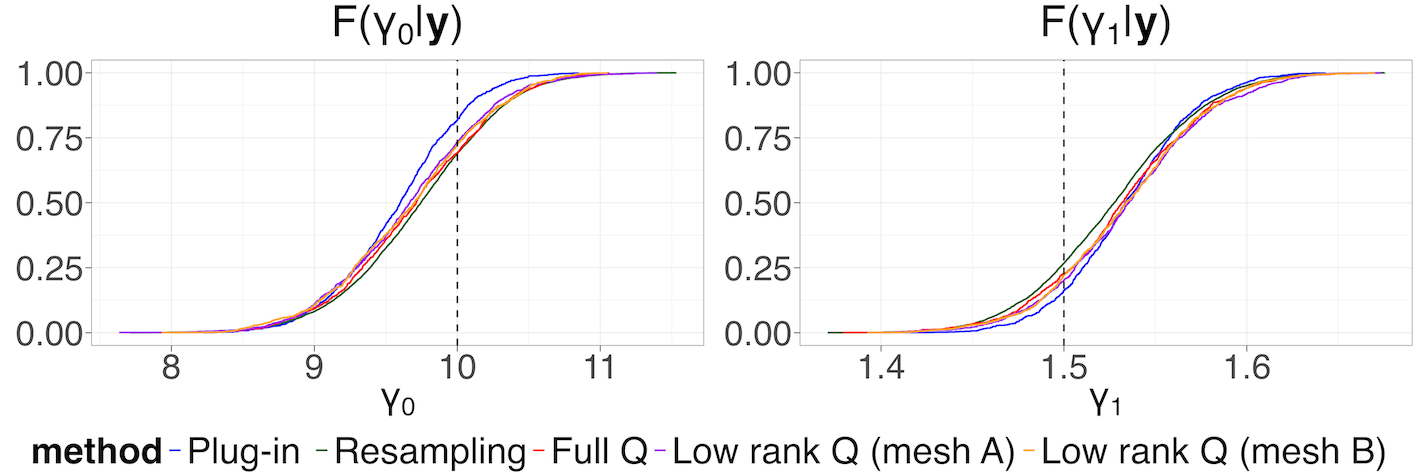}
    \caption{Estimated marginal posterior CDFs of $\gamma_0$ and $\gamma_1$ for a simulated dataset from the two-stage Gaussian model in Section \ref{subsec:twostage_spatial_Gaussian} using four methods of uncertainty propagation: plug-in, resampling, full $\mathbf{Q}$ method, low rank $\mathbf{Q}$ (mesh A), and low rank $\mathbf{Q}$ (mesh B)}
    \label{fig:res_spatial_gaussian_twostages2params_cdf}
\end{figure}

Figure \ref{fig:res_spatial_gaussian_twostages2params_cdf} shows the marginal posterior CDFs of $\gamma_0$ and $\gamma_1$ from the same simulated data. The plug-in method has the smallest posterior uncertainty for both parameters, but the difference is more apparent for $\gamma_0$. The resampling method has the highest posterior uncertainty, while the proposed methods provide a middle ground between the plug-in and resampling method. The posterior estimates from the full $\mathbf{Q}$ method and the low rank $\mathbf{Q}$ method are very similar.

\begin{figure}[t]
    \centering
    \subfigure[Comparison with high log precision values]{\includegraphics[width=.45\linewidth]{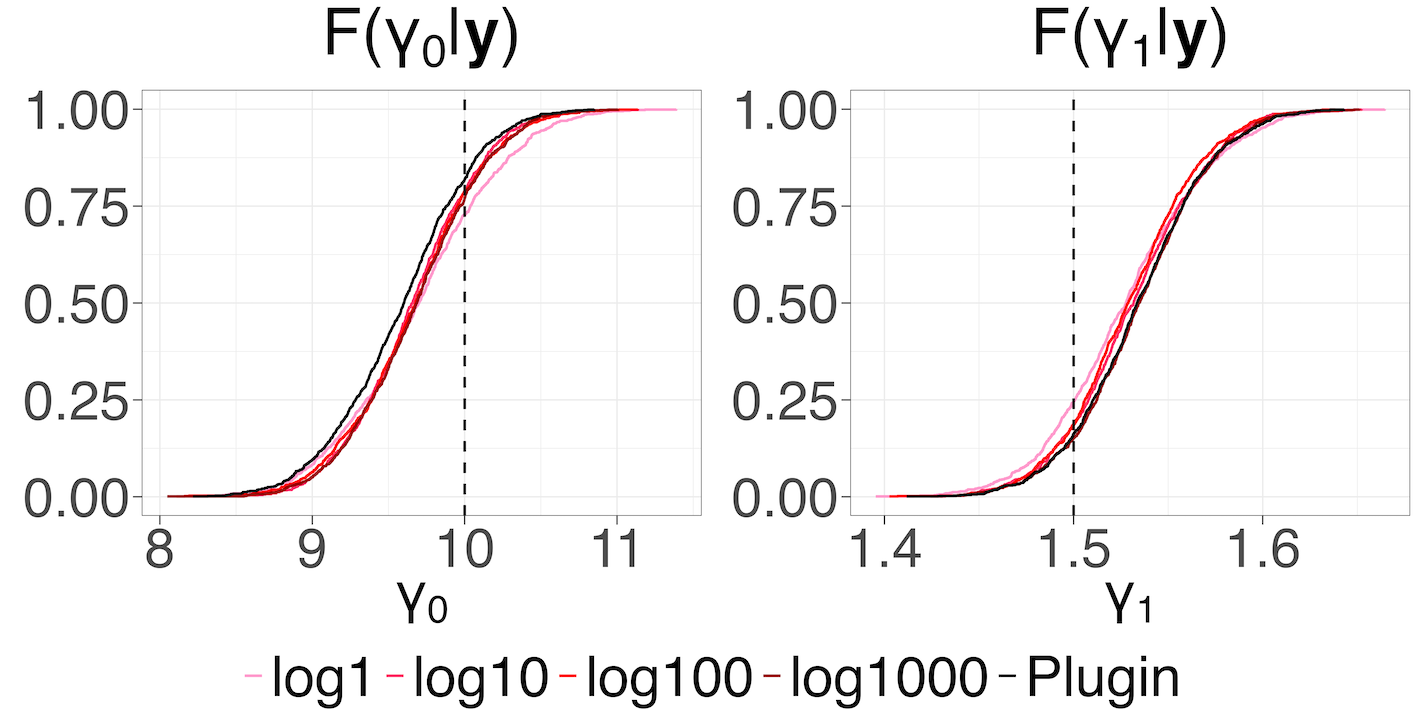}} \hspace{5mm}
    \subfigure[Comparison with low log precision values]{\includegraphics[width=.45\linewidth]{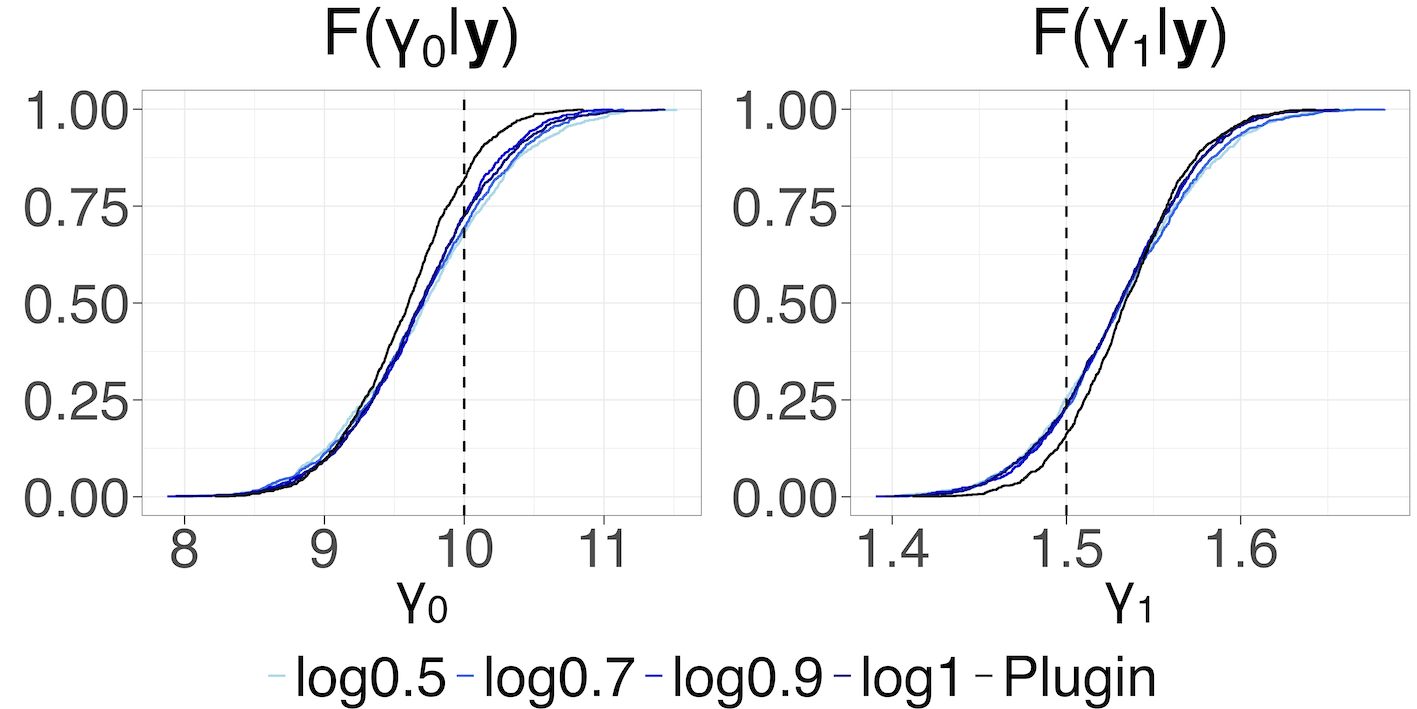}} 
    \caption{Comparison of the estimated marginal posterior CDFs of $\gamma_0$ and $\gamma_1$ for different fixed values of the log precision of the error component with the full $\mathbf{Q}$ uncertainty method using the simulated data example in Section \ref{subsec:twostage_spatial_Gaussian}}
\label{fig:res_spatial_gaussian_twostage_ecdfs_varylogprec_fullQ}
\end{figure}

Figures \ref{fig:res_spatial_gaussian_twostage_ecdfs_varylogprec_fullQ}a and \ref{fig:res_spatial_gaussian_twostage_ecdfs_varylogprec_fullQ}b show a comparison of the estimated posterior CDFs of $\gamma_0$ and $\gamma_1$, respectively, for different fixed values of the scaling parameter $\log(\tau_{\epsilon})$ of the error component using the full $\mathbf{Q}$ method, as discussed in Section \ref{subsec:proposedmethods}. The results show that as the log precision becomes larger, the estimated CDFs for $\gamma_0$ and $\gamma_1$ approaches the estimated CDFs of the plug-in method. Also, as the log precision value becomes smaller, the estimated CDFs deviate more from the estimated CDFs of the plug-in method, i.e., the posterior uncertainty becomes larger. The same insights are true from the results of the low rank $\mathbf{Q}$ method (see Section 2.5 of the Supplementary Material).


In terms of computational time, the crude plug-in method took 2.98 seconds to fit the second-stage model, while the full $\mathbf{Q}$ method took 9.82 seconds. The low rank $\mathbf{Q}$ approach took 13.97 seconds with mesh A, and 5.80 seconds with mesh B. This suggests that using a coarse mesh with the low rank $\mathbf{Q}$ does not always lead to reduced computational time. 
A plausible reason for this is that the linear predictor of the low rank $\mathbf{Q}$ method, as shown in Equations \eqref{eq:lowrankQuncertainty} and \eqref{eq:lowrankQuncertainty2}, is more complex and involves additional operations compared to the linear predictor of the full $\mathbf{Q}$ method in Equation \eqref{eq:fullQuncertainty}. We have the computational advantage of the low rank $\mathbf{Q}$ approach from using a coarse enough mesh, similar to mesh B in Figure \ref{fig:res_twostage_spatial_Gaussian_meshes}c. 
Lastly, the resampling method with $J = 30$, took 36.12 seconds using parallel computing with the \texttt{mclapply()} function in \texttt{R}.

The results from the SBC provide evidence that the  plug-in method underestimates the posterior uncertainty, while the resampling method is correct. The full $\mathbf{Q}$ and the low rank $\mathbf{Q}$ methods are also expected to give correct posteriors for $\gamma_0$ and $\gamma_1$. The computational benefits from the low rank $\mathbf{Q}$ method potentially depends on the coarseness of the mesh for the error component.

\subsection{A two-stage spatial model with Poisson observations}\label{subsec:spatial_twostage_poisson}

In this section, we perform the SBC in a two-stage spatial model with Poisson observations. We consider two model specifications: the first one, called the \textit{classical specification}, is similar to Equations \eqref{eq:motivatingeq3} and \eqref{eq:motivatingeq4} in Section \ref{sec:intro}. Here, the second-stage model specifies the log mean of the Poisson counts in each block as linear with respect to the block averages of $\mu(\mathbf{s})$. This approach is often used in such spatial misalignment problems since it is straightforward to implement (\citet{blangiardo2016two,lee2017rigorous,lee2021impact,liu2017incorporating, villejo2023data,cameletti2019bayesian}, and \citet{zhu2003hierarchical}). The second, named \textit{new specification}, introduces a spatially continuous latent intensity field $\lambda(\mathbf{s})$ for the Poisson counts, which is then linked to the log mean in a nonlinear way (\citet{lindgren2024inlabru}). This approach better represents the physical process by assuming that the observed Poisson counts are function of the averages of a latent intensity field over the areas. Although this results in a highly nonlinear model, it can be efficiently fitted using an approximate iterative method with INLA. This method extends the applicability of INLA beyond the linear predictor framework to accommodate more complex functional relationships and can be implemented with the \texttt{inlabru} library in \texttt{R} (\citet{lindgren2024inlabru}).



\subsubsection{Classical specification}\label{subsubsec:spatial_twostage_poisson_classical}

The first-stage latent model is similar to the one in Section \ref{subsec:twostage_spatial_Gaussian}, so that $\mu(\mathbf{s}) = \beta_0 + \beta_1 z(\mathbf{s})+\xi(\mathbf{s})$ and the observed data $\bm{\mathcal{D}}_1=\{\text{w}(\mathbf{s}_i), i=1,\ldots,n_{\text{w}}$\} also follows the classical error model. The latent process $\mu(\mathbf{s})$ is an input in the second-stage model as follows:
\begin{align*}
    &\text{y}(B) \sim \text{Poisson}\Big(\mu_{\text{y}}(B)\Big)  \label{eq:poisson} \;\;\;  \\
    & \mu_{\text{y}}(B)= \mathbb{E}[\text{y}(B)] = \text{E}(B)\times \lambda(B)\\
    &\log\Big(\lambda(B)\Big) =\gamma_0 + \gamma_1 \dfrac{1}{|B|}\int_B \mu(\mathbf{s}) d\mathbf{s} 
\end{align*}

The above model is closely related to the joint model in Equations \eqref{eq:motivatingeq1} -- \eqref{eq:motivatingeq4}, but here we have additional quantities $\text{E}(B)$ which are introduced as an \textit{offset} in order to account for the different sizes of the blocks $B$. For example, in spatial epidemiology where $\text{y}(B)$ is the observed disease count, $\text{E}(B)$ is the expected number of cases and are computed using the size and demographic structure of the population in block $B$ (\citet{lee2011comparison}). In this specification, $\lambda(B)$ is interpreted as the disease rate or risk (\citet{lee2017rigorous,blangiardo2016two}). We used the INLA-SPDE approach to fit the model. The second-stage data is $\bm{\mathcal{D}}_2=\{\text{y}(B),\; \forall B\}$. Similar to Section \ref{subsec:twostage_spatial_Gaussian}, the first-stage and second-stage model latent parameters are $\bm{x}_1=\{\beta_0,\beta_1,\omega_1,\omega_2,\ldots,\omega_K\}$ and $\bm{x}_2=\{\gamma_0,\gamma_1\}$, respectively. The first-stage model hyperparameters are  $\bm{\theta}_1 = \{\sigma_{e_1},\sigma_{\xi},\rho_{\xi}\}$. There are no second-stage model hyperparameters.

We simulate $\xi(\mathbf{s})$ and $z(\mathbf{s})$  as in Section \ref{subsec:twostage_spatial_Gaussian}. The spatial locations of the first-stage observations and the meshes for the full $\mathbf{Q}$ and low rank $\mathbf{Q}$ are also as in Section \ref{subsec:twostage_spatial_Gaussian}.
The configuration of the Poisson blocks is shown in Figure \ref{fig:res_spatial_twostage_poisson_classical_illustration}d. 
We used the following priors for the fixed effects: $\beta_0 \sim \mathcal{N}(0,10^2), \beta_1 \sim \mathcal{N}(0,5^2), \gamma_0 \sim \mathcal{N}(-2,1.5^2), \gamma_1 \sim \mathcal{N}(0,0.1^2).$ We used the PC prior for $\sigma_{e_1}$, particularly $\mathbb{P}(\sigma_{e_1} > 1)=0.5$, and the same joint log Gaussian prior for the Mat\'ern parameters from Section \ref{subsec:twostage_spatial_Gaussian}. 
Again, we consider four methods of uncertainty propagation: plug-in method, resampling method, full $\mathbf{Q}$ method, and the low rank $\mathbf{Q}$ method. 

\begin{figure}[t]
    \centering
    \subfigure[Plug-in method]{\includegraphics[width=.32\linewidth]{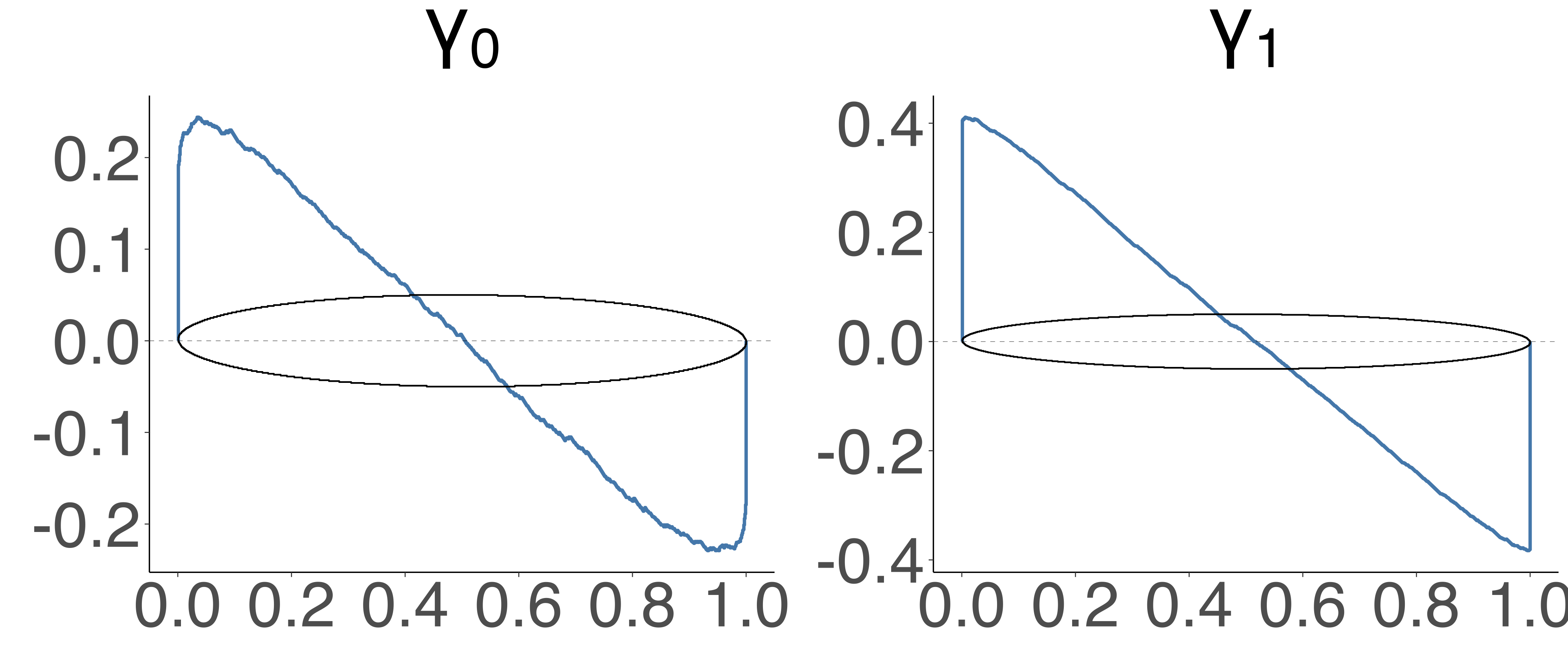}}     
    \subfigure[Resampling method]{\includegraphics[width=.32\linewidth]{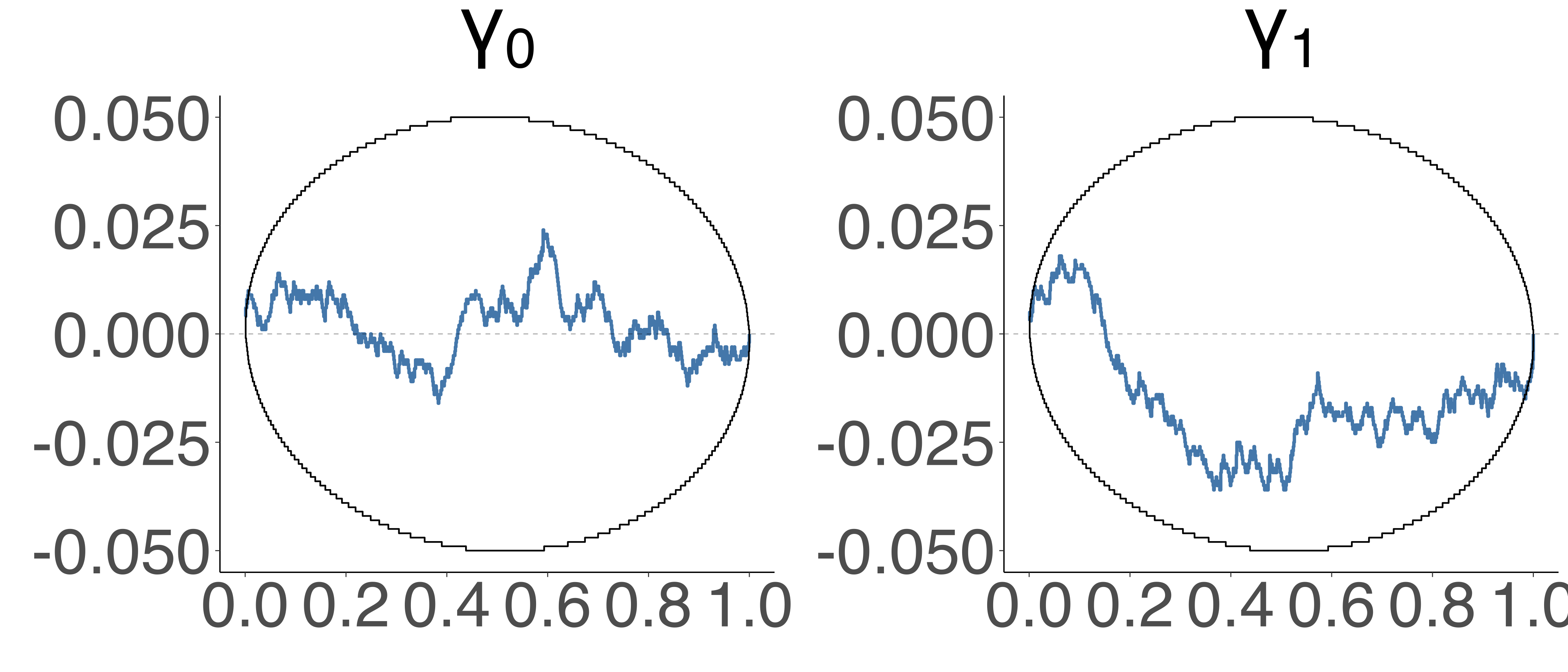}} 
    
    \subfigure[Full $\mathbf{Q}$ uncertainty method]{\includegraphics[width=.32\linewidth]{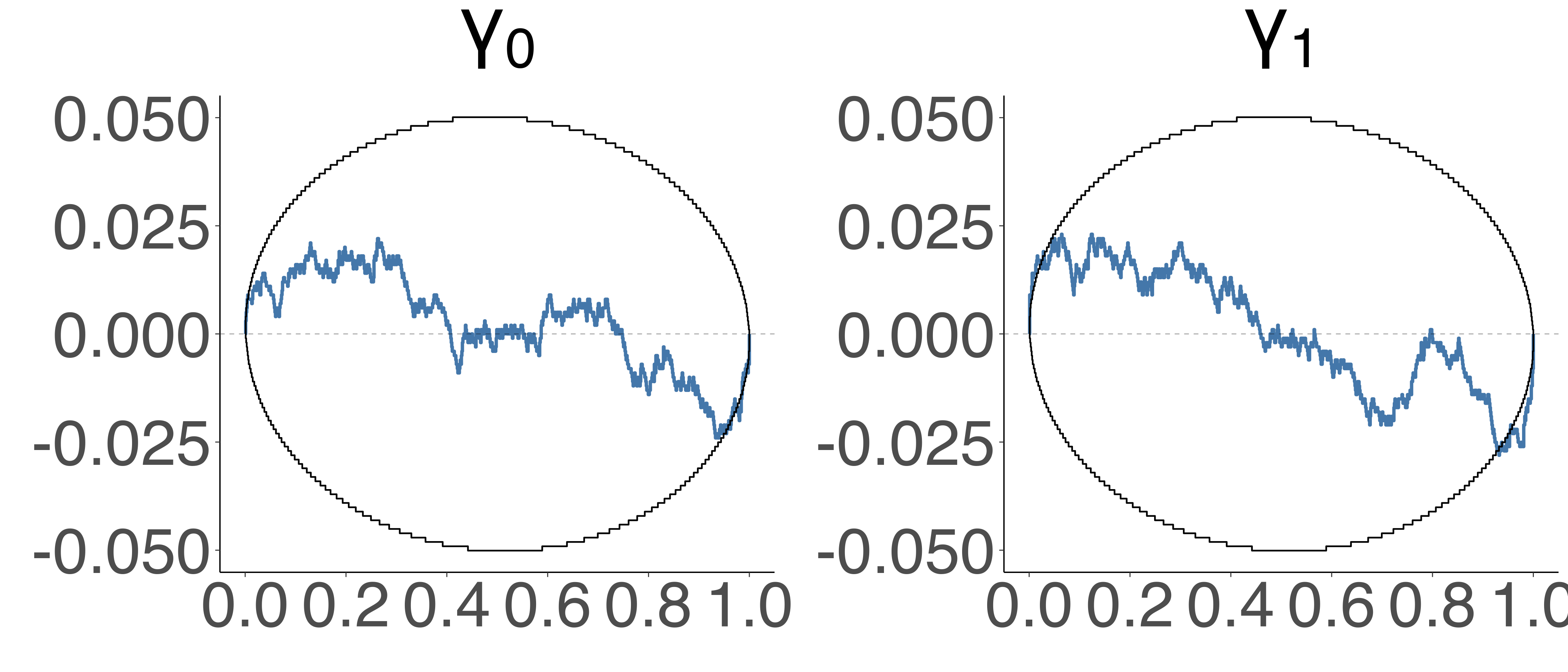}}
    \subfigure[Low rank $\mathbf{Q}$ (mesh A)]{\includegraphics[width=.32\linewidth]{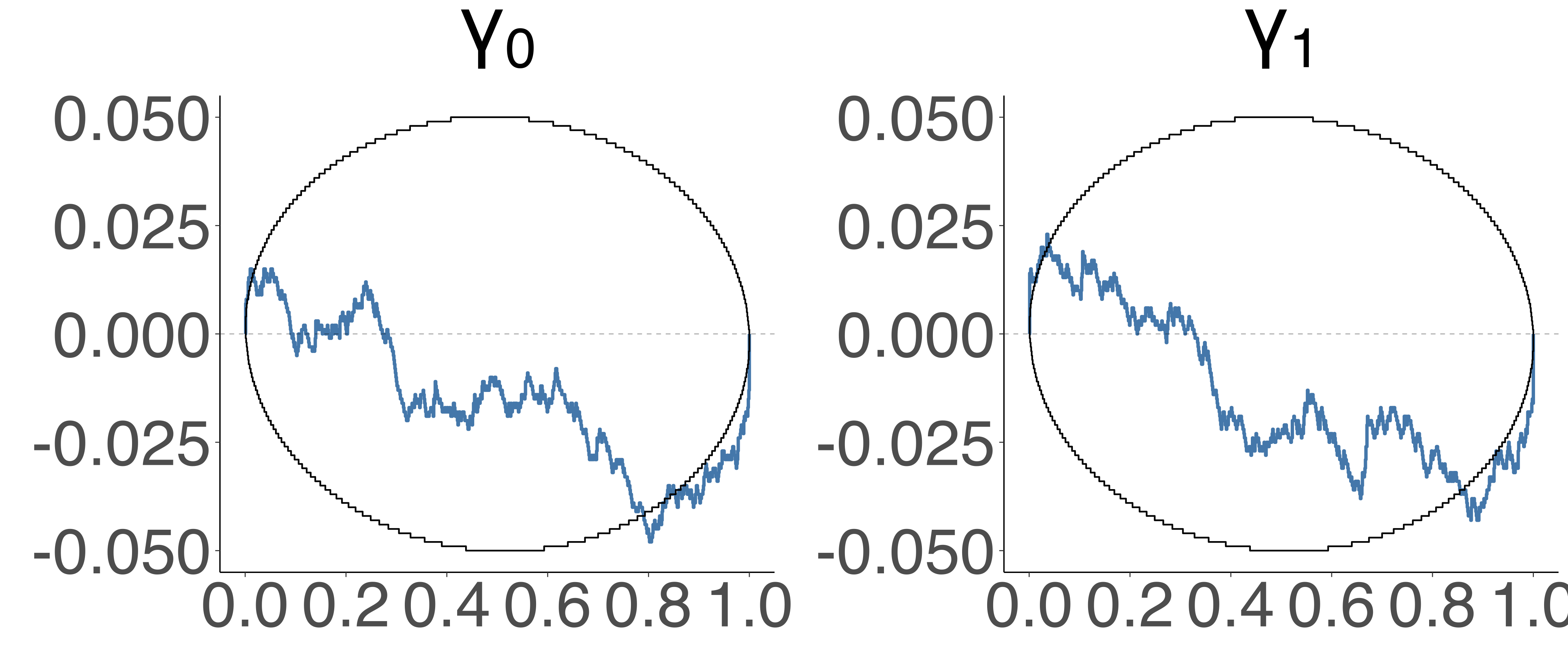}}
    \subfigure[Low rank $\mathbf{Q}$ (mesh B)]{\includegraphics[width=.32\linewidth]{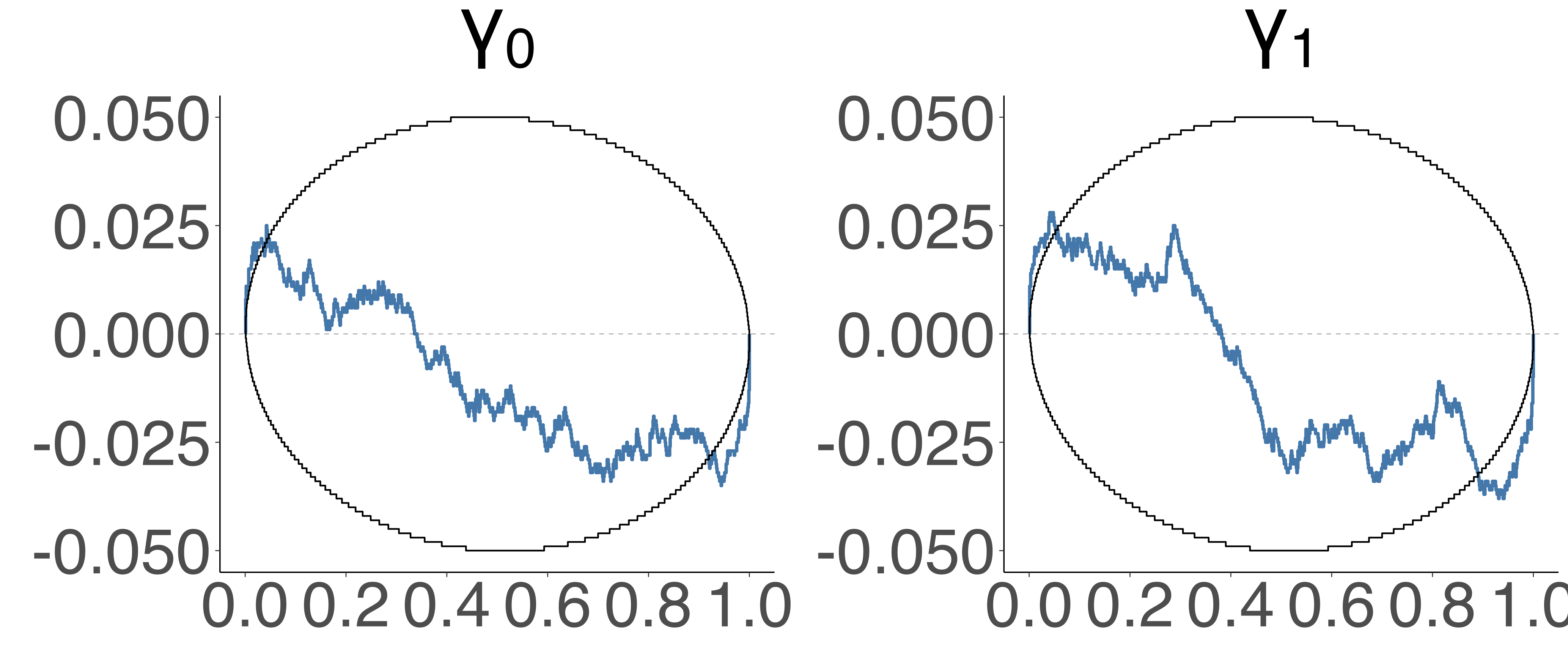}}
    \caption{ECDF difference plot of $p_k$ for $\gamma_0$ and $\gamma_1$ using Algorithm \ref{alg:sbc} out of 1000 data replicates for the classical specification of the two-stage Poisson spatial model (Section \ref{subsubsec:spatial_twostage_poisson_classical}) and using different approaches: (a) plug-in method (b) resampling method (c) full $\mathbf{Q}$  method (d) low rank $\mathbf{Q}$ (mesh A) method (e) low rank $\mathbf{Q}$ (mesh B) method} 
\label{fig:res_twostage_poisson_p_k_classical_ecdfdiff}
\end{figure}

Figure \ref{fig:res_twostage_poisson_p_k_classical_ecdfdiff}
shows the plot of the ECDF differences of $p_k$ for $\gamma_0$ and $\gamma_1$ using Algorithm \ref{alg:sbc} with 1000 data replicates (corresponding histograms are in Section 3.3 of the Supplementary Material). Again, the plug-in method appears to underestimate the true posterior uncertainty for both parameters, while the resampling and the full $\mathbf{Q}$ uncertainty methods do not show deviations from uniformity. The two versions of the low rank $\mathbf{Q}$ approach  (mesh A and mesh B) show slight deviation from uniformity, but not as bad as the plug-in method. Results for the first-stage model parameters using Algorithm \ref{alg:sbc} are shown in Section 3.1 of the Supplementary Material. As in Section \ref{subsec:twostage_spatial_Gaussian}, the histogram of the normalized ranks $p_k$ for $\sigma_{\xi}$ is $\cap$-shaped. Moreover, there are some mesh nodes which also fail the uniformity test using the KS test at 10\% significance level. 
Results using Algorithm \ref{alg:sbcconditional} are shown in Section 3.2 and Section 3.4 in the Supplementary material for the first-stage and second-stage model parameters, respectively. The results are coherent with those from Algorithm \ref{alg:sbc}.

Similar to Section \ref{subsec:twostage_spatial_Gaussian}, initial validation was done for a non-spatial two-stage model, i.e., without the spatial field $\xi(\mathbf{s})$ in the first stage. The results for both INLA and NUTS are in Section 5.2 of the Supplementary Material. The results are consistent with previous results that the plug-in method underestimates the posterior uncertainty of both $\gamma_0$ and $\gamma_1$, while the resampling and the proposed method are correct.

\subsubsection{New specification}\label{subsubsec:spatial_twostage_poisson_logsumexp}

For the new model specification, the second-stage model is as follows:
\begin{align*}
    &\text{y}(B) \sim \text{Poisson}\Big(\mu_{\text{y}}(B)\Big) \;\;\;  \\
    &\mu_{\text{y}}(B)= \mathbb{E}\big[\text{y}(B)\big] = \text{E}(B)\times \lambda(B)\\
    &\lambda(B) = \dfrac{1}{|B|} \int_B \lambda(\mathbf{s}) d\mathbf{s} = \dfrac{1}{|B|}\int_B  \exp\big\{\gamma_0 + \gamma_1\mu(\mathbf{s})\big\}d\mathbf{s}
\end{align*}

The first-stage model is the same as the one used for the classical specification; but here we assume that $\mu(\mathbf{s})$ is linked to another latent intensity field which we denote by $\lambda(\mathbf{s})$. We use the same covariate $z(\mathbf{s})$, the same spatial locations for the first-stage data, and the same configuration of the block/areas for the Poison outcomes as the classical specification. We also use the same priors for all model parameters, and compare the same uncertainty propagation methods. 
 
\begin{figure}[t]
    \centering
    \subfigure[Plug-in method]{\includegraphics[width=.32\linewidth]{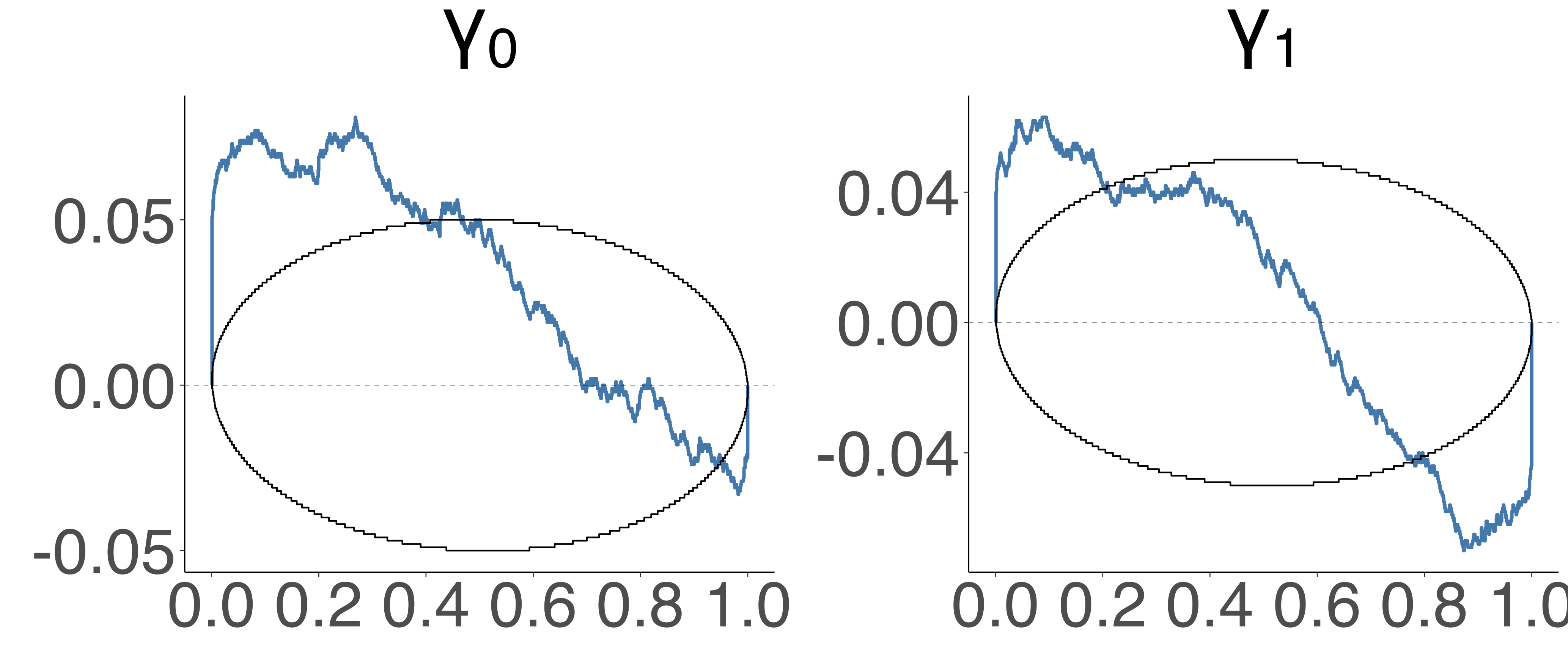}}     
    \subfigure[Resampling method]{\includegraphics[width=.32\linewidth]{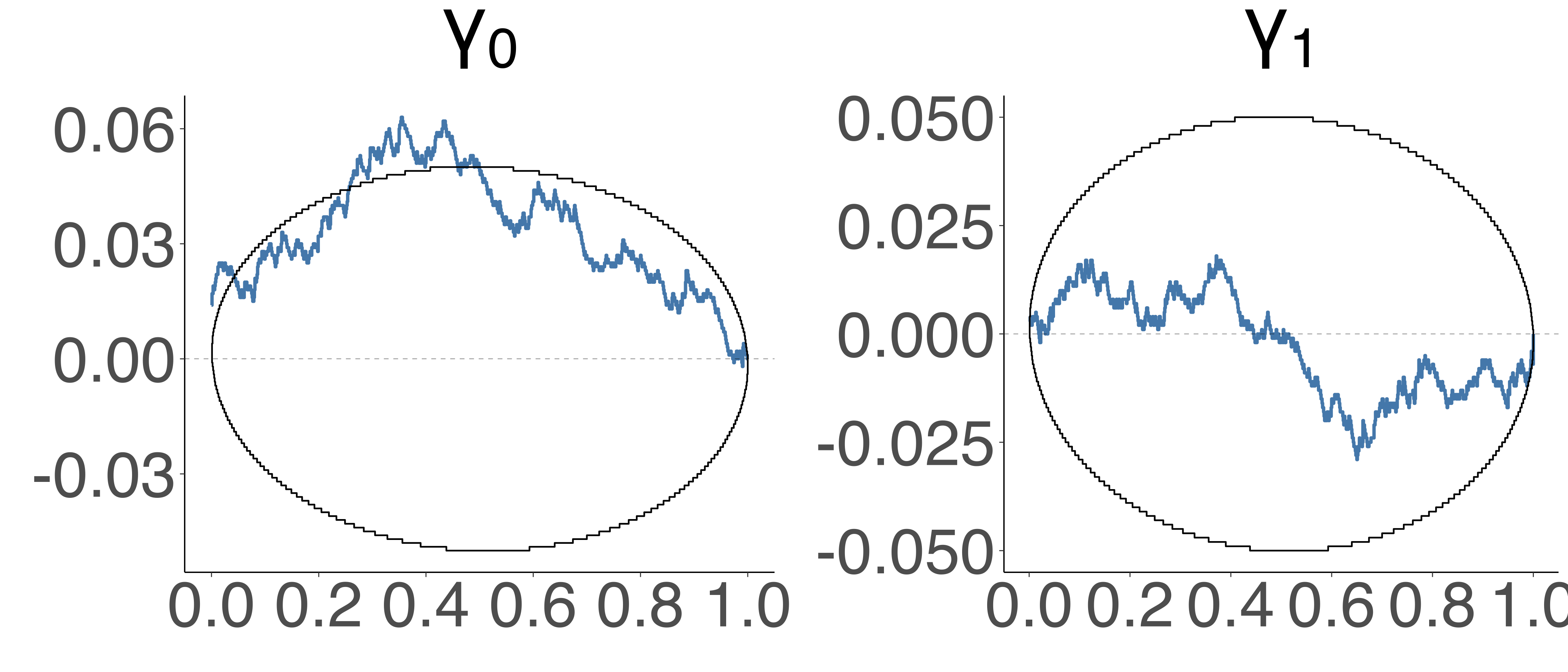}} 
    
    \subfigure[Full $\mathbf{Q}$ uncertainty method]{\includegraphics[width=.32\linewidth]{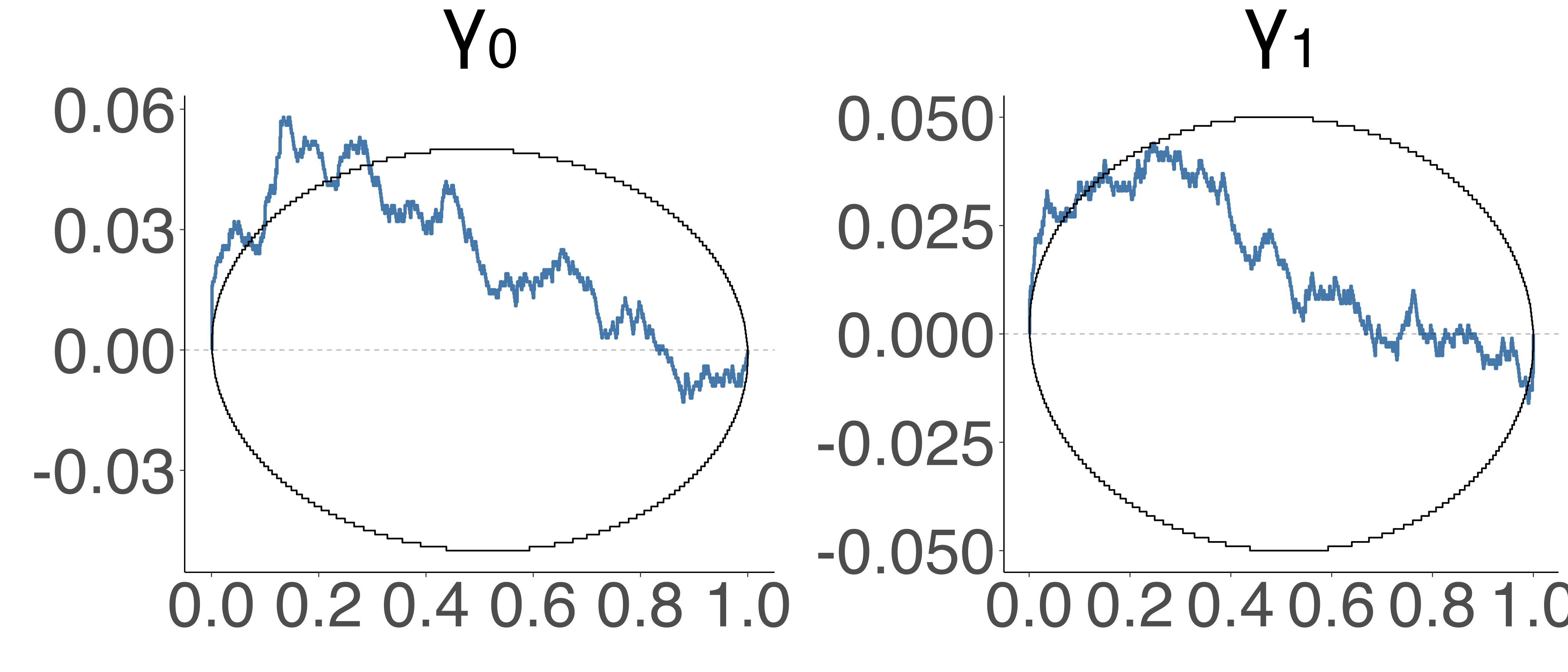}}
    \subfigure[Low rank $\mathbf{Q}$ (mesh A)]{\includegraphics[width=.32\linewidth]{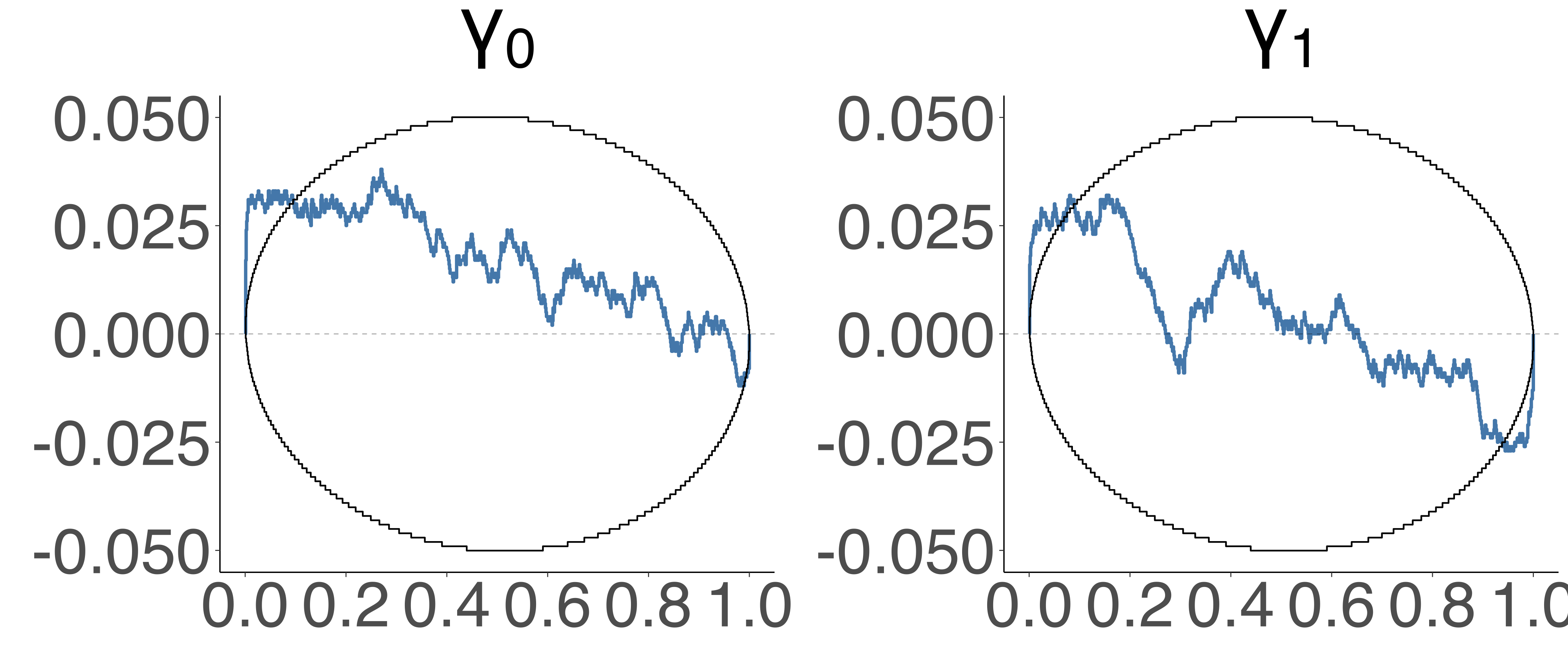}}
    \subfigure[Low rank $\mathbf{Q}$ (mesh B)]{\includegraphics[width=.32\linewidth]{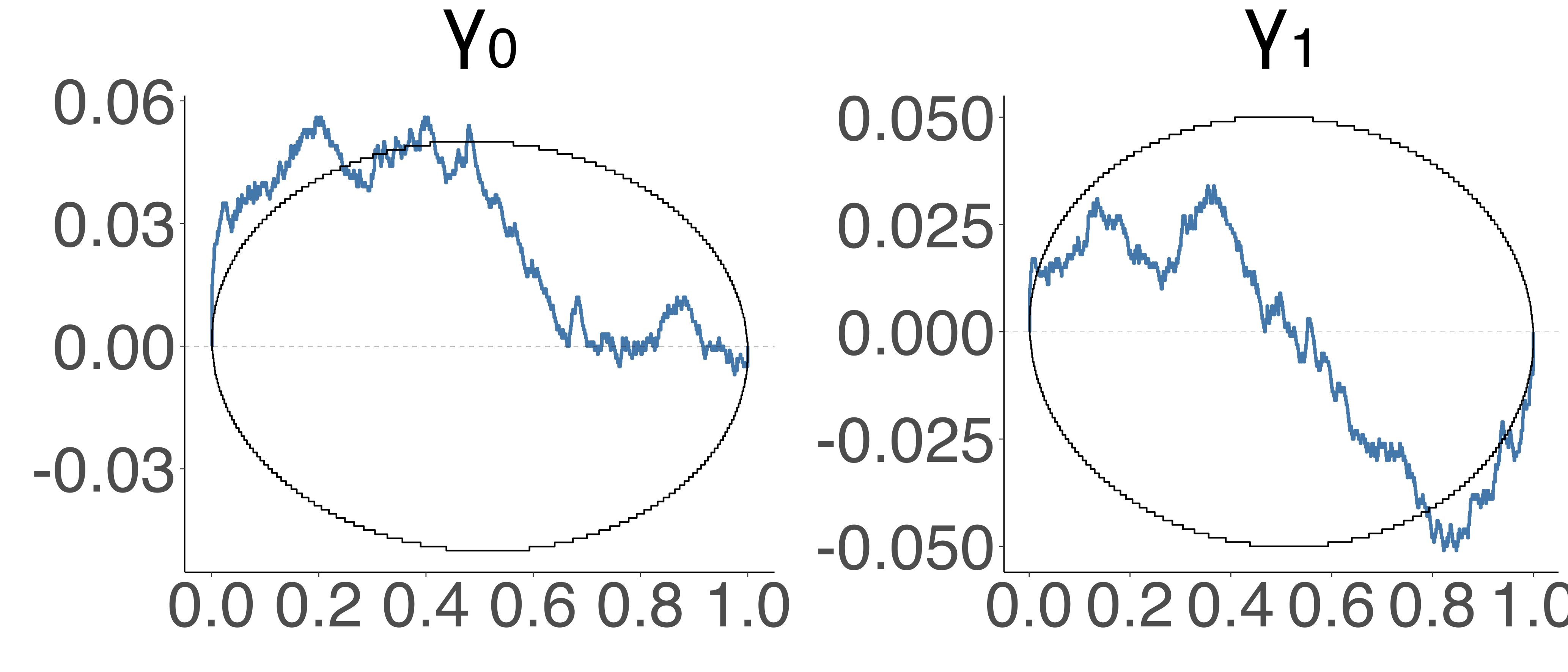}}
    \caption{ECDF difference plot of $p_k$ for $\gamma_0$ and $\gamma_1$ using Algorithm \ref{alg:sbc} out of 1000 data replicates for the new specification of the two-stage Poisson spatial model (Section \ref{subsubsec:spatial_twostage_poisson_logsumexp}) and using different approaches: (a) plug-in method (b) resampling method (c) full $\mathbf{Q}$  method (d) low rank $\mathbf{Q}$ (mesh A) method (e) low rank $\mathbf{Q}$ (mesh B) method} 
\label{fig:res_twostage_poisson_p_k_logsumexp_ecdfdiff}
\end{figure}

Figure \ref{fig:res_twostage_poisson_p_k_logsumexp_ecdfdiff}
shows the plot of the ECDF differences of the normalized ranks $p_k$ for $\gamma_0$ and $\gamma_1$ using Algorithm \ref{alg:sbc} and from 1000 data replicates. Histograms are provided in Section 3.5 of the Supplementary Material. The results show that the plug-in method underestimates the true posterior uncertainty for both parameters and introduces bias in the posterior distribution of $\gamma_0$. The resampling method correctly captures the posterior for $\gamma_1$, but  shows some bias for $\gamma_0$, though less severe than the plug-in method. The full $\mathbf{Q}$ and low rank $\mathbf{Q}$ methods also show potential bias for $\gamma_0$ on the same direction as the plug-in and resampling method. Moreover, the ECDF difference plot reveals a slight deviation from uniformity for $\gamma_1$, though less pronounced than that of the plug-in method. This suggests that the $\mathbf{Q}$-based methods strike a balance between the plug-in and the resampling method. Results from Algorithm \ref{alg:sbcconditional}, shown in Section 3.6 of the Supplementary Material, align with the insights from Algorithm \ref{alg:sbc}.


\subsubsection{Illustration with simulated data}

To illustrate the previous insights from the SBC, we simulate a data from both model specifications. We set the true values of the parameters as follows: $\beta_0=10, \beta_1=3, \gamma_0=-3, \gamma_1=0.15,\sigma_{e_1}^2=1,\sigma_{\xi}=0.6,\rho_{\xi}=4$. 
Moreover, we keep the spatial locations of the first-stage observations as in Section \ref{subsec:twostage_spatial_Gaussian} (Figure \ref{fig:res_twostage_gaussian_simdata}a), and  set $\text{E}(B) = 100$ for all blocks.  

\begin{figure}[t]
    \centering
    \subfigure[$\mu(\mathbf{s})$]{\includegraphics[width=.22\linewidth]{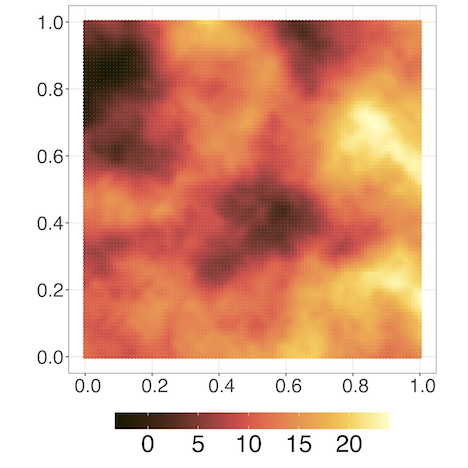}}
     \subfigure[$\mu(B)$]{\includegraphics[width=.22\linewidth]{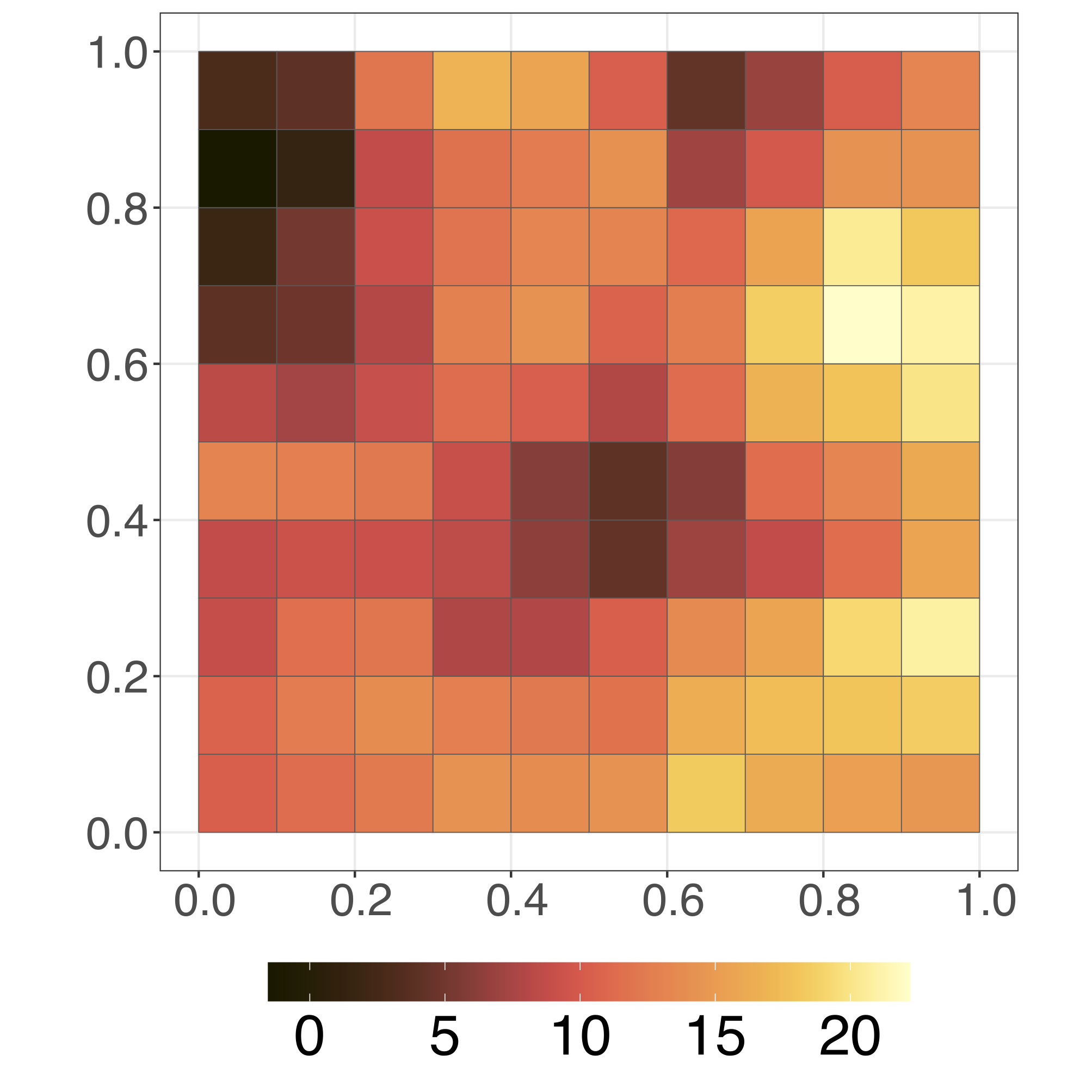}}
    \subfigure[$\lambda(B)$]{\includegraphics[width=.22\linewidth]{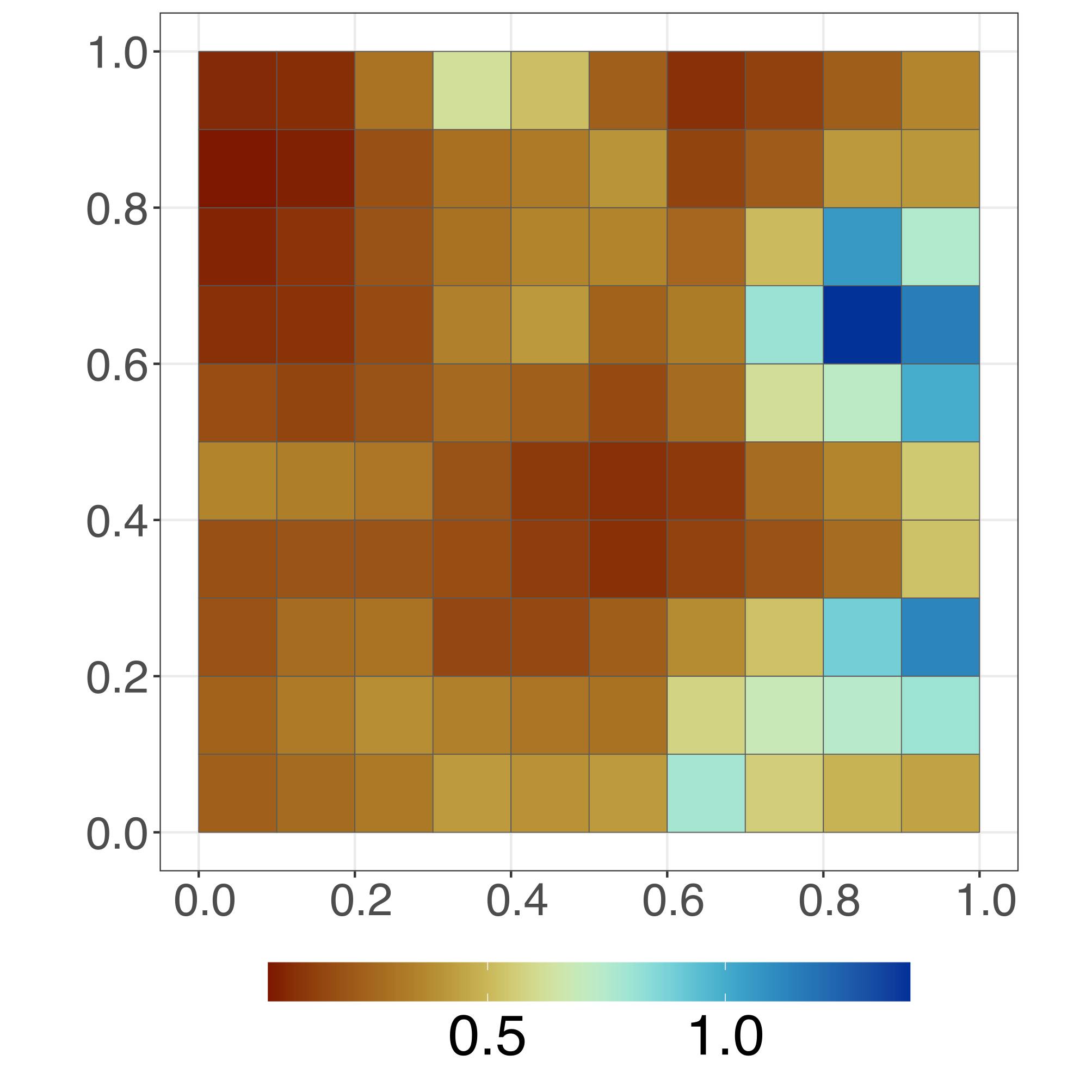}}
    \subfigure[$\text{y}(B)$]{\includegraphics[width=.22\linewidth]{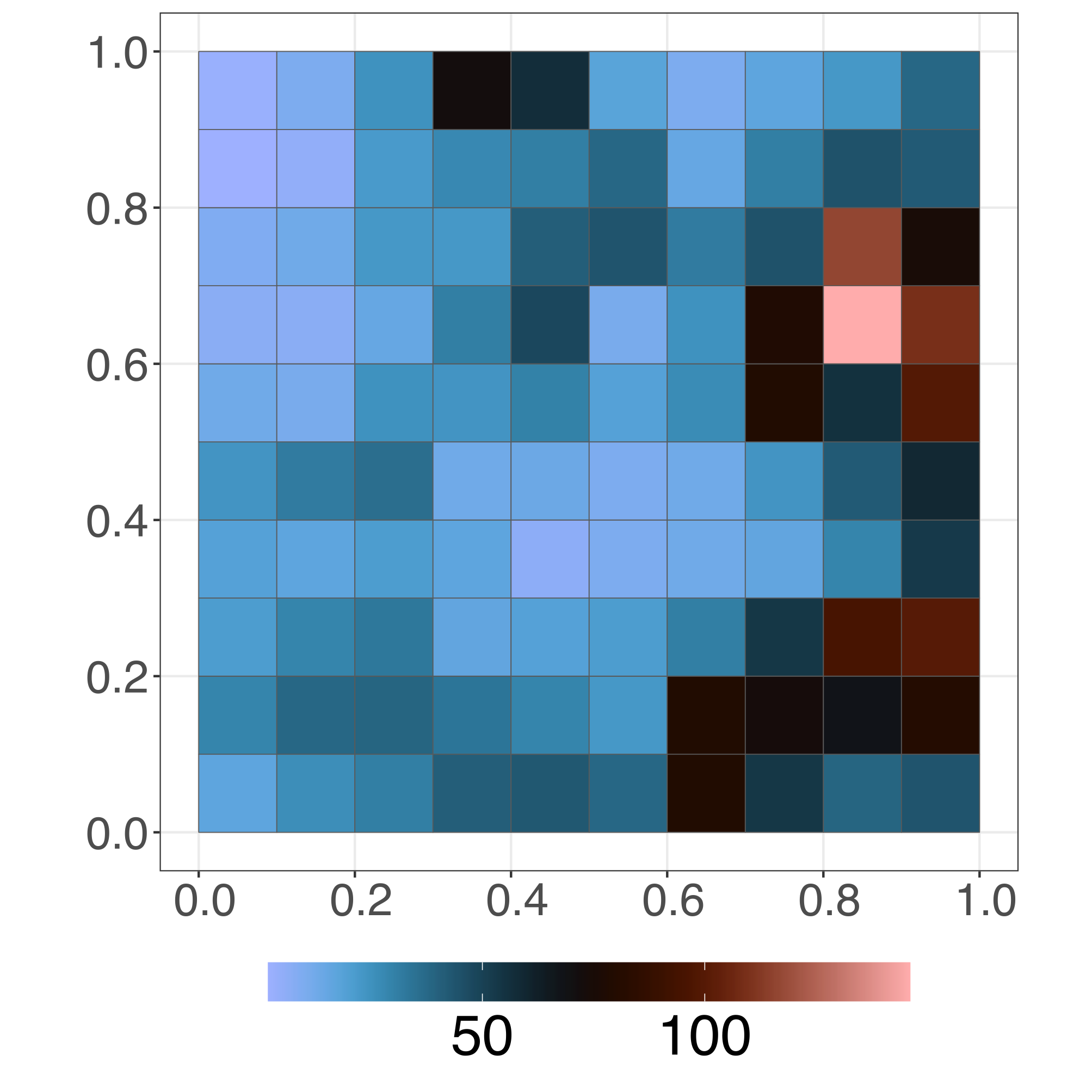}}    \caption{Simulated quantities from the classical model specification of the two-stage Poisson model in Section \ref{subsubsec:spatial_twostage_poisson_classical}}
\label{fig:res_spatial_twostage_poisson_classical_illustration}
\end{figure}

Figure \ref{fig:res_spatial_twostage_poisson_classical_illustration}a shows the simulated $\mu(\mathbf{s})$. 
The classical specification aggregates $\mu(\mathbf{s})$ over the blocks, which is shown in Figure \ref{fig:res_spatial_twostage_poisson_classical_illustration}b. The corresponding $\lambda(B)$ are in Figure \ref{fig:res_spatial_twostage_poisson_classical_illustration}c, while the simulated Poisson outcomes are in Figure \ref{fig:res_spatial_twostage_poisson_classical_illustration}d. For the new specification, we first compute the latent intensity field $\lambda(\mathbf{s})$ which is then aggregated over the blocks to yield $\lambda(B)$. The simulated $\lambda(\mathbf{s}),\lambda(B),$ and $\text{y}(B)$ for the new specification are shown in Figure 21 in the Supplementary Material.

Figures \ref{fig:res_spatial_twostage_poisson_estimates}a and \ref{fig:res_spatial_twostage_poisson_estimates}b show the estimated marginal posterior CDFs of $\gamma_0$ and $\gamma_1$ for the classical specification and new specification, respectively. For both model specifications, the plug-in method evidently has the smallest posterior uncertainty among the four approaches. The resampling method and the $\mathbf{Q}$-based methods have very similar posterior results. The posterior median for $\gamma_0$ using the new model specification is slightly overestimated, but is well within the 95\% credible interval. The results from this specific simulated data are consistent with the SBC results. 

\begin{figure}[t]
    \centering
    \subfigure[Classical specification]{\includegraphics[trim=0mm 0 0mm 0,clip,width=.48\linewidth]{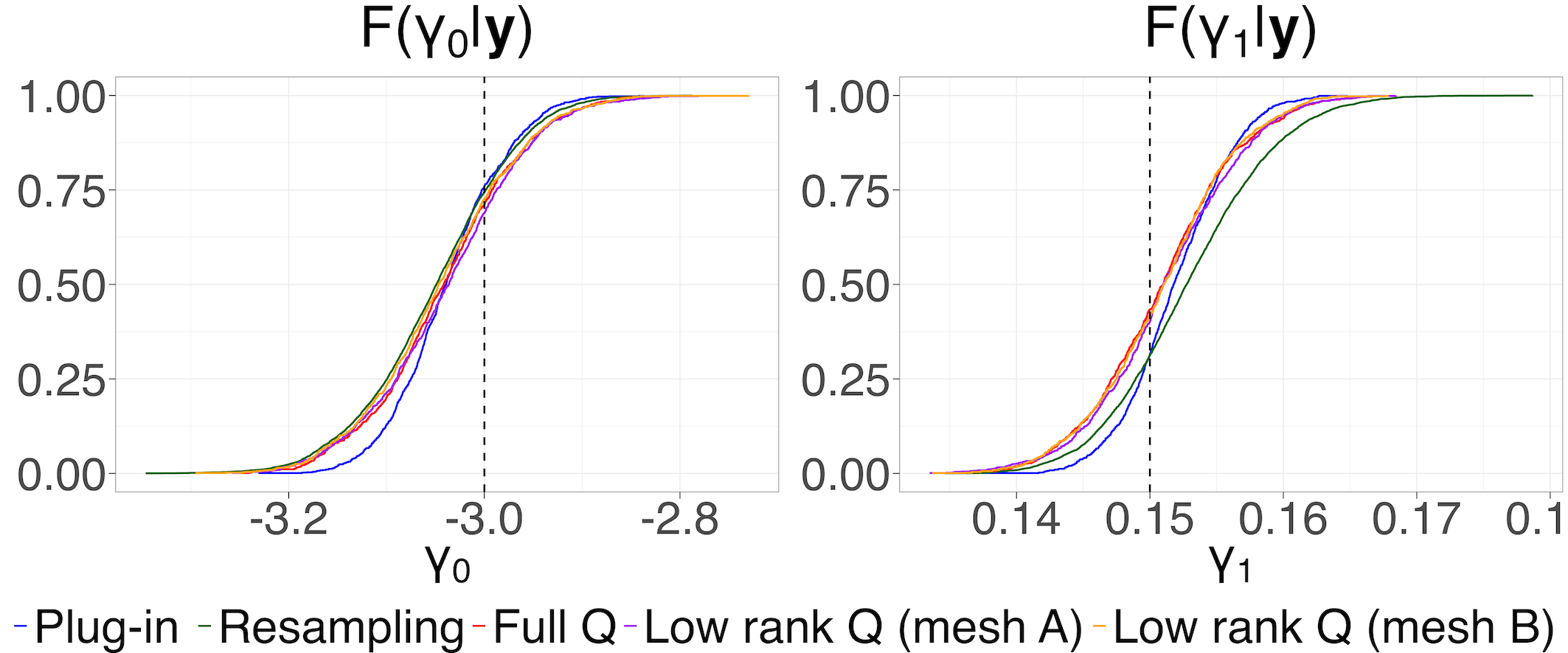}} \hspace{3mm}
    \subfigure[New specification]{\includegraphics[trim=0mm 0 0mm 0,clip,width=.48\linewidth]{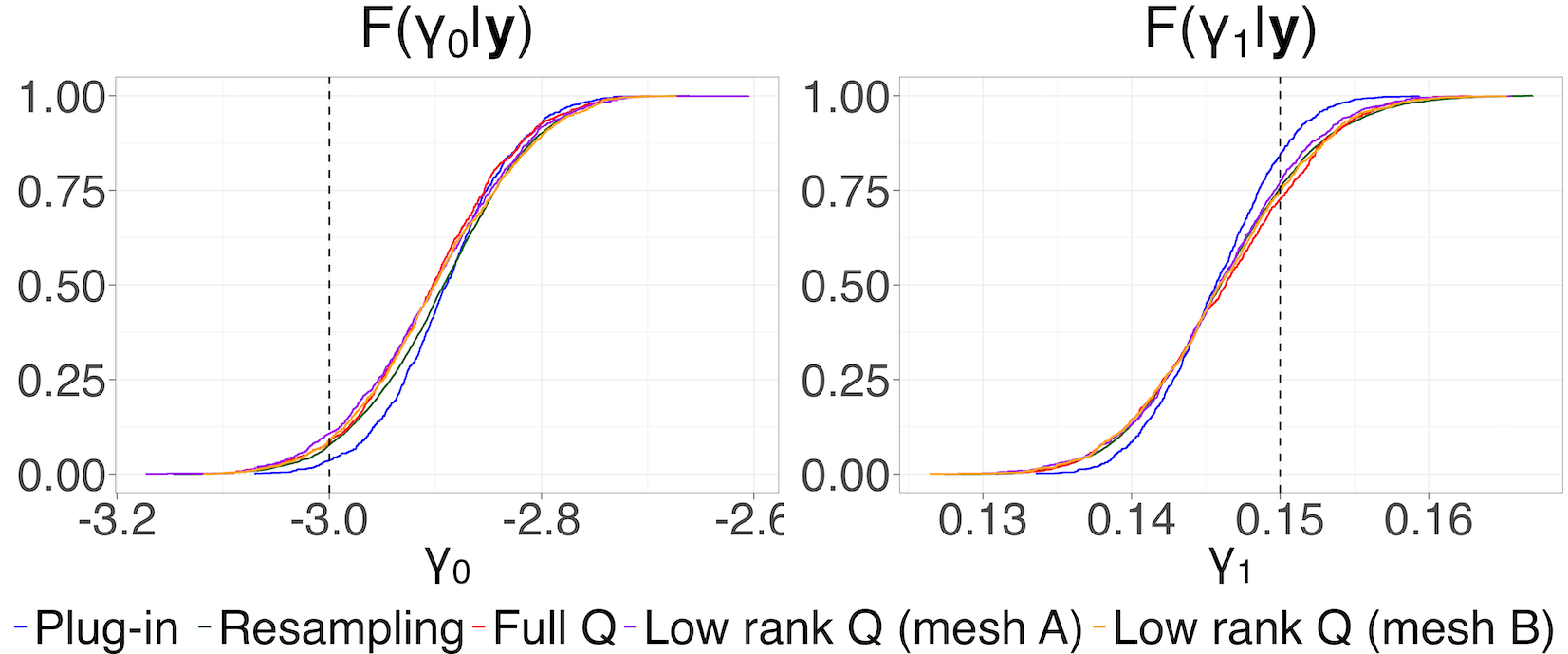}} 
    \caption{Marginal posterior CDFs of $\gamma_0$ and $\gamma_1$ for a simulated dataset from the two-stage Poison spatial model: (a) classical specification and (b) new specification; and using different estimation approaches: plug-in, resampling method, full $\mathbf{Q}$ method, low rank $\mathbf{Q}$ (mesh A) method, and low rank $\mathbf{Q}$ (mesh B) method}
\label{fig:res_spatial_twostage_poisson_estimates}
\end{figure}

\begin{figure}[t]
    \centering
    \subfigure[SD of $\lambda(\mathbf{s})$]{\includegraphics[scale=.32,trim=0mm 0 0mm 0,clip]{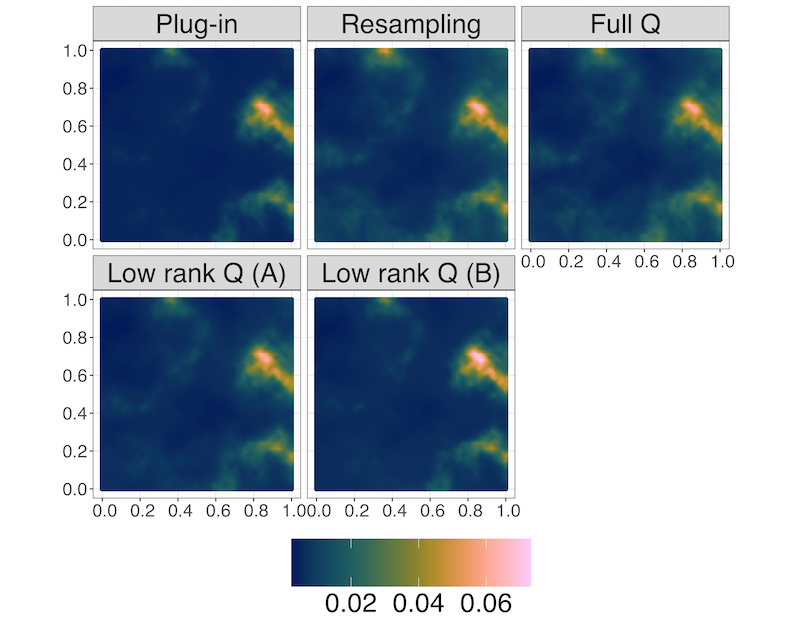}} 
    \subfigure[SD of $\lambda(B)$]{\includegraphics[scale=.32,trim=0mm 0 0mm 0,clip]{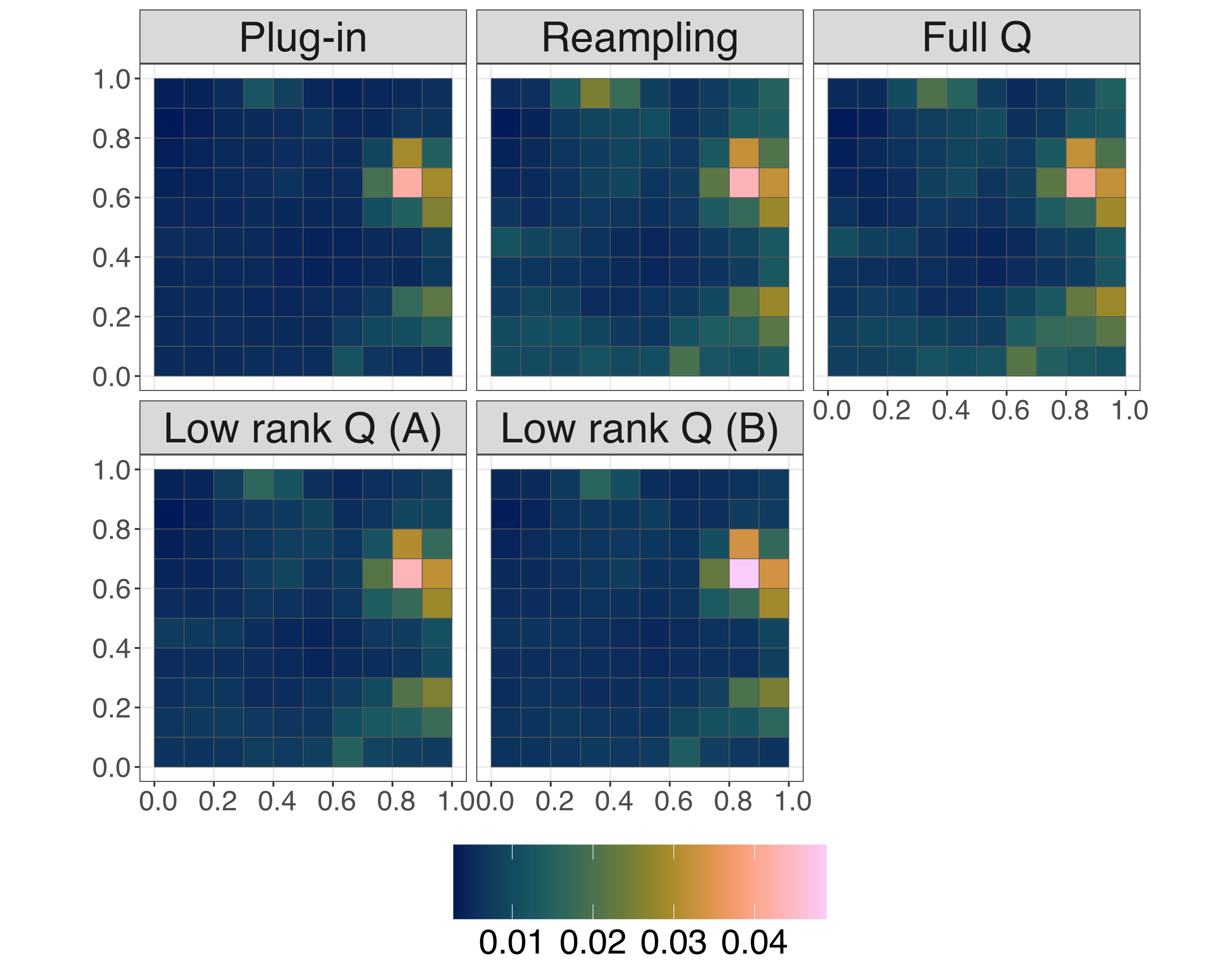}} 
    \caption{Comparison of the posterior uncertainty in (a) $\lambda(\mathbf{s})$ and (b) $\lambda(B)$ from a simulated data of the two-stage Poisson spatial model (new specification) using different approaches: plug-in method, resampling method, full $\mathbf{Q}$ method, low rank $\mathbf{Q}$ (mesh A) method, and low rank $\mathbf{Q}$ (mesh B) method}
\label{fig:res_spatial_twostage_poisson_logsumexp_pred_lambda_sd}
\end{figure}

Figures \ref{fig:res_spatial_twostage_poisson_logsumexp_pred_lambda_sd}a and \ref{fig:res_spatial_twostage_poisson_logsumexp_pred_lambda_sd}b show the posterior standard deviations of $\lambda(\mathbf{s})$ and $\lambda(B)$, respectively, using the new model specification and for the different uncertainty propagation approaches. It is evident that the plug-in approach generally has the smallest posterior uncertainty. The resampling method and the $\mathbf{Q}$-based methods have quite similar results, although the low rank $\mathbf{Q}$ method with a very coarse mesh has slightly smaller uncertainty estimates in some areas. The corresponding posterior means of $\lambda(\mathbf{s})$ and $\lambda(B)$ are shown in Figure 23 of Section 3.7 in the Supplementary Material. The posterior means of both $\lambda(\mathbf{s})$ and $\lambda(B)$ from the four uncertainty propagation approaches are very similar to the simulated truth. Moreover, the results from the classical model specification are shown in Figure 22 in the Supplementary Material. The results also show the same insights as the new model specification, i.e., the plug-in method has the smallest posterior uncertainty, while the resampling method and the $\mathbf{Q}$-based method have similar results.

Table \ref{tab:time_Poisson} shows the computational time (in seconds) for the different estimation approaches on the simulated data examples. For both the classical and new model specification, the plug-in method has the fastest computing time. The resampling method took the longest time for the classical model, while the full $\mathbf{Q}$ approach had a significant reduction in the computational time. In addition, the low rank $\mathbf{Q}$ (mesh A) took longer to run than the full $\mathbf{Q}$ method, but the low rank $\mathbf{Q}$ (mesh B) was faster than the previous two. This is consistent with the results from the simulated data example in Section \ref{subsec:twostage_spatial_Gaussian}, which show that the coarseness of the mesh for the error component is crucial in terms of the reduction in the computational time. On the other hand, for the new model specification, the full $\mathbf{Q}$ method took the longest computational time. The new model specification is a highly non-linear model, and introducing an error component all the more increases the model complexity; hence, making it plausible for the model fitting to even take longer. On the other hand, the low rank $\mathbf{Q}$ approach (both mesh A and mesh B) significantly reduced the computational time.  

\begin{table}[ht]
\centering
\begin{tabular}{lrr}
  \hline\hline
 \textbf{Method} & \textbf{Classical specification} & \textbf{New specification} \\ 
  \hline\hline
 Plug-in & 3.04 & 4.64 \\ 
  Resampling & 44.11 & 95.18 \\ 
  Full $\mathbf{Q}$ & 14.28 & 110.76 \\ 
 Low rank $\mathbf{Q}$ (mesh A) & 25.20 & 76.43 \\ 
  Low rank $\mathbf{Q}$ (mesh B) & 6.78 & 20.38 \\ 
   \hline\hline
\end{tabular}
\caption{Summary of computational time (in seconds) for the different approaches on the data illustration for the two-stage Poisson model}
\label{tab:time_Poisson}
\end{table}

The results for the two-stage Poisson models show that the plug-in method is expected to underestimate the posterior uncertainty in $\gamma_0$ and $\gamma_1$. On the other hand, the resampling approach is expected to be correct. However, there is a potential bias for the intercept $\gamma_0$ with the new model specification. The $\mathbf{Q}$-based methods provide a middle ground between the plug-in method and the resampling method, but the gain in the computational time depends on the coarseness of the mesh for the error component. For the new model specification, which is a highly non-linear model, using a very fine mesh for the error component may not be recommended since doing model fitting could potentially take a longer time than the resampling approach.   

\section{Real data application}\label{sec:data_application}

This section illustrates the proposed method in a real data application, which aims to link relative humidity (RH) and Dengue fever cases in the Philippines for August 2018. Dengue fever is an infectious disease common in tropical countries, and is caused by a virus transmitted by \textit{Aedes aegypti} and \textit{Aedes albopictus}, also known as yellow fever mosquito and Asian tiger mosquito, respectively. The association between climate variables and Dengue has been extensively studied (\citet{murray2013epidemiology, naish2014climate, lee2021impact, stolerman2019forecasting}). In particular, relative humidity is known to increase the risks of Dengue since high humidity enhances reproduction and breeding, and increases survival and lifespan of mosquitoes (\citet{thu1998effect, murray2013epidemiology, naish2014climate}).

\subsection{Data}

\begin{figure}[t]
    \centering
    \subfigure[Weather stations]{\includegraphics[width=.31\linewidth]{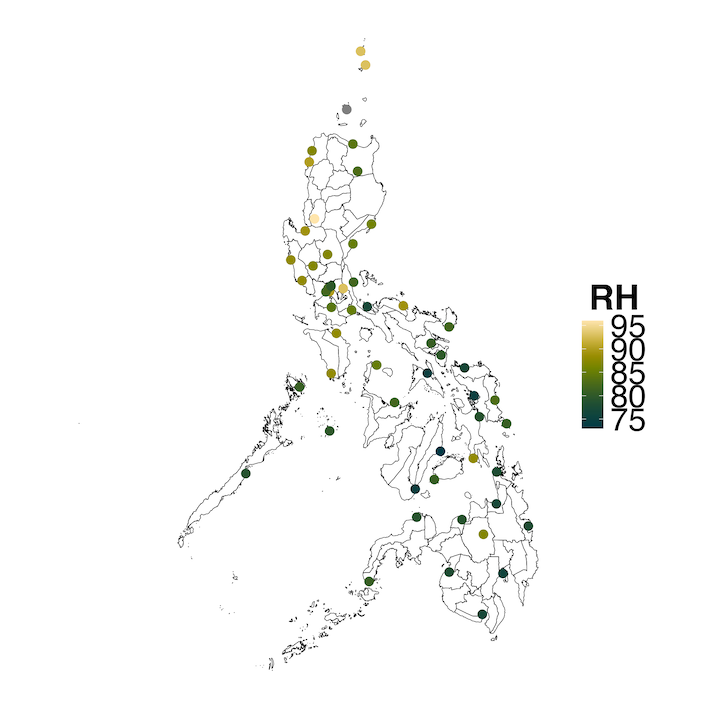}} 
    \subfigure[Dengue cases]{\includegraphics[width=.31\linewidth]{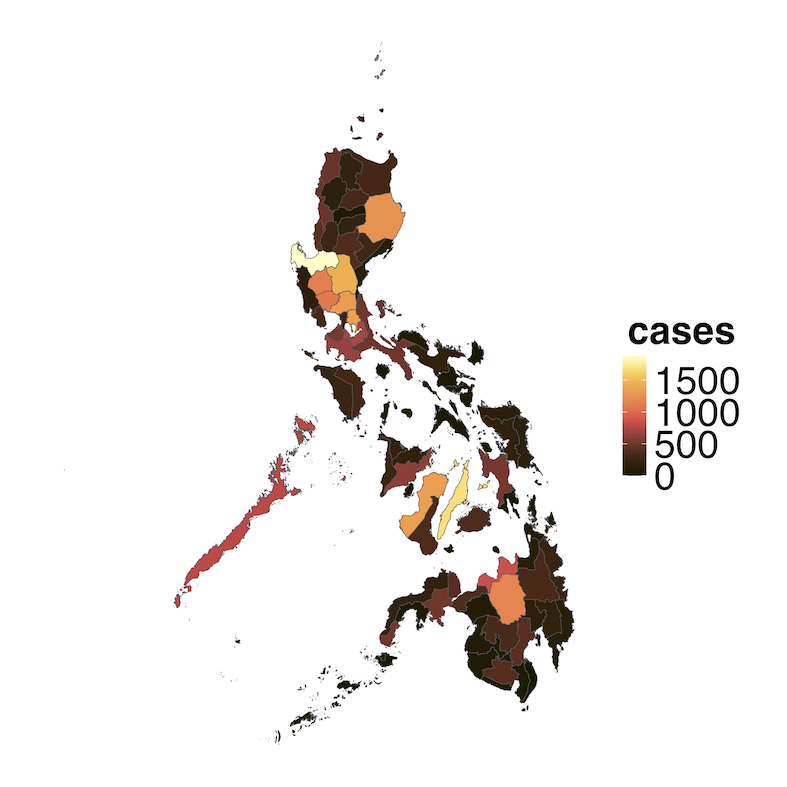}} 
    \subfigure[Dengue SIRs]{\includegraphics[width=.31\linewidth]{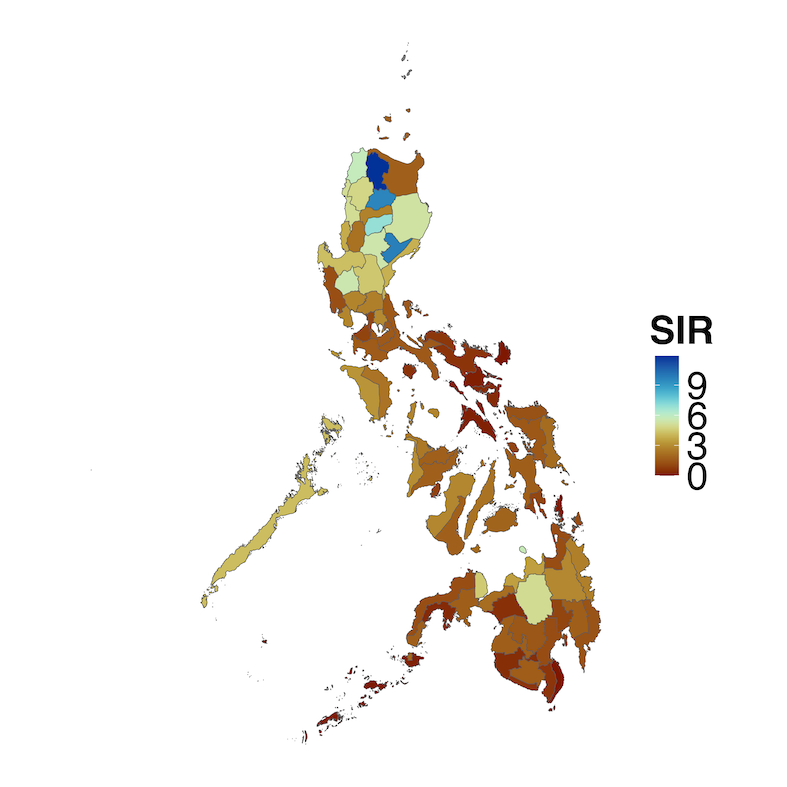}} 
    \caption{(a) weather stations in the Philippines (b) plot of Dengue cases by province for August 2018 (c) plot of the standardized incidence ratios (SIR) of Dengue by province for August 2018}
\label{fig:data_application}
\end{figure}
Monthly averages of relative humidity for August 2018 from 56 weather synoptic stations in the Philippines were provided by the Philippine Atmospheric, Geophysical and Astronomical Services Administration (PAGASA) (Figure \ref{fig:data_application}a). For each of the 82 provinces, the Epidemiological Bureau of the Department of Health provided the counts of Dengue fever (Figure \ref{fig:data_application}b).
The standardized incidence ratios (SIR), computed as the observed divided by the expected cases, are shown in Figure \ref{fig:data_application}c. The expected cases are derived using national rates and via internal standardization (\citet{waller2010disease}). In particular, if $r$ and $n(B)$ is the national rate and size of province $B$, respectively, then the expected cases for the province $B$ is $\text{E}(B)=r\times n(B)$. The SIR indicates the relative excess in the incidence of the disease with respect to what might have been expected based on the reference national rates (\citet{schoenbach2000understanding}).

\subsection{Model}

We perform the inference in a two-stage modelling framework. The first stage models RH, while the second stage models the Dengue health counts using information from the first-stage model as an input. 

\begin{itemize}
    \item \textbf{First-stage model} -- Suppose $\mu(\mathbf{s})$ is the log true relative humidity level at an arbitrary spatial location $\mathbf{s}$. We assume the following latent process:
\begin{equation*}
    \mu(\mathbf{s}) = \beta_0 + \beta_1\text{Elevation}(\mathbf{s}) + \beta_2\log\text{Temperature}(\mathbf{s}) + \beta_3\big(\log\text{Temperature}(\mathbf{s})\big)^2 + \xi(\mathbf{s}),
\end{equation*}
where $\xi(\mathbf{s})$ is a Mat\'ern field and $\bm{\beta} = \begin{pmatrix} \beta_0 & \beta_1 & \beta_2 & \beta_3
\end{pmatrix}^\intercal$ are fixed effects. We assume that the observed values at the weather stations follow the classical error model, i.e., $\text{w}(\mathbf{s}_i) = \mu(\mathbf{s}_i) + e(\mathbf{s}_i)$ and $e(\mathbf{s}_i) \overset{\text{iid}}{\sim} \mathcal{N}(0,\sigma^2_e), i=1,\ldots, 56.$ The temperature field is assumed to be known using the predicted values from the climate data fusion models in \citet{villejo2024data}.

\item \textbf{Second-stage model} -- We consider both the classical and new specification in Section \ref{subsec:spatial_twostage_poisson}. Suppose $\text{y}(B)$ and $\text{E}(B)$ are the observed and expected Dengue cases in province $B$, respectively. We assume that $\text{y}(B) \sim \text{Poisson}\big(\mu_{\text{y}}(B)\big), \mu_{\text{y}}(B)= \mathbb{E}[\text{y}(B)] = \text{E}(B)\times \lambda(B)$. This implies that $\lambda(B)=\dfrac{\text{y}(B)}{\text{E}(B)}$, so that the disease risk $\lambda(B)$ is also interpreted as the model-based estimate of the SIR. For the classical specification, we assume that 
\begin{align*}
    &\log\Big(\lambda(B)\Big) =\gamma_0 + \gamma_1 \dfrac{1}{|B|}\int_B \mu(\mathbf{s}) d\mathbf{s} + \phi(B),
\end{align*}
where $\phi(B)$ is an area-specific effect which we model as $\phi(B) \overset{\text{iid}}{\sim} \mathcal{N}(0,\sigma^2_B)$. For the new specification, we have:
\begin{align*}
  &\log\Big(\lambda(B)\Big) = \log\Bigg(\dfrac{1}{|B|}\int_B \lambda(\mathbf{s}) d\mathbf{s} +\phi(B)\Bigg), \;\;\; \lambda(\mathbf{s}) = \exp\big\{\gamma_0 + \gamma_1\mu(\mathbf{s})\big\}.
\end{align*} 
\end{itemize}

We use the INLA-SPDE approach to fit the model. The mesh, with 1077 nodes,  used to estimate the  the Mat\'ern field $\xi(\mathbf{s})$ is shown in Figure \ref{fig:application_stageone_fields_spde_highprec}a. We use vague priors for the fixed effects and $\sigma_{e}$, and a joint normal prior (Equations \eqref{eq:jointnormalpriorA} and \eqref{eq:jointnormalpriorB}) for the Mat\'ern parameters.
In particular, we use $\begin{bmatrix}
    \log(\tau) \\ \log(\kappa) 
\end{bmatrix} \sim \mathcal{N}\Bigg(\begin{bmatrix}
    2.71 \\ -4.66
\end{bmatrix} \Big), \begin{bmatrix}
   4 & 0 \\ 0 & 1
\end{bmatrix}\Bigg)$. This implies that a plausible range of values for the range parameter is from 110 km to 800 km, which are consistent with the estimates in \citet{villejo2024data}. Moreover, we set the plausible range of values for the marginal standard deviation as 0.2683 to 14.65, based on the empirical standard deviation of RH which is 5.37.


We compare the posterior estimates of $\gamma_0$ and $\gamma_1$ from the four uncertainty propagation approaches under consideration: plug-in, resampling, full $\mathbf{Q}$, and low rank $\mathbf{Q}$ approach. In this case study, there is no strong motivation for the low rank $\mathbf{Q}$ approach since the dimension of the latent parameters is not large. Nonetheless, we  explore the low rank $\mathbf{Q}$ approach in order to have a full comparison with the other approaches. Figure \ref{fig:application_stageone_fields_spde_highprec}b shows the mesh for the error component of the low rank $\mathbf{Q}$ approach, which has  546  nodes. 

\subsection{Results}


\subsubsection{First-stage model results}

\begin{figure}[t]
    \centering
    \subfigure[Full $\mathbf{Q}$ mesh]{\includegraphics[width=.25\linewidth]{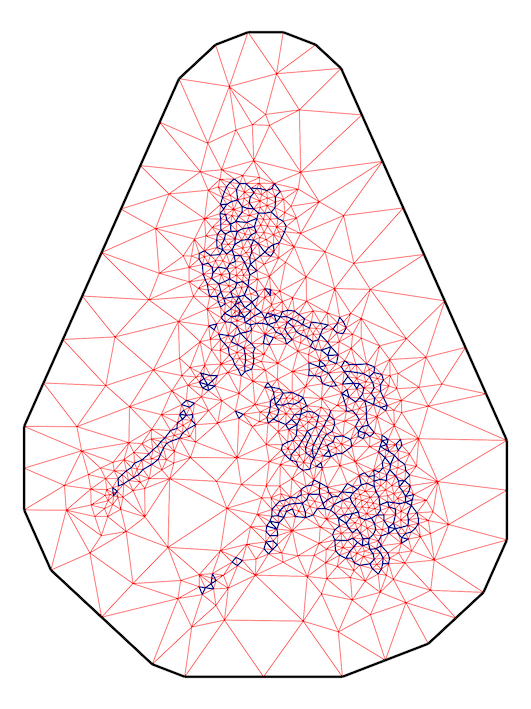}}
    \subfigure[Low rank $\mathbf{Q}$ mesh]{\includegraphics[ width=.25\linewidth]{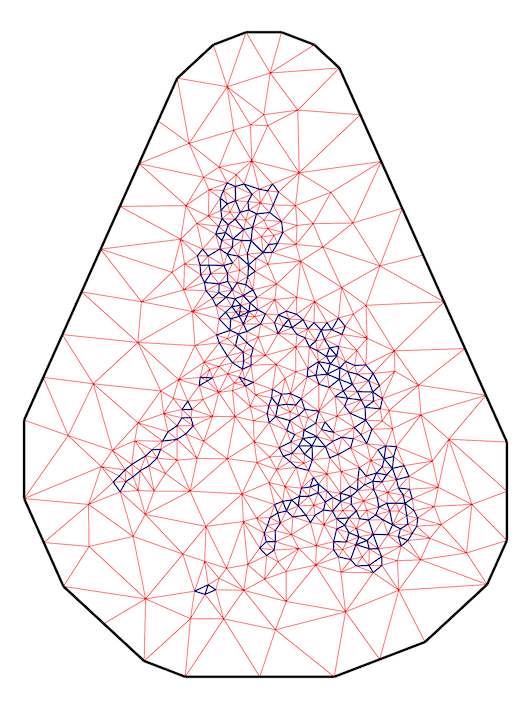}}\hspace{5mm}
    \subfigure[Posterior mean of RH]{\includegraphics[trim={7cm 1cm 7cm 0cm},clip,width=.21\linewidth]{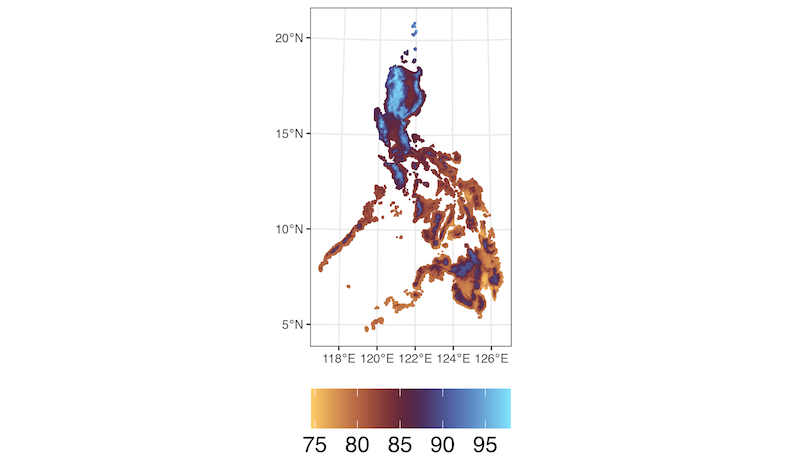}}\hspace{2mm} 
    \subfigure[Posterior SD of RH]{\includegraphics[trim={7cm 1cm 7cm 0cm},clip, width=.21\linewidth]{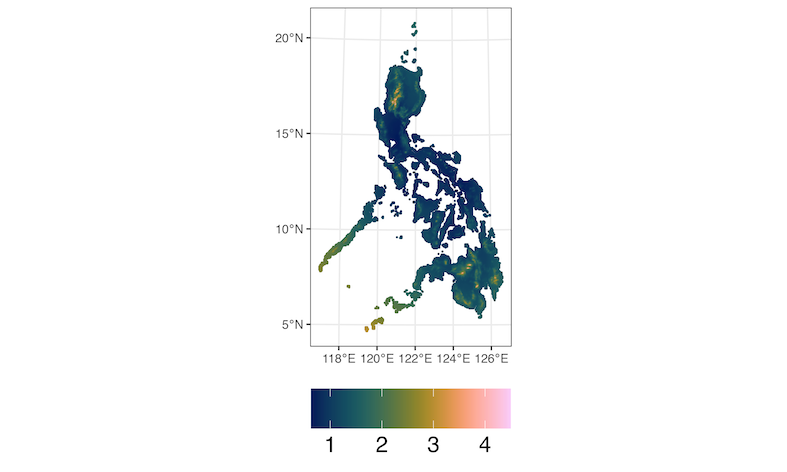}}
    \caption{(a) mesh for the full $\mathbf{Q}$ method (b) mesh for the low rank $\mathbf{Q}$ method (c) estimated RH field (d) posterior uncertainty of RH field}
\label{fig:application_stageone_fields_spde_highprec}
\end{figure}

Figures \ref{fig:application_stageone_fields_spde_highprec}c and \ref{fig:application_stageone_fields_spde_highprec}d show the estimated posterior mean and standard deviation of the relative humidity field. The posterior estimates of the parameters are reported in Table 1 of Section 4 in the Supplementary Material. Both elevation and temperature appear to be significant predictors in the model. The northwestern section of the country has high relative humidity levels compared to most of the eastern section. This is consistent with the climate dynamics of the country (\citet{floresbalagotClimatePH, kintanarClimatePH}).

\subsubsection{Second-stage model results}

\begin{figure}[h!]
    \centering
    \subfigure[95\% CI of RR with 1 SD change in RH]{\includegraphics[width=.49\linewidth]{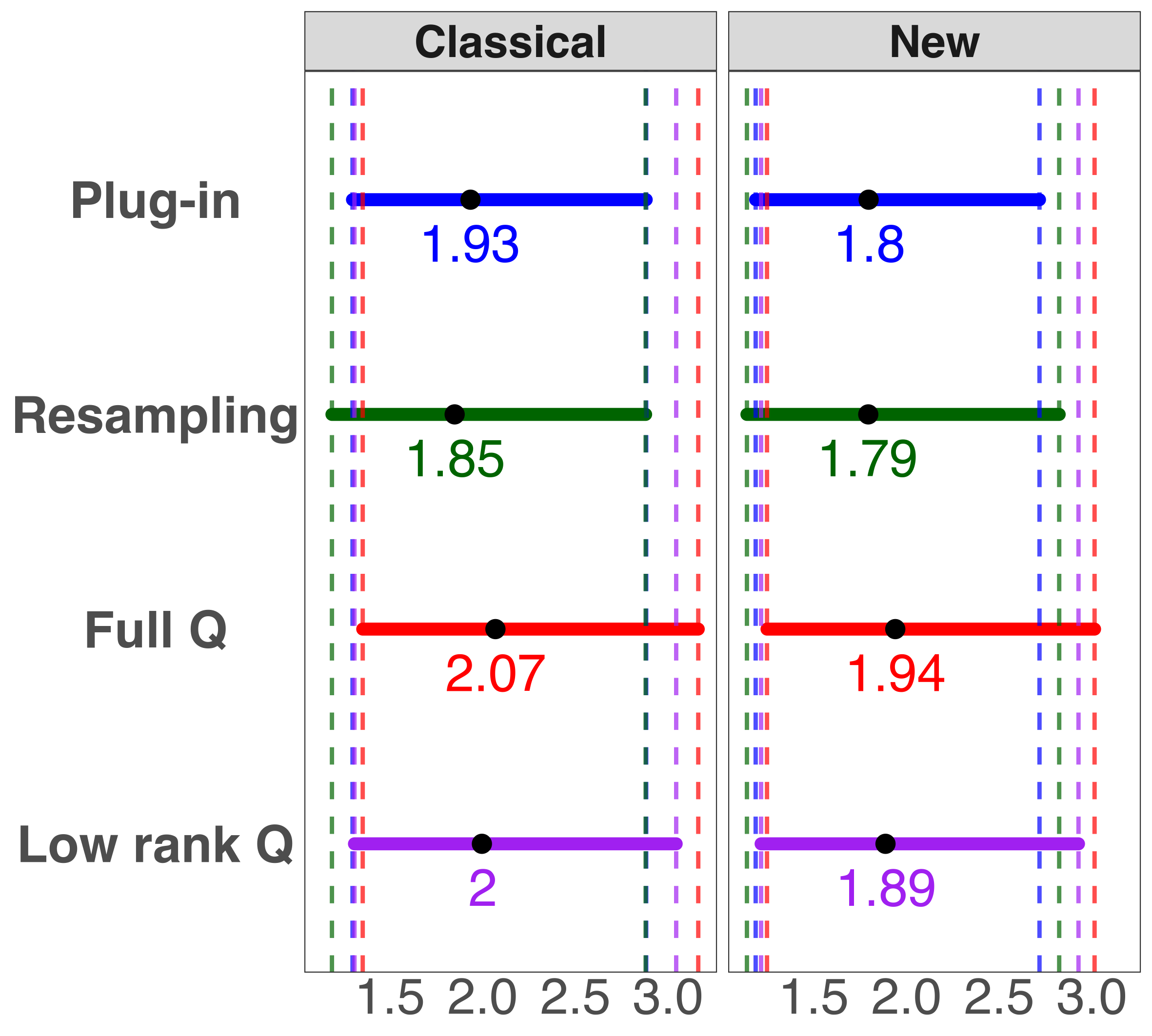}} 
    \hspace{0mm}
    \subfigure[95\% CI for $\gamma_0$]{\includegraphics[width=.49\linewidth]{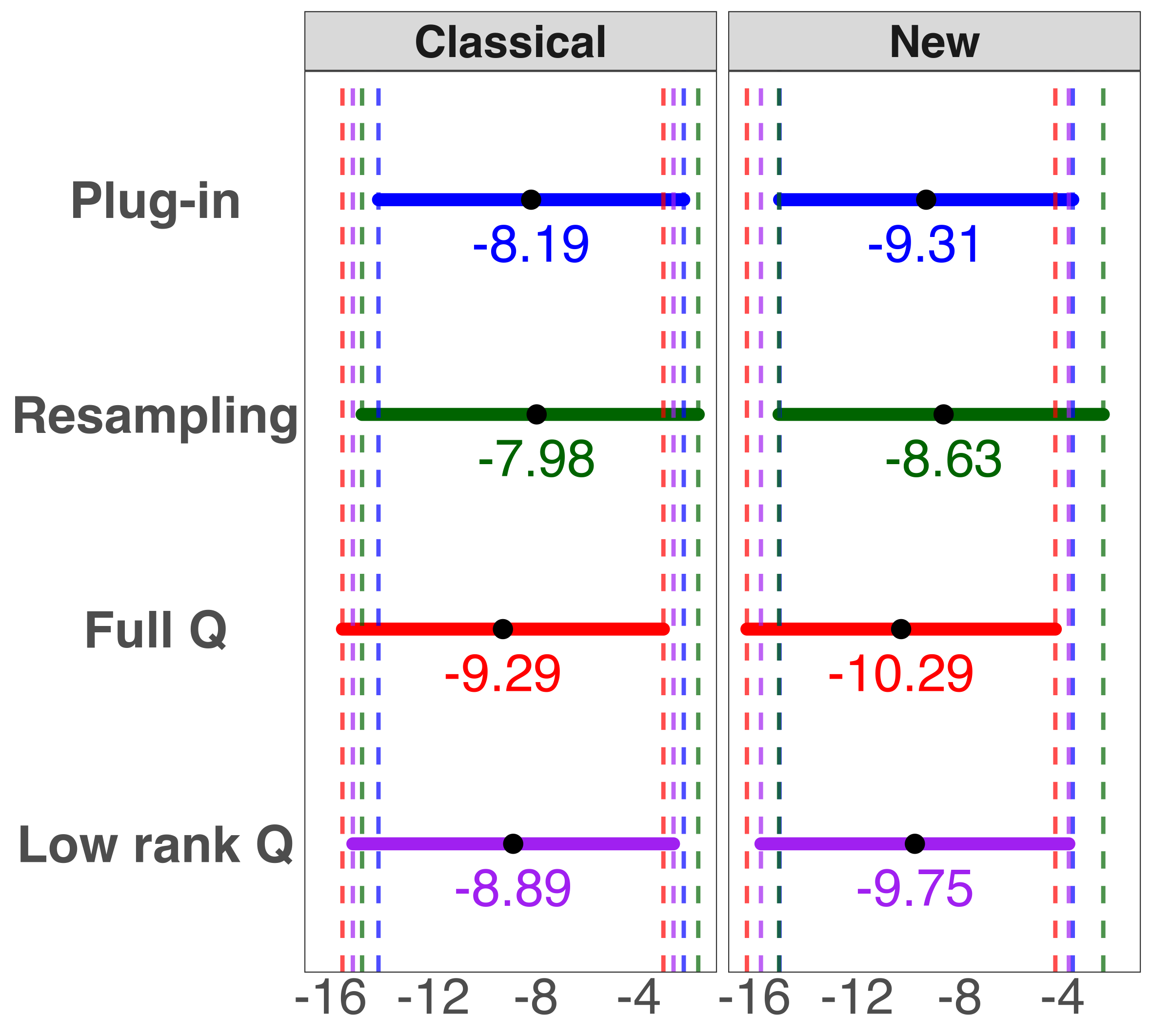}} 
    \caption{(a) 95\% CI of RR associated with 1 standard deviation change in relative humidity (b) 95\% CI for $\gamma_0$. Shown in broken lines (\sampleline{dashed}) are the lower and upper limit of the 95\% CI. The black dot ($\boldsymbol{\cdot}$) is the posterior mean} 
\label{fig:CIwidth_dataapplication2}
\end{figure}

\begin{figure}[h!]
    \centering
    \subfigure[95\% CI widths of RR for different $\omega$-units]{\includegraphics[width=.43\linewidth]{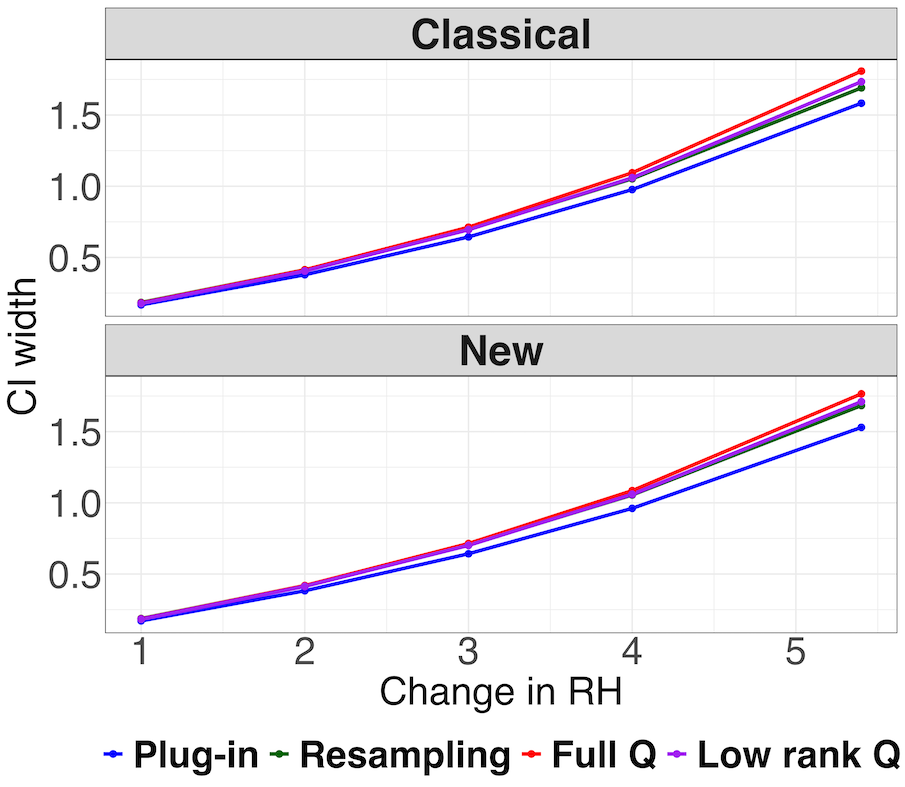}} 
    \hspace{2mm}
    \subfigure[95\% CI width for $\gamma_0$]{\includegraphics[width=.43\linewidth]{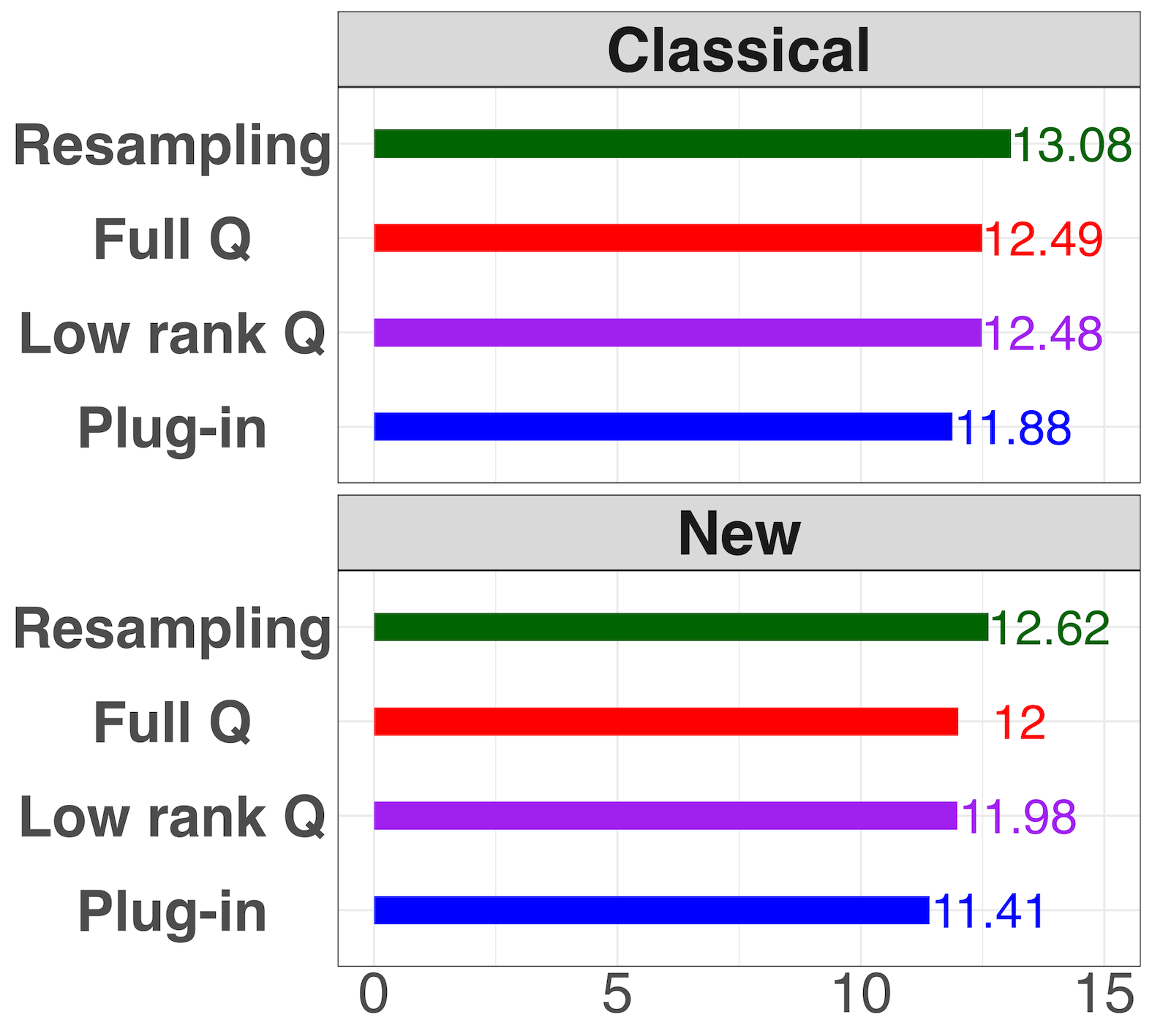}} 
    \caption{(a) 95\% CI width of RR associated with  $\omega$-units change in RH (b) 95\% CI width for $\gamma_0$} 
\label{fig:CIwidth_dataapplication}
\end{figure}

In spatial epidemiology, it is usually more useful to interpret the multiplicative change in the disease risk, referred to as a rate ratio (RR) or relative risk, associated with one SD change in the exposure variable, which is relative humidity in this application. Figure \ref{fig:CIwidth_dataapplication2}a shows the estimated RR and the 95\% CI (in dashed lines) associated with one SD change in RH. The actual 95\% CIs are shown in Figure 25 of the Supplementary Material. The results show that for a one SD change in RH, the risk of Dengue approximately doubles. This is consistent with several studies which have shown a positive association between RH and the risk of Dengue (\citet{thu1998effect, murray2013epidemiology, naish2014climate}). The width of the CIs from Figure \ref{fig:CIwidth_dataapplication2} does not seem to be too different, but this is the typical length of the CIs in spatial epidemiology literature (\citet{blangiardo2016two,liu2017incorporating,lee2017rigorous}). Also, both $\gamma_0$ and $\gamma_1$ are in the log risk scale, so that these differences are not negligible. We also computed the RR for other $\omega$-units change in RH. In particular,  we considered $\omega=\{1,2,3,4,5.4\}$, where 5.4 corresponds to one SD in RH. The results are shown in Figure \ref{fig:CIwidth_dataapplication}a. The figure shows that the CI widths in the RR are not too different when the magnitude of the change in the RH is small, but the difference in the CIs become more apparent when the change in RH is large. The posterior estimates for $\gamma_0$ and the 95\% CI widths for the different uncertainty propagation approaches are shown in Figure \ref{fig:CIwidth_dataapplication2}b and \ref{fig:CIwidth_dataapplication}b, respectively. Figures 24a and 24b in the Supplementary Material show the estimated marginal CDFs of $\gamma_0$ and $\gamma_1$ for the classical and new specification, respectively, while the point estimates and 95\% CI are shown in Tables 2 and 3 in Section 4 of the Supplementary Material. The resampling method and the two proposed methods have slightly larger posterior uncertainty than the plugin method for $\gamma_1$. The differences in the posterior uncertainty for $\gamma_0$ are more apparent.  

Figure \ref{fig:CIwidth_dataapplication2} shows some differences in the posterior results among the four uncertainty propagation approaches. The posterior mean and the lower limit of the 95\% CI of the RR for the resampling method is the lowest among the four uncertainty propagation approaches. This attenuation to the null risk of one is also observed in \citet{lee2017rigorous} and \citet{liu2017incorporating}, where they argue that it is due to the posterior predictive distribution of the first-stage model outweighing the spatial (or spatio-temporal) variation in the data, which results in the estimated effects being washed away. Even so, we also see that the posterior mean for $\gamma_0$ from the resampling method is the highest, so that both parameters balance each other out when calculating the log risks, $\log\big(\lambda(B)\big)$. This is the same observation, although in the opposite direction, from the results of the $\mathbf{Q}$-based methods, where the posterior mean of $\gamma_1$ is relatively high, but the posterior mean for $\gamma_0$ is relatively low. This observed push and pull between the two parameters explains why the estimated posterior means of $\lambda(B)$ for the four uncertainty propagation methods are very similar, as shown in Figure \ref{fig:predicted_modelSIRs}a for the classical specification and Figure 26 of Section 4 in the Supplementary Material for the new specification. Moreover, the estimated disease risks look very similar to the computed SIRs in Figure \ref{fig:data_application}c. The corresponding posterior SDs are shown in Figure \ref{fig:predicted_modelSIRs}b for the classical specification and Figure 27 in the Supplementary Material for the new specification. The differences in the posterior SDs are more apparent in the northern part of the country where the estimated risks are very high. The results show that the two proposed methods gave the highest posterior uncertainty, while the plug-in method has the lowest posterior uncertainty.

\begin{figure}[h!]
    \centering
    \subfigure[Posterior mean of $\lambda(B)$]{\includegraphics[width=.48\linewidth]{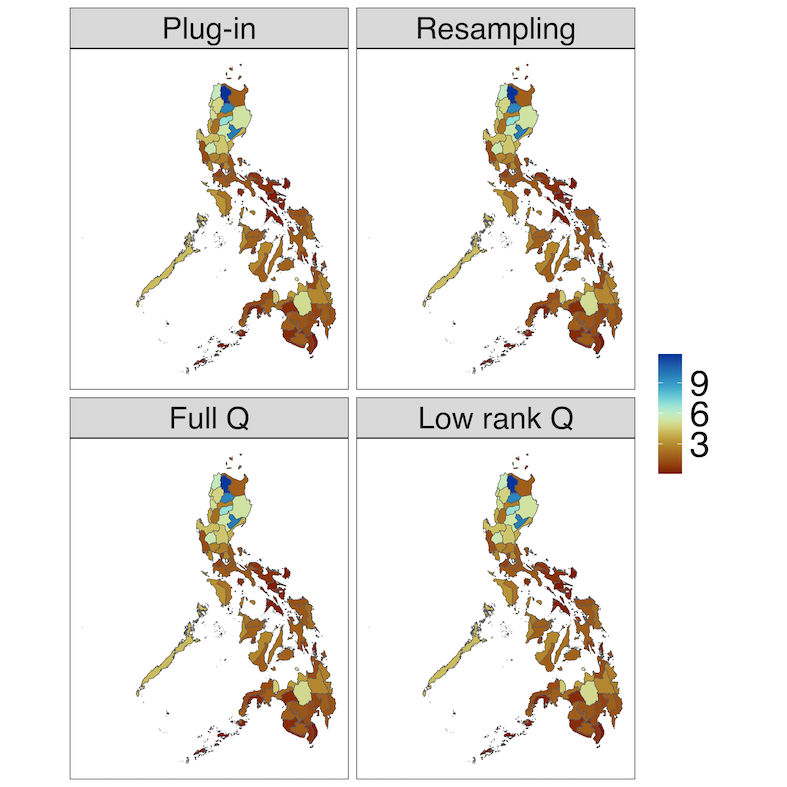}} 
    \hspace{2mm}
    \subfigure[Posterior SD of $\lambda(B)$]{\includegraphics[width=.48\linewidth]{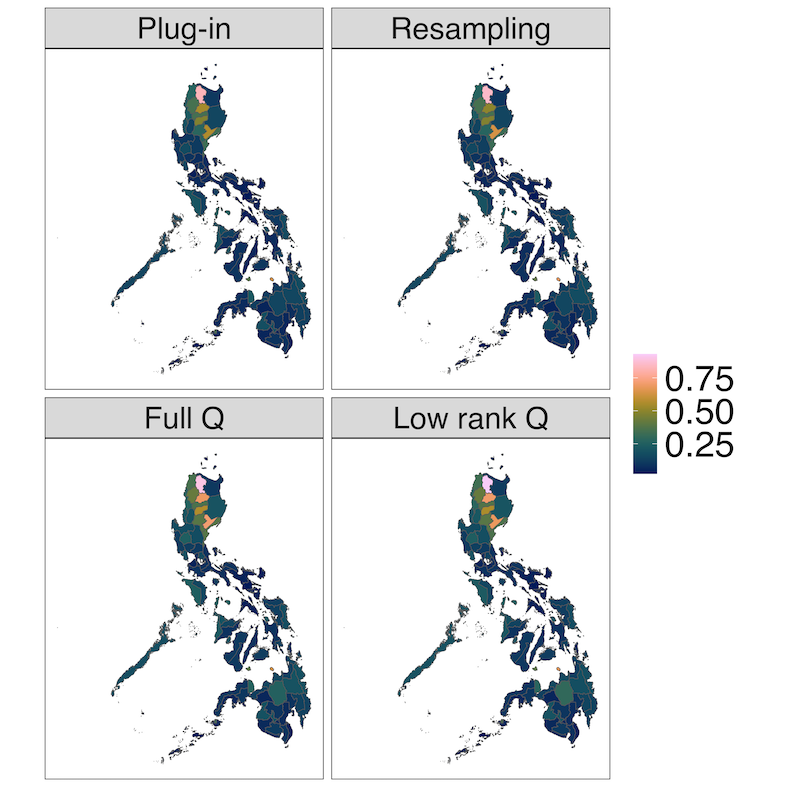}} 
    \caption{(a) Posterior mean of $\lambda(B)$ from the classical model specification (b) Posterior SD of $\lambda(B)$ from the classical model specification} 
\label{fig:predicted_modelSIRs}
\end{figure}

\section{Conclusions}\label{sec:conclusions}

This work addresses the problem of uncertainty propagation in two-stage Bayesian models. This approach is appropriate for scenarios when there is a clear one-directional physical relationship between the two models, e.g., climate affects Dengue cases or air pollution affects respiratory diseases. Also, it is a  practical approach when the first-stage model is already complex in itself; for example, it might involve fitting a complex data fusion model (\citet{villejo2023data}). In addition, a two-stage modeling framework avoids the potential unwanted feedback effects present in  fully Bayesian approaches (\citet{wakefield2006health,shaddick2002modelling,gryparis2009measurement}). 
The drawback of the two-stage modeling framework is that uncertainty is not automatically propagated between the two models.

In this paper, we validate different uncertainty propagation approaches for two-stage models by testing the self-consistency property of Bayesian models using the simulation-based calibration (SBC) method of \citet{talts2018validating}. In particular, we investigated the correctness of two commonly used methods for two-stage modeling: the plug-in method and the resampling method. In addition, we also explored a new method called the $\mathbf{Q}$ uncertainty method. This introduces a new model component, called an \textit{error component}, in the second-stage model. The error component is given a Gaussian prior with mean zero and precision matrix $\mathbf{Q}$, derived from the Gaussian approximation of the latent parameters of the first-stage model. 
The $\mathbf{Q}$ matrix can be of high dimension (for example in large spatio-temporal applications); hence, we also proposed a low rank approximation of the $\mathbf{Q}$ matrix. We then have two versions of the proposed method: the full $\mathbf{Q}$ method and the low rank $\mathbf{Q}$ method. The $\mathbf{Q}$ uncertainty method
is implemented using the INLA methodology, but we believe that the same idea can  be used in other Bayesian inference approaches like MCMC. We plan to explore this in a future work.


We also proposed a modification of the SBC method of \citet{talts2018validating} to address challenges specific to two-stage Bayesian models.  Here, the SBC algorithm is implemented conditional on fixed values of certain first-stage hyperparameters. This approach ensures that the evaluation focuses on the second-stage model parameters, avoiding the influence of first-stage parameters that may violate the self-consistency property.
Results from both the original SBC and the conditional SBC confirm that the plug-in method underestimates the posterior uncertainty of the second-stage model parameters, while the resampling method provides correct uncertainty estimates. The proposed $\mathbf{Q}$-based methods  also produce correct posterior uncertainty estimates, although  results indicate that the coarseness of the $\mathbf{Q}$ approximation can affect the accuracy of the approximate posteriors and the computational cost. In this work, we considered the individual parameters as test functions. The use of test functions which are data-dependent can be done in future work.


The computational efficiency of the $\mathbf{Q}$-based methods depends on the coarseness of the mesh for the error component. For the new specification of the Poisson model, the full $\mathbf{Q}$ method required longer computational time compared to the resampling method. This is because the $\mathbf{Q}$-based methods introduce an additional model component, significantly increasing model complexity and making fitting more computationally intensive, particularly for the highly nonlinear Poisson model. Nevertheless, a sufficiently coarse mesh in the low rank $\mathbf{Q}$ method can address this challenge while maintaining correctness.
Moreover, the low rank $\mathbf{Q}$ approach may also take longer to run compared to the full $\mathbf{Q}$ method if the resolution of the $\mathbf{Q}$ approximation is not coarse enough. The reason for this is that the predictor expression of the second-stage model which implements the low rank $\mathbf{Q}$ approach involves more matrix operations, which may potentially increase the computational requirements for model estimation. 


Some aspects of the $\mathbf{Q}$ uncertainty method require further investigation. Firstly, we fixed the scaling parameter of the $\mathbf{Q}$ matrix to 1, but this choice may not be optimal. 
When the scaling parameter is estimated rather than fixed, the results showed that its estimated value tends to be very large.
This behavior implies an increased confidence in the first-stage posterior estimates, effectively reducing the uncertainty in the error component, and consequently producing narrower uncertainty estimates for the second-stage model parameters. Our simulation experiments revealed that as the fixed value of the scaling parameter decreases, the posterior uncertainty of the second-stage model parameters widens and deviates more significantly from the crude plug-in method. Conversely, when the scaling parameter increases, the posterior uncertainty narrows, approaching the uncertainty estimates produced by the plug-in method. Despite this, the SBC results indicated that fixing the scaling parameter to 1 appears appropriate, as it did not violate the self-consistency property of the model. Nevertheless, further work is needed to determine whether this choice is indeed optimal.

Secondly, the low rank $\mathbf{Q}$ method requires a more thorough investigation. We hypothesize that the coarser the approximation to the $\mathbf{Q}$ matrix, the more likely it becomes that the self-consistency of the model is violated. We have not properly explored and investigated the breakdown point of the low rank approximation of $\mathbf{Q}$. This breakdown point is likely influenced by several factors, including the smoothness of the random field and the relative proportion of variability in the response variable explained by the fixed covariates versus the random field. A deeper understanding of these dependencies is essential to ensure the robustness of the low rank approximation.
Thirdly, we used the empirical Bayes approach to fit the models in both the simulation experiments and  data application.
This implies that the $\mathbf{Q}$ matrix is computed at the mode of the first-stage model hyperparameters. If another numerical integration strategy is chosen by the user when implementing INLA, such as a grid approach, there will be several $\mathbf{Q}$ matrices, one for each of the integration point for the model hyperparameter. In this scenario, we propose the use of the weighted average of the $\mathbf{Q}$ matrices, where the weights are the same integration weights from the numerical integration used to compute the approximated posteriors of the latent parameters.

The SBC method is a computational method for testing the self-consistency property of Bayesian models. It is implemented in a specific Bayesian model, prior specification, and Bayesian inference algorithm. For this work, we validated simple two-stage spatial models. However, in practice, the models we investigate are more complex. For the toy models considered in this work, we have shown and illustrated that the crude plug-in method indeed underestimates the posterior uncertainty of the second-stage model parameters. For more complex models, this underestimation of the posterior uncertainty will also be highly likely true. Moreover, our results also showed that the resampling method is correct. We also think that the resampling method should also be able to compute the correct posterior uncertainty for more complex models. However, the only way to exactly know this is to implement the SBC method for every new Bayesian model specification, new prior specification, and new Bayesian algorithm. This aligns with the proposal in \citet{talts2018validating} that the SBC should be an integral part of a robust Bayesian workflow (\citet{gelman2020bayesian}). However, the SBC method is very computationally expensive and might not be a practical route in many contexts. Therefore, we propose that a more practical approach to perform a two-stage model analysis is to implement different uncertainty propagation approaches and compare the obtained posterior uncertainties. The crude plug-in method is definitely the easiest strategy, but the results of which should only be taken as an initial understanding of the model. A more comprehensive analysis should involve doing resampling and other approaches, such as the proposed $\mathbf{Q}$ uncertainty method when the Bayesian inference is done using INLA. Implementing different uncertainty propagation strategies allows an objective comparison of the estimated posterior uncertainties, which would then help uncover interesting model insights and guide both the statistical and practical interpretation of the results.

\vspace{5mm}

\section*{Statements and Declarations}
\textbf{Competing Interests}: The authors have no competing interests to declare that are relevant to the content of this article. All authors certify that they have no affiliations with or involvement in any organization or entity with any financial interest or non-financial interest in the subject matter or materials discussed in this manuscript.

\textbf{Funding}: No funding was received for conducting this study.








\bibliographystyle{plainnat}
\bibliography{sample}

\end{document}


\maketitle


\section{Formal proof}

\subsection{Proof of Theorem 3.1}

Let $\pi(\bm{\theta}_{1})=\pi(\bm{\theta}_{1,\bm{x}})\pi(\bm{\theta}_{1,\bm{\mathcal{D}}})$ be the prior model, $\pi(\bm{x}_1|\bm{\theta}_1)$ be the latent model, and $\pi(\bm{\mathcal{D}}_1|\bm{x}_1,\bm{\theta}_1)$ be the observation density or probability mass function. Let $\bm{\chi}_1$ be the latent space, $\bm{\Theta}_{1,\bm{x}}$ be the $\bm{\theta}_{1,\bm{x}}$-space, and $\bm{\Theta}_{1,\bm{\mathcal{D}}}$ be the $\bm{\theta}_{1,\bm{\mathcal{D}}}$-space, all continuous. 

Suppose $\bm{\theta}_{1,\bm{x}}\in \bm{\Theta}_{1,\bm{x}}$ is fixed. Let $\tilde{\bm{\theta}}_{1,\bm{\mathcal{D}}}$ be a sample from the $\tilde{\bm{\theta}}_{1,\bm{\mathcal{D}}}$-space, i.e., $\tilde{\bm{\theta}}_{1,\bm{\mathcal{D}}}\sim \pi(\bm{\theta}_{1,\bm{\mathcal{D}}})$, $\tilde{\bm{x}}_1$ be a sample from the latent field model, i.e., $\tilde{\bm{x}}_1\sim \pi(\bm{x}_1|\bm{\theta}_{1,\bm{x}},\tilde{\bm{\theta}}_{1,\bm{\mathcal{D}}})$, and $\tilde{\bm{\mathcal{D}}}_1$ a sample from the observation model, i.e., $\tilde{\bm{\mathcal{D}}}_1 \sim \pi(\bm{\mathcal{D}}_1|\tilde{\bm{x}}_1,\bm{\theta}_{1,\bm{x}},\tilde{\bm{\theta}}_{1,\bm{\mathcal{D}}})$. 

Suppose that the posteriors from applying the Bayesian algorithm are $\pi(\bm{x}_1|\tilde{\bm{\mathcal{D}}}_1)$ and $\pi(\bm{\theta}_{1,\bm{\mathcal{D}}}|\tilde{\bm{\mathcal{D}}}_1)$. Let $\{\bm{x}_{1,\ell}\}, \ell=1,\ldots,L$ be independent samples from the posterior distribution, i.e., $\begin{pmatrix}
    \bm{x}_{1,1} & \bm{x}_{1,2} & \ldots \bm{x}_{1,L}
    \end{pmatrix} \overset{\text{iid}}{\sim}\hat{\pi}(\bm{x}_1|\tilde{\bm{\mathcal{D}}}_1)$. Also, let $\begin{pmatrix}
    \bm{\theta}_{1,\bm{\mathcal{D}},1} & \bm{\theta}_{1,\bm{\mathcal{D}},2} & \ldots \bm{\theta}_{1,\bm{\mathcal{D}},L}
\end{pmatrix} \overset{\text{iid}}{\sim}\hat{\pi}(\bm{\theta}_{1,\bm{\mathcal{D}}}|\tilde{\bm{\mathcal{D}}}_1)$.

\subsubsection{Result (1)}

Let $f:\bm{\chi}_1 \rightarrow \mathbb{R}$.  We define the rank statistic for a specific data outcome of the Bayesian model as
\begin{equation*}
    r = \sum_{\ell=1}^L \mathbb{I}[f(\bm{x}_{1,\ell}) < f(\tilde{\bm{x}}_1)], \;\;\; \mathbb{I}[f(\bm{x}_{1,\ell}) < f(\tilde{\bm{x}}_1)]=
    \begin{cases}
        1 & \text{if } f(\bm{x}_{1,\ell})< f(\tilde{\bm{x}}_1)\\
        0 & \text{if } f(\bm{x}_{1,\ell})\geq f(\tilde{\bm{x}}_1)
    \end{cases}
\end{equation*}

For conciseness, let $f_{\ell} \equiv f(\bm{x}_{1,\ell})$ and $f \equiv f(\tilde{\bm{x}}_1)$. Also, let $\pi(f)$ and $\pi(f|\bm{\mathcal{D}}_1)$ be the pushforward probability density function of $\pi(\bm{x}_1|\bm{\theta}_{1,\bm{x}})$ and $\pi(\bm{x}_1|\bm{\mathcal{D}}_1)$, respectively. Suppose $p_{\ell}=\mathbb{P}(f_\ell < f), \ell = 1,\ldots,L$. Also, we assume the ordering $f_1 \leq f_2 \leq \cdots \leq f_L$.
We then have:
\begin{align*}
    \pi(r) &= \int d\bm{\theta}_{1,\bm{\mathcal{D}}}dfd\bm{\mathcal{D}}_1\pi(\bm{\mathcal{D}}_1,f,\bm{\theta}_{1,\bm{\mathcal{D}}}|\bm{\theta}_{1,\bm{x}})\binom{L}{r} \prod_{\ell=1}^r p_{\ell}\prod_{\ell=r+1}^L (1-p_{\ell}) \\
    &=\binom{L}{r} \int d\bm{\theta}_{1,\bm{\mathcal{D}}}dfd\bm{\mathcal{D}}_1\pi(\bm{\mathcal{D}}_1,f,\bm{\theta}_{1,\bm{\mathcal{D}}}|\bm{\theta}_{1,\bm{x}})\prod_{\ell=1}^r\Bigg[\int_{-\infty}^{f}\pi(f_{\ell}|\bm{\mathcal{D}}_1,f,\bm{\theta}_{1,\bm{\mathcal{D}}},\bm{\theta}_{1,\bm{x}}) df_{\ell}\Bigg] \times \\ &\;\;\;\;\;\;\;\;\;\;\;\;\;\;\;\;\;\;\;\;\;\;\;\;\;\;\;\;\;\;\;\;\;\;\;\;\;\;\;\;\;\;\;\;\;\;\;\;\;\;\;\;\;\;\;\prod_{\ell=r+1}^L\Bigg[\int_{f}^{\infty}\pi(f_{\ell}|\bm{\mathcal{D}}_1,f,\bm{\theta}_{1,\bm{\mathcal{D}}},\bm{\theta}_{1,\bm{x}}) df_{\ell}\Bigg]
\end{align*}

The probability measure for generating $f_{\ell}$ depends only on $\bm{\mathcal{D}}_1$ and is independent of the conditioning model configuration. Hence, we can write $\pi(f_{\ell}|\bm{\mathcal{D}}_1,f,\bm{\theta}_{1,\bm{\mathcal{D}}},\bm{\theta}_{1,\bm{x}})=\pi(f_{\ell}|\bm{\mathcal{D}}_1)=\pi(f_{\ell}|\bm{\mathcal{D}}_1,\bm{\theta}_{1,\bm{x}})$. Further, since the model used to simulate data and construct posterior distributions is the same, then we have $\pi(f_{\ell}|\bm{\mathcal{D}}_1,\bm{\theta}_{1,\bm{x}})=\pi(f'|\bm{\mathcal{D}}_1,\bm{\theta}_{1,\bm{x}}), \ell=1,\ldots,L$. This implies that
\begin{align*}
    \pi(r)
    &=\binom{L}{r} \int d\bm{\theta}_{1,\bm{\mathcal{D}}}dfd\bm{\mathcal{D}}_1\pi(\bm{\mathcal{D}}_1,f,\bm{\theta}_{1,\bm{\mathcal{D}}}|\bm{\theta}_{1,\bm{x}})\Bigg[\int_{-\infty}^{f}\pi(f'|\bm{\mathcal{D}}_1,\bm{\theta}_{1,\bm{x}}) df'\Bigg]^r\Bigg[1-\int_{-\infty}^{f}\pi(f'|\bm{\mathcal{D}}_1,\bm{\theta}_{1,\bm{x}}) df'\Bigg]^{L-r}
\end{align*}

We use the fact that 
\begin{align}
\pi(\bm{\mathcal{D}}_1,f,\bm{\theta}_{1,\bm{\mathcal{D}}}|\bm{\theta}_{1,\bm{x}})&=\pi(f,\bm{\theta}_{1,\bm{\mathcal{D}}}|\bm{\mathcal{D}}_1,\bm{\theta}_{1,\bm{x}})\pi(\bm{\mathcal{D}}_1|\bm{\theta}_{1,\bm{x}}) \\  &=\pi(f|\bm{\mathcal{D}}_1,\bm{\theta}_{1,\bm{x}})\pi(\bm{\theta}_{1,\bm{\mathcal{D}}}|\bm{\mathcal{D}}_1,\bm{\theta}_{1,\bm{x}})\pi(\bm{\mathcal{D}}_1|\bm{\theta}_{1,\bm{x}}) \label{eq:appendix_thm3.1_res1}\\
&=\pi(f|\bm{\mathcal{D}}_1,\bm{\theta}_{1,\bm{x}})\pi(\bm{\mathcal{D}}_1,\bm{\theta}_{1,\bm{\mathcal{D}}}|\bm{\theta}_{1,\bm{x}}).
\end{align}
Equation \eqref{eq:appendix_thm3.1_res1} is true since $f$ and $\bm{\theta}_{1,\bm{\mathcal{D}}}$ are $d$-separated given $\bm{\mathcal{D}}_1$. This yields
\begin{align*}
   \pi(r)&= \binom{L}{r} \int d\bm{\theta}_{1,\bm{\mathcal{D}}}dfd\bm{\mathcal{D}}_1 \pi(f|\bm{\mathcal{D}}_1,\bm{\theta}_{1,\bm{x}})\pi(\bm{\mathcal{D}}_1,\bm{\theta}_{1,\bm{\mathcal{D}}}|\bm{\theta}_{1,\bm{x}})\Bigg[\int_{-\infty}^{f}\pi(f'|\bm{\mathcal{D}}_1,\bm{\theta}_{1,\bm{x}}) df'\Bigg]^r\times\\
   &\quad\quad\quad\quad\quad\quad\quad\quad\quad\quad\quad\quad\quad\quad\quad\quad\quad\quad \Bigg[1-\int_{-\infty}^{f}\pi(f'|\bm{\mathcal{D}}_1,\bm{\theta}_{1,\bm{x}}) df'\Bigg]^{L-r}  \\
    &=  \binom{L}{r} \int d\bm{\theta}_{1,\bm{\mathcal{D}}}d\bm{\mathcal{D}}_1 \pi(\bm{\mathcal{D}}_1,\bm{\theta}_{1,\bm{\mathcal{D}}}|\bm{\theta}_{1,\bm{x}})\int df \pi(f|\bm{\mathcal{D}}_1,\bm{\theta}_{1,\bm{x}})\Bigg[\int_{-\infty}^{f}\pi(f'|\bm{\mathcal{D}}_1,\bm{\theta}_{1,\bm{x}}) df'\Bigg]^r\times\\
    &\quad\quad\quad\quad\quad\quad\quad\quad\quad\quad\quad\quad\quad\quad\quad\quad\quad\quad \Bigg[1-\int_{-\infty}^{f}\pi(f'|\bm{\mathcal{D}}_1,\bm{\theta}_{1,\bm{x}}) df'\Bigg]^{L-r}
\end{align*}
Let $u = \int_{-\infty}^{f}\pi(f'|\bm{\mathcal{D}}_1,\bm{\theta}_{1,\bm{x}})df'$, so that $du=\pi(f|\bm{\mathcal{D}}_1,\bm{\theta}_{1,\bm{x}})df$. Thus,
\begin{align*}
    \pi(r) &= \binom{L}{r} \int d\bm{\theta}_{1,\bm{\mathcal{D}}}d\bm{\mathcal{D}}_1 \pi(\bm{\mathcal{D}}_1,\bm{\theta}_{1,\bm{\mathcal{D}}}|\bm{\theta}_{1,\bm{x}})\int du (u)^r(1-u)^{L-r}\\
    &= \binom{L}{r} B(r+1,L-r+1)\\
    &= \dfrac{1}{L+1}
\end{align*}
\qedsymbol

\subsubsection{Result (2)}

Let $f:\bm{\Theta}_{1,\bm{\mathcal{D}}} \rightarrow \mathbb{R}$.  We define the rank statistic for a specific data outcome of the Bayesian model as
\begin{equation*}
    r = \sum_{\ell=1}^L \mathbb{I}[f(\bm{\theta}_{1,\bm{\mathcal{D}},\ell}) < f(\tilde{\bm{\theta}}_{1,\bm{\mathcal{D}}})], \;\;\; \mathbb{I}[f(\bm{\theta}_{1,\bm{\mathcal{D}},\ell}) < f(\tilde{\bm{\theta}}_{1,\bm{\mathcal{D}}})]=
    \begin{cases}
        1 & \text{if } f(\bm{\theta}_{1,\bm{\mathcal{D}},\ell})< f(\tilde{\bm{\theta}}_{1,\bm{\mathcal{D}}})\\
        0 & \text{if } f(\bm{\theta}_{1,\bm{\mathcal{D}},\ell})\geq f(\tilde{\bm{\theta}}_{1,\bm{\mathcal{D}}})
    \end{cases}
\end{equation*}

For conciseness, let $f_{\ell} \equiv f(\bm{\theta}_{1,\bm{\mathcal{D}},\ell})$ and $f \equiv f(\tilde{\bm{\theta}}_{1,\bm{\mathcal{D}}})$. Also, let $\pi(f)$ and $\pi(f|\bm{\mathcal{D}}_1)$ be the pushforward probability density function of $\pi(\bm{\theta}_{1,\bm{\mathcal{D}}}|\bm{\theta}_{1,\bm{x}})=\pi(\bm{\theta}_{1,\bm{\mathcal{D}}})$ and $\pi(\bm{\theta}_{1,\bm{\mathcal{D}}}|\bm{\mathcal{D}}_1)$, respectively. Suppose $p_{\ell}=\mathbb{P}(f_\ell < f), \ell = 1,\ldots,L$. Also, we assume the ordering $f_1 \leq f_2 \leq \cdots \leq f_L$.

We can write the density of the rank statistic as
\begin{align*}
    \pi(r) &= \int dfd\bm{x}_1d\bm{\mathcal{D}}_1\pi(\bm{\mathcal{D}}_1,\bm{x}_1,f|\bm{\theta}_{1,\bm{x}})\binom{L}{r} \prod_{\ell=1}^r p_{\ell}\prod_{\ell=r+1}^L (1-p_{\ell}).
\end{align*}

We use the fact that
\begin{align}
    \pi(\bm{\mathcal{D}}_1,\bm{x}_1,f|\bm{\theta}_{1,\bm{x}})&=\pi(f,\bm{x}_1|\bm{\mathcal{D}}_1,\bm{\theta}_{1,\bm{x}})\pi(\bm{\mathcal{D}}_1|\bm{\theta}_{1,\bm{x}})\\
    &=\pi(f|\bm{\mathcal{D}}_1,\bm{\theta}_{1,\bm{x}})\pi(\bm{x}_1|\bm{\mathcal{D}}_1,\bm{\theta}_{1,\bm{x}})\pi(\bm{\mathcal{D}}_1|\bm{\theta}_{1,\bm{x}}) \label{eq:appendix_thm3.1_res2_a}\\
    &=\pi(f|\bm{\mathcal{D}}_1,\bm{\theta}_{1,\bm{x}})\pi(\bm{\mathcal{D}}_1,\bm{x}_1|\bm{\theta}_{1,\bm{x}}) \label{eq:appendix_thm3.1_res2_b}
\end{align}

Equation \eqref{eq:appendix_thm3.1_res2_a} is true since $f$ and $\bm{x}_1$ are $d$-separated given $\bm{\mathcal{D}}_1$.


This yields
\begin{align*}
   \pi(r) &=  \int \binom{L}{r}dfd\bm{x}_1d\bm{\mathcal{D}}_1\pi(f|\bm{\mathcal{D}}_1)\pi(\bm{\mathcal{D}}_1,\bm{x}_1|\bm{\theta}_{1,\bm{x}})\prod_{\ell=1}^r\Bigg[\int_{-\infty}^{f}\pi(f_{\ell}|\bm{\mathcal{D}}_1,f,\bm{x}_1,\bm{\theta}_{1,\bm{x}}) df_{\ell}\Bigg] \times  \\
   &\;\;\;\;\;\;\;\;\;\;\;\;\;\;\;\;\;\;\;\;\;\;\;\; \prod_{\ell=r+1}^L\Bigg[\int_{f}^{\infty}\pi(f_{\ell}|\bm{\mathcal{D}}_1,f,\bm{x}_1,\bm{\theta}_{1,\bm{x}})  df_{\ell}\Bigg]
\end{align*}
Similar to the argument in Result 1, we have $\pi(f_{\ell}|\bm{\mathcal{D}}_1,f,\bm{x}_1,\bm{\theta}_{1,\bm{x}})=\pi(f_{\ell}|\bm{\mathcal{D}}_1)=\pi(f_{\ell}|\bm{\mathcal{D}}_1,\bm{\theta}_{1,\bm{x}}), \ell=1,\ldots,L$. Also, we have $\pi(f_{\ell}|\bm{\mathcal{D}}_1,\bm{\theta}_{1,\bm{x}})=\pi(f'|\bm{\mathcal{D}}_1,\bm{\theta}_{1,\bm{x}}), \ell=1,\ldots,L$. This yields 
\begin{align*}
   \pi(r) &= \binom{L}{r}\int dfd\bm{x}_1d\bm{\mathcal{D}}_1\pi(f|\bm{\mathcal{D}}_1,\bm{\theta}_{1,\bm{x}})\pi(\bm{\mathcal{D}}_1,\bm{x}_1|\bm{\theta}_{1,\bm{x}})\prod_{\ell=1}^r\Bigg[\int_{-\infty}^{f}\pi(f'|\bm{\mathcal{D}}_1,\bm{\theta}_{1,\bm{x}}) df'\Bigg]^r \times\\
   &\quad\quad\quad\quad\quad\quad\quad\quad\quad\quad\quad\quad\quad\quad\quad\quad\quad\quad \Bigg[1-\int_{-\infty}^{f}\pi(f'|\bm{\mathcal{D}}_1,\bm{\theta}_{1,\bm{x}}) df'\Bigg]^{L-r} \\
    &=  \binom{L}{r} \int d\bm{x}_1d\bm{\mathcal{D}}_1\pi(\bm{\mathcal{D}}_1,\bm{x}_1|\bm{\theta}_{1,\bm{x}})\int df \pi(f|\bm{\mathcal{D}}_1,\bm{\theta}_{1,\bm{x}})\Bigg[\int_{-\infty}^{f}\pi(f'|\bm{\mathcal{D}}_1,\bm{\theta}_{1,\bm{x}}) df'\Bigg]^r \times \\
    &\quad\quad\quad\quad\quad\quad\quad\quad\quad\quad\quad\quad\quad\quad\quad\quad\quad\quad \Bigg[1-\int_{-\infty}^{f}\pi(f'|\bm{\mathcal{D}}_1,\bm{\theta}_{1,\bm{x}}) df'\Bigg]^{L-r}
\end{align*}

Let $u = \int_{-\infty}^{f}\pi(f'|\bm{\mathcal{D}}_1,\bm{\theta}_{1,\bm{x}})df'$, so that $du=\pi(f|\bm{\mathcal{D}}_1,\bm{\theta}_{1,\bm{x}})df$. Thus,
\begin{align*}
    \pi(r) &= \binom{L}{r} \int d\bm{x}_1d\bm{\mathcal{D}}_1\pi(\bm{\mathcal{D}}_1,\bm{x}_1|\bm{\theta}_{1,\bm{x}})\int du (u)^r(1-u)^{L-r}=\binom{L}{r} B(r+1,L-r+1)= \dfrac{1}{L+1}
\end{align*}
\qedsymbol

\subsection{Theorem 3.2}

\begin{theorem*}\label{thm:SBChierarhicalmodels2}
    Let $\pi(\bm{\theta}_1)$ be the prior model, $\pi(\bm{x}_1|\bm{\theta}_1)$ be the latent model, and $\pi(\bm{\mathcal{D}}_1|\bm{x}_1,\bm{\theta}_1)$ be the observation density or probability mass function. Let $\bm{\chi}_1$ be the latent space, and $\bm{\Theta}_1$ be the $\bm{\theta}_1$-space, and that both are continuous. Suppose $\bm{\theta}_1\in\bm{\Theta}_1$ is fixed. Let $\tilde{\bm{x}}_1$ be a sample from the latent field model, i.e., $\tilde{\bm{x}}_1\sim \pi(\bm{x}_1|\bm{\theta}_1)$, and $\tilde{\bm{\mathcal{D}}}_1$ a sample from the observation model, i.e., $\tilde{\bm{\mathcal{D}}}_1 \sim \pi(\bm{\mathcal{D}}_1|\tilde{\bm{x}}_1,\bm{\theta}_1,\mathbf{Z}_1)$. Suppose that the approximate posterior from applying the Bayesian algorithm is $\hat{\pi}(\bm{x}_1|\tilde{\bm{\mathcal{D}}}_1)$. Let $\{\bm{x}_{1,\ell}\}, \ell=1,\ldots,L$ be independent samples from the posterior distribution, i.e., $\begin{pmatrix}
    \bm{x}_{1,1} & \bm{x}_{1,2} & \ldots \bm{x}_{1,L}
\end{pmatrix} \overset{\text{iid}}{\sim}\hat{\pi}(\bm{x}_1|\tilde{\bm{\mathcal{D}}}_1)$. For any unidimensional function $f: \bm{\chi}_1\rightarrow \mathbb{R}$, the distribution of the rank statistic is given by
\begin{equation*}
    r = \sum_{\ell=1}^L \mathbb{I}\big[f(\bm{x}_{1,\ell}) < f(\tilde{\bm{x}}_1)\big], \;\;\; \mathbb{I}\big[f(\bm{x}_{1,\ell}) < f(\tilde{\bm{x}}_1)\big]=
    \begin{cases}
        1 & \text{if } f(\bm{x}_{1,\ell})< f(\tilde{\bm{x}}_1)\\
        0 & \text{if } f(\bm{x}_{1,\ell})\geq f(\tilde{\bm{x}}_1)
    \end{cases}
\end{equation*}
is $\mathcal{U}(0,1,\ldots,L)$.

\end{theorem*}

\textbf{Proof of Theorem 3.2}
 
{Let $\pi(\bm{\theta}_1)$ be the prior model, $\pi(\bm{x}_1|\bm{\theta}_1)$ be the latent model, and $\pi(\bm{\mathcal{D}}_1|\bm{x}_1,\bm{\theta}_1)$ be the observation density or proability mass function. Let $\bm{\chi}_1$ be the latent space, and $\bm{\Theta}_1$ be the $\bm{\theta}_1$-space, and that both are continuous.

Suppose $\bm{\theta}_1\in\bm{\Theta}_1$ is fixed. Let $\tilde{\bm{x}}_1$ be a sample from the latent field model, i.e., $\tilde{\bm{x}}_1\sim \pi(\bm{x}_1|\bm{\theta}_1)$, and $\tilde{\bm{\mathcal{D}}}_1$ a sample from the observation model, i.e., $\tilde{\bm{\mathcal{D}}}_1 \sim \pi(\bm{\mathcal{D}}_1|\tilde{\bm{x}}_1,\bm{\theta}_1)$. Note that we hold $\bm{\theta}_1$ fixed when generating the data replicates, but we fit the model assuming that both $\bm{\theta}_1$ and $\bm{x}_1$ are unknown. Let $\{\bm{x}_{1,\ell}\}, \ell=1,\ldots,L$ be independent samples from the posterior distribution $\pi(\bm{x}_1|\bm{\mathcal{D}}_1)$, i.e., $\begin{pmatrix}
    \bm{x}_{1,1} & \bm{x}_{1,2} & \ldots \bm{x}_{1,L}
\end{pmatrix} \overset{\text{iid}}{\sim}\pi(\bm{x}_1|\bm{\mathcal{D}}_1)$. Let $f:\bm{\chi}_1 \rightarrow \mathbb{R}$.  We define the rank statistic for a specific data outcome of the Bayesian model as
\begin{equation*}
    r = \sum_{\ell=1}^L \mathbb{I}\big[f(\bm{x}_{1,\ell}) < f(\tilde{\bm{x}}_1)\big], \;\;\; \mathbb{I}\big[f(\bm{x}_{1,\ell}) < f(\tilde{\bm{x}}_1)\big]=
    \begin{cases}
        1 & \text{if } f(\bm{x}_{1,\ell})< f(\tilde{\bm{x}}_1)\\
        0 & \text{if } f(\bm{x}_{1,\ell})\geq f(\tilde{\bm{x}}_1)
    \end{cases}
\end{equation*}
For conciseness, let $f_{\ell} \equiv f(\bm{x}_{1,\ell})$ and $f \equiv f(\tilde{\bm{x}}_1)$. Also, let $\pi(f)$ and $\pi(f|\bm{\mathcal{D}}_1)$ be the pushforward probability density function of $\pi(\bm{x}_1|\bm{\theta}_1)$ and $\pi(\bm{x}_1|\bm{\mathcal{D}}_1)$, respectively. Suppose $p_{\ell}=\mathbb{P}(f_\ell < f), \ell = 1,\ldots,L$. Also we assume the ordering $f_1 \leq f_2 \leq \cdots \leq f_L$.
We then have:
\begin{align*}
    \pi(r) &= \int dfd\bm{\mathcal{D}}_1\pi(\bm{\mathcal{D}}_1,f|\bm{\theta}_1)\binom{L}{r} \prod_{\ell=1}^r p_{\ell}\prod_{\ell=r+1}^L (1-p_{\ell}) \\
    &= \binom{L}{r} \int dfd\bm{\mathcal{D}}_1\pi(\bm{\mathcal{D}}_1|\bm{\theta}_1) \pi(f|\bm{\mathcal{D}}_1,\bm{\theta}_1)\prod_{\ell=1}^r\Bigg[\int_{-\infty}^{f}\pi(f_{\ell}|\bm{\mathcal{D}}_1,f,\bm{\theta}_1)  df_{\ell}\Bigg]\prod_{\ell=r+1}^L\Bigg[1-\int_{-\infty}^{f}\pi(f_{\ell}|\bm{\mathcal{D}}_1,f,\bm{\theta}_1)df_{\ell} \Bigg].
\end{align*}
The probability measure for generating $f_{\ell}$ depends only on $\bm{\mathcal{D}}_1$ and is independent of the conditioning model configuration. Hence we can write $ \pi(f_{\ell}|\bm{\mathcal{D}}_1,f,\bm{\theta}_1) = \pi(f_{\ell}|\bm{\mathcal{D}}_1)= \pi(f_{\ell}|\bm{\mathcal{D}}_1,\bm{\theta}_1),\ell=1,\ldots,L.$ This implies that 
\begin{align*}   
\pi(r)&= \binom{L}{r} \int dfd\bm{\mathcal{D}}_1\pi(\bm{\mathcal{D}}_1|\bm{\theta}_1) \pi(f|\bm{\mathcal{D}}_1,\bm{\theta}_1)\prod_{\ell=1}^r\Bigg[\int_{-\infty}^{f}\pi(f_{\ell}|\bm{\mathcal{D}}_1,\bm{\theta}_1) df_{\ell}\Bigg]\prod_{\ell=r+1}^L\Bigg[1-\int_{-\infty}^{f}\pi(f_{\ell}|\bm{\mathcal{D}}_1 ,\bm{\theta}_1)df_{\ell} \Bigg].
\end{align*}
Further, since the model used to simulate data and construct posterior distributions is the same, then we have $\pi(f_{\ell}|\bm{\mathcal{D}}_1,\bm{\theta}_1)=\pi(f'|\bm{\mathcal{D}}_1,\bm{\theta}_1), \ell=1,\ldots,L$. Consequently, we have
\begin{align*}
\pi(r)&= \binom{L}{r} \int dfd\bm{\mathcal{D}}_1 \pi(\bm{\mathcal{D}}_1|\bm{\theta}_1)\pi(f|\bm{\mathcal{D}}_1,\bm{\theta}_1)\prod_{\ell=1}^r\Bigg[\int_{-\infty}^{f}\pi(f'|\bm{\mathcal{D}}_1,\bm{\theta}_1) df'\Bigg]\prod_{\ell=r+1}^L\Bigg[1-\int_{-\infty}^{f}\pi(f'|\bm{\mathcal{D}}_1,\bm{\theta}_1)df' \Bigg]\\
    &=  \binom{L}{r} \int d\bm{\mathcal{D}}_1\pi(\bm{\mathcal{D}}_1|\bm{\theta}_1)  \int df\pi(f|\bm{\mathcal{D}}_1,\bm{\theta}_1)\Bigg[\int_{-\infty}^{f}\pi(f'|\bm{\mathcal{D}}_1,\bm{\theta}_1) df'\Bigg]^r\Bigg[1-\int_{-\infty}^{f}\pi(f'|\bm{\mathcal{D}}_1,\bm{\theta}_1) df'\Bigg]^{L-r}.
\end{align*}
Let $u = \int_{-\infty}^{f}\pi(f'|\bm{\mathcal{D}}_1,\bm{\theta}_1)df'$, so that $du=\pi(f|\bm{\mathcal{D}}_1,\bm{\theta}_1)df$. 
This yields  
\begin{align*}
    \pi(r) &= \binom{L}{r}\int d\bm{\mathcal{D}}_1\pi(\bm{\mathcal{D}}_1|\bm{\theta}_1)\int du (u)^r(1-u)^{L-r}\\
    &= \binom{L}{r} B(r+1,L-r+1)=\dfrac{L!}{r!(L-r)!}\dfrac{r!(L-r)!}{(L+1)!}=\dfrac{1}{L+1}
\end{align*}}
\qedsymbol

\newpage

\section{Results of SBC for the Gaussian model (Section 4.1)}

\subsection{SBC results for the first-stage model (Section 4.1) using Algorithm 2}

\begin{figure}[H]
    \centering
    \includegraphics[width=.32\linewidth]{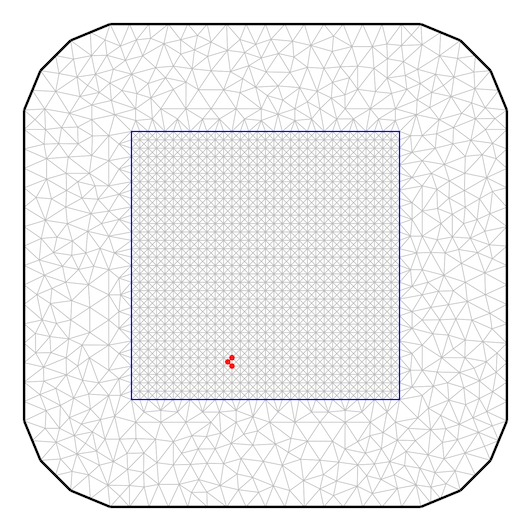}
    \caption{Results of the KS goodness-of-fit test for uniformity (at 10\% significance level) of the normalized ranks $p_k$ of the SPDE (mesh nodes) weights out of 1000 data replicates and using Algorithm 2. The red points show the mesh nodes which fail the KS test for uniformity}
\end{figure}

\begin{figure}[H]
    \centering
    \includegraphics[width=\linewidth]{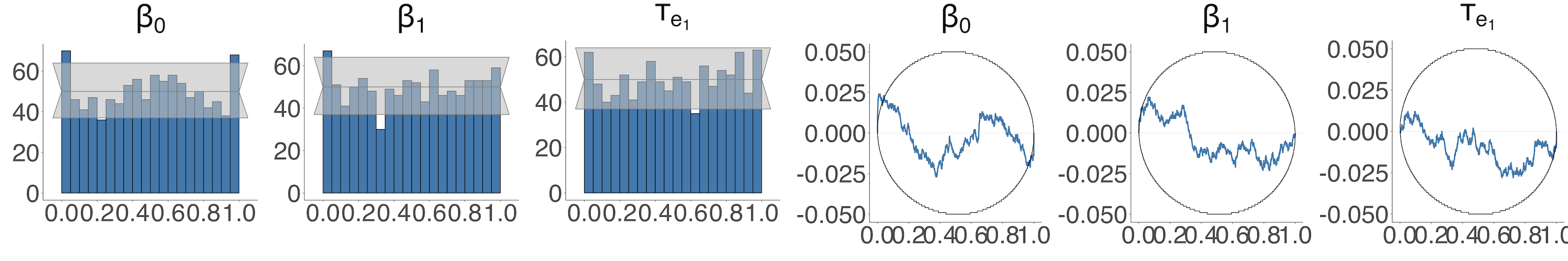}
    \caption{Histogram and ECDF difference plot of the normalized ranks $p_k$  for  $\beta_0$, $\beta_1$, and $\tau_{e_1}=1/\sigma^2_{e_1}$ out of 1000 data replicates using Algorithm 2}
  
\end{figure}

\begin{figure}[H]
    \centering
    \includegraphics[width=\linewidth]{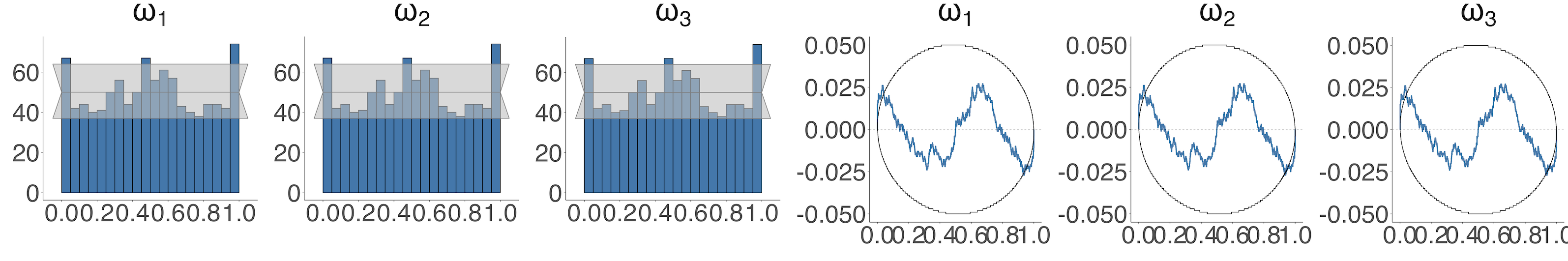}
    \caption{Histogram and ECDF difference plot of the normalized ranks $p_k$ for $\omega_1$, $\omega_2$, and $\omega_3$ out of 1000 data replicates using Algorithm 2}

\end{figure}

\begin{figure}[H]
    \centering
    \includegraphics[width=\linewidth]{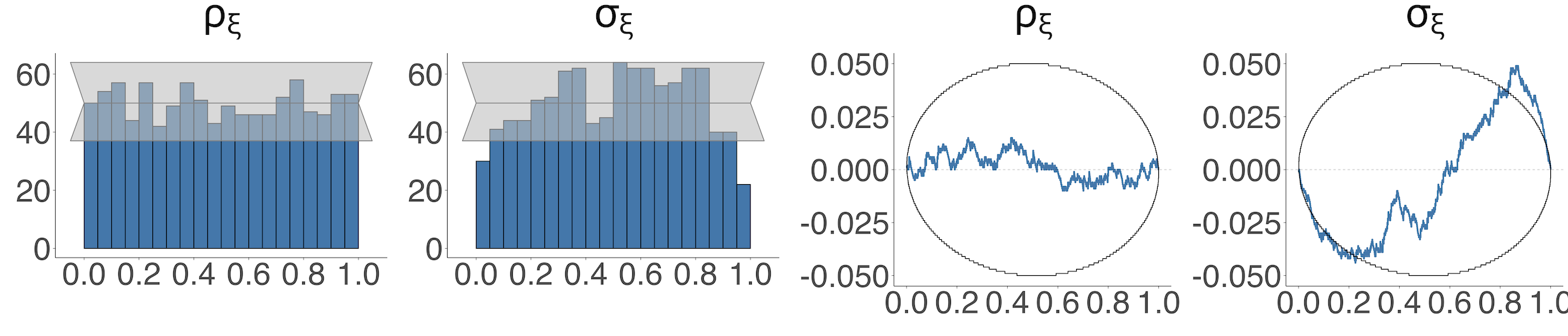}
    \caption{Histogram and ECDF difference plot of the normalized ranks $p_k$ for  $\rho_{\xi}$ and $\sigma_{\xi}$ out of 1000 data replicates using Algorithm 2}
\end{figure}


\subsection{SBC results for the first-stage model (Section 4.1) using Algorithm 3}

\begin{figure}[H]
    \centering
    \includegraphics[width=\linewidth]{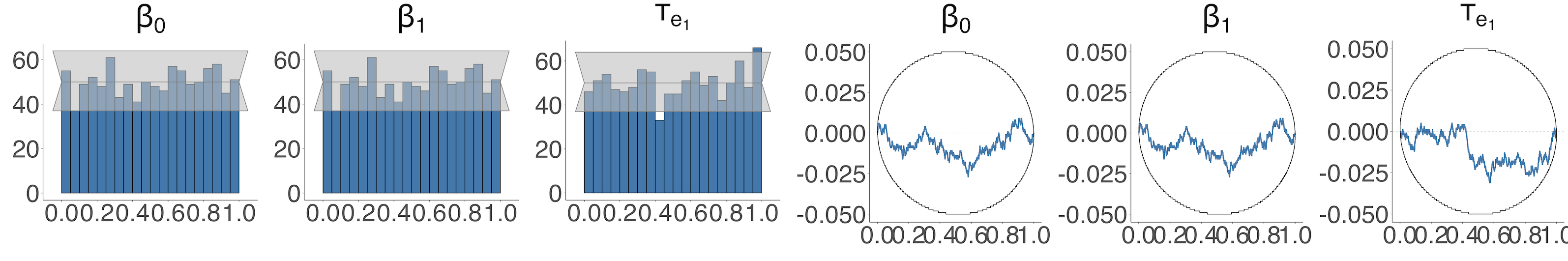}
    \caption{Histogram and ECDF difference plot of the normalized ranks $p_k$  for  $\beta_0$, $\beta_1$, and $\tau_{e_1}=1/\sigma^2_{e_1}$ out of 1000 data replicates using Algorithm 3 and using PC prior for the Mat\'ern parameters}
\end{figure}

\begin{figure}[H]
    \centering
    \includegraphics[width=\linewidth]{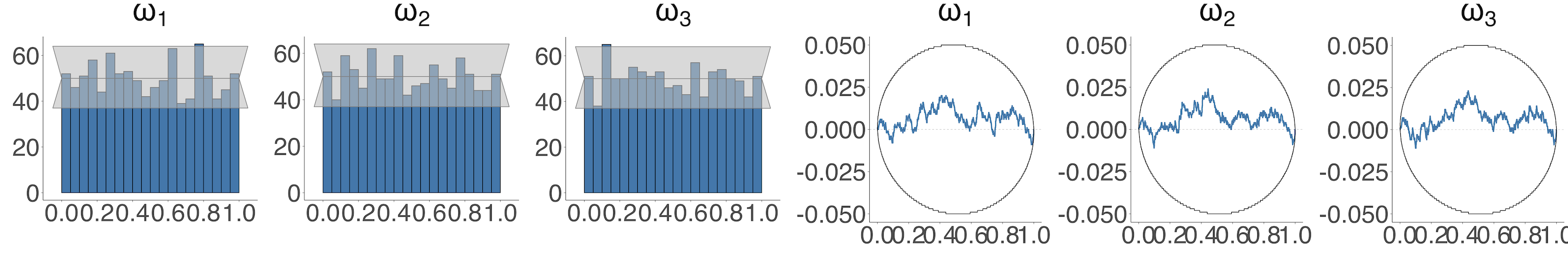}
    \caption{Histogram and ECDF difference plot of the normalized ranks $p_k$ for $\omega_1$, $\omega_2$, and $\omega_3$ out of 1000 data replicates using Algorithm 3 and using PC prior for the Mat\'ern parameters}
\end{figure}

\subsection{SBC results for $\gamma_0$ and $\gamma_1$ (Section 4.1) using Algorithm 2}

\begin{figure}[H]
    \centering
    \subfigure[Plug-in method]{\includegraphics[width=.71\linewidth]{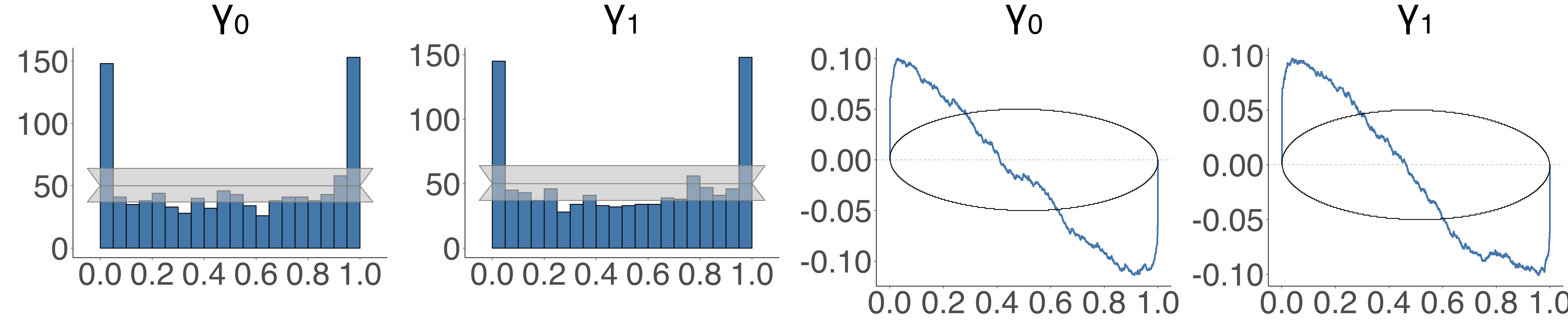}} 
    \subfigure[Resampling method]{\includegraphics[width=.71\linewidth]{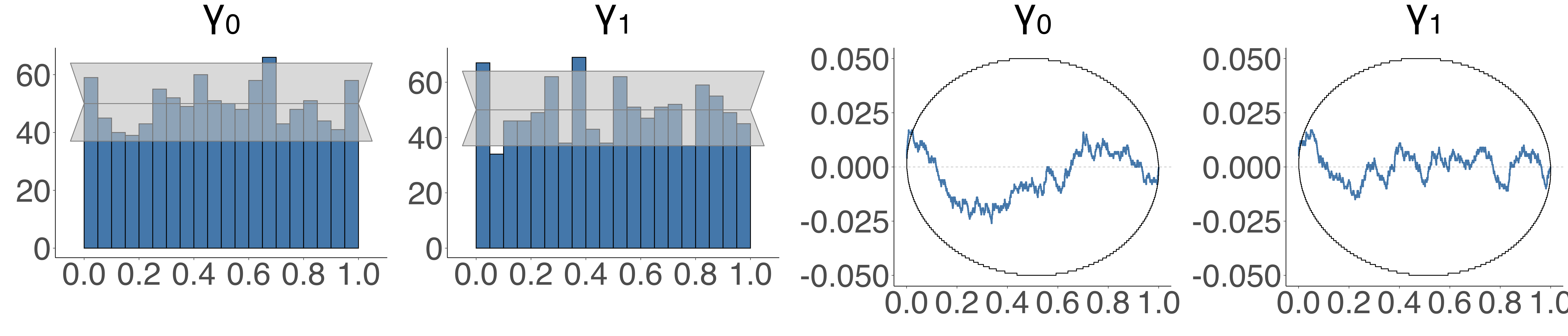}}
    \subfigure[Full $\mathbf{Q}$ uncertainty method]{\includegraphics[width=.71\linewidth]{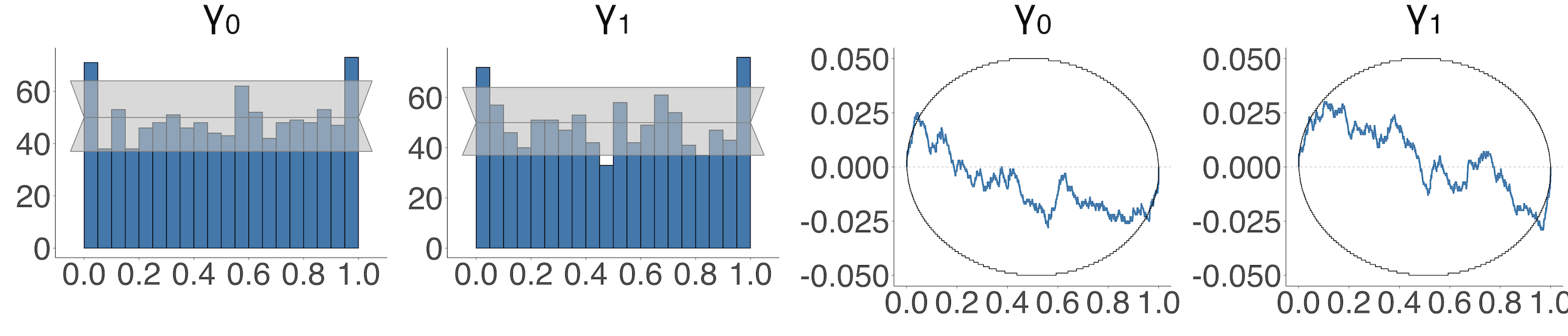}}
    \subfigure[Low rank $\mathbf{Q}$ uncertainty method (mesh A)]{\includegraphics[width=.71\linewidth]{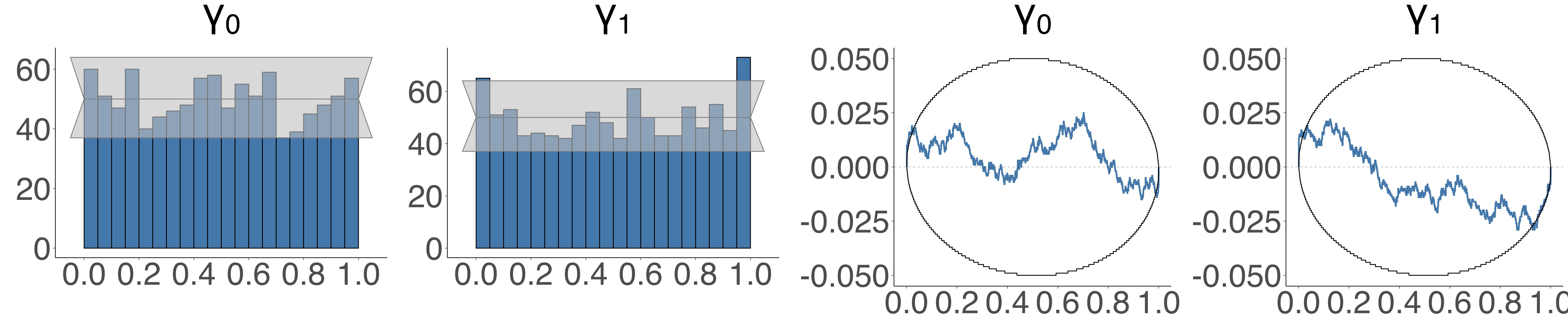}}
    \subfigure[Low rank $\mathbf{Q}$ uncertainty method (mesh B)]{\includegraphics[width=.71\linewidth]{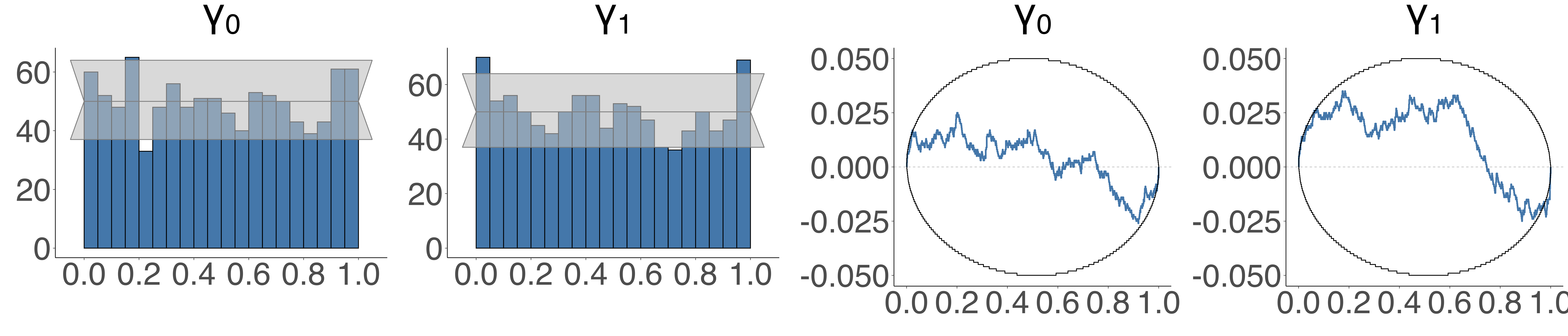}}
    \caption{Histogram and ECDF difference plot of the normalized ranks $p_k$ using Algorithm 2 for the second-stage model parameters $\gamma_0$ and $\gamma_1$ out of 1000 data replicates for the two-stage Gaussian spatial model (Section 4.1) using INLA-SPDE and with different approaches: (a) plug-in method (b) resampling method (c) full $\mathbf{Q}$ method (d) low rank $\mathbf{Q}$ method (mesh A) (e) low rank $\mathbf{Q}$ method (mesh B)}
\end{figure}

\subsection{SBC results for $\gamma_0$ and $\gamma_1$ (Section 4.1), Algorithm 3}

\begin{figure}[H]
    \centering
    \subfigure[Plug-in method]{\includegraphics[width=.71\linewidth]{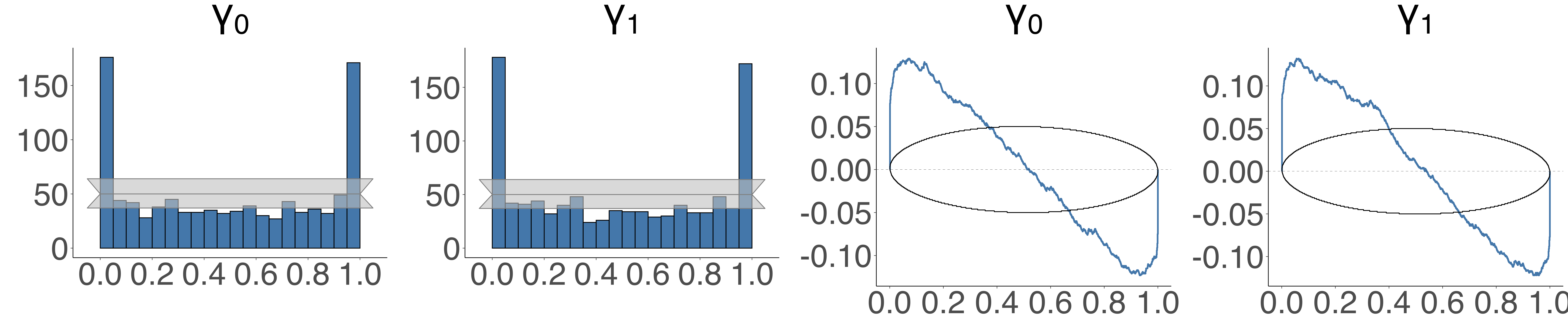}} 
    \subfigure[Resampling method]{\includegraphics[width=.71\linewidth]{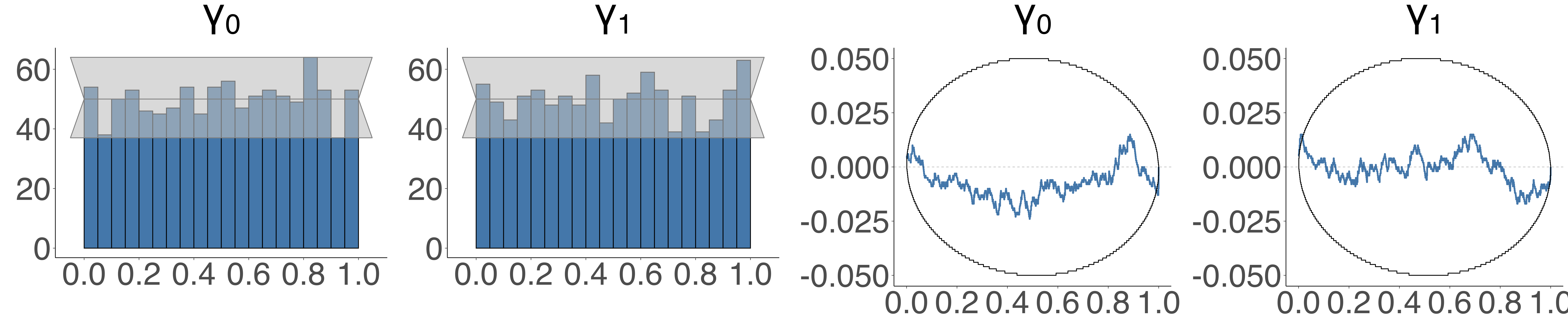}}
    \subfigure[Full $\mathbf{Q}$ uncertainty method]{\includegraphics[width=.71\linewidth]{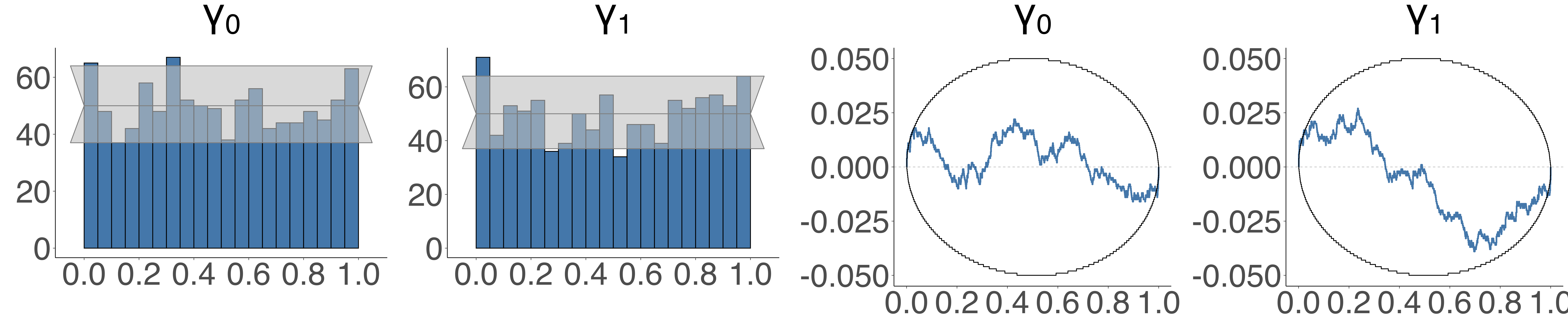}}
    \subfigure[Low rank $\mathbf{Q}$ uncertainty method (mesh A)]{\includegraphics[width=.71\linewidth]{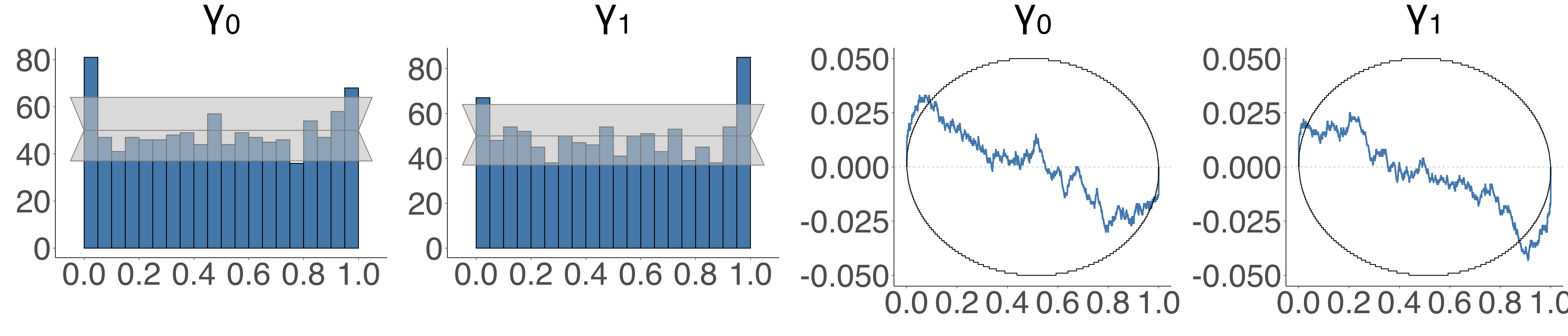}}
    \subfigure[Low rank $\mathbf{Q}$ uncertainty method (mesh B)]{\includegraphics[width=.71\linewidth]{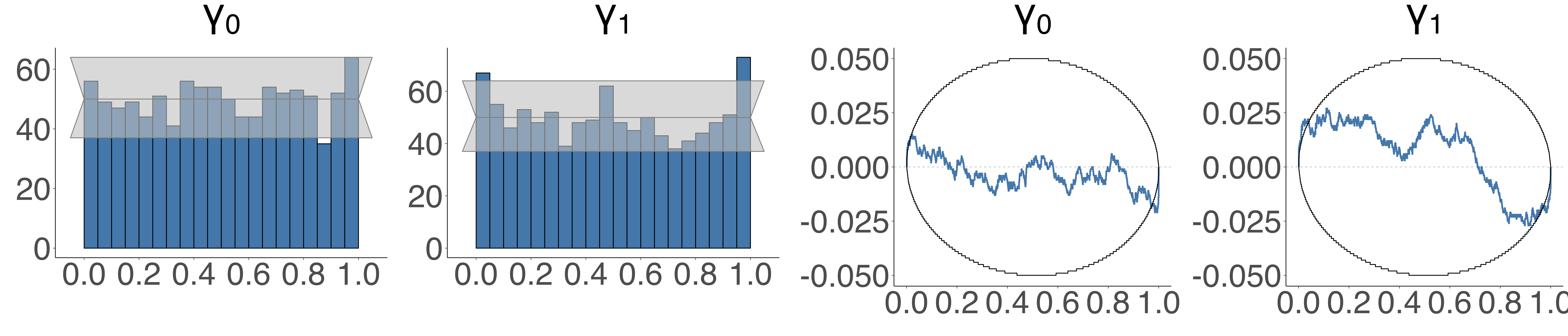}}
    \caption{Histogram and ECDF difference plot of the normalized ranks $p_k$ using Algorithm 3 for the second-stage model parameters $\gamma_0$ and $\gamma_1$ out of 1000 data replicates for the two-stage Gaussian spatial model (Section 4.1) using INLA-SPDE and with different approaches: (a) plug-in method (b) resampling method (c) full $\mathbf{Q}$ method (d) low rank $\mathbf{Q}$ method (mesh A) (e) low rank $\mathbf{Q}$ method (mesh B)}
\end{figure}

\subsection{Illustration with a simulated data }\label{supp:twostage_spatial_gaussian}

\begin{figure}[H]
    \centering
    \subfigure[Comparison with high log precision values]{\includegraphics[width=.49\linewidth]{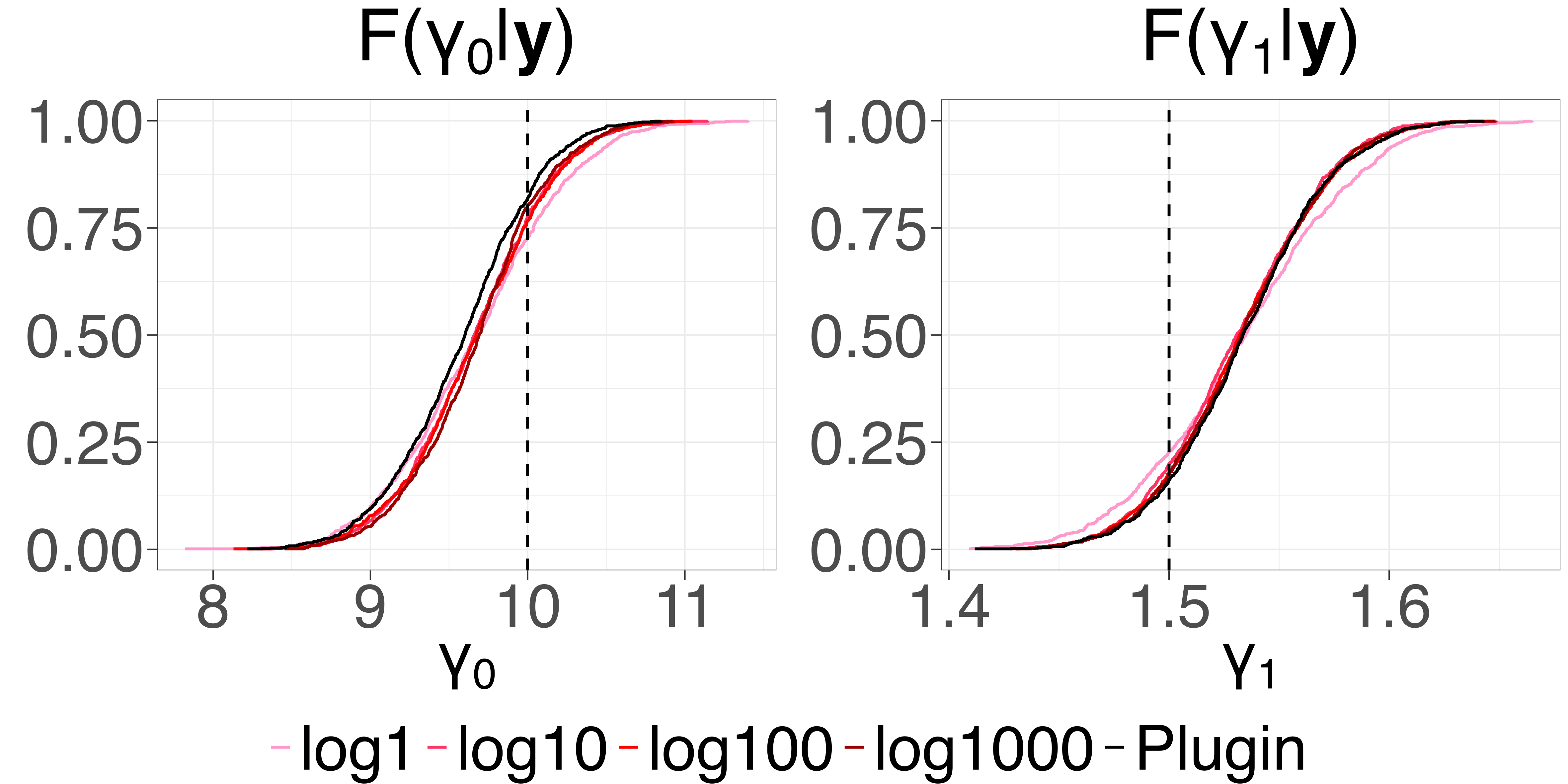}} 
    \subfigure[Comparison with low log precision values]{\includegraphics[width=.49\linewidth]{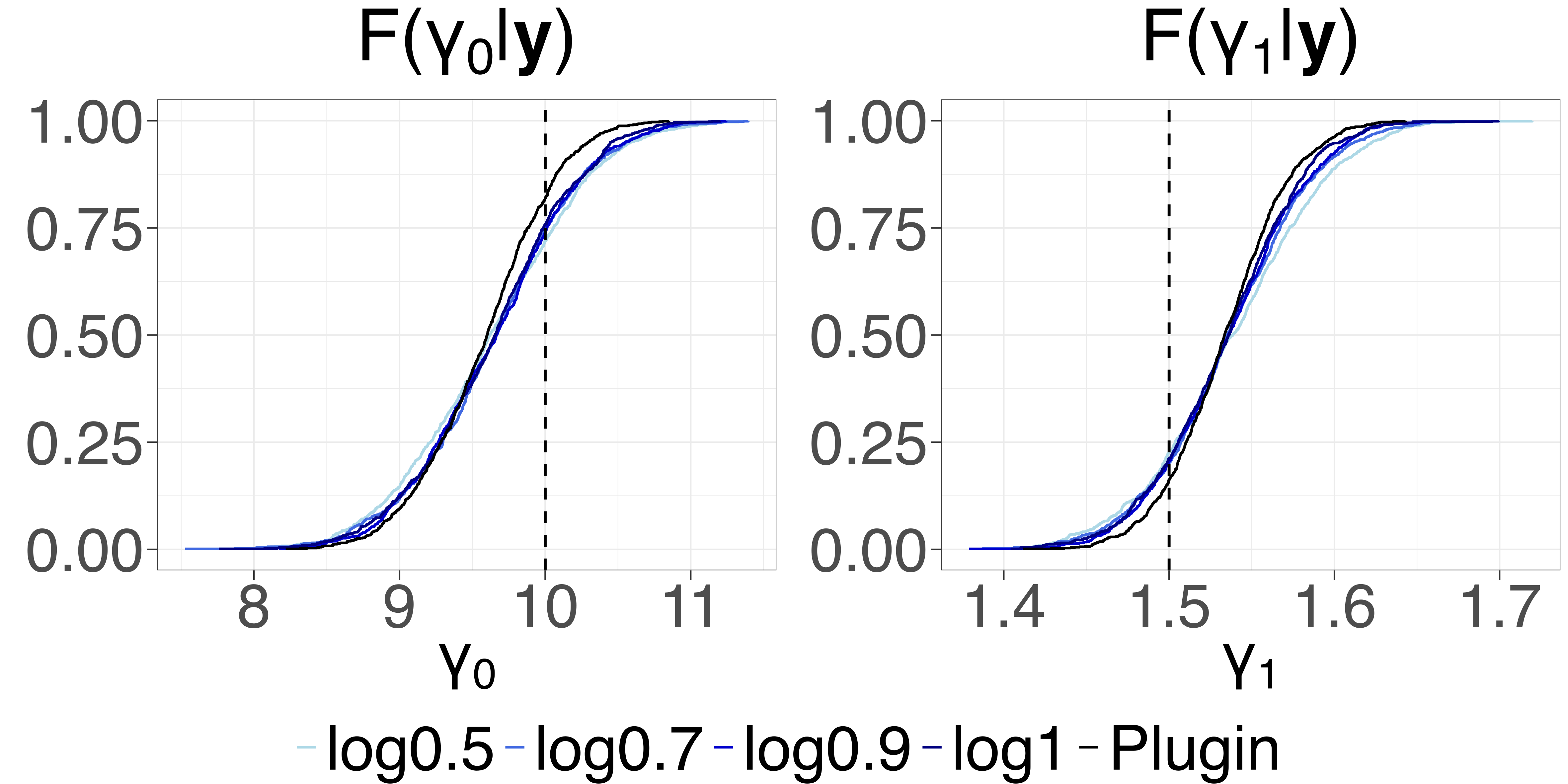}} 
    \caption{Comparison of the estimated posterior CDFs of the second-stage model parameters $\gamma_0$ and $\gamma_1$ for different values of the log precision of the error component in the low rank $\mathbf{Q}$ uncertainty method (mesh A) using the simulated data example in Section 4.1}
\end{figure}

\begin{figure}[H]
    \centering
    \subfigure[Comparison with high log precision values]{\includegraphics[width=.49\linewidth]{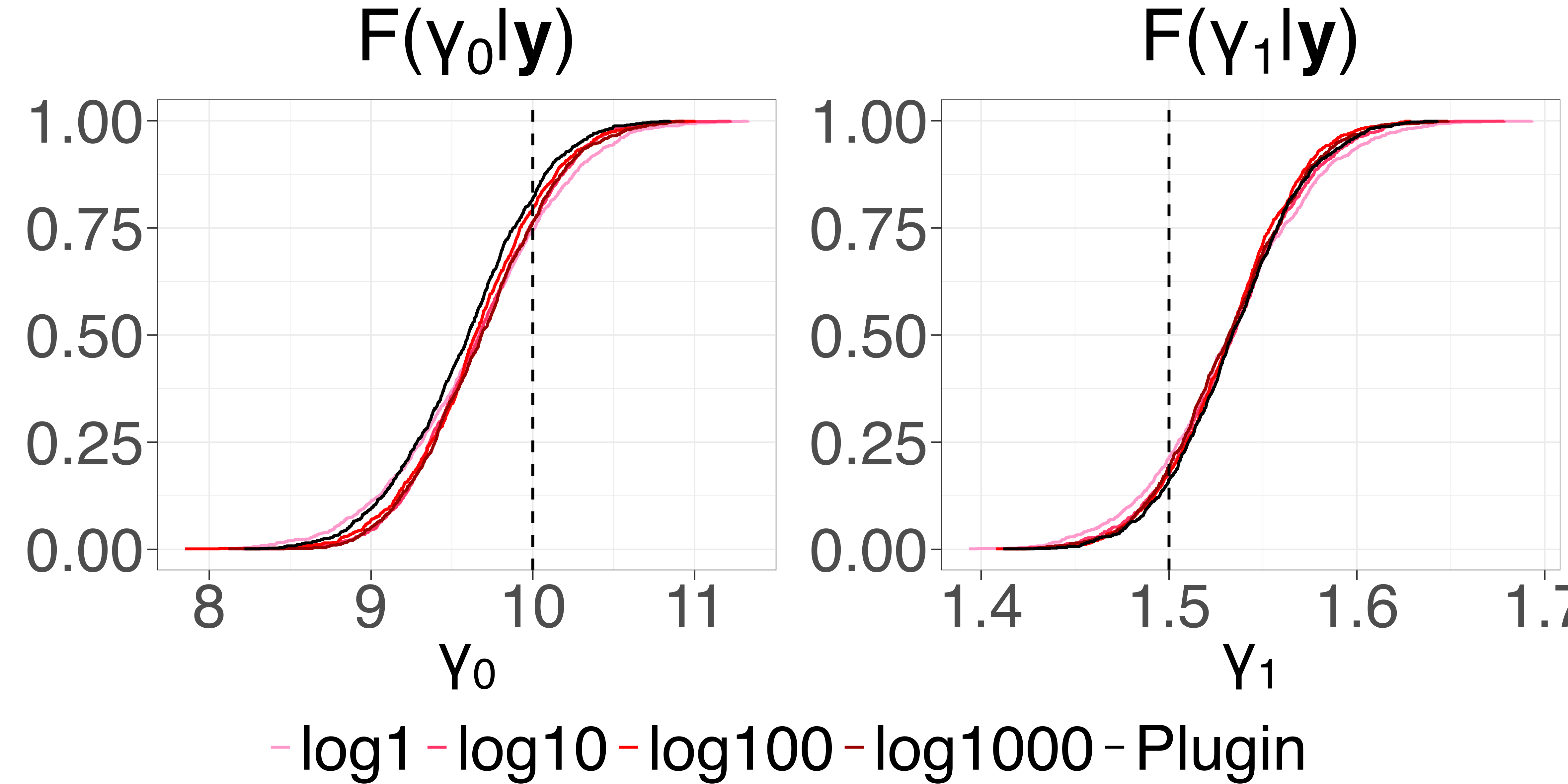}} 
    \subfigure[Comparison with low log precision values]{\includegraphics[width=.49\linewidth]{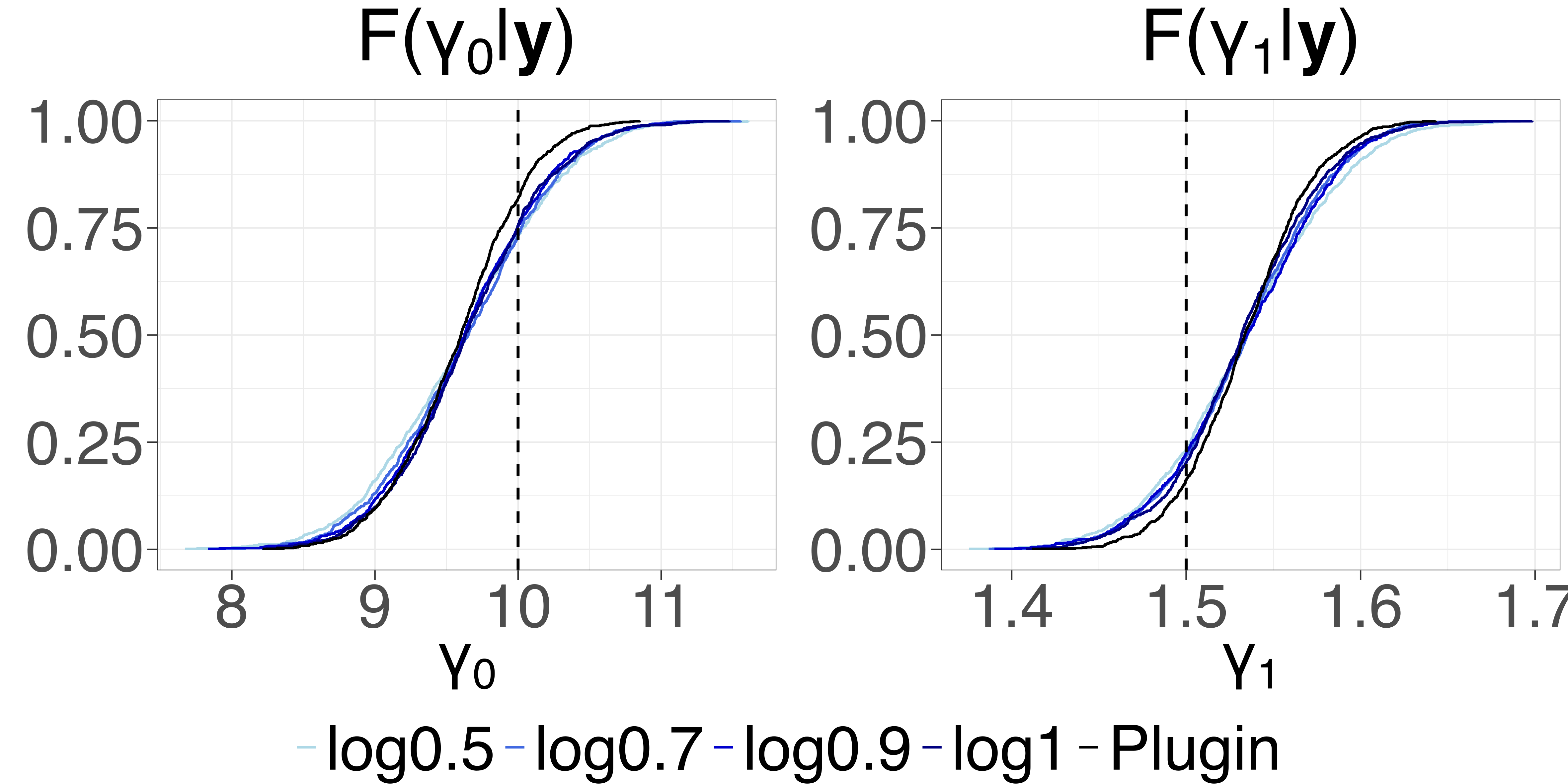}} 
    \caption{Comparison of the estimated posterior CDFs of the second-stage model parameters $\gamma_0$ and $\gamma_1$ for different values of the log precision of the error component in the low rank $\mathbf{Q}$ uncertainty method (mesh B) using the simulated data example in Section 4.1}
\end{figure}

\newpage

\section{Results of SBC for the Poisson model (Section 4.2)}

\subsection{SBC results for the first-stage model (Section 4.2) using Algorithm 2}

\begin{figure}[H]
    \centering
    \includegraphics[width=.32\linewidth]{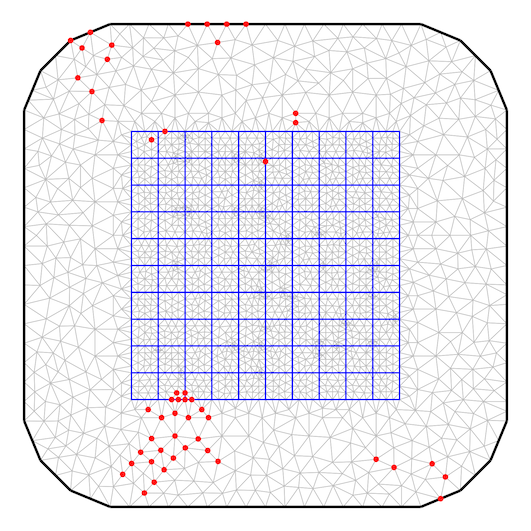}
    \caption{Results of the KS goodness-of-fit test for uniformity (at 10\% significance level) of the normalized ranks $p_k$ of the SPDE (mesh nodes) weights out of 1000 data replicates and using Algorithm 2. The red points show the mesh nodes which fail the KS test for uniformity}

\end{figure}

\begin{figure}[H]
    \centering
    \includegraphics[width=\linewidth]{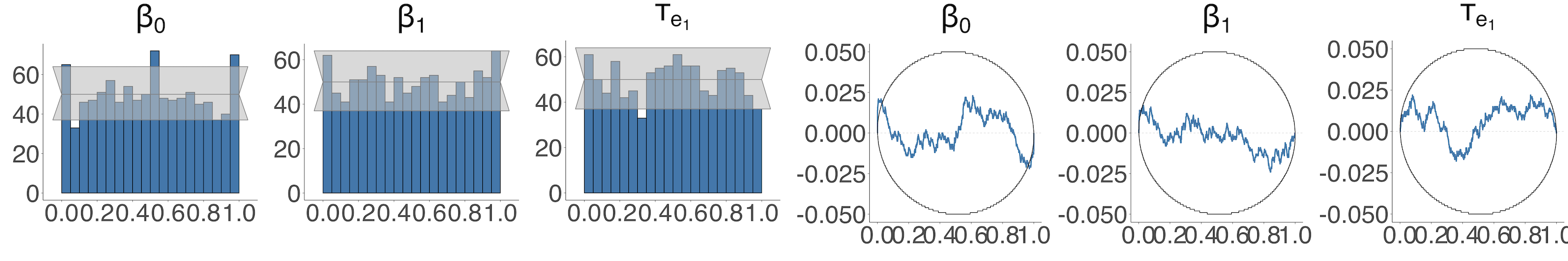}
    \caption{Histogram and ECDF difference plot of the normalized ranks $p_k$ for  $\beta_0$, $\beta_1$, and $\tau_{e_1}=1/\sigma^2_{e_1}$ out of 1000 data replicates using Algorithm 2}
\end{figure}

\begin{figure}[H]
    \centering
    \includegraphics[width=\linewidth]{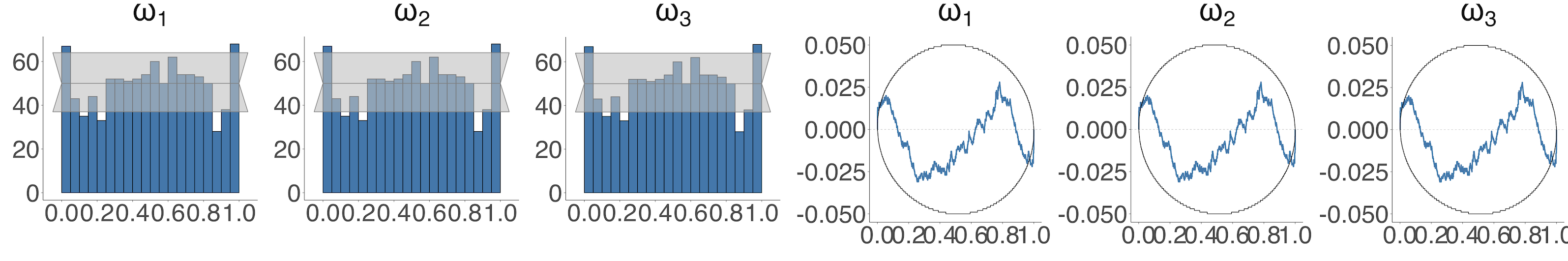}
    \caption{Histogram and ECDF difference plot of the normalized ranks $p_k$ for  $\omega_1$, $\omega_2$, and $\omega_3$ out of 1000 data replicates using Algorithm 2}
\end{figure}

\begin{figure}[H]
    \centering
    \includegraphics[width=\linewidth]{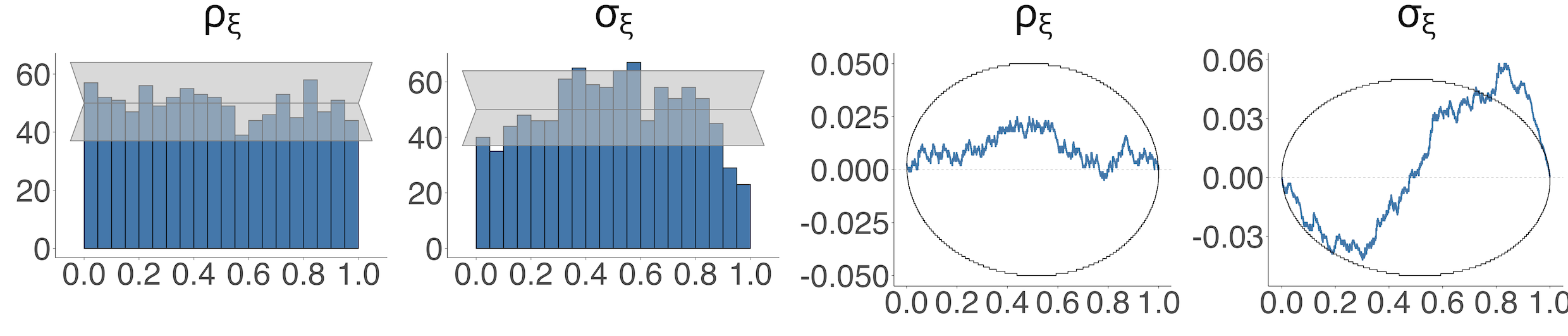}
    \caption{Histogram and ECDF difference plot of the normalized ranks $p_k$ for  $\rho_{\xi}$ and $\sigma_{\xi}$ out of 1000 data replicates using Algorithm 2}
\end{figure}


\subsection{SBC results for first-stage model (Section 4.2) using Algorithm 3}

\begin{figure}[H]
    \centering
    \includegraphics[width=\linewidth]{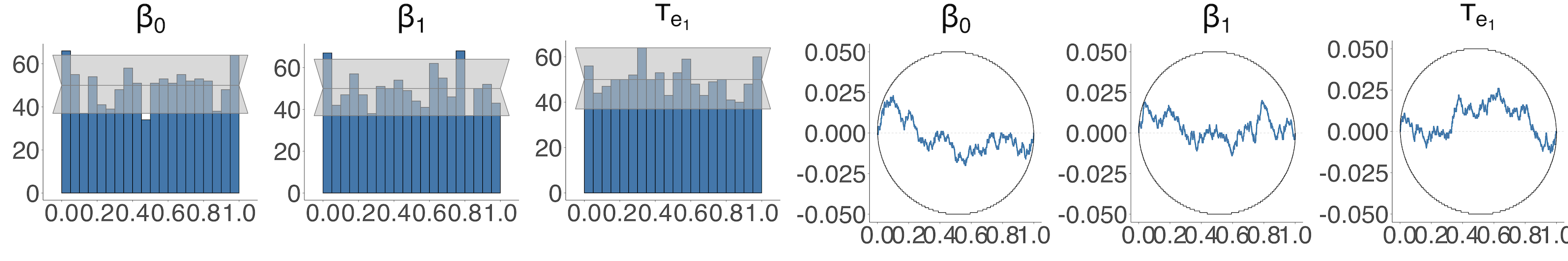}
    \caption{Histogram and ECDF difference plot of the normalized ranks $p_k$ for  $\beta_0$, $\beta_1$, and $\tau_{e_1}=1/\sigma^2_{e_1}$ out of 1000 data replicates and using PC prior for the Mat\'ern parameters using Algorithm 3}
\end{figure}

\begin{figure}[H]
    \centering
    \includegraphics[width=\linewidth]{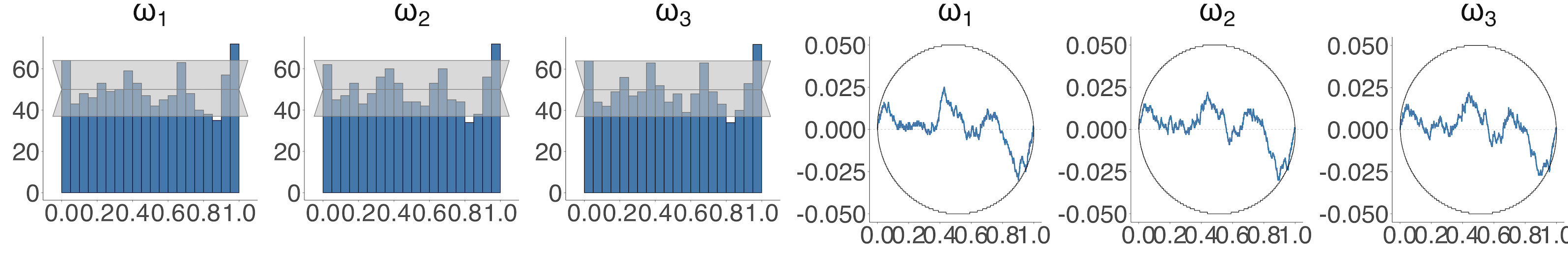}
    \caption{Histogram and ECDF difference plot of the normalized ranks $p_k$ for  $\omega_1$, $\omega_2$, and $\omega_3$ out of 1000 data replicates and using PC prior for the Mat\'ern parameters using Algorithm 3}
\end{figure}

\subsection{SBC results for $\gamma_0$ and $\gamma_1$ of the classical model specification (Section 4.2.1) using Algorithm 2}

\begin{figure}[H]
    \centering
    \subfigure[Plug-in method]{\includegraphics[width=.71\linewidth]{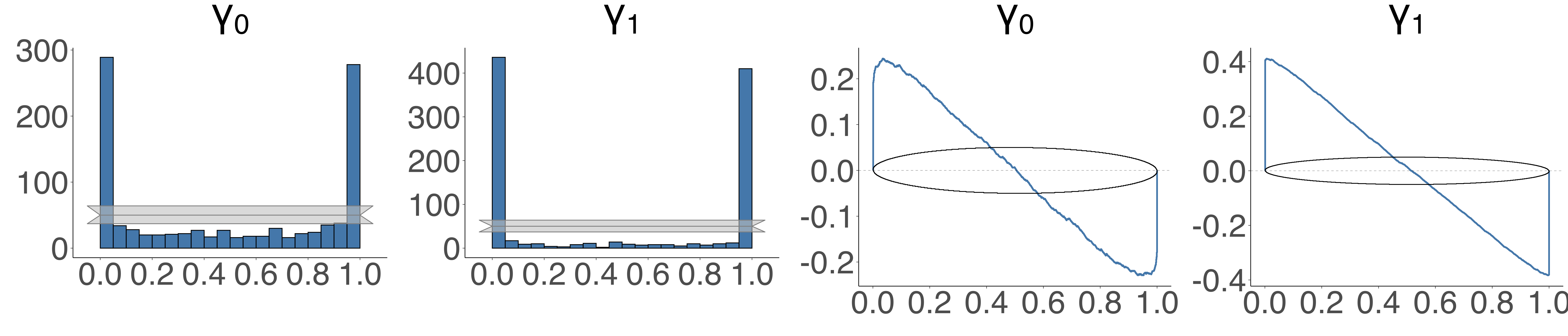}} 
    \subfigure[Resampling method]{\includegraphics[width=.71\linewidth]{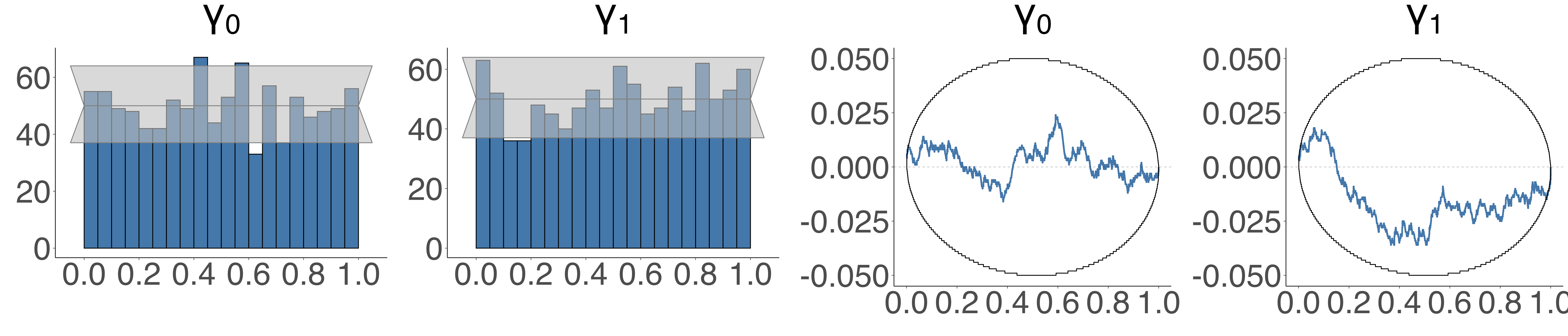}}
    \subfigure[Full $\mathbf{Q}$ uncertainty method]{\includegraphics[width=.71\linewidth]{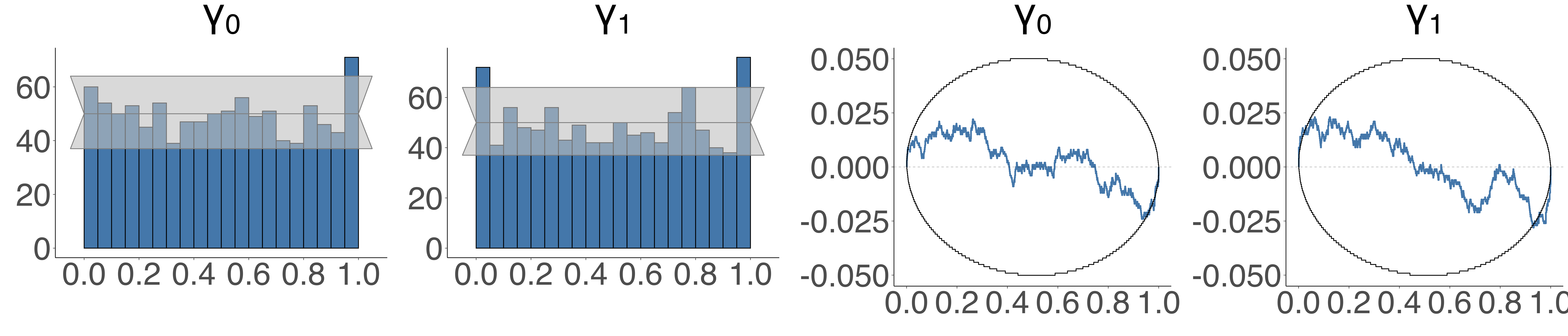}}
    \subfigure[Low rank $\mathbf{Q}$ uncertainty method (mesh A)]{\includegraphics[width=.71\linewidth]{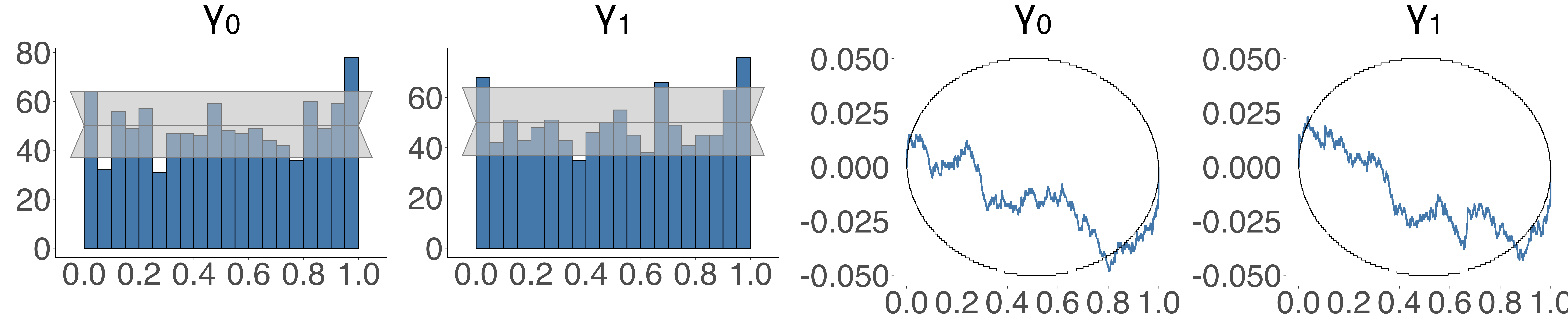}}
    \subfigure[Low rank $\mathbf{Q}$ uncertainty method (mesh B)]{\includegraphics[width=.71\linewidth]{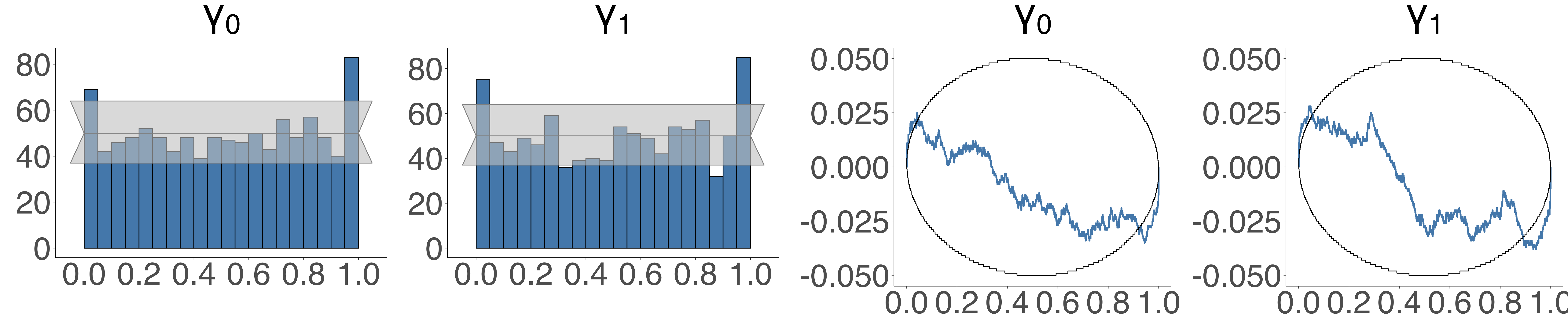}}
    \caption{Histogram and ECDF difference plot of the normalized ranks $p_k$ using Algorithm 2 for the second-stage model parameters $\gamma_0$ and $\gamma_1$ out of 1000 data replicates for the classical specification of the two-stage Poisson spatial model (Section 4.2.1) using INLA-SPDE and with different approaches: (a) plug-in method (b) resampling method (c) full $\mathbf{Q}$ method (d) low rank $\mathbf{Q}$ method (mesh A) (e) low rank $\mathbf{Q}$ method (mesh B)}
\end{figure}

\subsection{SBC results for $\gamma_0$ and $\gamma_1$ of the classical model specification (Section 4.2.2) using Algorithm 3}

\begin{figure}[H]
    \centering
    \subfigure[Plug-in method]{\includegraphics[width=.71\linewidth]{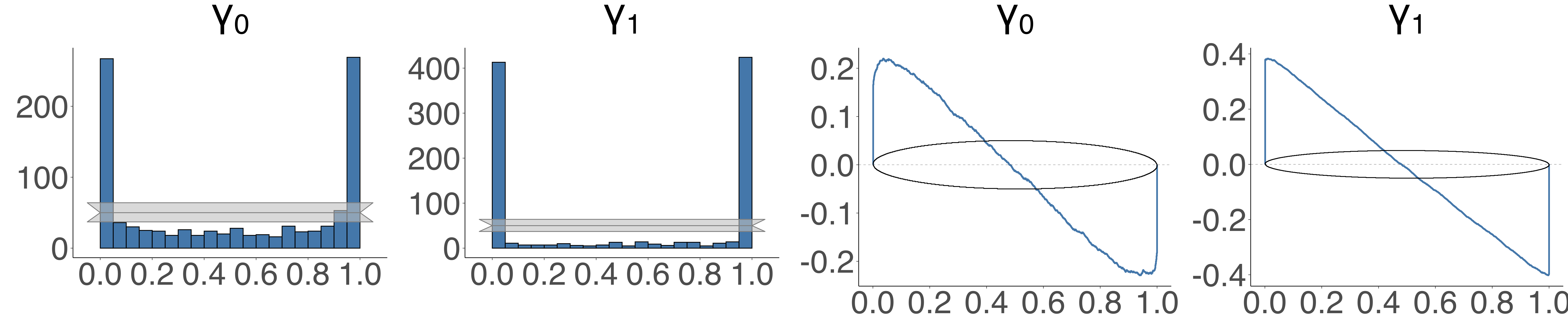}} 
    \subfigure[Resampling method]{\includegraphics[width=.71\linewidth]{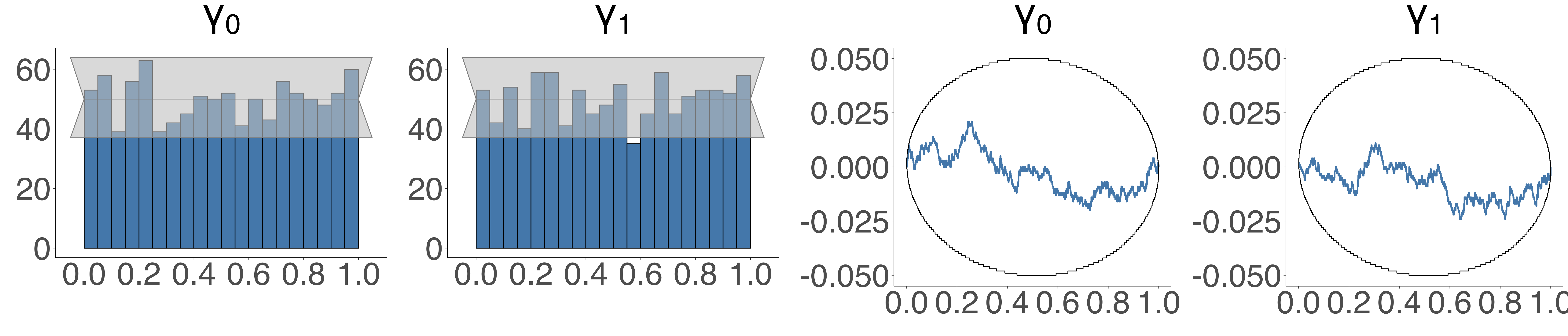}}
    \subfigure[Full $\mathbf{Q}$ uncertainty method]{\includegraphics[width=.71\linewidth]{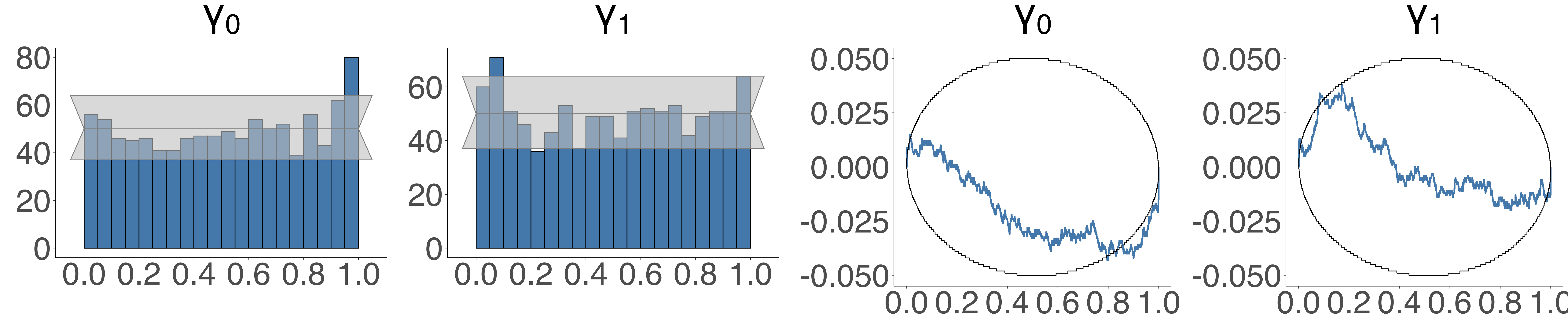}}
    \subfigure[Low rank $\mathbf{Q}$ uncertainty method (mesh A)]{\includegraphics[width=.71\linewidth]{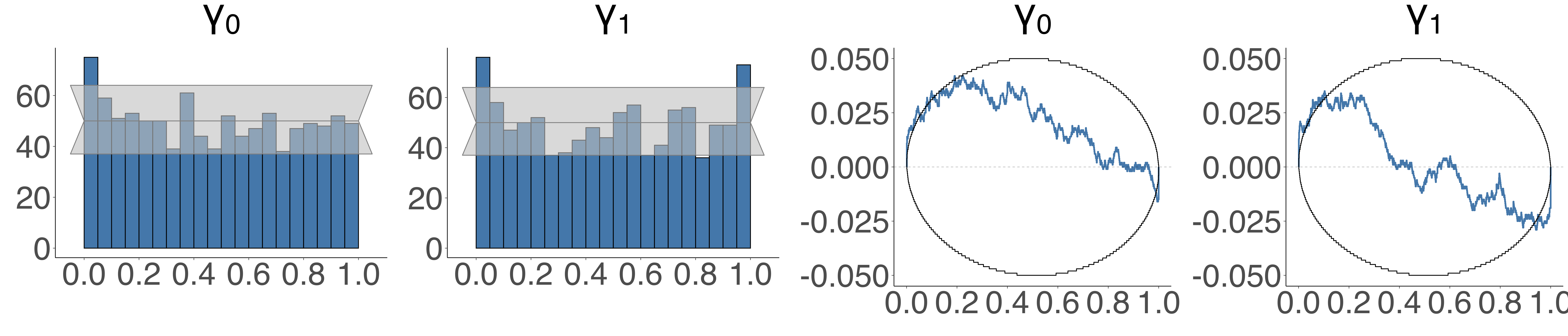}}
    \subfigure[Low rank $\mathbf{Q}$ uncertainty method (mesh B)]{\includegraphics[width=.71\linewidth]{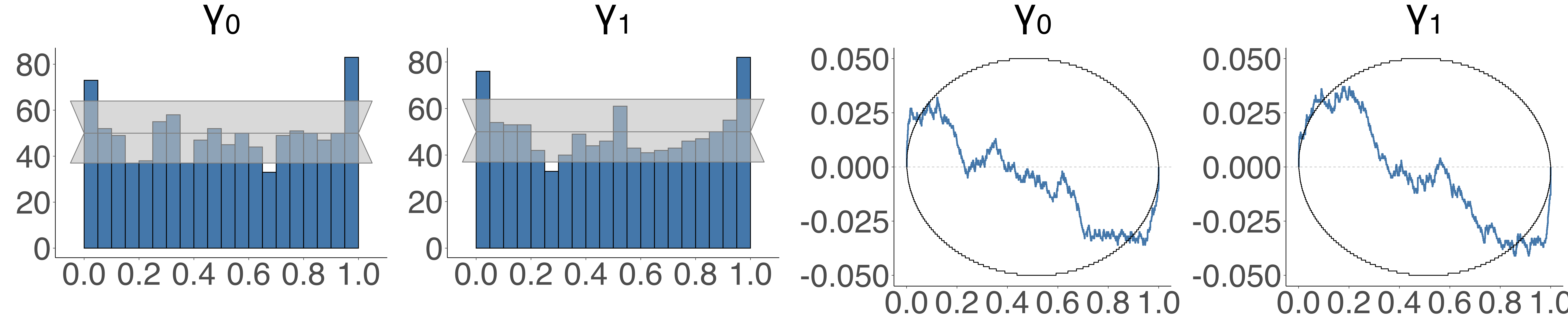}}
    \caption{Histogram and ECDF difference plot of the normalized ranks $p_k$ using Algorithm 3 for the second-stage model parameters $\gamma_0$ and $\gamma_1$ out of 1000 data replicates for the classical specification of the two-stage Poisson spatial model (Section 4.2.1) using INLA-SPDE and with different approaches: (a) plug-in method (b) resampling method (c) full $\mathbf{Q}$ method (d) low rank $\mathbf{Q}$ method (mesh A) (e) low rank $\mathbf{Q}$ method (mesh B)}
\end{figure}

\subsection{SBC results for $\gamma_0$ and $\gamma_1$ of the new model specification (Section 4.2.2) using Algorithm 2}

\begin{figure}[H]
    \centering
    \subfigure[Plug-in method]{\includegraphics[width=.71\linewidth]{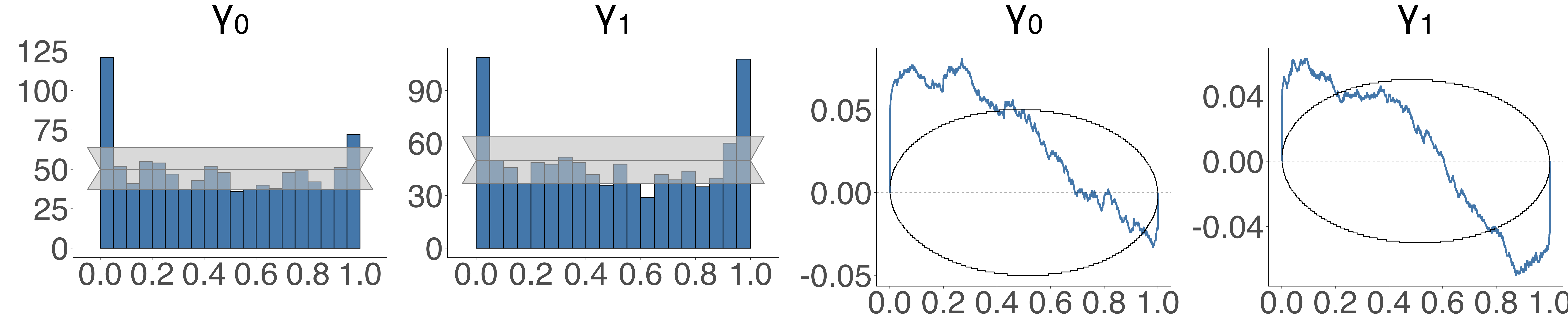}} 
    \subfigure[Resampling method]{\includegraphics[width=.71\linewidth]{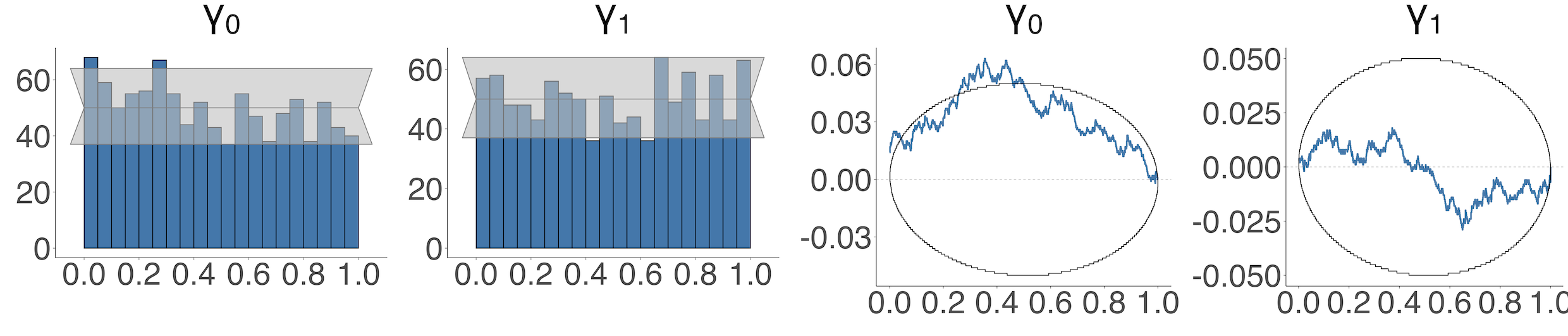}}
    \subfigure[Full $\mathbf{Q}$ uncertainty method]{\includegraphics[width=.71\linewidth]{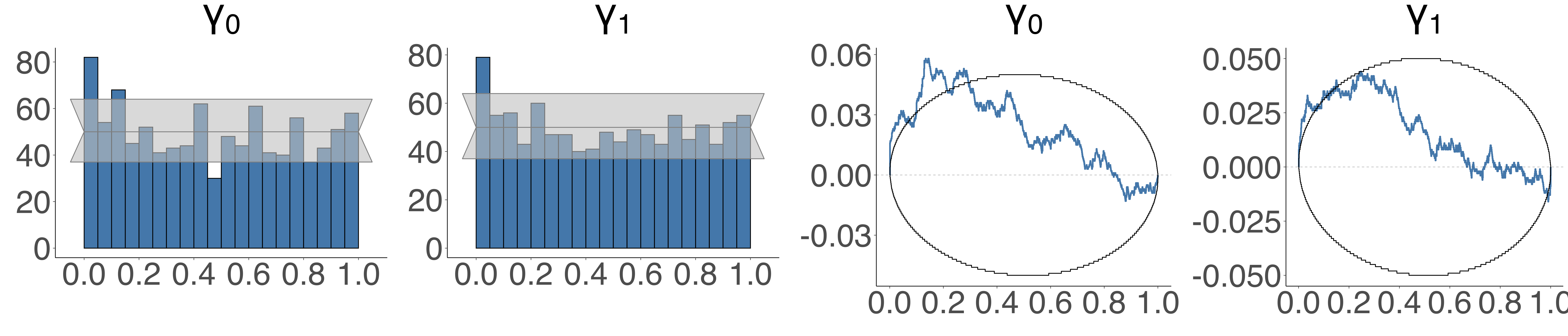}}
    \subfigure[Low rank $\mathbf{Q}$ uncertainty method (mesh A)]{\includegraphics[width=.71\linewidth]{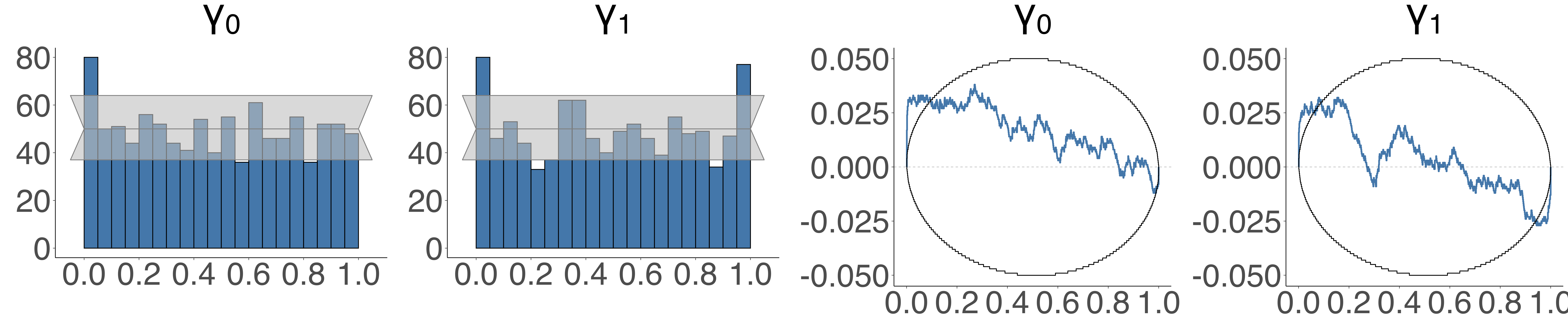}}
    \subfigure[Low rank $\mathbf{Q}$ uncertainty method (mesh B)]{\includegraphics[width=.71\linewidth]{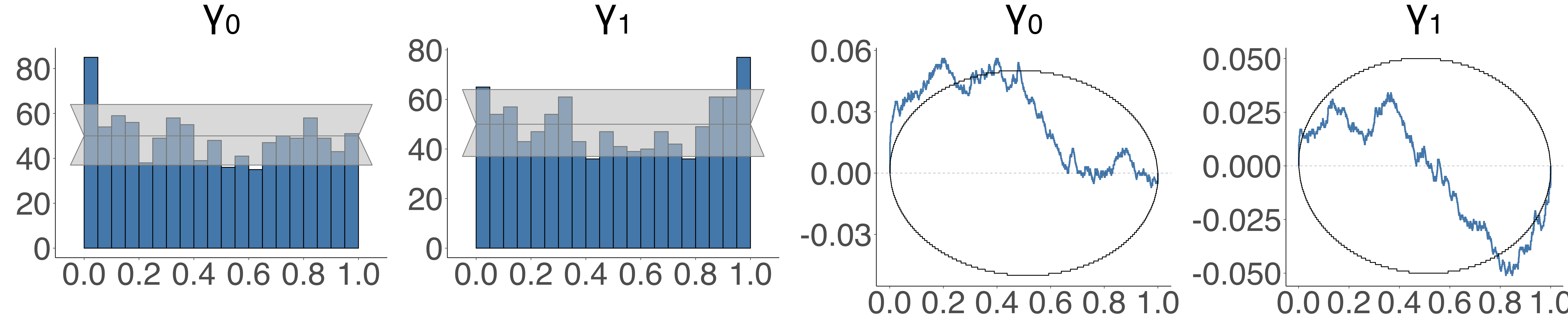}}
    \caption{Histogram and ECDF difference plot of the normalized ranks $p_k$ using Algorithm 2 for the second-stage model parameters $\gamma_0$ and $\gamma_1$ out of 1000 data replicates for the new specification of the two-stage Poisson spatial model (Section 4.2.2) using INLA-SPDE and with different approaches: (a) plug-in method (b) resampling method (c) full $\mathbf{Q}$ method (d) low rank $\mathbf{Q}$ method (mesh A) (e) low rank $\mathbf{Q}$ method (mesh B)}
\end{figure}

\subsection{SBC results for $\gamma_0$ and $\gamma_1$ of the new model specification (Section 4.2.2) using Algorithm 3}

\begin{figure}[H]
    \centering
    \subfigure[Plug-in method]{\includegraphics[width=.71\linewidth]{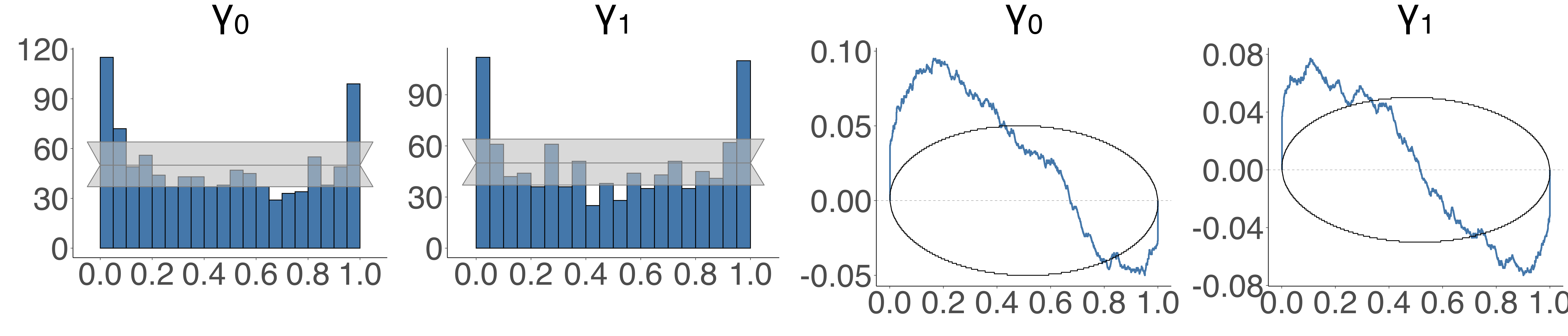}} 
    \subfigure[Resampling method]{\includegraphics[width=.71\linewidth]{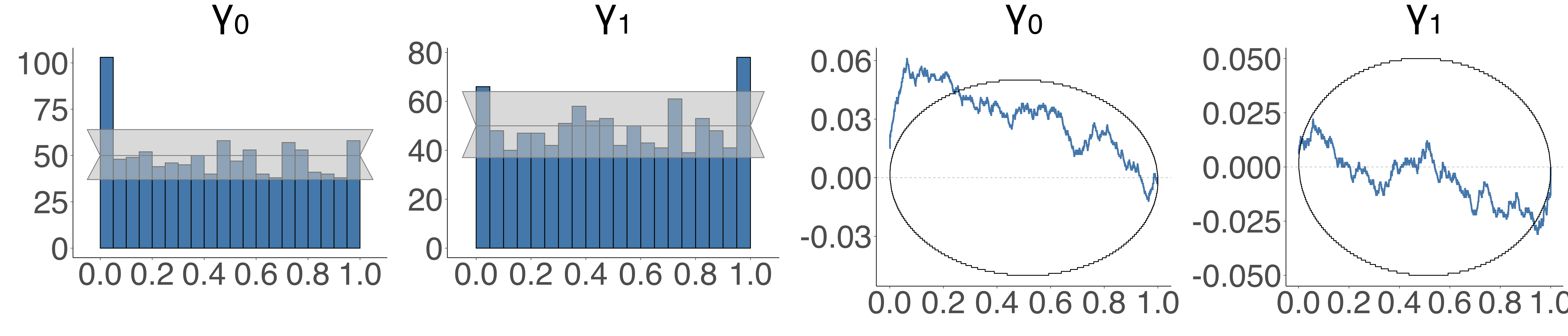}}
    \subfigure[Full $\mathbf{Q}$ uncertainty method]{\includegraphics[width=.71\linewidth]{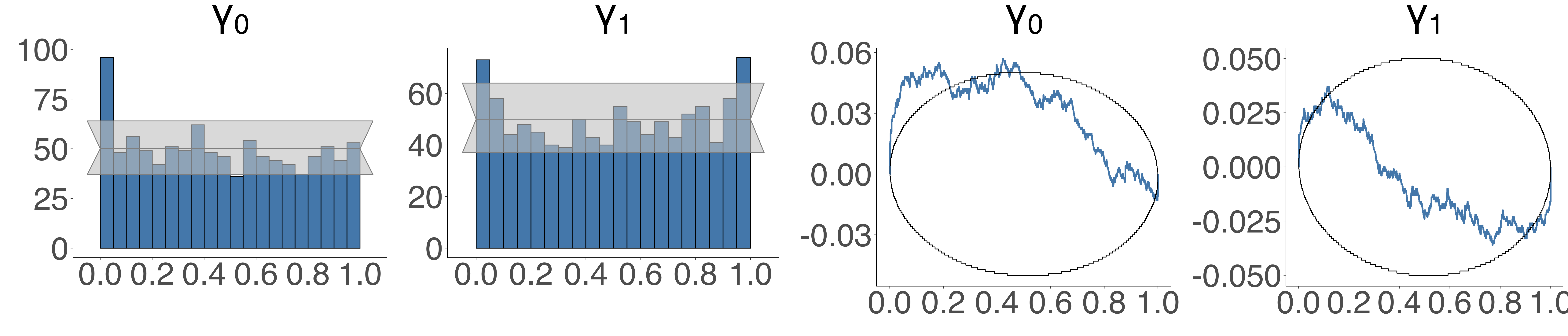}}
    \subfigure[Low rank $\mathbf{Q}$ uncertainty method (mesh A)]{\includegraphics[width=.71\linewidth]{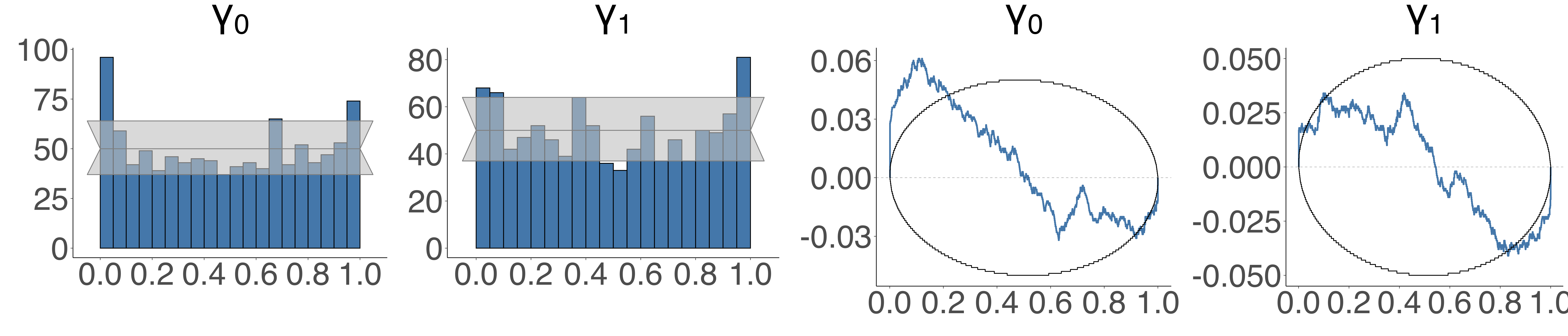}}
    \subfigure[Low rank $\mathbf{Q}$ uncertainty method (mesh B)]{\includegraphics[width=.71\linewidth]{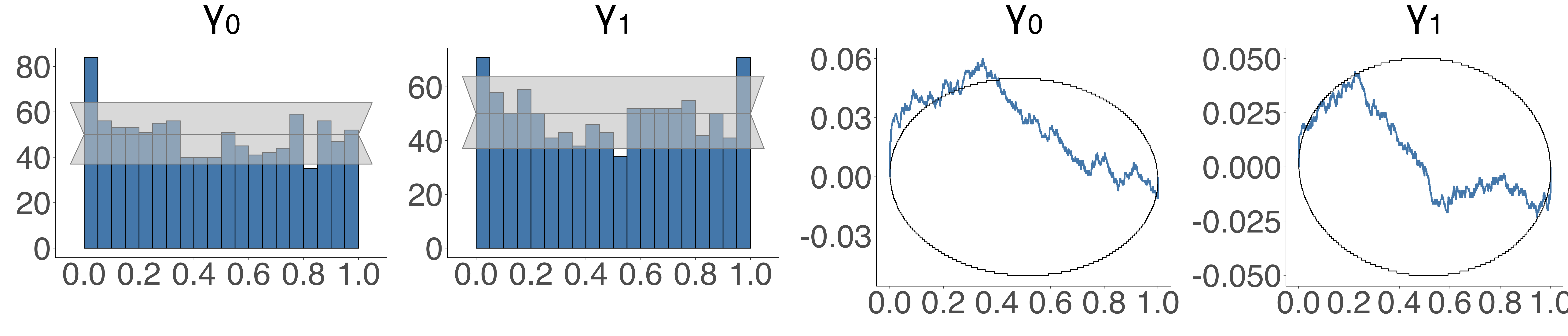}}
    \caption{Histogram and ECDF difference plot of the normalized ranks $p_k$ using Algorithm 3 for the second-stage model parameters $\gamma_0$ and $\gamma_1$ out of 1000 data replicates for the new specification of the two-stage Poisson spatial model (Section 4.2.2) using INLA-SPDE and with different approaches: (a) plug-in method (b) resampling method (c) full $\mathbf{Q}$ method (d) low rank $\mathbf{Q}$ method (mesh A) (e) low rank $\mathbf{Q}$ method (mesh B)}
\end{figure}

\subsection{Illustation with a simulated data}

\begin{figure}[H]
    \centering
    \subfigure[$\lambda(\mathbf{s})$]{\includegraphics[width=.2\linewidth]{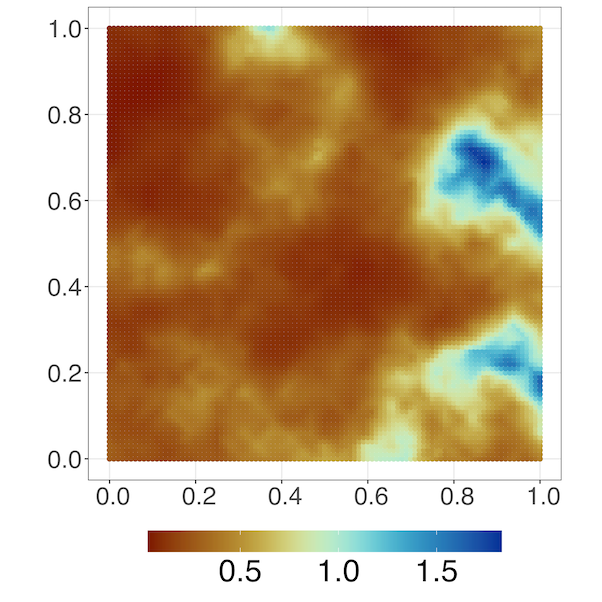}}
    \subfigure[$\lambda(B)$]{\includegraphics[width=.2\linewidth]{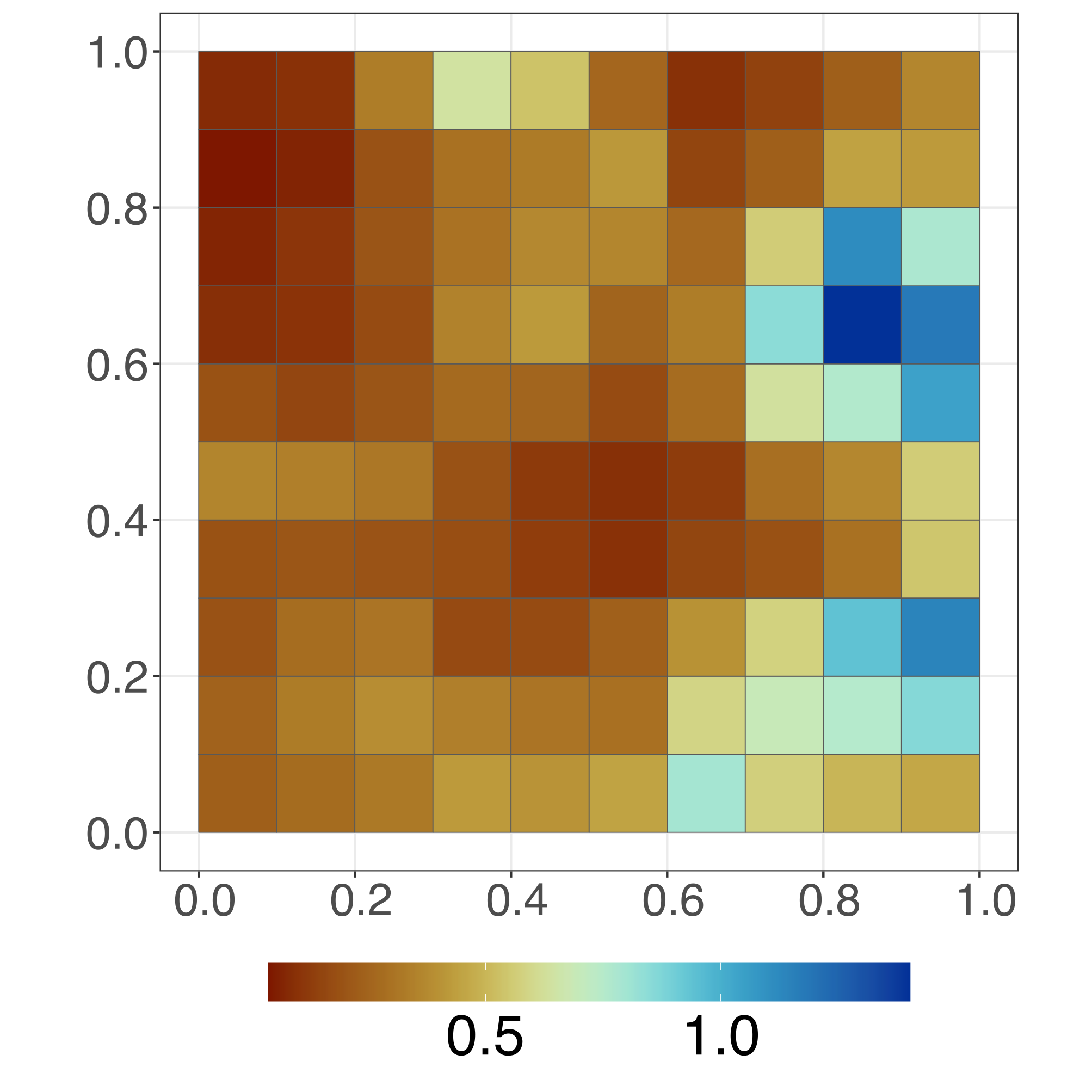}}
    \subfigure[$Y(B)$]{\includegraphics[width=.2\linewidth]{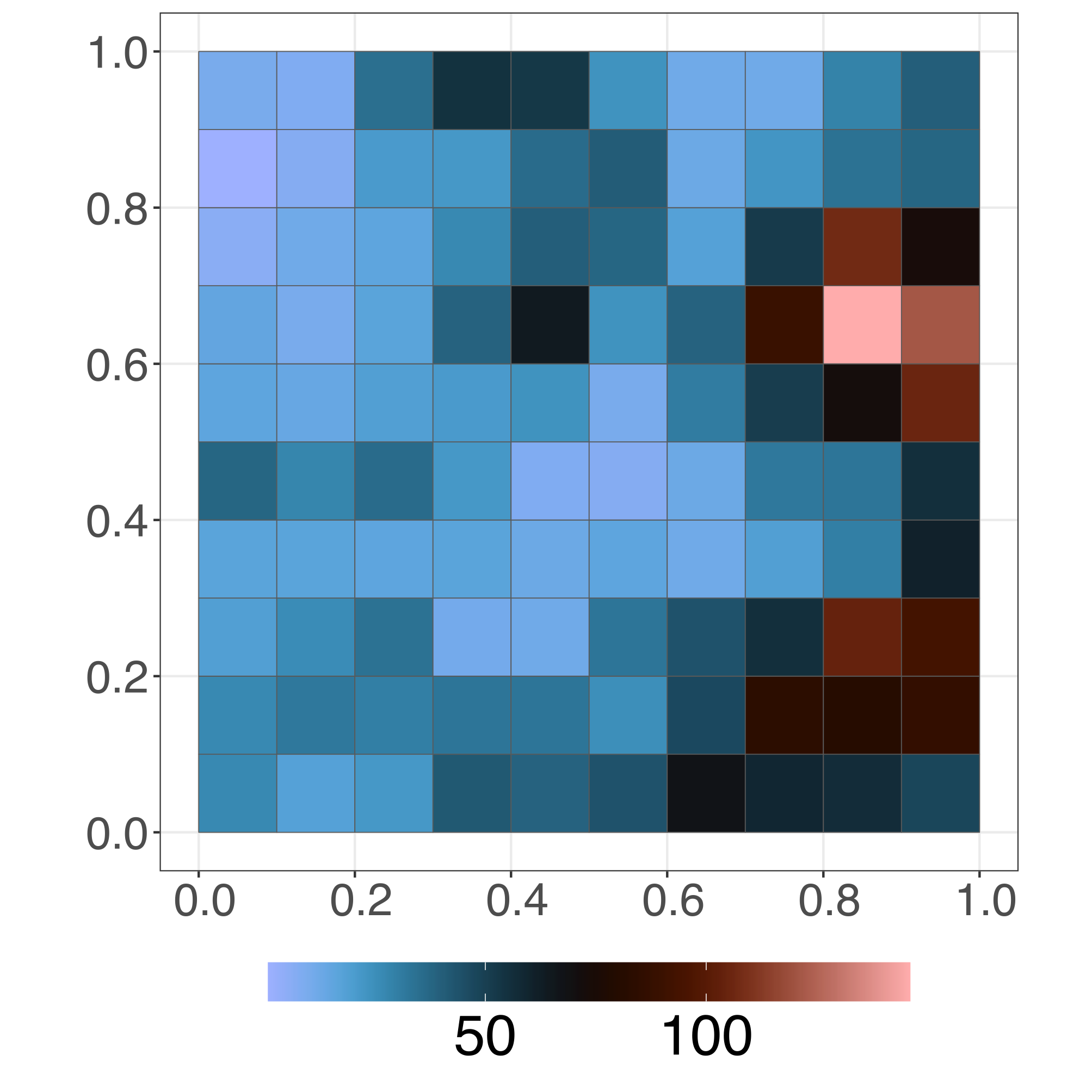}}
    \caption{Simulated quantities from the new specification of the two-stage Poisson spatial model in Section 4.2.2}
\label{fig:res_spatial_twostage_poisson_logsumexp_illustration}
\end{figure}

\begin{figure}[H]
    \centering
    \subfigure[Posterior mean]{\includegraphics[trim=0mm 0 0mm 0,clip,width=.45\linewidth]{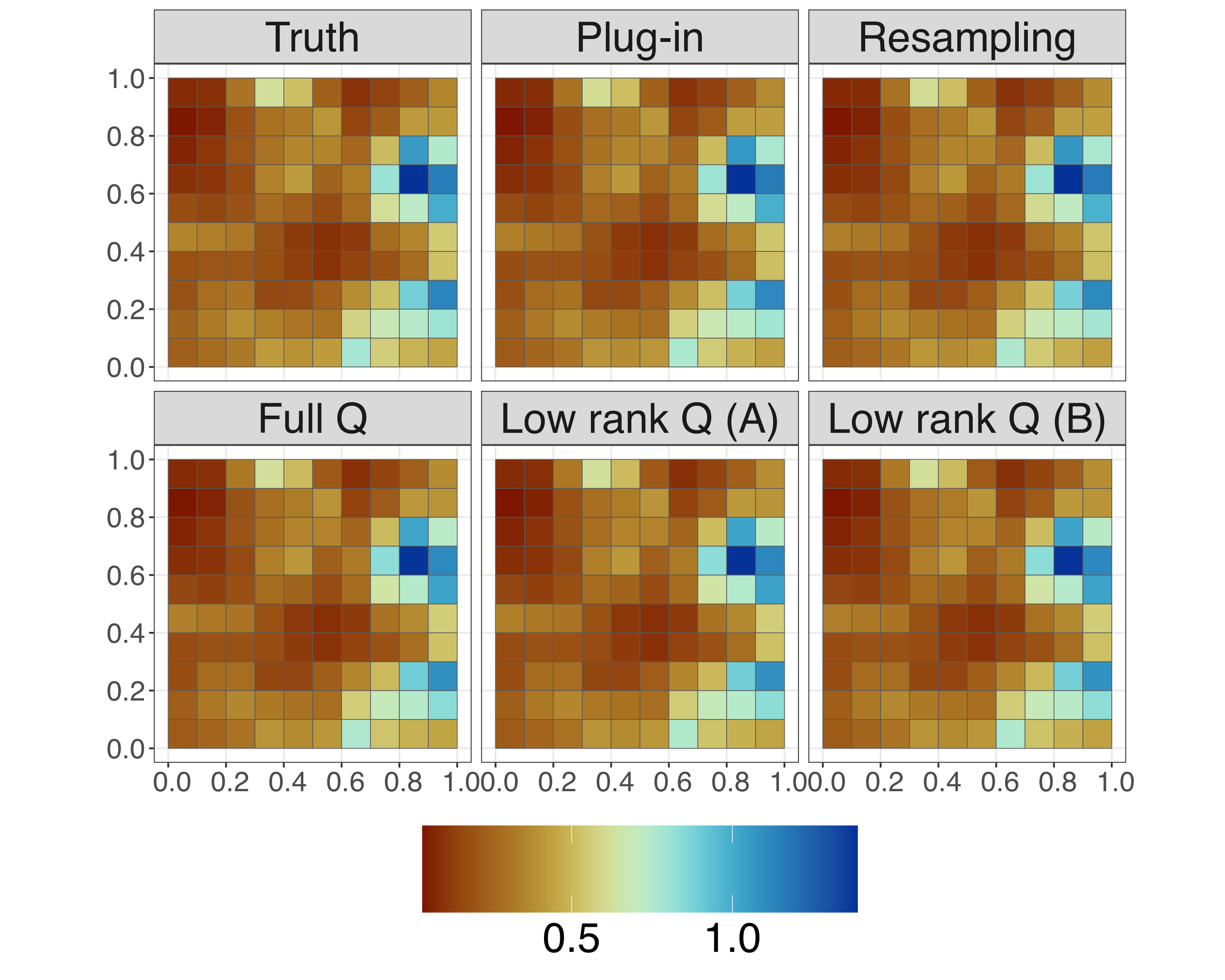}} 
    \subfigure[Posterior standard deviation]{\includegraphics[trim=0mm 0 0mm 0,clip,width=.45\linewidth]{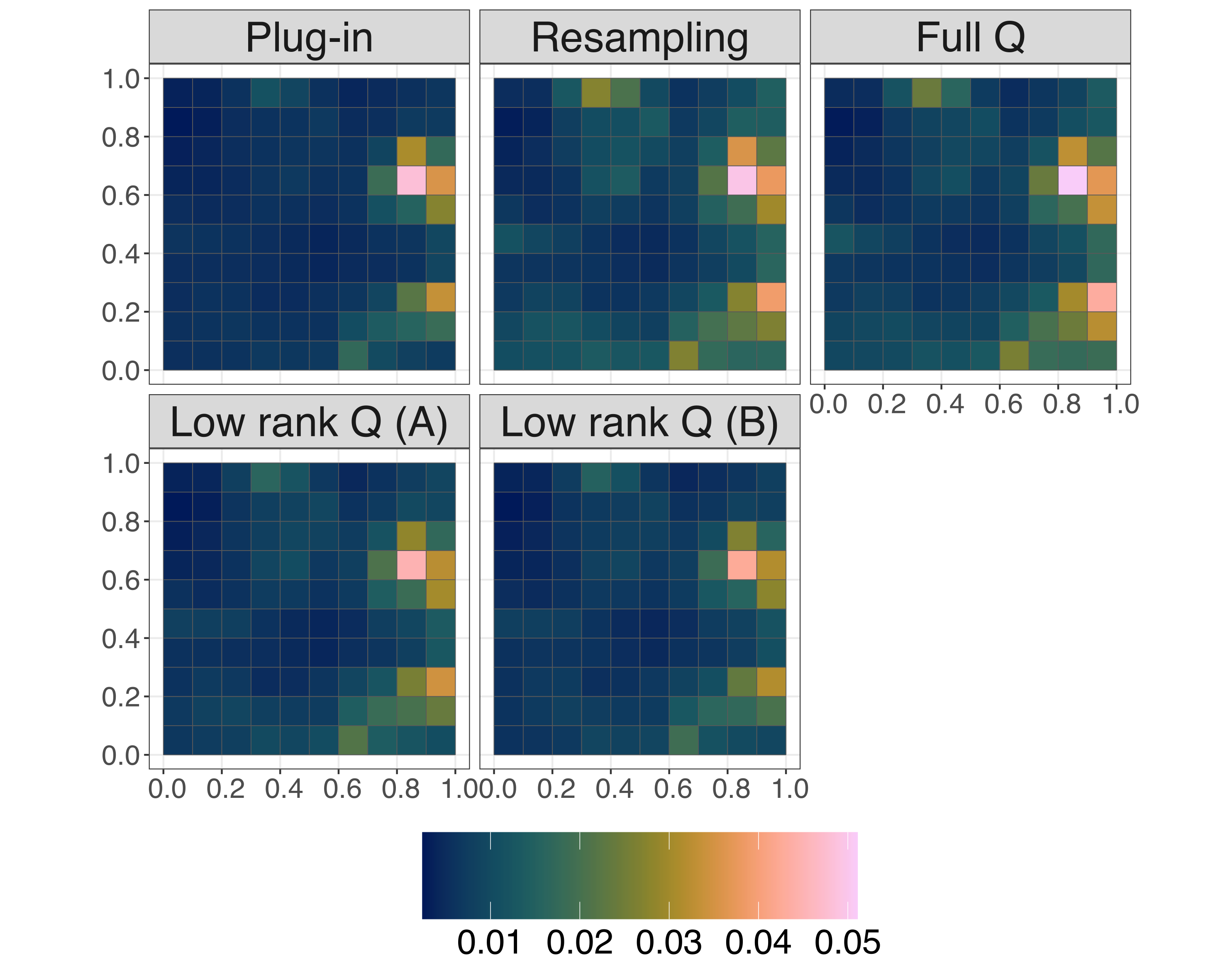}} 
    \caption{Comparison of (a) the posterior mean and (b) posterior standard deviation of $\lambda(B)$ from a simulated data of the two-stage Poisson spatial model (classical specification) in Section 4.2.1 using different approaches: the plug-in method, resampling method, full $\mathbf{Q}$ method, low rank $\mathbf{Q}$ (mesh A) method, low rank $\mathbf{Q}$ (mesh B) method}
\label{fig:res_toy_twostage_poisson_classical_pred_lambdaB}
\end{figure}

\begin{figure}[H]
    \centering
    \subfigure[$\lambda(\mathbf{s})$]{\includegraphics[trim=0mm 0 0mm 0,clip,width=.45\linewidth]{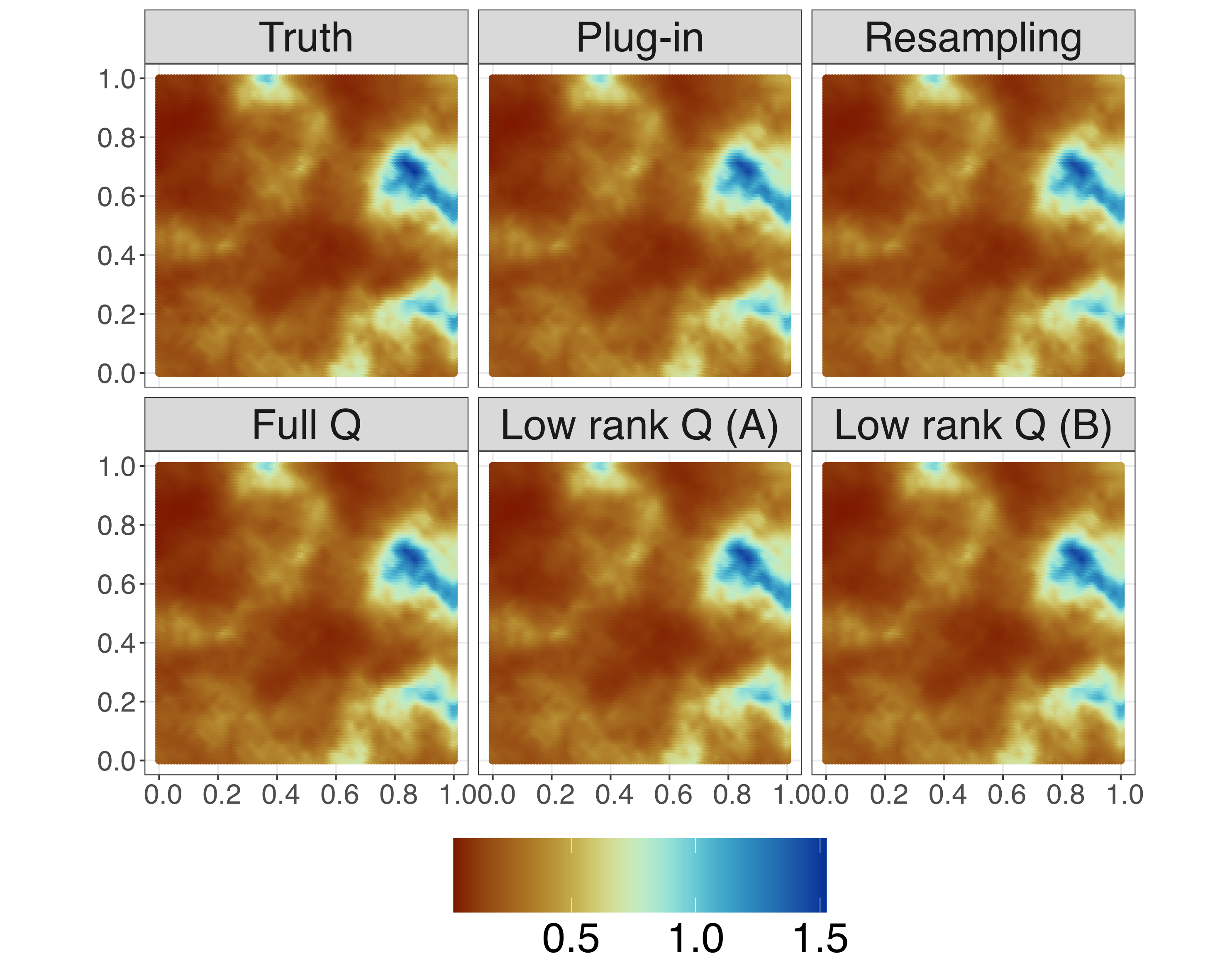}} 
    \subfigure[$\lambda(B)=\dfrac{1}{|B|}\int_B\lambda(\mathbf{s})d\mathbf{s}$]{\includegraphics[trim=0mm 0 0mm 0,clip,width=.45\linewidth]{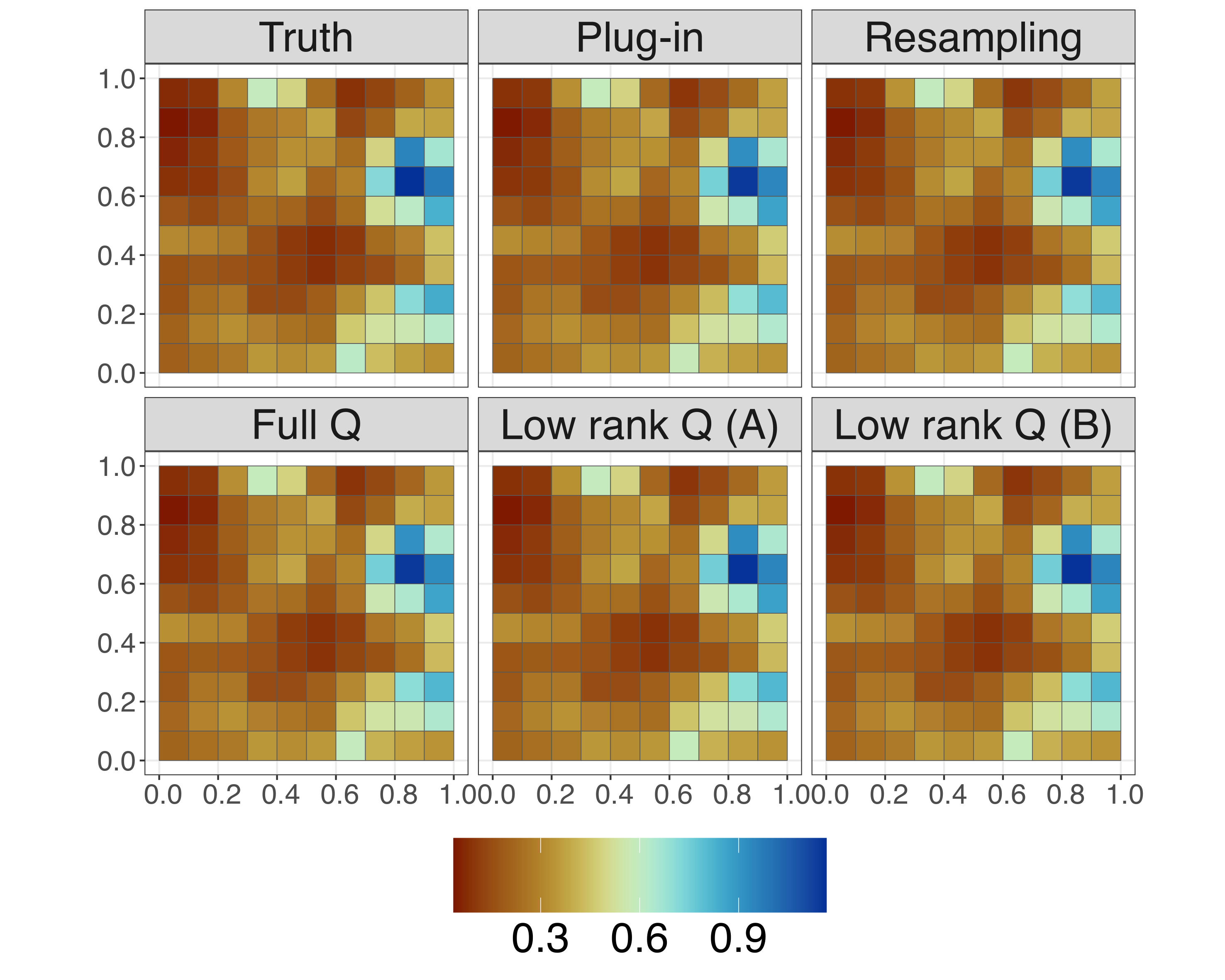}} 
    \caption{Comparison of the posterior mean for (a) $\lambda(\mathbf{s})$ and (b) $\lambda(B)$ from a simulated data of the two-stage Poisson model (new specification) in Section 4.2.2 using different approaches: the plug-in method, resampling method, full $\mathbf{Q}$ method, low rank $\mathbf{Q}$ (mesh A) method, and low rank $\mathbf{Q}$ (mesh B) method}
\label{fig:res_spatial_twostage_poisson_logsumexp_pred_lambda_mean}
\end{figure}

\section{Data Application}\label{supp:data_application}

\begin{table}[H]
\centering
\scalebox{.95}{\begin{tabular}{rrrrr}
  \hline\hline
 & Mean & SD & P2.5 & P97.5 \\ 
  \hline\hline
$\beta_0$ & 78.1553 & 4.1435 & 70.0341 & 86.2765 \\ 
  $\beta_1$ & -0.0058 & 0.0033 & -0.0122 & 0.0006 \\ 
  $\beta_2$ & 3.2959 & 0.6142 & 2.0920 & 4.4998 \\ 
  $\beta_3$ & -0.1124 & 0.0210 & -0.1535 & -0.0713 \\ 
  $1/\sigma^2_{e_1}$ & 0.1448 & 0.0325 & 0.0910 & 0.2181 \\ 
  $\log(\tau)$ & 3.8219 & 0.6807 & 2.4272 & 5.1048 \\ 
  $\log(\kappa)$ & -6.1369 & 0.6244 & -7.3248 & -4.8675 \\ 
   \hline\hline
\end{tabular}}
\caption{Posterior estimates of first-stage model parameters: posterior mean, posterior standard deviation (SD), and 95\% credible intervals}
\label{tab:res_dataapplication_firststage_spde}
\end{table}

\begin{minipage}{\textwidth}
  \begin{minipage}[]{0.48\textwidth}
    \centering
    \scalebox{.9}{
\centering
\begin{tabular}{lrrrrr}
  \hline\hline
 Method &  & Mean & SD & P2.5 & P97.5 \\ 
  \hline\hline
Plug-in&$\gamma_0$ & -8.1913 & 3.0307 & -14.1313 & -2.2513 \\ 
  &$\gamma_1$ & 0.1006 & 0.0353 & 0.0313 & 0.1698 \\ \hline
  Resampling&$\gamma_0$ & -7.9764 & 3.3459 & -14.7697 & -1.6871 \\ 
  &$\gamma_1$ & 0.0981 & 0.0388 & 0.0240 & 0.1758 \\ \hline
  Full $\mathbf{Q}$&$\gamma_0$ & -9.2889 & 3.1863 & -15.5341 & -3.0438 \\ 
  &$\gamma_1$ & 0.1131 & 0.0371 & 0.0403 & 0.1859 \\ \hline
  Low rank $\mathbf{Q}$&$\gamma_0$ & -8.8892 & 3.1837 & -15.1291 & -2.6493 \\ 
  &$\gamma_1$ & 0.1086 & 0.0371 & 0.0357 & 0.1814 \\ 
   \hline\hline
\end{tabular}}
\captionof{table}{Posterior estimates of second-stage model (classical specification)}
\label{tab:application_gammaestimates_classical_spde_highprec}
  \end{minipage}
  \hfill
  \begin{minipage}[]{0.48\textwidth}
    \centering
\scalebox{.9}{
\begin{tabular}{lrrrrr}
  \hline\hline
 Method&& Mean & SD & P2.5 & P97.5 \\ 
  \hline\hline
Plug-in&$\gamma_0$ & -9.3117 & 2.9108 & -15.0168 & -3.6067 \\ 
  &$\gamma_1$ & 0.1131 & 0.0336 & 0.0473 & 0.1790 \\ \hline
 Resampling &$\gamma_0$ & -8.6326 & 3.2423 & -15.0377 & -2.4195 \\ 
  &$\gamma_1$ & 0.1048 & 0.0372 & 0.0314 & 0.1788 \\ \hline
 Full $\mathbf{Q}$ &$\gamma_0$ & -10.2863 & 3.0618 & -16.2873 & -4.2852 \\ 
  &$\gamma_1$ & 0.1240 & 0.0354 & 0.0546 & 0.1933 \\ \hline
  Low rank $\mathbf{Q}$&$\gamma_0$ & -9.7508 & 3.0567 & -15.7418 & -3.7597 \\ 
  &$\gamma_1$ & 0.1180 & 0.0354 & 0.0487 & 0.1874 \\ 
   \hline\hline
\end{tabular}
}
      \captionof{table}{Posterior estimates of second-stage model (New specification)}
      \label{tab:application_gammaestimates_new_spde_highprec}
    \end{minipage}
  \end{minipage}



\begin{figure}[t]
    \centering
    \subfigure[Classical model specification]{\includegraphics[width=.48\linewidth]{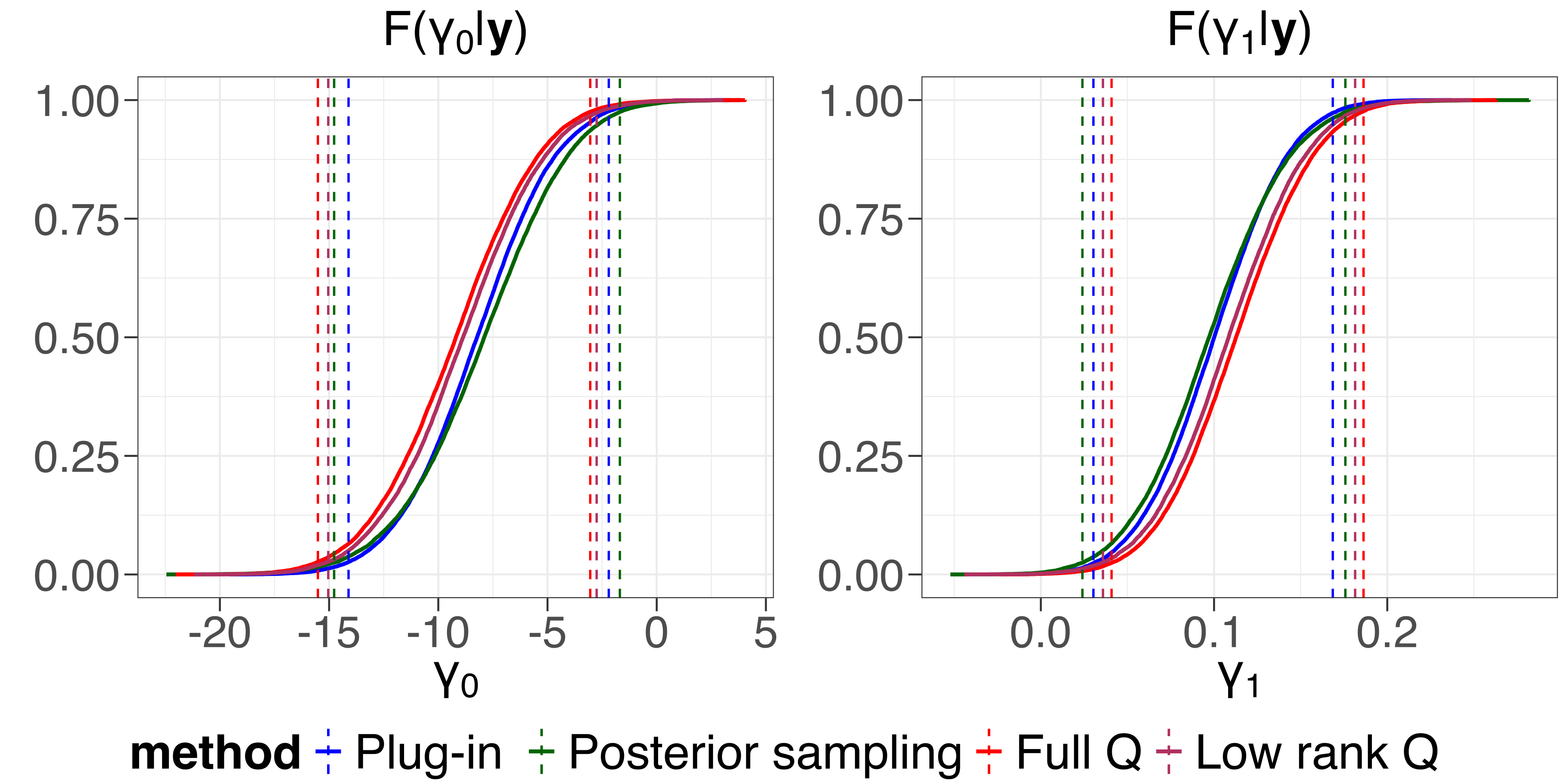}} 
    \subfigure[New model specification]{\includegraphics[width=.48\linewidth]{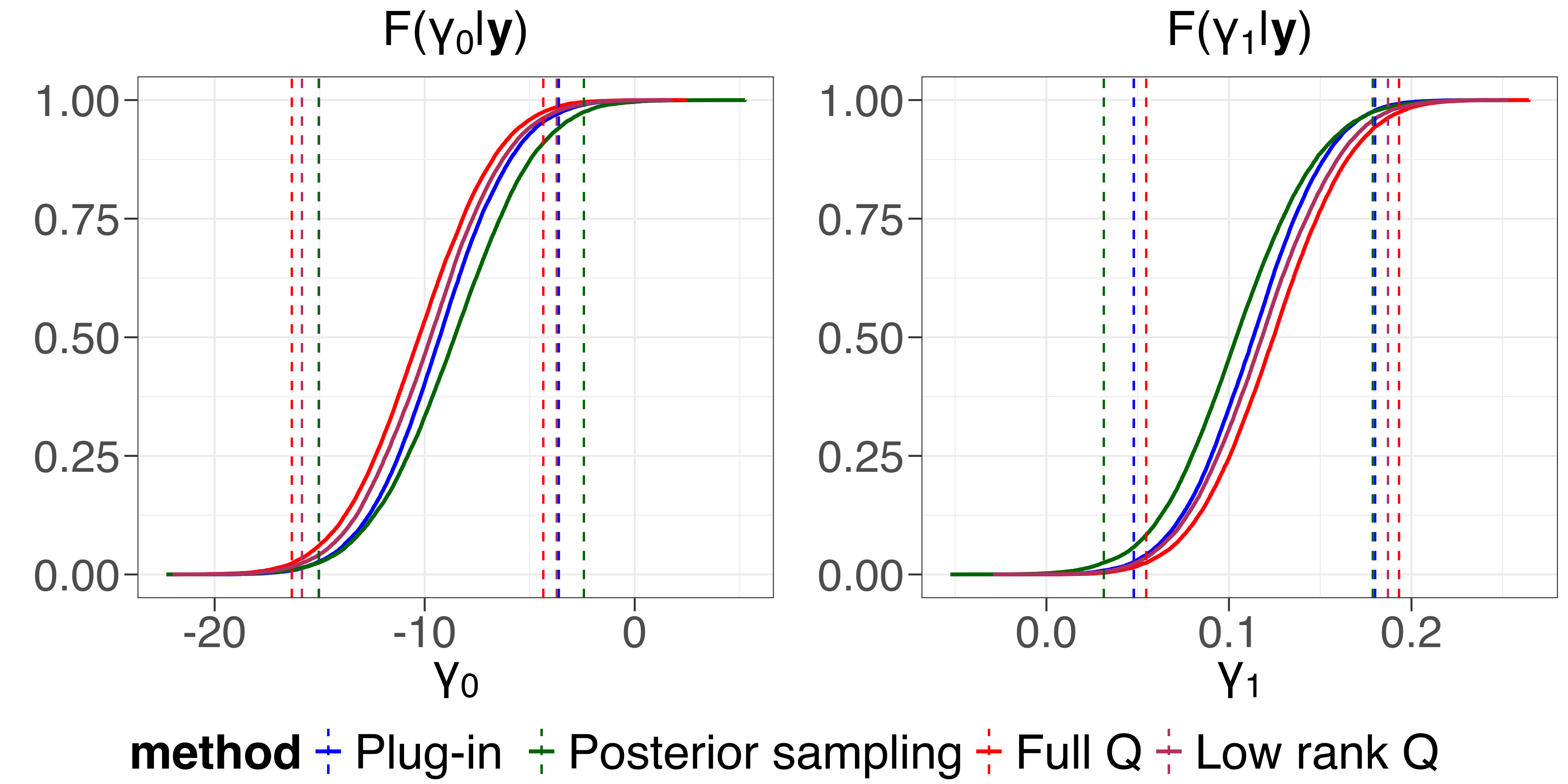}} 
    \caption{Comparison of marginal posteriors of $\gamma_0$ and $\gamma_1$ using four uncertainty propagation approaches: (a) classical model specification (b) new model specification}
\label{fig:application_gammaestimates_spde_highprec}
\end{figure}

\begin{figure}[h!]
    \centering
    \subfigure[95\% CI of RR with 1 SD change in RH]{\includegraphics[width=.49\linewidth]{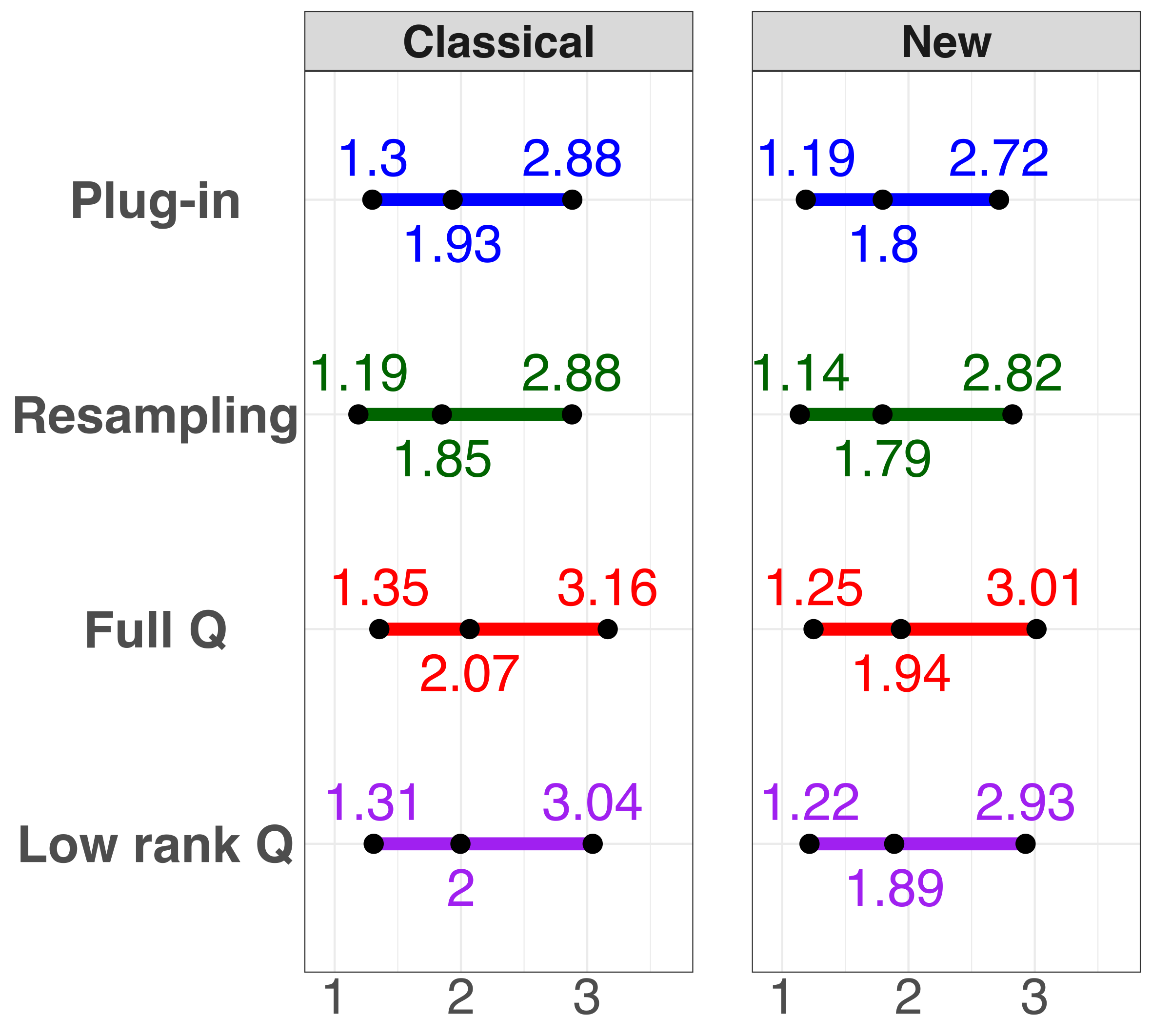}} 
    \hspace{0mm}
    \subfigure[95\% CI for $\gamma_0$]{\includegraphics[width=.49\linewidth]{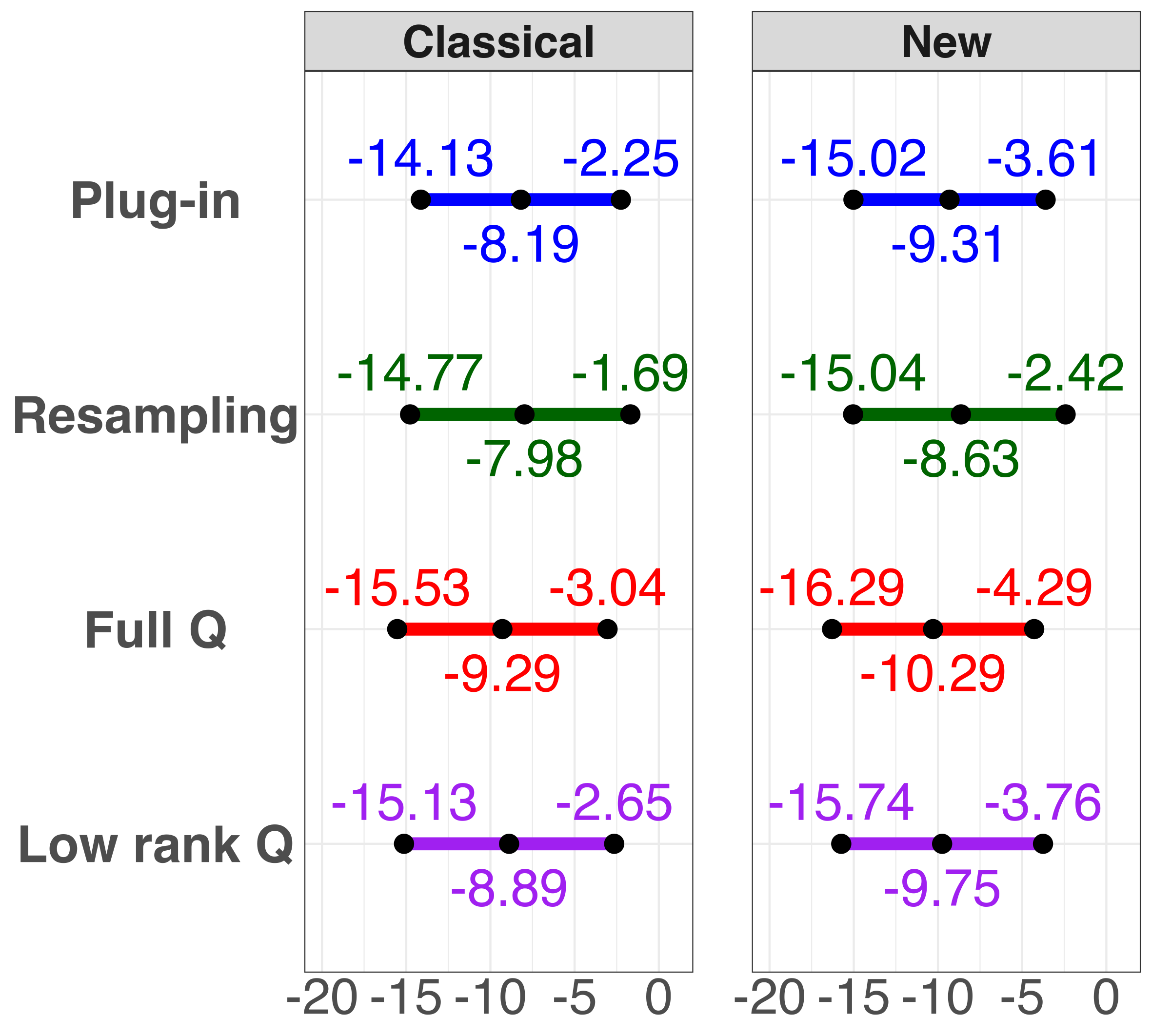}} 
    \caption{(a) 95\% CI of RR associated with 1 SD change in relative humidity (b) 95\% CI for $\gamma_0$} 
\end{figure}


\begin{figure}[H]
    \centering
    \includegraphics[trim={0cm 3cm 0cm 3cm},clip,scale=.45]{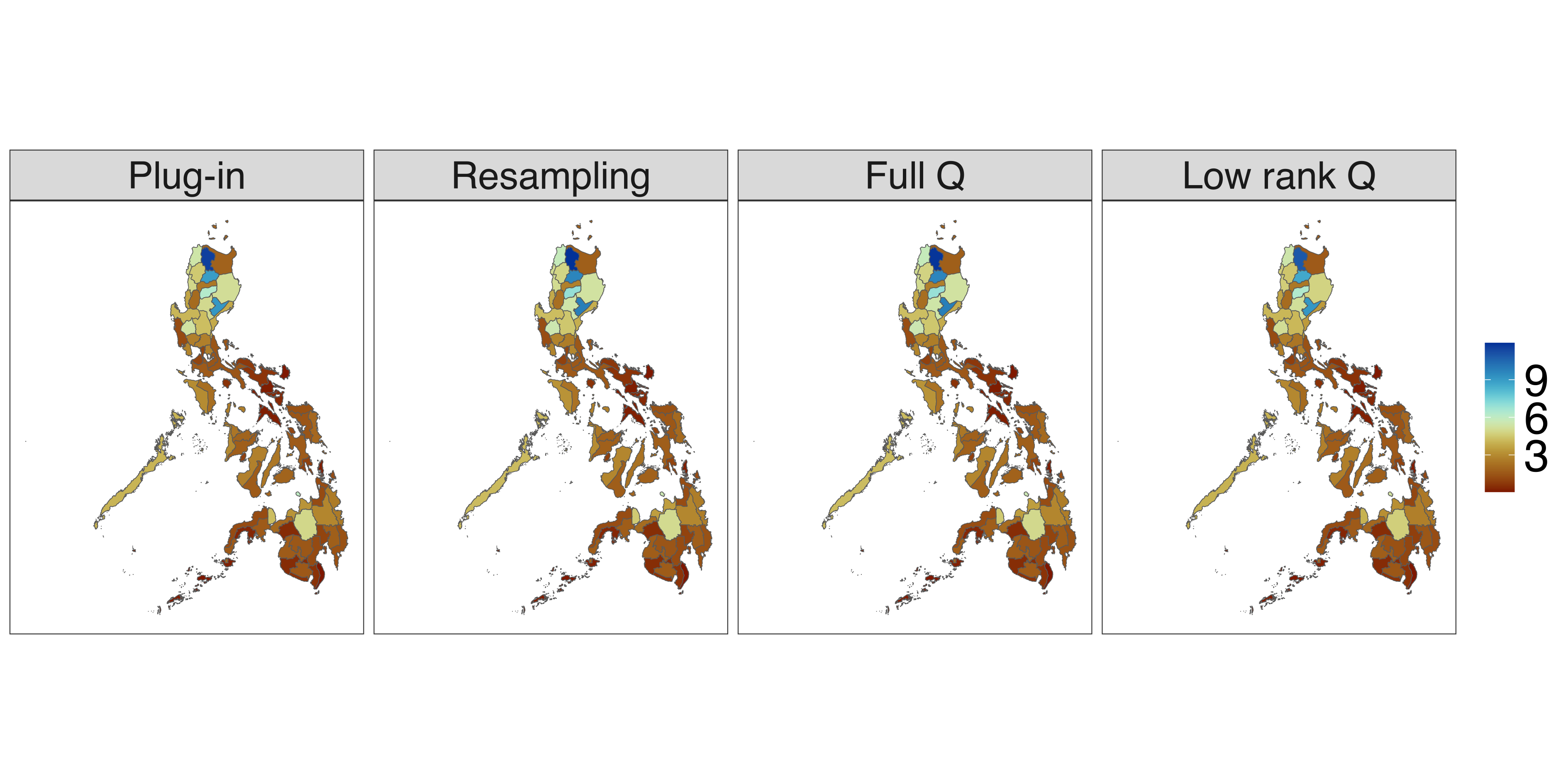}
    \caption{Posterior means of $\lambda(B)$ using the new specification of the Poisson model}
    \label{fig:posteriormean_lambdaB_logsumexp_spde_highprec}
\end{figure}

\begin{figure}[H]
    \centering
    \includegraphics[trim={0cm 3cm 0cm 3cm},clip,scale=.45]{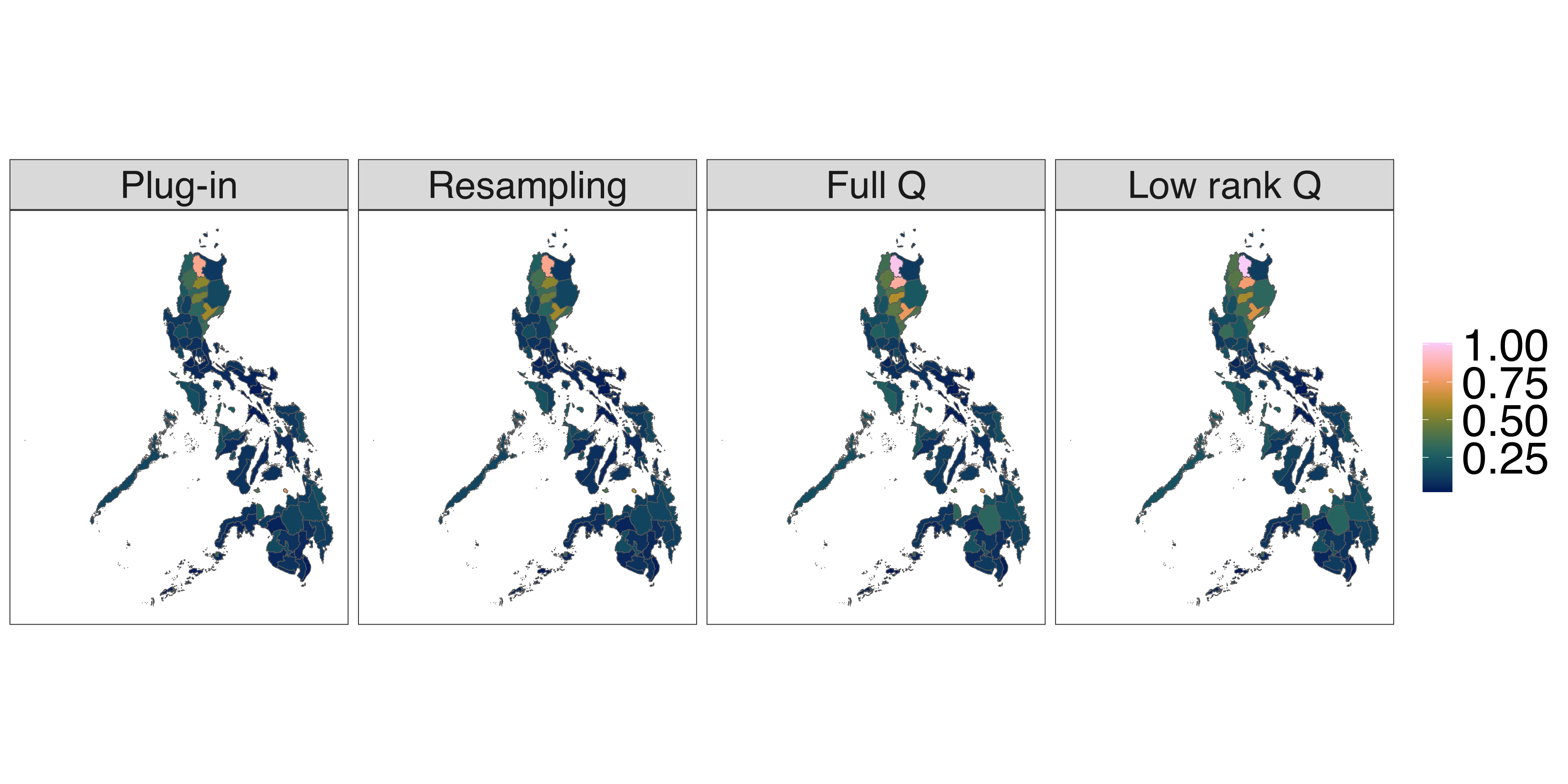}
    \caption{Posterior standard deviations of $\lambda(B)$ using the new specification of the Poisson model}
    \label{fig:posteriorSD_lambdaB_logsumexp_spde_highprec}
\end{figure}

\section{Results of SBC for non-spatial two-stage models}

\subsection{Gaussian model}

\begin{figure}[H]
    \centering
    \includegraphics[width=\linewidth]{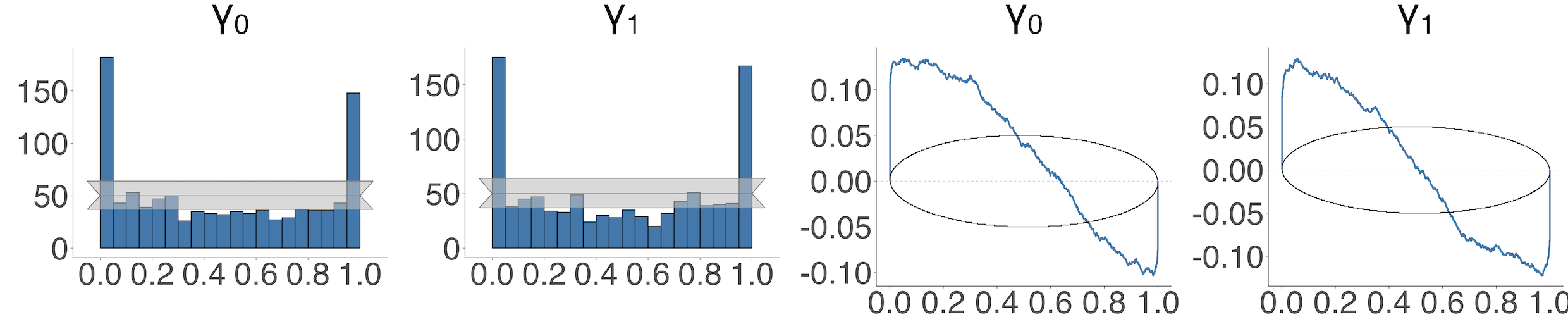}
    \caption{Histogram and ECDF difference plot of the normalized ranks $p_k$ for  $\gamma_0$ and $\gamma_1$ out of 1000 data replicates using  INLA-SPDE and the plug-in method for the two-stage Gaussian model}
\end{figure}

\begin{figure}[H]
    \centering
    \includegraphics[width=\linewidth]{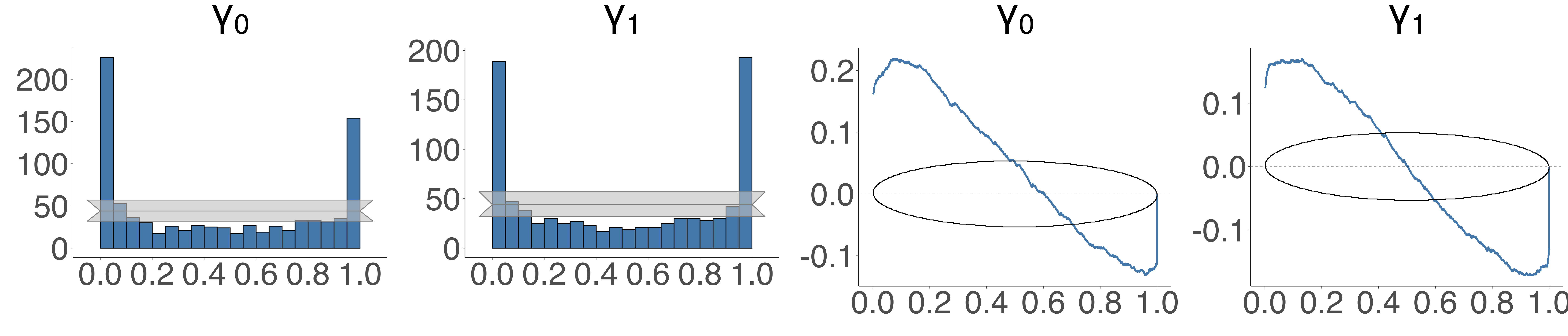}
    \caption{Histogram and ECDF difference plot of the normalized ranks $p_k$ for  $\gamma_0$ and $\gamma_1$ out of 1000 data replicates using NUTS and the plug-in method for the two-stage Gaussian model}
\end{figure}

\begin{figure}[H]
    \centering
    \includegraphics[width=\linewidth]{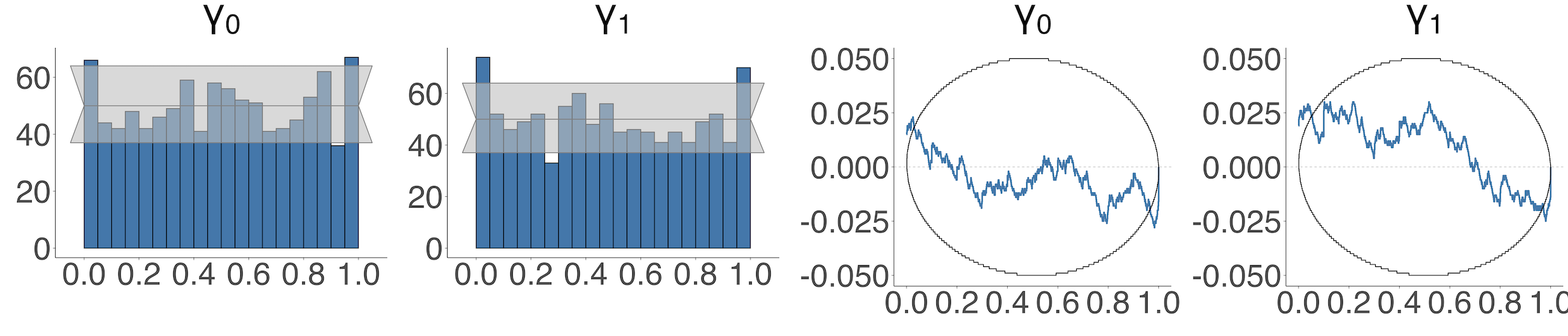}
    \caption{Histogram and ECDF difference plot of the normalized ranks $p_k$ for  $\gamma_0$ and $\gamma_1$ out of 1000 data replicates using  INLA-SPDE and the resampling method for the two-stage Gaussian model}
\end{figure}

\begin{figure}[H]
    \centering
    \includegraphics[width=\linewidth]{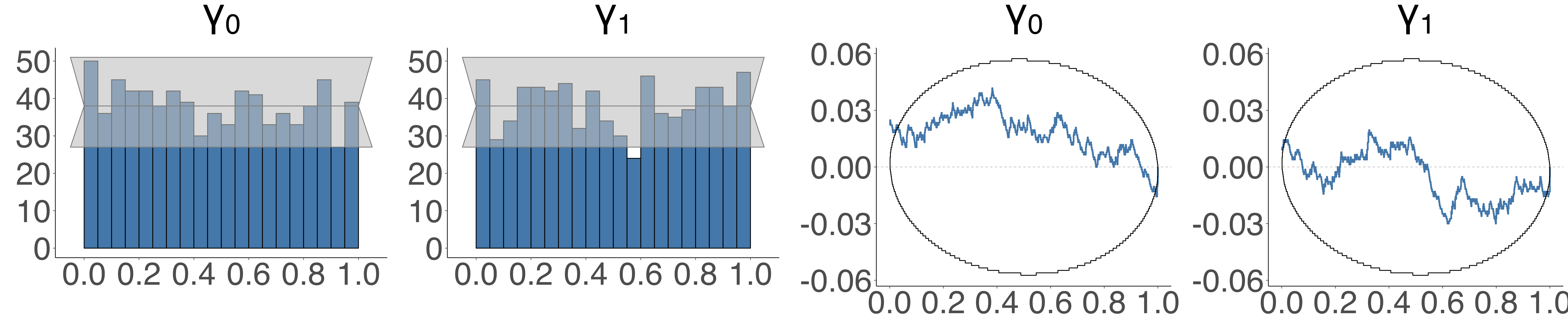}
    \caption{Histogram and ECDF difference plot of the normalized ranks $p_k$ for  $\gamma_0$ and $\gamma_1$ out of 1000 data replicates using NUTS and the resampling method for the two-stage Gaussian model}
\end{figure}

\begin{figure}[H]
    \centering
    \includegraphics[width=\linewidth]{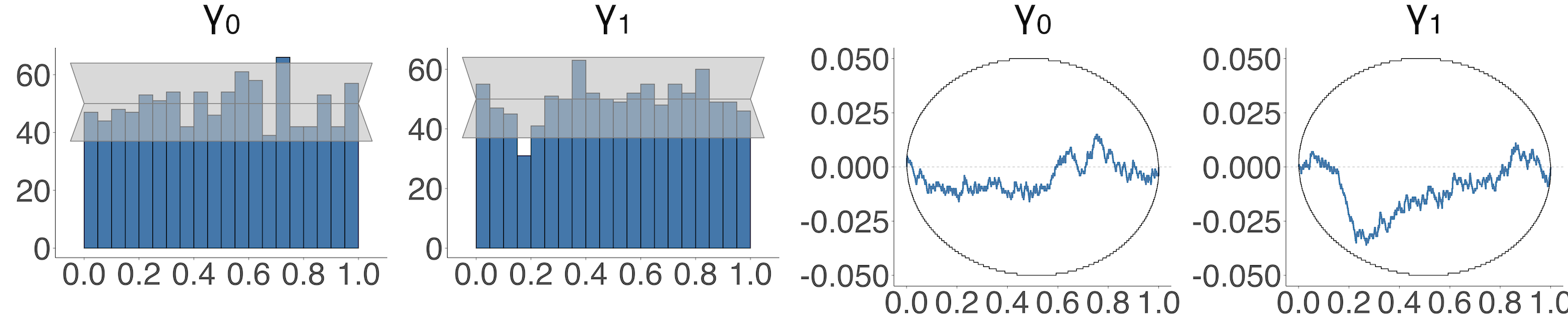}
    \caption{Histogram and ECDF difference plot of the normalized ranks $p_k$ for  $\gamma_0$ and $\gamma_1$ out of 1000 data replicates using the full $\mathbf{Q}$ method for the two-stage Gaussian model}
\end{figure}

\subsection{Poisson model - classical specification}

\begin{figure}[H]
    \centering
    \includegraphics[width=\linewidth]{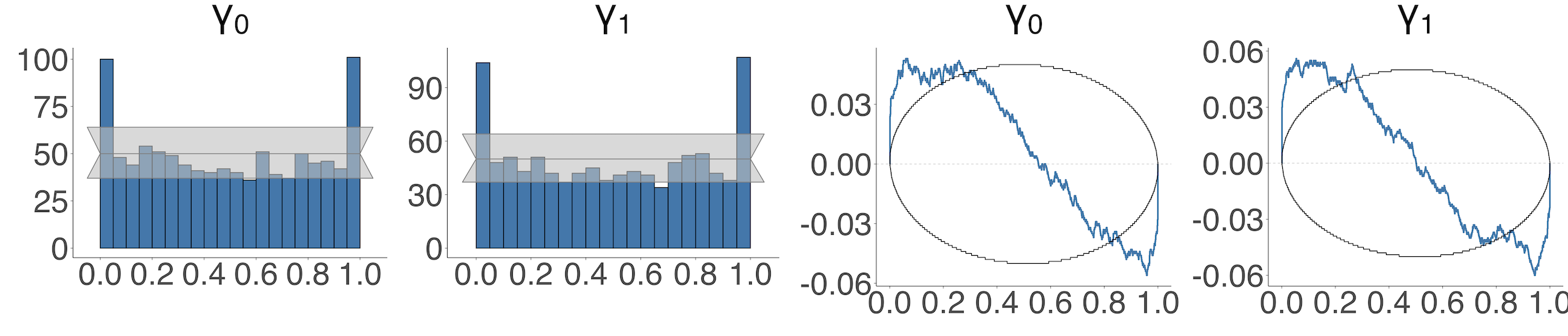}
    \caption{Histogram and ECDF difference plot of the normalized ranks $p_k$ for  $\gamma_0$ and $\gamma_1$ out of 1000 data replicates using  INLA-SPDE and the plug-in method for the two-stage Poisson model (classical specification)}
\end{figure}

\begin{figure}[H]
    \centering
    \includegraphics[width=\linewidth]{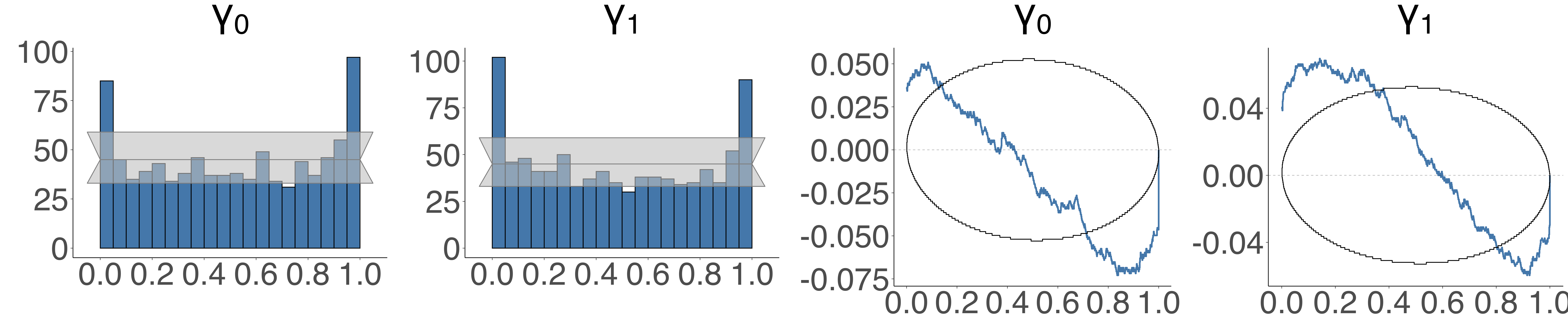}
    \caption{Histogram and ECDF difference plot of the normalized ranks $p_k$ for  $\gamma_0$ and $\gamma_1$ out of 1000 data replicates using NUTS and the plug-in method (classical specification)}
\end{figure}

\begin{figure}[H]
    \centering
    \includegraphics[width=\linewidth]{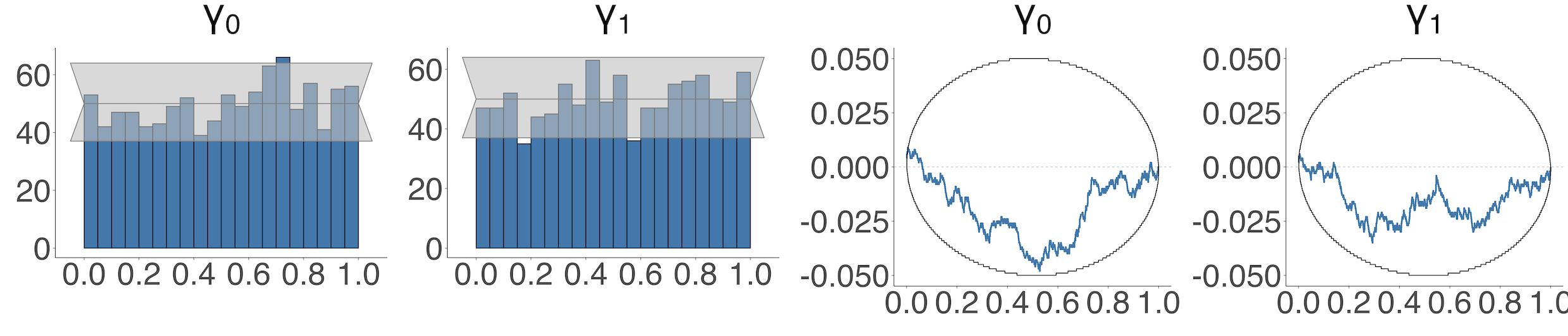}
    \caption{Histogram and ECDF difference plot of the normalized ranks $p_k$ for  $\gamma_0$ and $\gamma_1$ out of 1000 data replicates using  INLA-SPDE and the resampling method for the two-stage Poisson model (classical specification)}
\end{figure}

\begin{figure}[H]
    \centering
    \includegraphics[width=\linewidth]{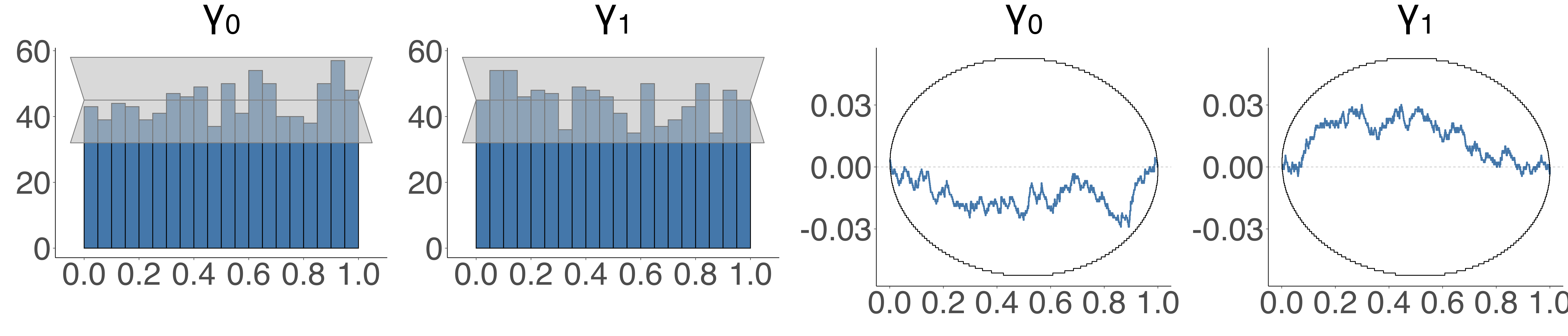}
    \caption{Histogram and ECDF difference plot of the normalized ranks $p_k$ for  $\gamma_0$ and $\gamma_1$ out of 1000 data replicates using NUTS and the resampling method for the two-stage Poisson model (classical specification)}
\end{figure}

\begin{figure}[H]
    \centering
    \includegraphics[width=\linewidth]{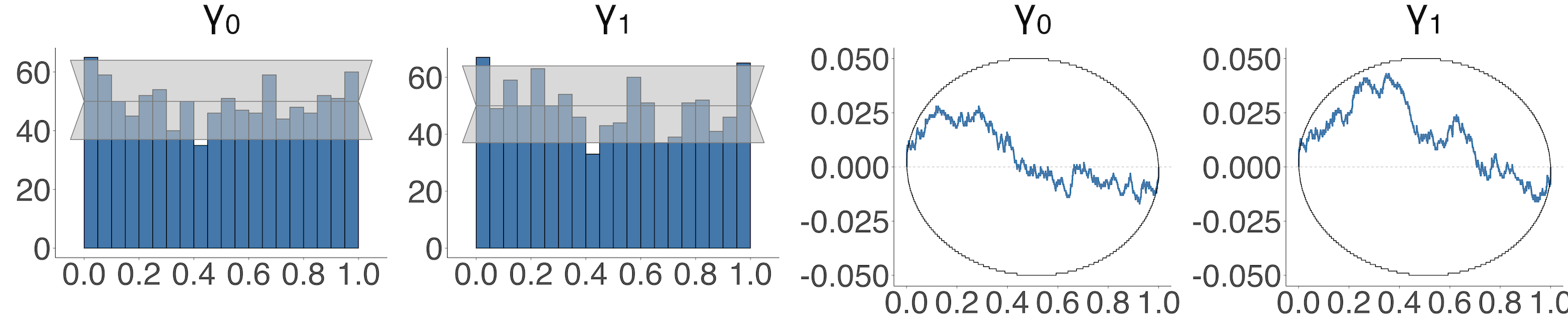}
    \caption{Histogram and ECDF difference plot of the normalized ranks $p_k$ for  $\gamma_0$ and $\gamma_1$ out of 1000 data replicates using the full $\mathbf{Q}$ method for the two-stage Poisson model}
\end{figure}